\documentclass[12pt]{puthesis}
\newcommand{\proquestmode}{}
\title{On the typical properties of inverse problems in statistical mechanics}

\submitted{September 2012}  % degree conferral date (January, April, June, September, or November)
\copyrightyear{2012}  % year in which the copyright is secured by publication of the dissertation.
\author{Iacopo Mastromatteo}
\adviser{Professor Matteo Marsili}  %replace with the full name of your adviser
\department{Statistical Physics}

\usepackage{graphicx}
\usepackage{verbatim}

\usepackage{multirow}
\usepackage{longtable}

\usepackage{booktabs}

\usepackage{bm}

\usepackage{sidecap}

\usepackage{amsthm}
\theoremstyle{definition}
\newtheorem{definition}{Definition}[chapter]
\theoremstyle{plain}
\newtheorem{proposition}{Proposition}[chapter]
\newtheorem{remark}{Remark}[chapter]
\newtheorem{theorem}{Theorem}[chapter]
\newtheorem{corollary}{Corollary}[chapter]

\usepackage{fancyhdr}

\usepackage{amssymb}
\usepackage{amsmath}

\usepackage{color}
\usepackage[usenames,dvipsnames,svgnames,table]{xcolor}

\ifdefined\printmode

\usepackage{url}

\else

\ifdefined\proquestmode
\usepackage{hyperref}
\hypersetup{bookmarksnumbered}

\makeatletter
\hypersetup{pdftitle=\@title,pdfauthor=\@author}
\makeatother

\else
\usepackage{hyperref}
\hypersetup{colorlinks,bookmarksnumbered}

\makeatletter
\hypersetup{pdftitle=\@title,pdfauthor=\@author}
\makeatother

\fi % proquest or online formatting
\fi % printed or online formatting

\newcommand{\beq}{\begin{equation}}
\newcommand{\eeq}{\end{equation}}
\newcommand{\beqa}{\begin{eqnarray}}
\newcommand{\eeqa}{\end{eqnarray}}

\newcommand{\arctanh}{\textrm{arctanh }}
\newcommand{\expect}[1]{\mathbb{E}\left[ #1 \right]}

\newcommand{\bchi}{{\bm{\chi}}}
\newcommand{\bpsi}{{\bm{\psi}}}
\newcommand{\bp}{{\bm{p}}}
\newcommand{\bq}{{\bm{q}}}
\newcommand{\bphi}{{\bm{\phi}}}
\newcommand{\bg}{{\bm{g}}}
\newcommand{\bx}{{\bm{x}}}
\newcommand{\by}{{\bm{y}}}
\newcommand{\bX}{{\bm{X}}}
\newcommand{\bV}{{\bm{V}}}
\newcommand{\bh}{{\bm{h}}}
\newcommand{\bJ}{{\bm{J}}}
\newcommand{\bmg}{{\bm{m}}}
\newcommand{\bc}{{\bm{c}}}
\newcommand{\bT}{{\bm{T}}}
\newcommand{\bK}{{\bm{K}}}
\newcommand{\bGam}{{\bm{\gamma}}}
\newcommand{\bLam}{{\bm{\lambda}}}
\newcommand{\bMu}{{\bm{\mu}}}
\newcommand{\bAl}{{\bm{\alpha}}}
\newcommand{\bDl}{{\bm{\delta}}}
\newcommand{\bs}{{\bm{s}}}

\newcommand{\defin}[1]{
\begin{definition} #1 \end{definition}
}

\newcommand{\prop}[1]{
\begin{proposition} #1 \end{proposition} 
}
\newcommand{\rmk}[1]{
\begin{remark} #1 \end{remark}
}
\newcommand{\theor}[1]{
\begin{theorem} #1 \end{theorem}
}
\newcommand{\corol}[1]{
\begin{corollary} #1 \end{corollary}
}
\newcommand{\prf}[1]{
\begin{proof} #1 \end{proof}
}

\ifodd 0
\renewcommand{\maketitlepage}{}
\renewcommand*{\makecopyrightpage}{}
\renewcommand*{\makeabstract}{}

\else

\abstract{
In this work we consider the problem of extracting a set of interaction parameters from an high-dimensional dataset describing $T$ independent configurations of a complex system composed of $N$ binary units. This problem is formulated in the language of statistical mechanics as the problem of finding a family of couplings compatible with a corresponding set of empirical observables in the limit of large $N$. We focus on the typical properties of its solutions and highlight the possible spurious features which are associated with this regime (model condensation, degenerate representations of data, criticality of the inferred model). We present a class of models (complete models) for which the analytical solution of this inverse problem can be obtained, allowing us to characterize in this context the notion of stability and locality. We clarify the geometric interpretation of some of those aspects by using results of differential geometry, which provides means to quantify consistency, stability and criticality in the inverse problem. In order to provide simple illustrative examples of these concepts we finally apply these ideas to datasets describing two stochastic processes (simulated realizations of a Hawkes point-process and a set of time-series describing financial transactions in a real market) .
}

\acknowledgements{
I am truly indebted with my advisor M. Marsili for encouraging me to spend three years studying beautiful and challenging problems. I have been consistently borrowing his advices and profiting of his skills, which he never refused to share.
I thank A.C. Barato, C. Battistin, M.Bardoscia, E.Zarinelli and P.Zoi, with whom I had the pleasure to collaborate with during the course of the PhD.
I especially thank Andr\'e for his efforts in coercing this thesis into an almost readable form.
I acknowledge the former and the current members of M. Marsili group (M. Bardoscia, F. Caccioli, L. Caniparoli, L. Dall'Asta, G. De Luca, D. De Martino, G. Gori, G. Livan, P. Vivo), for all the interesting discussions we had and for the exceedingly long time we spent in the ICTP cafeteria. I thank my classmates A. De Luca and J. Viti (with whom I'm going to share another part of academic life), together with F. Buccheri, L. Foini, F. Mancarella and X. Yu. I thank M. Masip for his constant disposability and his wise advices. I would also like to thank all the persons from outside SISSA and ICTP with whom I had the opportunity to have valuable and stimulating interactions all over these years: M. Alava, F. Altarelli, E. Aurell, J.P. Bouchaud, A. Braunstein, S. Cocco, A. Codello, S. Franz, A. Kirman, F. Lillo, Y. Roudi, B. T\'oth, R. Zecchina. \\ \\
On a personal note, I would like to thank my parents for all their support.  Finally, I thank Najada for the years we spent together, and for her decision to follow me during the forthcoming ones.
}
\published{

\begin{enumerate}

\item{
\textsc{Mastromatteo, I.}\\
Beyond inverse Ising model: structure of the analytical solution for a class of inverse problems.
\emph{Arxiv preprint: arXiv:1209.1787} (2012).
}

\item{
\textsc{Barato A.C., Mastromatteo I., Bardoscia M. and Marsili M.}\\
Impact of meta-order in the minority game. Submitted to \emph{Quant. Finance}.
\emph{Arxiv preprint: arXiv:1112.3908} (2012).
}

\item{
\textsc{Mastromatteo I., Zarinelli E. and Marsili M.}\\
Reconstruction of financial network for robust estimation of systemic risk.
 \emph{J. Stat. Mech.} \textbf{P03011} (2011).
}

\item{
\textsc{Mastromatteo I. and Marsili M.}\\
On the criticality of inferred models.
\emph{J. Stat. Mech.} \textbf{P10012} (2011).
}

\item{
\textsc{Mastromatteo I., Marsili M. and Zoi P.} \\
Financial correlations at ultra-high frequency: theoretical models and empirical estimation. 
\emph{Eur. Phys. J. B}, \textbf{80} (2) 243-253 (2011).
}

\end{enumerate}
Chapter 2 has an introductory purpose, and contains mainly non-original work. Chapter 3 presents original results not yet published. Chapter 4 discusses the ideas behind article number 1, while Chapter 5 covers exhaustively the content of article number 4. The subject of the remaining manuscripts has not been included in this thesis.
}

\fi  % disable frontmatter

\begin{document}

\makefrontmatter

\fancyhead{} % clear all header fields
\fancyhead[RO,LE]{\slshape \leftmark}
\pagestyle{fancy}
%\renewcommand{\sectionmark}[1]{\markright{\MakeUppercase{Section \thesection.\ } }}

% If you've disabled frontmatter, you can insert the toc manually
%\tableofcontents\clearpage
% \include lets us split up the document (and each include starts a new page):
\chapter{Introduction\label{ch:intro}}
The generations living during the last twenty or thirty years witnessed a huge scientific revolution which has been, essentially, technology driven. An impressive amount of computational power became cheaply available for people and institutions, while at the same time the quantity of data describing many aspects of our world started to grow in a seemingly unbound fashion: the human genoma can be efficiently sequenced in some days \cite{Shendure:2008qf,Wheeler:2008ve}, the interactions among proteins in a human body can in principle be enumerated one-by-one \cite{Rual:2005bh}, financial transactions are recorded with resolutions well below one second \cite{SEC2010:dq}, the dynamics of networks of all kinds (social, economics, neural, biological) can be tracked in real-time. 
Parallel to this, the widely accepted scientific paradigm according to which it is necessary to ground reliable models on solid first principles started to crumble: promising results evidenced that it is possible to extract accurate statistical models from empirical datasets without even trying to guess what is their underlying structure, nor to characterize which input-output relations govern their behavior. Large datasets can be automatically and faithfully \emph{compressed} in small sets of coefficients \cite{Donoho:2006ly}, their features can be \emph{described} accurately with unsupervised algorithms, new data can be \emph{predicted} with a given degree of accuracy on the basis of the older one (see for example \cite{:ly}). Google uses pattern recognition and Bayesian techniques to translate from one language to the other regardless of the formal rules of the grammar \cite{:zr}, and Netflix can predict how much you will rate a movie (one to five) with an error around 0.85 without knowing anything about you but a few of your former preferences \cite{:uq}. The embarrassing success of this approach compels a basic epistemological question about modeling: does an approach based solely on statistical learning lead to any actual understanding? What does one learn about a system when processing data in this way? \\
This problem is particularly relevant when dealing with the task of high-dimensional inference, in which a typically large set of parameters is extracted from an even larger dataset of empirical observations. What meaning has to be associated with each of the many parameters extracted from data? Are there \emph{combinations} of such numbers describing global, macroscopic features of the system? A prototypical example is provided by the study of networks of neurons, in which one would like to understand how the brain works (e.g., the presence of collective states of the network, the possibility to store and retrieve informations) by processing data describing the behavior of a huge set of elementary units (the neurons). This task can be thought of as a seemingly hopeless one: in a way it is similar to reverse-engineering how a laptop works by probing the electric signal propagating through its circuitry. A modern answer to this type of arguments is the idea that if data is sufficient and the inference algorithm is good enough, some of the actual features of the system will eventually be detected. In the case of a laptop, one can think to extract from data not only the wiring pattern a set of cables, but to detect collective features such as the fact that a computer is an essentially deterministic object (in contrast to biological networks, where fluctuations are essential), or that it possesses multiple collective states (say, switched-on, switched-off or sleepy). \\
Physics, and in particular statistical mechanics, has much to do with all of this picture for two main reasons. The first one is technical: while the high-dimensional limit is a relatively new regime in the field of statistical inference, statistical mechanics has since long developed mathematical descriptions of systems composed by a very large (or better, infinite) number of interacting components \cite{Huang:1987nx}. Hence, mapping problems of statistical inference onto problems of statistical mechanics opens the way to a remarkable amount of mathematical machinery which can be used to solve quickly and accurately problems which become very complicated for large systems \cite{Kappen:1998dt,Tanaka:1998tg}. This is even more true since the study of heterogeneous and glassy materials produced sophisticated tools (replica trick, cavity methods) suitable to study systems in which no apparent symmetry or regularity is present, as often found in data describing complex systems \cite{Mezard:1987fk}.
The second, and more philosophical, reason is that statistical mechanics is naturally built to explain collective behaviors on the basis of individual interactions. Just as the ideal gas can be understood by studying the aggregate behavior of many non-interacting particles, or the emergence of spontaneous magnetization can be derived by studying the interactions of single spins, statistical mechanics can be used to predict the collective behavior of biological, social and economic systems starting from a given set of rules describing the interaction of some fundamental units \cite{De-Martino:2006os}. In 1904 Ludwig Boltzmann, almost a century before anyone could take him literally, anticipated that
\begin{quote}
\emph{
``The wide perspectives opening up if we think of applying this science to the statistics of living beings, human society, sociology and so on, instead of only to mechanical bodies, can here only be hinted at in a few words.''}
\end{quote}
Hence, from the perspective of (large-scale) statistical learning, it is natural to use statistical mechanics methods to study the emergence of collective properties of a system once the microscopic interactions of the fundamental units have been reconstructed through a careful analysis of empirical data. \\
Unfortunately, even if one is able to do that, it is not always easy to understand how much of the inferred model faithfully describes the system: it is possible, and it is often the case, that the procedure which is used to perform the data analysis influences so much the outcome that the actual properties of the system get lost along the way, and the inferred model shows a spurious behavior determined just by the fitting procedure. For example, models with binary interactions may describe very well systems in which the interaction is actually multi-body \cite{Hinton:2010vn}, just as critical models (strongly fluctuating statistical systems) may fit random set of observables much better than ordinary ones \cite{Mastromatteo:2011ly}. Noise itself may be fitted very well by sophisticated models, while non-stationary systems might be accurately described by using equilibrium distributions \cite{Tyrcha:2012fk}. In all of these cases, it is important to develop quantitative tools which allow to distinguish between spurious features of the inferred model and genuine ones. \\
The purpose of this work is precisely to inquire some of those aspects in the simpler setting in which we consider a statistical system consisting in a string of $N$ binary variables, used to model $T$ independently drawn configurations. We will show that, while the small $N$ regime the problem of inference can be completely controlled (chapter \ref{ch:Inference}), the large $N$ regime becomes computationally intractable and non-trivial collective properties may emerge (chapter \ref{ch:HighDimInf}). Such features be observed independently of the data, and have to be associated uniquely with the properties of the model which is used to perform the inference, regardless of the system which one is trying to describe. In chapter \ref{ch:Completeness} we will show under which conditions the problem of inferring a model is easy, showing in some cases its explicit solution. We will also evidence the limits of non-parametric inference, highlighting that for under-sampled systems correlations might be confused with genuine interactions. In chapter \ref{ch:Geometry} we will provide a geometric interpretation for the problem of inference, showing a metric which can be used to meaningfully assess the collective phase of an inferred system. We will apply these ideas to two datasets, describing extensively the results of their analysis in the light of our approach.
\chapter{Binary Inference\label{ch:Inference}}
In this chapter we will describe the problem of extracting information from empirical datasets describing a stationary system composed of a large number of interacting units. Interestingly, this problem has almost simultaneously received a great deal of attention from the literature of diverse communities (biology \cite{Socolich:2005vy,Weigt:2009on}, genetics \cite{Braunstein:2008sa}, neuroscience \cite{Schneidman:2006vg,Shlens:2006uq,Cocco:2009mb}, economy, finance \cite{Lillo:2008ht,Moro:2009wy,Lachapelle:2010kv}, sociology). This can be traced back to two main reasons: first, it is now possible across many fields to analyze the synchronous activity of the components of a complex system (e.g., proteins in a cell, neurons in the brain, traders in a financial market) due to technological advantages either in the data acquisition procedures or in the experimental techniques used to probe the system. Secondly, data highly resolved in time is often available, which (beyond implying that finer time-scales can be explored) provides researchers with a large number of observations of the system. Defining as $N$ the number of components of the system and as $T$ the number of available samples, these last observations can be summarized by asserting that the limit of large $N$ and large $T$ can be accessed for a large number of complex systems. In this work we will restrict ourselves to the more specific case in which such systems are described by binary units, reminding to the reader that (i) most of what will be shown can be generalized to the case of non-binary (Potts) or continuous variables \cite{Wainwright:2008kx} and (ii) the binary case already allows to describe in detail several systems \cite{Schneidman:2006vg,Shlens:2006uq,Cocco:2009mb}.
In section \ref{sec:DirectProblem} we describe the models that we consider, which usually go under the name of \emph{exponential families} and are justified on the basis of the maximum entropy principle (appendix \ref{app:MaxEntPr}), and state the \emph{direct problem}, alias the calculation of the observables given the model. In section \ref{sec:InvProblem} we present the problem of inferring a model from data (the \emph{inverse problem}) and characterize it as the Legendre-conjugated of the direct one. In section \ref{sec:Regularization} we present the \emph{regularization} techniques which can be used to cure the pathological behavior of some inverse problems and improve their generalizability.
Although the results presented in this chapter are far from being original, we aim to show as transparently as possible the deep connections between information theory and statistical mechanics, emphasizing the strong analogy between direct and inverse problems.

\section{The direct problem \label{sec:DirectProblem}}
We introduce in this section the \emph{direct problem} -- which deals with finding the observables associated with a given statistical model -- as a preliminary step towards the formulation of an inference problem. This is the problem typically considered by statistical mechanics, hence we will adopt most of the terminology and the notation from this field. The main results that we will present are associated with the \emph{free energy} -- which we    use in order to generate the averages and the covariances of the model -- and to its relations with the notion of Kullback-Leibler divergence and the one of Shannon entropy. Finally, we will characterize the large and small deviation properties of the empirical averages of the observables under the model.

\subsection{Statistical model}
We consider a system of $N$ binary spins $ s = (s_1,\dots , s_N) \in \{-1,1\}^N = \Omega$, indexed by $i \in V = \{1,\dots, N\}$. A \emph{probability density} $\bp$ is defined as any positive function $\bp : \Omega \to \mathbb{R}$ such that $\sum_s p(s) = 1$, while the space of all possible probability densities on $\Omega$ is denoted as $\mathcal M(\Omega)$.
We also consider a families of real-valued functions $\bphi : \Omega \to \mathbb{R}^{|\bphi|}$ with components $\bphi (s) = (\phi_1 (s), \dots , \phi_{|\bphi|} (s))$, which will be referred as \emph{binary operators}, and are more  commonly known in the literature of statistical learning as \emph{sufficient statistics} or \emph{potential functions} \cite{Wainwright:2008kx}, and will be used in order to construct a probability density on the configuration space of the system.
\defin{
Given a set of binary operators $\bphi =\{ \phi_\mu \}_{\mu=1}^M$ and a vector of real numbers $\bg=\{ g_\mu \}_{\mu=1}^M$ a \emph{statistical model} is defined as the pair $(\bphi,\bg)$. Its associated probability density $\bp = (p_s)_{s\in \Omega}$ is given by
\beq
p( s) = \frac{1}{Z(\bg)} \exp \left( \sum_{\mu=1}^M g_\mu  \phi_\mu( s) \right) \; , \label{eq:ProbDensity}
\eeq
whereas the normalization constant $Z(\bg)$ is defined as
\beq
Z(\bg) = \sum_{ s} \exp \left( \sum_{\mu=1}^M g_\mu \phi_\mu( s)  \right) \label{eq:PartFunction}
\eeq
and is referred as the \emph{partition function} of the model. The \emph{free energy} $F(\bg)$ is defined as $F(\bg) = -\log Z(\bg)$. 
}
For conciseness, the identity operator will always be labeled as the zero operator $\phi_0( s) = 1$, in order to reabsorb the normalization constant $Z(\bg)$ into its conjugated coupling $g_0$. The probability density will be written as $p( s) = p_s$, so that (\ref{eq:ProbDensity}) will be compactly written as
\beq
p_s = \exp \left( \sum_{\mu=0}^M g_\mu  \phi_{\mu,s}  \right)  \label{eq:ProbDensityCompact} \; .
\eeq
With these definitions, the coupling $g_0$ results equal to the free energy $g_0 = - \log Z( \bg) = F(\bg)$. Given a family of operators $\bphi$, we also denote as $\mathcal M (\bphi)$ the set of all the statistical models of the form (\ref{eq:ProbDensity}) obtained by varying the coupling vector $\bg$.
Given the probability density (\ref{eq:ProbDensity}) and a generic subset $\Gamma \subseteq V$ (which we call a \emph{cluster}),  we also define the \emph{marginal} $p^\Gamma(s^\Gamma) $ as
\beq
p^\Gamma(s^\Gamma ) = \sum_{s_i | i \not\in \Gamma} p( s) \; , \label{eq:MarginalDef}
\eeq
which expresses the probability to find spins belonging to the $\Gamma$ in a given configuration once the degrees of freedom associated with spins outside such cluster have been integrated out (whereas $p^\emptyset = 1$ and $p^V (s) = p(s)$). \\
This construction will be used to study inference problems in which the $M$ operators $\phi ( s)$ are \emph{a priori} known. We will disregard for the moment the issue of optimally selecting the most appropriate operators in order to describe a given set of data, an important problem known as \emph{model selection}. Let us indeed remind that models of the form (\ref{eq:ProbDensity}) can be justified on the basis of the maximum entropy principle, which will be stated in appendix \ref{app:MaxEntPr}.
The next notions which will be defined are the one of ensemble average and the one of susceptibility which will be extensively used throughout our discussion.
\defin{
Given a statistical model $(\bphi,\bg)$ of the form (\ref{eq:ProbDensity}), we define the \emph{ensemble average} of an operator $\phi_\mu$ as the quantity
\beq
\langle \phi_\mu \rangle = \sum_{s} \phi_{\mu,s} p_s \; ,
\eeq
while the \emph{generalized susceptivity} matrix $\hat \bchi$ is defined as the covariance matrix whose elements are given by
\beq
\chi_{\mu, \nu} = \langle \phi_\mu \phi_\nu \rangle - \langle \phi_\mu \rangle \langle \phi_\nu \rangle  \; . \label{eq:GenSusc}
\eeq
}
Beyond describing fluctuations around the ensemble average of the $\bphi$ operators, the generalized susceptibility $\hat \bchi$ is a fundamental object in the field of information theory \cite{Cover:1991fk}, in whose context is more often referred as \emph{Fisher information}, and is more commonly defined as
\beq
\chi_{\mu, \nu} = - \left< \frac{\partial^2 \log p_s}{\partial g_\mu \partial g_\nu} \right> \; .
\eeq
Its relevance in the field of information theory and statistical learning will later be elucidated by properties (\ref{eq:MeanEmpAvg}) and (\ref{eq:CovEmpAvg}) which concern with the direct problem. Sanov thorem (\ref{eq:Sanov}), Cram\'er-Rao bound (\ref{eq:CRBound}), together with equations (\ref{eq:InvMeanMaxLik}) and (\ref{eq:InvCovMaxLik}), clarify its role in the context of the inverse problem.
\prop{The free energy function enjoys the properties
\beq
\langle \phi_\mu \rangle = - \frac{\partial F}{\partial g_\mu}  \label{eq:GenFuncAvg}
\eeq
and
\beq
\chi_{\mu,\nu} = - \frac{\partial^2 F}{\partial g_\mu \partial g_\nu}   \; , \label{eq:GenFuncSusc}
\eeq
thus it is the generating function of the averages and of the fluctuations of the operators $\phi_\mu$ contained in the model.
}
Equation (\ref{eq:GenFuncSusc}) implies that covariances $\chi_{\mu,\nu}$ are related to the response of the ensemble averages with respect to changes of the couplings through
\beq
\chi_{\mu,\nu} = \langle \phi_\mu \phi_\nu \rangle - \langle \phi_\mu \rangle \langle \phi_\nu \rangle = \frac{\partial \langle \phi_\mu \rangle}{\partial g_\nu} \; ,
\eeq
a relation known as \emph{fluctuation-dissipation} relation, which is a direct consequence of the stationary nature of the probability distribution (\ref{eq:ProbDensity}).
Another fundamental property of the free energy function $F(\bg)$ is its \emph{concavity}, which will later allow us to relate the field of statistical inference with the one of convex optimization (appendix \ref{app:ConvexOpt}).
It can be shown (appendix \ref{app:FreeEnConc}) that:
\prop{ $\phantom{a}$
\begin{itemize}
\item{The susceptibility matrix $\hat \bchi$ is a positive semidefinite matrix, thus the free energy $F(\bg)$ is a concave function.}
\item{If the family of operators $\bphi$ is \emph{minimal} (i.e.\ it doesn't exist a non-zero vector $\bx$ such that $\sum_\mu x_\mu \phi_{\mu,s} $ is constant in $s$), then the susceptibility matrix $\hat \bchi$ is strictly positive definite and the free energy $F(\bg)$ is strictly concave.}
\end{itemize}
}

\defin{
Given a statistical model $(\bphi,\bg)$ of the form (\ref{eq:ProbDensity}), the \emph{direct problem} is defined as the calculation of the free energy $F(\bg)$, of the averages $\langle  \bphi \rangle$ and of the susceptibility matrix $\hat \bchi$ as functions of the coupling vector $\bg$.
}

\subsection{Entropy and Kullback-Leibler divergence}
In this section we will define the concept of Shannon entropy, which will be used as an information theoretic measure of the information content of a distribution.
\defin{Given a probability density $\bp$, we define the Shannon entropy $S(\bp)$ as the function
\beq
S(\bp) = - \sum_s p_s \log p_s \; \label{eq:ShannonEntropy}
\eeq
}
The quantity $S(\bp)$ measures the amount of disorder associated with the random variable $s$, and satisfies the following properties:
\begin{itemize}
\item{$0 \leq S(\bp) \leq \log |\Omega|$. In particular $ S(\bp) = 0$ for $p(s) = \delta_{s,s^\prime}$ (when the variable $s$ is maximally informative), while $ S(\bp) = \log |\Omega|$ for the flat case $p(s) = 1/|\Omega|$ (in which $s$ is maximally undetermined).}
\item{The function $S(\bp)$ is concave in $\bp$.}
\end{itemize}
They can be proven straightforwardly, as for example in \cite{Cover:1991fk}.
Another information-theoretic notion which will be extensively used is the Kullback-Leibler divergence $D_{KL}(\bp|\bq)$, which characterizes the distance between two probability distributions. Although it doesn't satisfy the symmetry condition nor the triangular inequality required to define a proper measure of distance, in chapter \ref{ch:Geometry} we will show that indeed a rigorous concept of distance can be extracted by means of the Kullback-Leibler divergence.
\defin{Given a pair of probability densities $\bp$ and $\bq$, the \emph{Kullback-Leibler} divergence $D_{KL}(\bp||\bq)$ is defined as
\beq
D_{KL}(\bp||\bq) = \sum_s p_s \log \frac{p_s}{q_s}
\eeq
}
Such quantity enjoys the following properties:
\begin{itemize}
\item{ $D_{KL} (\bp||\bq)\geq 0 $ for any pair of probability densities $\bp$, $\bq$.}
\item{$D_{KL} (\bp||\bq) = 0$ if and only if $\bp=\bq$.}
\item{$D_{KL} (\bp||\bq)$ is a convex function in both $\bp$ and $\bq$.}
\end{itemize}
These property justify the role played by the Kullback-Leibler divergence in information theory, and can be proven straightforwardly (see \cite{Cover:1991fk}). Notice indeed that given two statistical models $(\bphi, \bg)$ and $(\bphi,\bg^\prime)$ respectively associated with densities $\bp$ and $\bp^\prime$, the entropy and the Kullback-Leibler divergence can be written as
\beqa
S(\bp) &=& -F(\bg) - \sum_{\mu=1}^M g_\mu \langle \phi_\mu \rangle_{\bg} \label{eq:RelEntrFreeEn} \\
D(\bp || \bp^\prime) &=& F(\bg) - F(\bg^\prime) + \sum_{\mu=1}^M (g_\mu - g_\mu^\prime) \langle \phi_\mu \rangle_{\bg} \label{eq:RelDKLFreeEn} \; ,
\eeqa
so that the concavity properties of $S(\bp)$ and $D_{KL}(\bp || \bq)$ can be related to the ones of the free energy $F(\bg)$. These quantities will be relevant in order to characterize the large deviation properties both for the direct and of the inverse problem.

\subsection{Observables \label{sec:Observables}}
Throughout all our discussion, we will focus on the case in which $T$ independent, identically distributed (i.i.d.) configurations of the system denoted as $\hat \bs = \{  s^{(t)}\}_{t=1}^T$ are observed.
The joint probability of observing the dataset $\hat \bs$ (also called \emph{likelihood}) given a statistical model $(\bphi,\bg)$ is
\beq
P_T(\hat \bs| \bg) = \prod_{t=1}^T p( s^{(t)}) = \exp \left( T \sum_{\mu=0}^M g_\mu \bar \phi_\mu \right) \label{eq:Likelihood}
\eeq
where the quantities
\beq
\bar \phi_\mu = \frac{1}{T} \sum_{t=1}^T \phi_\mu ( s^{(t)})
\eeq
are called \emph{empirical averages}. It is worth remarking that $P_T(\hat \bs)$ depend on the observed configurations just through the empirical averages $\bar \bphi$. We will denote averages over the measure $P_T(\hat \bs | \bg)$ with the notation $\langle \dots \rangle_T$.
We also define the \emph{empirical frequencies} (also known as \emph{type}) $\bar \bp$ as the vector with components
\beq
\bar p_s = \frac{1}{T} \sum_{t=1}^T \delta_{s,s^{(t)}} \; ,
\eeq
which enjoys the following properties:
\begin{itemize}
\item{It is positive and normalized ($\sum_s \bar p_s =1$), thus it defines a probability density on $\Omega$ (i.e., $\bar \bp \in \mathcal M(\Omega)$).}
\item{The empirical averages $\bar \phi$ can be obtained as $\bar \phi_\mu = \sum_s \phi_{\mu,s} \bar p_s$ .}
\item{If the dataset $\hat \bs$ is generated by a probability distribution $\bp$, then $\bar \bp$ is distributed according to the multinomial distribution
\beq
P_T(\bar \bp | \bp) = \left( \prod_s  \frac{p_s^{T_s}}{ T_s !} \right) T! \; \delta \left(T - \sum_s T_s \right)  \; ,
\eeq
where $T_s = T \bar p_s$. Its first and second momenta are
\beqa
\langle \bar p_s \rangle_T &=& p_s \\
\langle \bar p_s  \bar p_{s^\prime} \rangle_T - \langle \bar p_s \rangle \langle \bar p_{s^\prime} \rangle_T &=& \frac{1}{T} \left( \delta_{s,s^\prime} p_s - p_s p_{s^\prime} \right) \; .
\eeqa
}
\end{itemize}
Finally, given a collection of operators $\bphi$ we will denote the set of all empirical averages $\bar \bphi$ that are compatible with at least one probability density in $\Omega$ with
\beq
\mathcal G (\bphi) = \Bigg\{ \bar \bphi \in \mathbb{R}^M \; \Bigg| \;  \exists \, \bar \bp \, \in \mathcal M (\Omega) \textrm{ s.t. } \, \bar \phi_\mu = \sum_s \phi_{\mu,s} \bar p_s \; \forall \mu \Bigg\} \; ,
\eeq
which is called in the literature \emph{marginal polytope} \cite{Wainwright:2008kx}. It can be proven that (see for example \cite{Wainwright:2008kx}):
\begin{itemize}
\item{
$\mathcal G (\bphi)$ is a convex set (i.e., given $\bar \bphi$, $\bar \bphi^\prime \in \mathcal G(\bphi)$, for any $\alpha \in [0,1]$ also \\ $\alpha \bar \bphi + (1-\alpha) \bar \bphi^\prime \in \mathcal G(\bphi)$).
}
\item{
$\mathcal G (\bphi) = \textrm{conv} \{ \bphi (s) \in \mathbb{R}^M \; | \; s \in \Omega \}$, where $\textrm{conv}\{ \cdot \}$ denotes the convex hull operation.
}
\item{
$\mathcal G(\bphi)$ is characterized by the Minkowski-Weyl theorem as a subset of $\mathbb{R}^M$ identified by a finite set of inequalities. More formally, one can find a set of vectors $\{ \bm{x}_a, y_a\}_{a=1}^d$ with $d$ finite such that
\beq
\mathcal G (\bphi) = \left\{ \bphi \in \mathbb{R}^M \; \Bigg| \;  \sum_{\mu=1}^M x_{\mu,a} \bar \phi_\mu \geq y_a \; \forall a \in \{ 1,\dots d\} \right\}
\eeq
}
\end{itemize}

\subsection{Small and large deviations \label{sec:DirSLDev}}
In the case of the direct problem it is natural to formulate the following questions:
\begin{enumerate}
\item{
What are the most likely values for the empirical averages $\bar \bphi$?
}
\item{
How probable it is to find rare instances $\hat \bs$?
}
\end{enumerate}
The first question is relatively easy to answer, and characterizes the role of the generalized susceptivity in the direct problem as ruling the convergence of the empirical averages to the ensemble averages\footnote{In the framework that we are considering (i.i.d.\ sampling of configurations drawn by the same distribution) empirical averages always converge to ensemble averages with an error scaling as $1/\sqrt{T}$. Indeed it makes sense to model the case in which the probability measure $\bp$ breaks into \emph{states}, so that for any finite length experiment, just samples belonging to the same state are observed. This is meant to model the phenomenon of ergodicity breaking, which we will comment about in section \ref{sec:ErgBreak}.}, as shown in the following and proven in appendix \ref{app:SmallDevEmpAv}.
\prop{
Given a statistical model $(\bphi,\bg)$, the empirical averages $\bar \bphi$ satisfy the relations
\beqa
\langle \bar \phi_\mu \rangle_T &=& \langle \phi_\mu \rangle \label{eq:MeanEmpAvg} \\ 
\langle \bar \phi_\mu \bar \phi_\nu \rangle_T -  \langle \bar \phi_\mu \rangle_T \langle \bar \phi_\nu \rangle_T  &=& \frac{\chi_{\mu,\nu}}{T}  \label{eq:CovEmpAvg} \; .
\eeqa
}
The explicit form of the likelihood function (\ref{eq:Likelihood}) allows to answer exhaustively also to the second question.
\prop{
Given a probability density $\bp$ defined by a statistical model $(\bphi,\bg)$, the function $I_\bp(\bar \bp) = -\frac{1}{T}\log P_T(\bar \bp | \bp) = -F(\bg) - \sum_{\mu=1}^M g_\mu \bar \phi_\mu$ is the \emph{large deviation function} for the direct problem.
}
This implies that the probability of observing dataset a generic $\hat \bs$ decays exponentially in $T$, with a non-trivial rate function $I_\bp(\bar \bp)$ determined by the empirical averages $\bar \bphi$ only. Also notice that the large deviation function can be expressed entirely in terms of the entropy and the Kullback-Leibler divergence as
\beq
I_{\bp}(\bar \bp) = D_{KL}(\bar \bp|| \bp) + S(\bar \bp) \; . \label{eq:DirLargeDev}
\eeq

\section{The inverse problem \label{sec:InvProblem}}
In this section we introduce the \emph{inverse problem} of extracting a coupling vector $\bg^\star$ given a set of operators $\bphi$ and a vector of empirical averages $\bar \bphi$. We will present this problem as dual with respect to the direct one, showing that just as the knowledge of the free energy $F(\bg)$ completely solves the direct  problem, the Legendre transform of $F(\bg)$ denoted as $S(\bar \bphi)$ and characterized as the Shannon entropy, analogously controls the inverse one.

\subsection{Bayesian formulation}
We will be interested in calculating the set of couplings $\bg^\star$ which best describes a given set of data $\hat \bs$ of length $T$ within the statistical model $(\bphi,\bg)$. Bayes theorem provides a mathematical framework in which the problem can be rigorously stated, by connecting the likelihood function $ P_T(\hat \bs |  \bg)$ described in section \ref{sec:Observables} to the \emph{posterior} of the model $P_T( \bg | \hat \bs)$, which specifies the probability that the data $\hat \bs$ has been generated by model $ \bg$. Bayes theorem states in fact that
\beq
P_T( \bg | \hat \bs) \propto P_T(\hat \bs |  \bg) P_0( \bg) \; , \label{eq:BayesTheor}
\eeq
where $P_0( \bg)$ is known as the \emph{prior}, and quantifies the amount of information which is \emph{a priori} available about the model by penalizing or enhancing the probability of models specified by $ \bg$ by an amount $P_0( \bg)$. Bayes theorem also links the concept of prior to the one of regularization which will be discussed in section \ref{sec:Regularization}, but for the moment we will consider the prior $P_0( \bg)$ to be uniform (i.e.\ a $ \bg$-independent constant), so that it can be reabsorbed into the pre factor of equation (\ref{eq:BayesTheor}).
In this case finding the best model to describe the empirical averages may mean:
\begin{itemize}
	\item{Finding the point in the space of couplings $ \bg$ in which the function $P_T( \hat \bs |  \bg)$ is maximum (\emph{maximum likelihood} approach).}
	\item{Finding  the region of the space of couplings in which such probability is high (\emph{Bayesian} approach).}
\end{itemize}
These two approaches lead to very similar results in the case in which the likelihood function is strictly concave, as one can prove by means of large deviation theory (see section \ref{sec:InvSLDev} and appendix \ref{app:ConvInfCoup}). Roughly speaking, when the number of observations $T$ is large, the posterior $P_T(\bg | \hat \bs)$ concentrates around the maximum likelihood parameter, being the rate of convergence  fixed by the stability matrix of the maximum and the number of samples $T$. Hence we will later  define as the inverse problem the characterization of the maximum likelihood parameters and of their linear stability, disregarding the detailed shape of the function $P_T(\bg | \hat \bs)$.

\subsection{Maximum likelihood criteria}
The maximum likelihood criteria requires to find the maximum of the likelihood function $P_T(\hat \bs |  \bg)$, whose solution is obtained by differentiation of equation (\ref{eq:Likelihood}) with respect to the couplings $g_\mu$, and reads for each $\mu$
\beq
\langle \phi_\mu \rangle = \bar \phi_\mu \; , \label{eq:MomentumMatching}
\eeq
a condition which will be referred as \emph{momentum matching} condition. Thus, the best parameters $ \bg^\star$ describing a set of data $\hat \bs$ under the model (\ref{eq:ProbDensity}) in absence of prior are the ones for which the ensemble averages of the model are matched with the empirical ones.
\rmk{It is easy to see that the matching condition (\ref{eq:MomentumMatching}) can alternatively be obtained by minimizing the Kullback-Leibler divergence $D_{KL}(\bar \bp | \bp)$ between the probability distribution defined by the empirical frequencies $\bar \bp$ and the probability density $\bp$ defined by the statistical model $(\bphi,\bg)$.
}

\subsection{Statement of the inverse problem}
The concavity properties of the likelihood function (or equivalently, of the free energy $F(\bg)$), allow for a characterization of the problem of inferring the maximum likelihood parameters $\bg^\star$ given data $\hat \bs$ in terms of a Legendre transform of $F(\bg)$. \\
\defin{
Given a minimal set of operators $\bphi$ and a set of empirical averages $\bar \bphi$, the function $S(\bar \bphi)$ is defined as the Legendre transform
\beq
-S( \bar \bphi) = \max_{ \bg} \left( \sum_{\mu=1}^M g_\mu \bar \phi_\mu + F( \bg) \right) \; . \label{eq:LegendreEntropy}
\eeq
We denote with $\bg^\star$ the (only) value of $\bg$ maximizing equation (\ref{eq:LegendreEntropy}). Such quantity satisfies
\beq
\bar \phi_\mu = - \frac{\partial F(\bg)}{\partial g_\mu} \Bigg|_{\bg=\bg^\star}\; . \label{eq:LegendreCondition}
\eeq
By construction the statistical model  $(\bphi,\bg^\star)$ verifies the matching condition (\ref{eq:MomentumMatching}).
}
By considering the Shannon entropy $S(\bp) = - \sum_s p_s \log p_s $ and by plugging probability density $\bp^\star$ inside its definition, one can see that it holds
\beq
S(\bp^\star) = -\sum_{\mu=1}^M g_\mu^\star \bar \phi_\mu - F( \bg^\star ) = S(\bar  \bphi) \; ,
\eeq
which characterizes the Legendre transformation (\ref{eq:LegendreEntropy}) of the free energy $F(\bg)$: $S(\bar \bphi)$ is the Shannon entropy of the distribution expressed as a function of the empirical averages.
\rmk{
The existence of a solution $\bg^\star (\bar \bphi)$ to the minimization problem defining the entropy $S(\bar \bphi)$ is guaranteed by a general result stating that given any operator set $\bphi$ defining a marginal polytope $\mathcal G (\bphi)$, the empirical averages $\bar \phi_\mu = \sum_s \phi_{\mu,s} \bar p_s$ can be matched by ensemble averages $\langle \bphi^\star \rangle$ associated with the statistical model $(\bphi,\bg^\star)$, with $\bg^\star \in (\mathbb{R} \cup \{ -\infty, +\infty\} )^M$. The interested reader is referred to \cite{Wainwright:2008kx} for the mathematical details.
}
\prop{
By differentiation of equation (\ref{eq:LegendreEntropy}) one finds that
\beq
- \frac{\partial S}{\partial \bar \phi_\mu } = g_\mu^\star\;, \label{eq:GenFuncCoupling}
\eeq
while by applying the chain rule to the equation $\delta_{\mu,\nu} = \partial g_\mu / \partial g_\nu$ one finds that
\beq
-\frac{\partial^2 S}{\partial \bar \phi_\mu  \partial \bar \phi_\nu} = \chi^{-1}_{\mu,\nu} \;. \label{eq:GenFuncInvFisher}
\eeq
}
Equations (\ref{eq:GenFuncCoupling}) and (\ref{eq:GenFuncInvFisher}) are analogous to equations (\ref{eq:GenFuncAvg}) and (\ref{eq:GenFuncSusc}) which relate to the direct problem. Just as the free energy $F(\bg)$ generates averages and susceptibilities in the direct problem, the entropy $S(\bar \bphi)$ is the generating function for the inverse one. Hence, an inference problem can be solved by explicitly computing the Shannon entropy $S(\bar \bphi)$ and finding its maximum (either analytically or numerically).

\defin{
The problem of determining the entropy $S(\bar \bphi)$, the inferred couplings $ \bg^\star$ and the inverse susceptibility $\hat \bchi^{-1}$ as functions of the averages $\bar  \bphi$ will be referred as the \emph{inverse problem}.
}
\subsection{Small and large deviations \label{sec:InvSLDev}}
Two questions analogous to the ones formulated in section \ref{sec:DirSLDev} in the case of the direct problem can be formulated for the inverse problem, namely: (i) what are the most likely values for the inferred coupling $\bg^\star$ obtained by a dataset $\hat \bs$ of length $T$? and (ii) how likely it is that such dataset has been generated by a model very different from the maximum likelihood one? In order to answer to those two questions we need to consider the large deviation function for the inverse problem. This can be obtained by noting that in absence of a prior, Bayes theorem and equation (\ref{eq:DirLargeDev}) imply that
\beq
P_T(\bp | \bar \bp) \propto P_T(\bar \bp | \bp) = e^{ -T (D_{KL}(\bar \bp|| \bp) + S(\bar \bp) )} \propto e^{ -T D_{KL}(\bar \bp|| \bp)}
\eeq
so that we can prove the following proposition.
\prop{
Given a vector of empirical frequencies $\bar \bp$, the large deviation function for the inverse problem $I_{\bar \bp}(\bp) \propto - \frac{1}{T} \log P_T(\bp|\bar \bp)$ is given by the Kullback-Leibler divergence
\beq
I_{\bar \bp}(\bp) = D_{KL}(\bar \bp|| \bp) \; .
\eeq
}
This implies that the probability for data $\bar \bp$ to be generated by any model $\bp$ decays exponentially fast in $T$ with a rate function given by the large deviation function $D_{KL}(\bar \bp || \bp)$. This result can be seen as a particular case of a more general theorem, which is known as Sanov theorem and whose proof can be found in appendix \ref{app:SanovTheor}.\footnote{We won't adopt the informal version of the theorem often found in literature (see for example \cite{Mezard:2009ko}), which doesn't require the introduction of the set $\mathcal M^\prime$. In such form the theorem is not valid when, for any value of $T$, $\mathcal M$ has empty intersection with the set of realizable empirical frequencies, as the probability for any point in $\mathcal M$ to be realized is strictly zero regardless of $T$.}
\theor{Consider a statistical model defined by a probability distribution $\bp$, and a (compact) set of probability densities $\mathcal M \subseteq \mathcal M (\Omega)$. Then if $\bar \bp$ is a vector of empirical frequencies sampled from the distribution $P_T(\hat \bs| \bp)$, it holds that
\beq
\lim_{\delta \to 0} \lim_{T\to\infty} - \frac{1}{T} \log \textrm{Prob} (\bar \bp \in \mathcal M^\prime) =  D_{KL} (\bq^\star || \bp) \; , \label{eq:Sanov}
\eeq
where $\bq^\star = \arg\min_{\bq \in \mathcal M} D_{KL}(\bq || \bp)$ and $\mathcal M^\prime$ is the compact set $\mathcal M^\prime = \{\bp + \delta \bp = \bp^\prime  \; \in \mathcal M(\Omega) \; |\;  \bp \in \mathcal M \,,\, ² \delta \bp \in [-\delta,\delta]^{|\Omega|} \; \}$.
}
Building on these results, we can provide an answer for our first question and find out what are the most likely distributions $\bp$ having generated data $\bar \bp$. In particular, it is possible to expand the Kullback-Leibler divergence around its minimum and perform a saddle-point estimation, obtaining the following result.
\prop{
Consider a generic dataset $\hat \bs$ defining the empirical distribution $\bar \bp \in \mathcal M(\Omega)$. Then, given a family of operators $\bphi$, the posterior probability (with uniform prior) $P_T(\bg | \bar \bp) \propto P_T(\bar \bp | \bg)$ defines a probability measure on space $\mathcal M (\bphi)$, parametrized by the coupling vector $\bg$ which defines the statistical model $\bp$. The averages and the covariances under this measure are given in the large $T$ limit by
\beqa
\frac{ \int d\bg \; g_\mu e^{-T D_{KL}(\bar \bp || \bp)} }{\int d\bg \;  e^{-T D_{KL}(\bar \bp || \bp)}} &\xrightarrow[T \to \infty]{}& g_\mu^\star \label{eq:InvMeanMaxLik} \\
\frac{ \int d\bg \; g_\mu g_\nu \, e^{-T D_{KL}(\bar \bp || \bp)} }{\int d\bg \;  e^{-T D_{KL}(\bar \bp || \bp)}} - g_\mu^\star g_\nu^\star &\xrightarrow[T \to \infty]{}& \frac{\chi^{-1}_{\mu,\nu}}{T} \label{eq:InvCovMaxLik} \; .
\eeqa
where $\bg^\star$ is the maximum likelihood estimator of $\bg$ and $\hat \bchi^{-1}$ is the inverse of the Fisher information matrix calculated in $\bg^\star$.
}
This result (proved in appendix \ref{app:ConvInfCoup}) characterizes the inverse of the generalized susceptibility as the matrix quantifying the speed in $T$ at which the probability measure on the inferred couplings concentrates around the maximum likelihood estimate. The centrality of this matrix in the inverse problem is also provided by a rigorous bound that can be proven for the covariance of any unbiased estimator, and known as Cram\'er-Rao bound. From this perspective, $\chi^{-1}_{\mu,\nu}$, can be seen as establishing a bound to the maximum rate of convergence for the estimator of a coupling.
\theor{
Consider a statistical model $(\bphi,\bg)$ with $F(\bg)$ strictly concave and an \emph{unbiased} estimator of the couplings $\bg^\star $ (i.e., such that $\langle g_\mu^\star \rangle_T= g_\mu$). Then the covariance matrix of $\bg^\star$ under the measure $\langle \dots \rangle_T $ satisfies
\beq
\left< (\bg^\star - \bg)(\bg^\star - \bg)^T \right>_T \succeq \frac{\hat \bchi^{-1}}{T} \label{eq:CRBound}
\eeq
where with $\hat{ \bm{X}} \succeq \hat{ \bm{Y}}$ we indicate that the matrix $\hat{ \bm{X}} - \hat{ \bm{Y}}$ is positive semidefinite.
}
The proof of this theorem is presented in the appendix \ref{app:CRBound}.

\subsection{Examples}
\subsubsection{Independent spins model\label{sec:IndepSpins}}
The simplest model of the form (\ref{eq:ProbDensity}) which can be considered is of the form
\beq
p( s) = \frac{1}{Z(\bh)} \exp \left( \sum_{i\in V} h_i s_i \right) \label{eq:IndepSpins}
\eeq
and will be called \emph{independent spin model}. The model contains $N$ operators of the form $\{\phi_{\{i\}} ( s) = s_i\}_{i\in V}$ (called in the following \emph{magnetizations}), whose conjugated couplings are denoted as $g_{\{ i\}} = h_i$ (and referred as \emph{external fields}). The empirical magnetizations will be denoted as $\bar s_i= m_i$. The direct problem can be solved by evaluating the partition function of the model, so that the free energy $F(\bh)$ results
\beq
F(\bh) = - N \log 2 - \sum_{i\in V} \log \cosh h_i \; .
\eeq
The ensemble averages and generalized susceptibilities can be obtained by differentiation, and are given by
\begin{eqnarray}
m_i &=& \tanh h_i \\
\chi_{i,j} &=& \frac{\delta_{i,j}}{\cosh^2 h_i}
\end{eqnarray}
The inverse problem is also easily solvable, as the Legendre transformation of $F(h)$ can explicitly be computed, and the entropy results
\beq
S(\bmg) = - \sum_{i\in V} \left( \frac{1+m_i}{2} \log  \frac{1+m_i}{2}   + \frac{1-m_i}{2} \log  \frac{1-m_i}{2} \right)
\eeq
while by differentiation one finds
\begin{eqnarray}
h_i^\star&=& \arctanh m_i \\
\chi^{-1}_{i,j} &=& \frac{\delta_{i,j} }{1-m_i^2} \label{eq:GenSuscIndepSpin}
\end{eqnarray}
The additivity both of the entropy and of the free energy, which are crucial in order to solve the model, descend directly by the independence of $p( s)$, which can be written as a product of single spin marginals
\beq
p( s) =\prod_{i\in V} p^{\{ i\}} (s_i) \; .
\eeq
Notice that the existence of the solution is guaranteed for any $\bmg$ in the hypercube $[-1,1]^N$, while its uniqueness is enforced by the minimality of the operator set $\{ s_i \}_{i=1}^N$ (which is additionally an orthogonal set in the sense that will be defined in (\ref{eq:Orthogonality})). As expected, for $m_i = \pm 1$, the estimator $h^\star_i$ is divergent, so that $h_i^\star (m_i = \pm 1)= \pm \infty$. 

\subsubsection{The pairwise model\label{sec:IsingModel}}
The next model that will be presented is known in a large variety of fields with different names (Ising model in physics, graphical model in the field of statistical learning), and is defined by the probability density
\beq
p( s) = \frac{1}{Z(\bh,\hat \bJ)} \exp \left( \sum_{i\in V} h_i s_i + \sum_{(i,j) \in E} J_{ij} s_i s_j \right) \; , \label{eq:IsingModel}
\eeq
where $E$ is a given set of \emph{edges}, that is, a given subset of $\{ (i,j) \in V  \times V\; | \; i < j\}$.
While in statistical mechanics it has been extensively used since 1925 as a prototypical model to study magnetic materials \cite{Ising:1925kx,Baxter:1982vn}, it has deserved a special interest in the field of statistical learning as it is the simplest model which is able to capture the correlation structure of a given dataset\footnote{This can be shown via the maximum entropy principle, which is presented and thoroughly commented in appendix \ref{app:MaxEntPr}.}.
The operator content of this model is a set of $N$ magnetizations, conjugated to their corresponding external fields (as in section \ref{sec:IndepSpins}), and a set of $|E| \leq \frac{N(N-1)}{2}$ operators $\{ \phi_{\{i,j\}}(s_i,s_j) = s_i s_j \}_{(i,j) \in E}$ conjugated to a set of \emph{pairwise couplings} $g_{\{i,j\}} = J_{ij}$. We will call \emph{empirical correlations} the averages $\overline{s_i s_j} = c_{ij}$.

\rmk{This direct problem for the pairwise model is \emph{hard} to solve in the general case for even moderate values of $N$, in the sense that the calculation of the partition function $Z(\bg)$ is a problem which is known to belong to the \#P-complete class \cite{Jerrum:1990zr,Jaeger:1990ys}.
Only for some subclasses of this general problem an exact, analytical solution for the partition function can be obtained (e.g., regular lattices, trees) and evaluated in polynomial time, while in general just approximate solutions can be obtained in polynomial time \cite{Jaeger:1990ys}. Another possible approach consists in finding approximated  expressions for the partition function $Z(\bh,\hat \bJ)$ which are proven to converge in the limit of large system size or weak interaction to the exact result for the free energy of the model (mean-field approximations).
}

In the next sections we will introduce specific versions of model (\ref{eq:IsingModel}) for which we will be able to solve the inverse problem, namely the fully connected ferromagnet (section \ref{sec:FCFerromagnet}) and the pairwise tree (section \ref{sec:InvPairTree}).

\section{The regularized inverse problem \label{sec:Regularization}}
The inverse problem described in section \ref{sec:InvProblem} may appear extremely easy to solve due to the concavity of the free energy $F(\bg)$. The optimization of concave functions is usually very easy because fast algorithms such as gradient ascents can find in short time a maximizer (if any) for $F(\bg)$ (appendix \ref{app:ConvexOpt}). Despite that, there are several cases in which this procedure may be problematic, so that the function $F(\bg)$ is often replaced by a modified function $F(\bg) - H_0(\bg)$ which enforces a better behavior for the inverse problem. In this case the function $H_0(\bg)$ is called a regularizer. In a Bayesian setting, regularization can be understood as an injection of \emph{a priori} information about a statistical model. Indeed the issue of regularization is a topic of fundamental importance in the field of statistical inference well beyond the need of enforcing mathematical tractability of the model.  In particular it can be used to deal with these cases:
\begin{itemize}
\item{
{\bf Divergencies:} Regularization can cure divergencies, by removing infinite couplings. A solution to any inverse problem is guaranteed to exist for any set of empirical averages $\bar \bphi \in \mathcal G(\bphi)$, but such solution may be located at the boundary of the coupling space, in whose case one or more couplings are divergent. Penalizing large couplings with a regularizer ensures that the inferred couplings attain a finite value. This is often the case for neurobiological or protein data and can be related to undersampling, as motivated in sections \ref{sec:CompleteInverseProblem}, \ref{sec:InvPairTree} and \ref{sec:Inv1DChain}  \cite{Cocco:2011fk,Schneidman:2006vg,Weigt:2009on,Cocco:2009mb,Cocco:2012uq}.
}
\item{
{\bf Uniqueness:} Regularization can enforce uniqueness for the solution of the inverse problem, by removing the zero models of the $\hat \bchi$ matrix. Such modes can arise if the family $\bphi$ is not minimal (appendix \ref{app:FreeEnConc}), or can be linked to the large $N$ limit (chapter \ref{ch:HighDimInf}).
}
\item{
{\bf Generalizability:} Regularization can be used to improve generalizability of a statistical model in the case of under sampling: if the inferred probability has a much smaller entropy with respect to the true one, an inferred model is likely not to be predictive. A compromise between faithfulness to the data and simplicity of the model can nevertheless be achieved by penalizing the complexity of the model with a regularization term, which is expected to lift the entropy of the inferred model. The balance between over and under fitting can be heuristically evaluated  by using cross-validation methods (e.g.,  by using one half of the data to calibrate the model and by computing the likelihood of the other half) or by using a complexity measure for the inferred model (such as the Akaike information criterium \cite{Akaike:1973uq} or the Bayesian information criterion \cite{Schwarz:1978kx}), in order to tune the regularizer to a correct value (see also section \ref{sec:Complexity}).
}
\item{
{\bf Model selection:} Finally, regularization can be used as a tool to perform model selection. In the case in which data are distributed according to a specific, unknown, statistical model, it is possible to perform inference by using a more general distribution which is likely to contain (or to be very close) to the true one. By adding a suitable regularizing term (such as an L-1 or L-0 norm) it is sometimes possible to recover the original model as a particular sub-class of a more general distribution. For example, this has been used in the context of graph reconstruction, where models defined by specific topologies have been successfully selected by a regularizer out of the space of all possible graph structures \cite{Ravikumar:2010ys,Wainwright:2007zr}.
}
\end{itemize}

\subsection{Bayesian formulation}
Consider an empirical dataset $\hat \bs$ and a model defined by a set of operators $\bphi$. Then the posterior of the model can be written as in (\ref{eq:BayesTheor}), in which it is $P_T( \bg | \hat \bs) \propto P_T(\hat \bs |  \bg) P_0( \bg)$, so that the problem of inference can be reformulated as the minimization of the function
\beq
H(\bg | \hat \bs) = - \log P_T(\hat \bs |  \bg) - \log P_0(\bg) = -T \sum_{\mu=0}^M g_\mu \bar \phi_\mu - \log P_0(\bg) \; . \label{eq:LogPosterior} 
\eeq
\defin{
Given a statistical model $(\bphi,\bg)$ and a positive prior function $P_0(\bg)$ we define a \emph{regularizer} as the function $H_0(\bg) = - \log P_0(\bg)$.
}
Notice that due to convexity of the $\hat \bchi$ matrix, if the regularizer $H_0(\bg)$ is (strictly) convex, also $H(\bg | \hat \bs)$ is (strictly) convex. Hence, the introduction of a strictly convex prior can be used to remove zero modes from the $\hat \bchi$ matrix thus enforcing a unique solution for the inverse problem. In our analysis we will restrict to the case of convex regularizers. Also notice that if $H_0(\bg) = +\infty$ when any component of $\bg^\star$ is divergent, the solution of the inverse problem is confined to a finite region of the coupling space.

\subsection{Two popular regularizers}
We present two known regularization schemes, with the purpose of providing simple examples of convex regularizers, while showing at the same time two widely used regularization mechanisms. The details about the properties and the implementation of the algorithms used to solve these regularized problems are reminded in appendix \ref{app:ConvexOpt}.

\subsubsection{L-2 regularization \label{sec:L2Regular}}
Given a statistical model $(\bphi, \bg)$, a set of empirical frequencies $\bar \bphi$ and a vector $\bm{\beta}$ such that it is component-wise $\beta_\mu > 0$, we consider the minimization problem
\beq
H(\bg) = -T \sum_{\mu=0}^M g_\mu \bar \phi_\mu + \sum_{\mu=1}^M \frac{\beta_\mu}{2} g_\mu^2 \, ,
\eeq
which we call the \emph{L-2 regularized} inverse problem. This choice for $H_0 (\bg)$ enforces strict concavity of the problem and finiteness of the values of $\bg^\star$, which should satisfy the set of equations
\beq
\bar \phi_\mu - \langle \phi_\mu \rangle - \frac{\beta_\mu}{T} \, g_\mu = 0\; .
\eeq
This regularization corresponds to the Gaussian prior $P_0(\bg) \propto \exp \left( - \sum_\mu \frac{\beta_\mu}{2} g_\mu^2 \right)$. Notice also that the regularizer is differentiable, so that a solution of this problem can be addressed efficiently by using techniques such as the ones described in the first part of appendix \ref{app:ConvexOpt}. As in the non-regularized case, the main computational limitation consists in calculating the gradient of the minus log-likelihood function $-\log P_T(\bar \bphi | \bg)$, which requires the knowledge of the averages $\langle \bphi \rangle$ as functions of the coupling vector $\bg$. This regularization procedure is typically used to remove infinite values arising in the solution of the non-regularized inverse problem.

\subsubsection{L-1 regularization \label{sec:L1Regular}}
We also present the \emph{L-1 regularized} inverse problem, which is defined by the minimization problem
\beq
H(\bg) = -T \sum_{\mu=0}^M g_\mu \bar \phi_\mu + \sum_{\mu=1}^M \beta_\mu |g_\mu| \, ,
\eeq
corresponding to the choice of an exponentially decaying prior $\exp \big( - \sum_\mu \beta_\mu |g_\mu| \big)$ $\propto P_0(\bg)  $ for the coupling vector $\bg$. Analogously to the L-2 case, this regularizer is convex and enforces a finite value for the inferred couplings $\bg^\star$. Unlike that, this regularizer is non-differentiable. This introduces some difficulties in the solution of the minimization problem, as shown in the second part of appendix \ref{app:ConvexOpt}, where it is shown that the inferred coupling vector should satisfy the equation
\beq
0 \in \bar \phi_\mu - \langle \phi_\mu \rangle - \beta_\mu \, \textrm{sgn} (g_\mu) \; ,
\eeq
where $\textrm{sgn} (x)$ is the set-valued function defined by equation (\ref{eq:SgnFunct}). The main interest in this regularizer arises from its efficacy as a feature-selector, as it is able to provide sparse solutions for $\bg^\star$, i.e., to put exactly to zero some components of the inferred couplings vector. Despite being first used in the field of \emph{compressed-sensing}, in which the use of the L-1 regularizer has been exploited to solve underconstrained sets of linear equations \cite{Donoho:2006ly}, this regularized has been successfully applied in the field of binary inference (also called \emph{logistic regression}), in which it has been useful to reconstruct the structure of an interaction network of the form (\ref{eq:IsingModel}) \cite{Ravikumar:2010ys,Wainwright:2007zr} and even in more general cases dealing with non-binary interaction \cite{Schmidt:2010hc}.

\rmk{The two regularizers presented so far are special cases of the L-$p$ regularization scheme, which is associated with the choice $H_0(\bg) \propto \sum_\mu \beta_\mu || g_\mu ||_p$, where $|| x ||_p = |x|^p$ is the L-$p$ norm of $x$. Notice that the L-$p$ regularizer is convex (hence leading to computationally tractable minimization problems) for $p \geq 1$, and is strictly so for $p>1$. In particular, the L-1 regularizer can be seen as the simple (\emph{alias}, convex) regularizer that is closer to the L-0 one, which is associated with the problem of minimizing the number of inferred parameters for a fixed value of the posterior, a criterium which one would think to use in order to minimize the complexity of the inferred model.
}

\subsection{Examples}
\subsubsection{Independent spins model}
\begin{figure}[h] %  figure placement: here, top, bottom, or page
   \centering
   \includegraphics[width=4in]{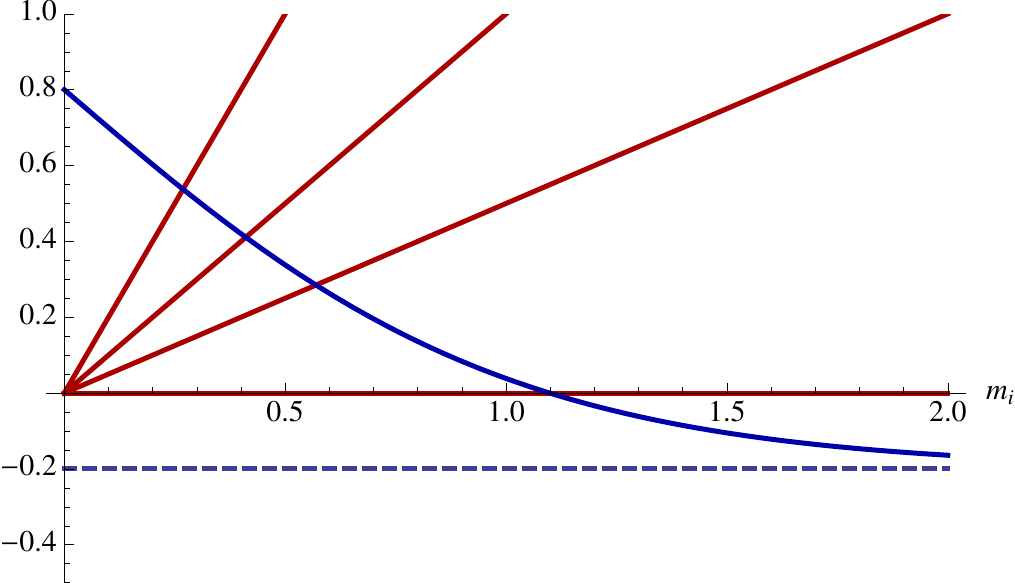} 
   \caption{Graphical solution of equation (\ref{eq:IndepSpinsL2Norm}) yielding the inferred field $h_i^\star$ for the L-2 regularized independent spin model. The blue curve displays the quantity $m_i - \tanh h_i$ in the case $m_i = 0.8$, while the red ones show the product $\beta_i h_i / T$ for $\beta_i/T = 0.5, 1 ,2$. The dashed line plotted for reference corresponds to the line $m_i -1$.}
   \label{fig:L2Solution}
\end{figure}

\begin{figure}[!t] %  figure placement: here, top, bottom, or page
   \centering
   \includegraphics[width=4in]{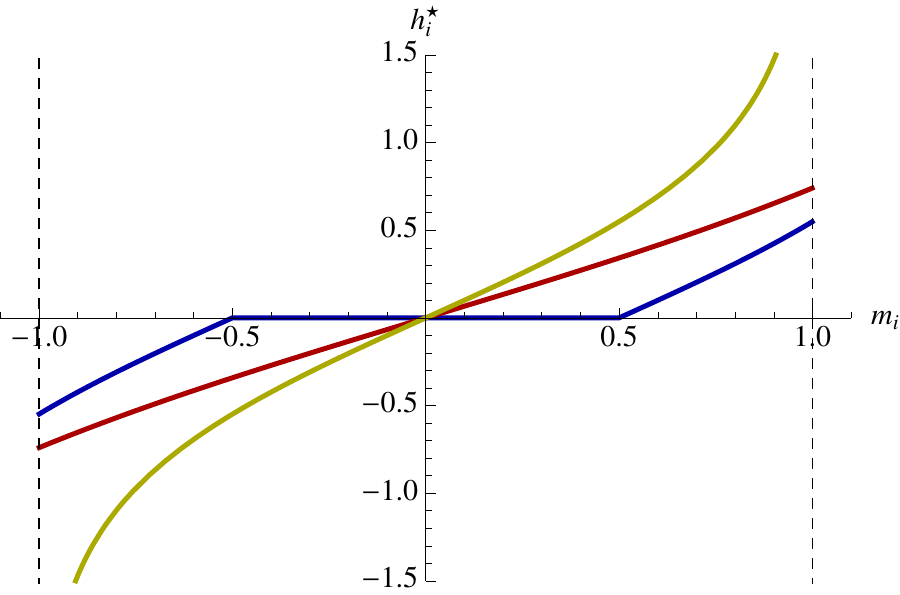} 
   \caption{Solution for the inferred field $h_i^\star$ for the L-1 (blue line) and L-2 (red line) regularized independent spin model as a function of the empirical magnetization for $\beta_i / T = 0.5$. The solution for the non-regularized problem is also plotted for comparison (yellow line).}
   \label{fig:L1vsL2}
\end{figure}

Consider the model defined by the probability density (\ref{eq:IndepSpins}). Then we can consider the regularized inverse problems in which one tries to minimize
\beq
H(\bh | \hat \bs) = -T \sum_{i \in V} \big(h_i m_i - N \log 2 - \log \cosh h_i \big) + H_0(\bh) \; ,
\eeq
in the two cases $H_0(\bh) = \sum_i \frac{\beta_i}{2} \, h_i^2$ and $H_0(\bh) = \sum_i \beta_i |h_i|$ corresponding respectively to the L-2 and L-1 norm.
In the first case the (decoupled) set of equations which has to be solved in order to find the vector $\bh$ is
\beq
m_i - \tanh (h_i) = \frac{\beta_i}{T} \, h_i \; , \label{eq:IndepSpinsL2Norm}
\eeq
whose graphical solution is depicted in figure \ref{fig:L2Solution}.
Such plot and equation (\ref{eq:IndepSpinsL2Norm}) also show that the inferred couplings $h_i$ attain a finite value for any of $-1 \leq m_i \leq 1$ and $0 \leq \beta_i < \infty$.
In the case of the L-1 norm, one has to solve the decoupled set of equations
\beq
m_i - \tanh h_i = \frac{\beta_i}{T} \, \textrm{sgn}(h_i) \; ,
\eeq 
whose solution is
\beq
h_i = \left\{ \begin{array}{ccc} 0  & \textrm{if} & \frac{\beta_i}{T} > |m_i| \\ \arctanh[m_i - \frac{\beta_i}{T} \, \textrm{sign} (m_i) ]  & \textrm{if} &  \frac{\beta_i}{T} \leq  |m_i| \end{array} \right. \; .
\eeq
The solution for $h_i$ in the two cases for a specific value of $\beta_i$ is plotted in figure \ref{fig:L1vsL2}.
Notice that:
\begin{itemize}
\item{
Both regularizations schemes produce a finite $h_i$ in the case $|m_i| = 1$.
}
\item{
The zero-field solutions of the L-1 regularized problem can be seen as arising from a complexity related criteria, stating that \emph{operators which do not add enough descriptive power to the model should be suppressed by the assignment of zero weight to their conjugated coupling}. In this example the notion of ``\emph{enough descriptive power}'' is quantified through the comparison of $\beta_i$ against the directional derivative of the log-likelihood $\partial_{h_i} \log P_T(m_i \gtrless 0| h_i) |_{h_i \to 0^{\pm}} = T m_i$.
}
\end{itemize}
Despite its trivial solution, we have chosen to present this problem as it shows with simplicity some basic features of the L-1 and L-2 regularizers which are retained even in more complicated scenarios.

\chapter{High-dimensional inference \label{ch:HighDimInf}}
Rigorous results in information theory -- such as the ones presented in section \ref{sec:InvSLDev} -- are able to provide both qualitative and quantitative understanding of the inverse problem in the regime of finite $N$ and large $T$, the case most of the literature on statistical learning deals with, while computational techniques such as the ones described in appendix \ref{app:ConvexOpt} provide efficient means to find its solution. Nevertheless, recent technological advances in several fields (such as biology, neuroscience, finance, economy) are pushing the fields of statistical learning towards a less trivial regime, in which both $N$ and $T$ are large, with a given relation among system size and number of samples keeping fixed their scaling. The reason for this change of perspective is that it is now possible for several complex systems to  record a large number of data samples describing simultaneous the activity of the many microscopic constituents \cite{Braunstein:2008sa,Schneidman:2006vg,Shlens:2006uq,Cocco:2009mb,Lillo:2008ht,Moro:2009wy,Lachapelle:2010kv}. The question that naturally arises in this case is whether it makes sense to consider a model with a large (possibly very large) number of parameters, if the data available is also very large. The answer is non-trivial, and requires the addition of some degree of complexity to the problem of inference. The first problem which has to be addressed (section \ref{sec:LimitationsHighN}) is of purely technical nature, and deals with the problem of finding the minimum of a convex function when its gradient is computationally intractable. Then, we will describe some interesting conceptual problems which arise when considering the large $N$ limit. For simplicity, we will consider initially the problem in which both $N$ and $T$ are large, but the number of inferred parameters $M$ is finite (section \ref{sec:LargeNFiniteM}). Discussing the case in which $M$ scales with $N$ as well will require the introduction of the notion of \emph{disorder}, which we will briefly comment about in section \ref{sec:LargeNLargeM}. 

\section{Computational limitations and approximate inference schemes \label{sec:LimitationsHighN}}
In appendix \ref{app:ConvexOpt} we show how it is possible to construct algorithms which are guaranteed to find a minimum (if any) for a convex function. Then the solution of the inverse problem can be written as a minimization problem over a convex function $H(\bg)$ of the form
\beq
H(\bg) = - \log P_T(\bar s | \bg) -\log P_0 (\bg) \; ,
\eeq
that problem is in principle solved. Indeed, the problem which often arises in many practical cases is that the naive minimization of this function can be extremely slow, and \emph{ad-hoc} techniques have to be implemented in order to overcome this problem.

\subsection{Boltzmann Learning}
One of the most intuitive algorithms to solve the inverse problem is provided by the \emph{Boltzmann learning} procedure \cite{Ackley:1985hc}, which consists in the application of algorithm \ref{app:GradDesc} to the inverse problem described in section \ref{sec:InvProblem}. In that case, the minimization procedure of $H(\bg)$ consists in constructing a succession $\{ \bg^{(k)}\}_{k=1}^K$ of the form
\beq
\bg^{(k+1)} = \bg^{(k)} - \epsilon_k \nabla H(\bg^{(k)})   \,
\eeq
where $\{\epsilon_k\}_{k=1}^K$ is a \emph{schedule} satisfying the set of conditions (\ref{eq:Armijo}) which enforce the convergence of $\bg^{(k)}$ to the minimum (if any) $\bg^\star$.
Indeed the computation of each of the $\bg^{(k)}$ requires the evaluation of $H(\bg^{(k)})$ and the calculation of a gradient of the form
\beq
\nabla H(\bg^{(k)}) = T \left( \langle \phi_\mu \rangle_{\bg^{(k)}} - \bar \phi_\mu \right) + \frac{\partial}{\partial g_\mu} H_0(\bg^{(k)})\; .
\eeq
The calculation of the gradient (or the sub-gradient) of $H(\bg)$ requires evaluating the ensemble averages of the operators $\bphi$, which is a computationally challenging task if $N$ is even moderately large. This is true even when the function $H$ and the ensemble averages $\langle \bphi \rangle$ are not computed via direct enumeration (which would in principle entail a summation over $2^N$ states for each of the $M$ operators plus the identity), and are instead calculated with Monte Carlo methods. The number of iterations required to calculate each of the gradients and the function $H$ with a controlled precision is in fact typically fast growing in $N$, being the quality of the approximation and the time computational power required to obtain it dependent on the algorithm which is adopted to compute the averages (see for example \cite{Ackley:1985hc,Mezard:2009ko,MacKay:2003vn,Krauth:1998fk}).
Summarizing:
\begin{itemize}
\item{
Boltzmann learning is able to solve with arbitrary precision any inverse problem.
}
\item{
The computational power required to solve the inverse problem through the Boltzmann learning procedure with a given degree of accuracy (i.e.\ $H(\bg^{(K)}) - H(\bg^\star)$ smaller than a fixed $\epsilon$) grows \emph{fast} in $N$.
}
\end{itemize}

\subsection{Mean field approaches for pairwise models \label{sec:MFApproach}}
An alternative approach to the Boltzmann learning procedure can be constructed by adopting so-called \emph{mean-field} techniques, which allow to obtain efficient approximations for the free energy $F(\bg)$ and the averages $\langle \bphi \rangle$ of a statistical model. Such techniques are suitable for systems whose partition function can be quickly, although approximately, evaluated with a precision which either increases with the system size $N$ or decreases with the magnitude of the interactions, so that in many practical applications the difference between the approximated observables and the exact ones is very small \cite{Roudi:2009li,Roudi:2009qm}.
For pairwise models of the form (\ref{eq:IsingModel}), mean-field approximations are well-known since long time in statistical physics. In particular we will consider approaches in which the free energy of the model (\ref{eq:IsingModel}) is expanded in a series around a non-interacting or a weakly correlated model (naive mean field, TAP approximation, Sessak-Monasson approximation), or obtained by assuming a factorization property of the probability distribution in terms of one and two body marginals (Bethe approximation). We will briefly describe these approximate inference schemes without providing explicit derivations, supplying the interested reader with the necessary references.\\
In order to motivate the mean-field approach, we first state the result \cite{Plefka:1982vn}.

\prop{Consider a pairwise model of the form
\beq
p(s) = \frac{1}{Z(\bh,\beta \hat \bJ)} \exp \left( \sum_{i \in V} h_i s_i + \beta \sum_{(i,j) \in E} J_{ij} s_i s_j \right) \; ,
\eeq
where $\beta > 0$ is an expansion parameter.
Then its free energy can be written as
\beq
F_\beta(\bh,\bJ) = \sum_{n=0}^\infty \beta^n \frac{\partial^n F_\beta}{\partial \beta^n}  \label{eq:Plefka}
\eeq
where the terms $\frac{\partial^n F_\beta}{\partial \beta^n} $ are functions such that: (i) depend only on the couplings $J_{ij}$ and the ensemble magnetizations $\langle s_i\rangle$ (ii) for $n \geq 1$ the $n$-th term involves $n$-th powers of $J_{ij}$ (iii) the ensemble magnetizations satisfy the self-consistency equations
\beq
\frac{\partial F_\beta(\bh , \beta \hat \bJ)}{\partial \langle s_i \rangle } = 0 \; .
\eeq.
}
Leaving aside the problem of convergence of the series (\ref{eq:Plefka}), the free energy for a generic pairwise model can in principle be obtained by setting $\beta=1$ in the above expansion.
\begin{itemize}
\item{{\bf Naive mean field:}
The naive mean field approximation can be obtained by truncating the series (\ref{eq:Plefka}) for $n=2$, thus obtaining the expression
\beqa
F_{nMF}(\bh,\bJ) &=& \sum_i \left[ \frac{1+\langle s_i\rangle}{2} \log \frac{1+\langle s_i\rangle}{2} +  \frac{1- \langle s_i\rangle}{2} \log \frac{1- \langle s_i\rangle}{2} \right] \nonumber \\
&-& \sum_{i \in V} h_i \langle s_i \rangle - \sum_{(i,j) \in E} J_{ij} \langle s_i \rangle \langle s_j \rangle \; , \label{eq:FreeEnergynMF}
\eeqa
while the self-consistency equations become
\beq
\langle s_i \rangle = \tanh \left( \sum_{(i,j) \in E} J_{ij} \langle s_j \rangle + h_i \right) \; .
\eeq
The solution of the inverse problem within this inference scheme can be obtained by inserting the momentum matching condition $\langle s_i \rangle = m_i$ in the previous expression, yielding a first set of relations among $\bh^\star$, $\hat \bJ^\star$ and $\bmg$. Matching the correlations $c_{ij}$ with the ensemble averages $\langle s_i s_j \rangle$ requires instead the use of linear response theory\footnote{Nor by using this inference scheme, nor by using TAP approximation one is able to enforce the momentum matching condition for the correlations without resorting to linear response. This is due to the decorrelation property of the mean-field approximation, which will be thoroughly commented for a simpler model in section \ref{sec:FCFerromagnet}.} \cite{Kappen:1998dt}, which can be used to to prove that
\beq
\chi_{\{i\},\{j\}} = \frac{\partial \langle s_i \rangle}{\partial h_j} = c_{ij} - m_i m_j \; .
\eeq
Putting those informations together, one finds that
\beqa
(\hat \bc - \bmg \bmg^T)^{-1}_{ij} &=&  \frac{\delta_{ij}}{1-m_i^2} - J_{ij}^\star \label{eq:MFInversionJ} \\
h_i^\star &=& \textrm{atanh} (m_i) - \sum_{i<j} J_{ij}^\star m_j \label{eq:MFInversionH}
\eeqa
}
\item{ {\bf TAP approximation:}
The Thouless-Anderson-Palmer (TAP) approximation can be obtained by considering an additional term in the expansion (\ref{eq:Plefka}), often denoted as \emph{Onsager reaction} \cite{Thouless:1977kx}, leading to the expression for the free energy
\beqa
F_{TAP}(\bh,\bJ) &=& \sum_i \left[ \frac{1+\langle s_i\rangle}{2} \log \frac{1+\langle s_i\rangle}{2} +  \frac{1- \langle s_i\rangle}{2} \log \frac{1- \langle s_i\rangle}{2} \right] \label{eq:FreeEnTAP}\\
&-& \sum_{i \in V} h_i \langle s_i \rangle - \sum_{(i,j) \in E} J_{ij} \langle s_i \rangle \langle s_j \rangle - \frac{1}{2} \sum_{(i,j) \in E} J_{ij}^2 (1-m_i^2)(1-m_j^2) \nonumber \; ,
\eeqa
and the self-consistency relation\footnote{Notice that the potential emergence of multiple solutions of equation (\ref{eq:TAPSelfCons}) is a known feature of several pairwise models, and is generally associated with the emergence of an instability linked with the presence of a glassy phase \cite{Almeida:1978uq}.}
\beq
\langle s_i \rangle = \tanh \left( \sum_{(i,j) \in E} J_{ij} \left[ \langle s_j \rangle - J_{ij} (1-\langle s_j \rangle^2) \langle s_i \rangle \right] + h_i \right) \; . \label{eq:TAPSelfCons}
\eeq
Also in this case, in order to apply this approximation to the inverse problem  \cite{Tanaka:1998tg}, one has to use the momentum matching conditions together with linear response theory, leading to the expression \cite{Ricci-Tersenghi:2011uq}
\beqa
(\hat \bc - \bmg \bmg^T)^{-1}_{ij} &=& \left[ \frac{1}{1-m_i^2} + \sum_{k\in V} J_{ik}^\star (1-m_k^2) \right] \delta_{ij} - J_{ij}^\star - 2 J_{ij}^{\star 2} m_i m_j \quad \phantom{.} \label{eq:TAPInversionJ}\\
h_i^\star &=& \textrm{atanh} (m_i) - \sum_{i < j } J_{ij}^\star \left[m_j - J_{ij}^\star (1- m_j^2) m_i \right] \; .  \label{eq:TAPInversionH}
\eeqa
}
\end{itemize}
While the expansion (\ref{eq:Plefka}) is a series for $F(\bh, \hat \bJ)$, and is hence associated with the direct problem, it is also possible to find an analogous expansion for the entropy $S(\bmg, \hat \bc)$ due to Sessak and Monasson which is more naturally associated with the inverse problem \cite{Sessak:2009lf}.
\prop{
Given a pairwise model of the form (\ref{eq:IsingModel}), the entropy $S(\bmg, \hat \bc)$ can be expanded as
\beq
S(\bmg, \beta \bm{\delta} \hat \bc ) \sum_{n=0}^\infty \beta^n \frac{\partial^n S(\bmg, \beta \bm{\delta} \hat \bc )}{\partial \beta^n} \label{eq:SMExpansion}
\eeq
where $\beta > 0$ is a parameter controlling the expansion and $\bm{\delta} \hat \bc = \hat \bc - \bmg \bmg^T $. One can see that (i) the terms $\frac{S(\bmg, \beta \bm{\delta} \hat \bc )}{\partial \beta^n}$ depend upon $\bmg$ and $\bm{\delta} \hat \bc$, (ii) for $n \geq 1$ the $n$-th term of the expansion contains powers of the connected correlation $c_{ij} - m_i m_j$ of order $n$.
}
By setting $\beta = 1$, it is also possible to use such an expansion to construct a mean field approximation: the terms in (\ref{eq:SMExpansion}) can be constructed explicitly through a recursion relation, and each of those can be represented by a diagram, converting the series (\ref{eq:SMExpansion}) into a diagrammatic expansion.
\begin{itemize}
\item{ {\bf Sessak-Monasson expansion}
An infinite number of terms of the expansion (\ref{eq:SMExpansion}) (which are associated with \emph{loop} diagrams and \emph{two-spin} diagrams) are analytically resumed in \cite{Sessak:2009lf}, where it is found that their contribution leads to
\beqa
J^\star_{ij}  &=&   \delta_{ij}(1-m_i^2)  - (\hat \bc - \bmg \bmg^T)^{-1}_{ij} \nonumber \\
&+& \frac{1}{4} \log \left[ \frac{(1+m_i + m_j + c_{ij})(1-m_i - m_j + c_{ij})}{(1+m_i - m_j - c_{ij})(1-m_i + m_j - c_{ij})} \right]  \\
&-& \frac{c_{ij} - m_i m_j}{(1-m_i^2)(1-m_j^2) - (c_{ij}-m_i m_j)^2} \nonumber \; ,
\eeqa
which is commonly referred as the Sessak-Monasson approximation.
}
\end{itemize}
Notice that the expansion (\ref{eq:SMExpansion}) automatically leads to a series expansion for the external fields and the couplings by using relation (\ref{eq:GenFuncCoupling}) and exploiting the linearity of the derivative, without the need of resorting to linear response theory. \\
A different type of approximation is the so-called Bethe approximation, in which the free energy is written as
\beqa
F_{BA}(\bh, \bJ) = &-& \sum_{(i,j) \in E} p^{\{ i,j \}}(m_i,m_j, c_{ij}) \log p^{\{ i,j \}}(\langle s_i \rangle,\langle s_j \rangle, \langle s_i s_j \rangle)  \nonumber \\
&-& \sum_{i \in V} (1- |\partial i|) p^{\{ i\}}(\langle s_i \rangle) \log p^{\{ i \}}(\langle s_i \rangle) \label{eq:BetheApprox} \\ 
&-& \sum_{i\in V} h_i \langle s_i \rangle - \sum_{(i,j) \in E} J_{ij} \langle s_i s_j \rangle \; , \nonumber
\eeqa
where $\partial i =\{(i,j) \in E \}$ and the averages $\langle s_i \rangle $ and $\langle s_i s_j \rangle$ are self-consistently chosen in order to minimize (\ref{eq:BetheApprox}). This approximate expression is exact whenever the probability distribution $p(s)$ can be written as a product of one and two body marginals, which is true in the case of trees (see section \ref{sec:InvPairTree} and appendix \ref{app:TreeFactorization}). Notice that for generic systems, the self-consistence equations are not guaranteed to yield a unique, stable solution, being the solutions to the minimization conditions associated with fixed points of the so-called Belief-Propagation (BP) algorithm for constraint satisfaction problems \cite{Mezard:2009ko}. The expression for the averages obtained by using the free-energy (\ref{eq:BetheApprox}) is given by \cite{Ricci-Tersenghi:2011uq}
\beq
\langle s_i \rangle = \tanh \left[ h_i + \sum_{j | (i,j) \in \partial i} \textrm{atanh} \, \left[ \tanh ( J_{ij}) f(\langle s_i \rangle, \langle s_j \rangle , \tanh J_{ij} )  \right] \right] \; ,
\eeq
where
\beq
f(m_1,m_2,t) = \frac{1-t^2 - \sqrt{(1-t^2)^2 -4t(m_1 -m_2 t )(m_2 -m_1 t)}}{2 t(m_2 - m_1 t)} \; . \label{eq:SelfConsistencyBA}
\eeq
\begin{itemize}
\item{{\bf Bethe approximation}
The use of linear response theory together with equation (\ref{eq:SelfConsistencyBA}) allows to find a solution of the inverse problem in Bethe approximation, yielding
\beqa
J_{ij}&=& \textrm{atanh} \Bigg[ m_i m_j - \frac{1}{2\left(\widehat{ \bm{\delta} \bc}^{-1}\right)_{ij}} \sqrt{1+ 4 (1-m_i^2)(1-m_j^2)\left(\widehat{ \bm{\delta} \bc}^{-1}\right)_{ij}^2} \\
&+& \frac{1}{\left(\widehat{ \bm{\delta} \bc}^{-1}\right)_{ij}}  \Bigg(  \frac{1}{4} - m_i m_j\left(\widehat{ \bm{\delta} \bc}^{-1}\right)_{ij}  \sqrt{1+ 4 (1-m_i^2)(1-m_j^2)\left(\widehat{ \bm{\delta} \bc}^{-1}\right)_{ij}^2}  \nonumber \\
&+&  (2 m_i^2 m_j^2 - m_i^2 - m_j^2 ) \left(\widehat{ \bm{\delta} \bc}^{-1}\right)_{ij}^2 \Bigg)^{1/2} \Bigg] \nonumber \\
h_i &=&\textrm{atanh} \, (m_i)   - \sum_{j \in V} \textrm{atanh} \, \left[ \tanh ( J_{ij}) f(m_i, m_j , \tanh (J_{ij}) )  \right] \quad \quad \quad \phantom{.}
\eeqa
where
$\widehat{ \bm{\delta} \bc }= \hat \bc - \bmg \bmg^T$. Notice that this equation describes the fixed point solution of the susceptibility propagation algorithm (SuscProp) \cite{Mezard:2009ul} without the need of numerically iterating the algorithm itself \cite{Ricci-Tersenghi:2011uq}.
}
\end{itemize}
\rmk{
The techniques described above have been extensively used in order to solve the inverse problem for the pairwise model. Indeed no general result for the quality of these approximations is rigorously known, thus it is worth remarking that (i) several approximations have been tested on synthetic and experimental data (see for example \cite{Cocco:2009mb,Roudi:2009li,Roudi:2009qm,Marinari:2010vn,Ricci-Tersenghi:2011uq,Aurell:2012ys,Cocco:2012uq}) in order to check their performance and (ii) those approximations describe the correct expression of the free energy for some specific models. In particular the free energy (\ref{eq:FreeEnergynMF}) is the exact free energy for the (either homogeneous or heterogeneous) Curie-Weiss model in the limit of large $N$, (\ref{eq:FreeEnTAP}) is the correct free energy for the Sherrington-Kirkpatrick model \cite{Plefka:1982vn} and the Bethe approximation is exact for loop-less graphs (appendix \ref{app:TreeFactorization}).
}

\section{The large $N$, finite $M$ regime \label{sec:LargeNFiniteM}}
We will be interested in sketching some features of the inverse problem which arise for large values of $N$ (a regime known in statistical mechanics as the \emph{thermodynamic limit}), and in commenting about their role in the solution of an inference problem such as the one described in section \ref{sec:InvProblem}. In particular we will consider the following issues:
\begin{itemize}
\item{
{\bf Loss of concavity:} A model defined by a strictly concave free-energy $F(\bg)$ may develop null-modes associated with the matrix $\hat \bchi$. This implies that the solution of the inverse problem may lose its uniqueness or, more precisely, large regions of the space $\mathcal M(\bphi)$ might be associated with similar sets of empirical averages $\bar \bphi$.
}
\item{
{\bf Model condensation:} Models undergoing a so-called \emph{second order phase transition} display a divergence of one or more components of the generalized susceptibility matrix $\hat \bchi$. This indicates that large portions of the marginal polytope $\mathcal G(\bphi)$ can be described by slightly shifting the values of $\bg$ around the \emph{critical point} in which $\hat \bchi$ diverges. More generally, even for non-critical points finite regions of the space of the empirical averages can be mapped by the inverse problem onto sets of apparently vanishing measure of the space $\mathcal M(\bphi)$. We call this behavior \emph{model condensation}, a phenomenon which will be discussed in great detail in chapter \ref{ch:Geometry}.
}
\item{
{\bf Ergodicity breaking:} The probability measure $\bp$ may break in a set of $P$ \emph{states}, each of them characterized by a different probability density $\bp^{(\alpha)}$ (with $\bp = \sum_{\alpha=1}^P q_\alpha \bp^{(\alpha)}$ and $\sum_{\alpha=1}^P q_\alpha$ = 1). If this is the case, empirical averages produced with a finite amount of data $T \ll |\Omega|$  by any realistic dynamics concentrate according to the measure $\bp^{(\alpha)}$ rather than the full measure $\bp$. Then, equation (\ref{eq:MeanEmpAvg}) fails to hold and the sampled averages are no longer representative of the global probability measure. Hence, the notion of ergodicity breaking deals with the direct problem more than with the inverse one, as it relates to the problem of the convergence of the averages $\bar \bphi$ to the empirical ensemble averages $\langle \bphi \rangle$. As the discussion of this phenomenon will require the addition of some structure to the direct problem, we will briefly comment its role in section \ref{sec:ErgBreak}.}
\end{itemize}
Those features are expected to be universal, i.e., present in several models in the limit $N \to \infty$ limit. Nevertheless, we will just study a single model known as the fully connected ferromagnet, and try to underline the characteristics which are expected to generalize also to other type of models.

\section{Fully-connected ferromagnet \label{sec:FCFerromagnet}}
We want to illustrate some of the features described above by discussing a completely solvable model. Such model is a particular case of the pairwise model (\ref{eq:IsingModel}), and is also known as the Curie-Weiss model of magnetism. It has been used as a prototypical model to study the emergence of a spontaneous magnetization in ferromagnetic materials, as it is one of the simplest statistical models which are able to describe a thermodynamic \emph{phase transition} between a non-ordered phase and an ordered one.

\defin{Consider the pair of operators $\bphi = \left( \sum_i s_i , \frac{1}{N} \sum_{i<j} s_i s_j\right)$, and the statistical model $(\bphi,\bg)$ defined by $\bg=(h,J)$, so that its associated probability density is given by
\beq
p(s) = \frac{1}{Z(h,J)} \exp \left(\frac{J}{N} \sum_{i<j} s_i s_j + h \sum_i s_i \right) \; . \label{eq:FCFerromagnet}
\eeq
We call this model a \emph{fully connected ferromagnet}. As for the pairwise model, we will write  $m = \frac{1}{N} \sum_i \bar s_i$ and $c = \frac{2}{N(N-1)} \sum_{i<j} \overline{s_i s_j} $.
}
Due to symmetry, we will consider without loss of generality the model in the region $h \geq 0$. The free energy of the model $F(h,J)$ can be calculated in the large $N$ limit using a saddle-point approximation, and can be written as
\beq
F(h,J) \xrightarrow[N\to \infty]{}  \frac{J}{2}  + F_{0}(h,J) + F_{fluct}(h,J) + F_{trans}(h,J) \; , \label{eq:SaddlePointFCFerro}
\eeq
where $F_{0}(h,J)$ is the leading term of the saddle point expansion, $F_{fluct}(h,J)$ describes the Gaussian fluctuations around the saddle point solution and $F_{trans}(h,J)$ accounts for the presence of multiple solutions (the details of the expansion and the definition of the terms can be found in appendix \ref{app:SPFerromagnet}). Due to linearity of the derivative, it is possible to solve the direct problem taking into account the contributions of those terms separately.
The phenomenology of the model is well-known, and can be roughly described keeping into account only the term $F_0(h,J)$. In particular for low values of $J$ the direct problem has only one stable solution (\emph{paramagnetic phase}), while for high values of $J$ two stable solutions for the empirical averages emerge (\emph{ferromagnetic} phase). In the case $h=0$ the two regimes are separated by a phase transition in which the fluctuations of the average magnetization diverge.

\subsection{The mean-field solution}
The solution of the direct problem considering only $F_0(h,J)$ will be called \emph{mean-field} solution. Notice that due to the scaling $F_0(h,J) \propto N$, for large values of $N$ this contribution dominates the free energy $F(h,J)$.
\prop{For all $i \neq j$ the mean-field solution for the fully connected ferromagnet is :
\beqa
\left< \sum_i s_i \right>_0 &=& N \, m_{s.p.}(h,J)  \label{eq:MFFerroAvg} \\
\left< \frac{1}{N} \sum_{i<j} s_i s_j\right>_0 &=& N \, \frac{m_{s.p.}^2 (h,J)}{2} \, \label{eq:MFFerroCorr}
\eeqa
while the susceptibility matrix is given by
\beq
\chi_{0} = N \, \chi_{s.p.} \left( \begin{array}{cc} 1 & m_{s.p.} \\ m_{s.p.} & m^2_{s.p.} \end{array} \right) \; ,
\eeq
where $m_{s.p.}$ is the absolute minimum of the function $f_{h,J}(m) = \frac{1+m}{2} \log \frac{1+m}{2} + \frac{1-m}{2} \log \frac{1-m}{2} - \frac{J m^2}{2} - hm$
and $\chi_{s.p.} = \partial m_{s.p.} / \partial h$.
}
\rmk{
It is easy to check that the mean-field solution describes independent spins. In fact equations (\ref{eq:MFFerroAvg}) and (\ref{eq:MFFerroCorr}) imply that for large $N$ and $i\neq j$
\beq
\langle s_i \rangle^2 = \langle s_i s_j \rangle \; .
\eeq
}
This fact is a consequence of the pathological behavior of the mean-field solution of this model. In particular this implies that the inverse problem has a solution just along the line $ (m,c) = (m,m^2)$, while it is easy to see(appendix \ref{app:FCMargPoly}) that for a generic distribution $\bar \bp \in \mathcal M(\Omega)$ the set of all possible empirical averages (i.e., the marginal polytope associated with the fully connected ferromagnet) is
\beq
\mathcal G (\bphi) = \Bigg\{ (m,c) \in \mathbb{R}^2 \Bigg| m \in [-1,1] \wedge c \in \left[\frac{m^2 - 1/N}{1-1/N} , 1 \right]\Bigg\}
\eeq
This implies the following fact concerning the inverse problem.
\prop{
The inverse problem for the fully connected ferromagnet has a mean-field solution if and only if $(m,c) = (m,m^2)$. In that case, the entropy is given by
\beq
S(m,m^2) = N \left( \frac{1+m}{2} \log \frac{1+m}{2} +  \frac{1-m}{2} \log \frac{1-m}{2}  \right)
\eeq
while the couplings belong to the space
\beqa
h^\star &=& \arctanh  m - \delta J \, m \\
J^\star &=&  \delta J
\eeqa
restricted to the region in which $\textrm{sign}(m) = \textrm{sign} (h)$. Finally, the inverse susceptibility matrix is divergent.
}
This last fact can be understood by checking that the matrix $\chi_0$ has eigenvalue decomposition $N \left(0, \frac{1-m_{s.p.}^4}{1-J + J m_{s.p.}^2}\right)$. In particular, the null eigenvalue has eigenvector $(-m,1)$ which  indicates that the mean field solution of the direct problem is invariant under the change of couplings
\beq
(h,J) \rightarrow (h-\delta J m_{s.p.}, J + \delta J) \; .
\eeq
Thus, the inverse problem maps all the points belonging to the one-dimensional region $(m,m^2)$ on the two-dimensional plane $(h,J)$.
This apparently contradicts the remark in section \ref{sec:InvProblem} about the existence of solutions to the inverse problem for any point belonging to the marginal polytope $\mathcal G (\bphi)$. Indeed, we will show in the next section that keeping properly into account the presence of the $h=0, J>1$ line allows to understand this discrepancy. Interestingly, the two-dimensional region $\mathcal G (\bphi) \backslash \{ (Nm ,\frac{N-1}{2} m^2) \; |\;  m \in [-1,1]\}$ is mapped on such one-dimensional line.

\subsection{Finite $N$ corrections}
Keeping into account the terms $F_{fluct}$ and $F_{trans}$ allows to describe the transition from the finite $N$ regime to the mean-field one. In particular, the Gaussian fluctuations around the mean-field solution extend the region in which the inverse problem is solvable to a strip of finite width in the space $\mathcal G (\bphi)$.
\prop{
Given $(Nm,\frac{N-1}{2} c) \in \mathcal G (\bphi)$, the inverse problem for a fully connected ferromagnet described by the terms $F_0$ and $F_{fluct}$ of equation (\ref{eq:SaddlePointFCFerro}) has solution if and only if $c = m^2 + \frac{\delta c}{N}$ with $\delta c$ finite, and reads\footnote{In the literature concerning the so-called inverse Ising model, this result is typically derived by differentiating the relation
\beq
\arctanh m_i = h_i + \frac{1}{N}\sum_{k} J_{ik} m_k
\eeq
with respect to $m_j$, and by recognizing that through linear response one can write $( \partial h / \partial m)_{ij}^{-1} = c_{ij} - m_i m_j \approx \delta c_{ij} / N$ \cite{Kappen:1998dt,Roudi:2009li}.
}
\beqa
h &=& \arctanh m - J m \label{eq:FCFerroInfField} \\
J &=& \frac{\delta c}{(1-m^2)(1-m^2 + \delta c)} \label{eq:FCFerroInfCoup} \; .
\eeqa
}
\prf{
This can easily be proved by keeping into account the contributions to the averages $\langle \dots \rangle_0$, $\langle \dots \rangle_{fluct}$ shown in appendix  \ref{app:SPFerromagnet} and imposing $m_{s.p.} = m + \delta m / N $, $c = m^2 + \delta c / N$ in the momentum matching condition.
}The null eigenvalue of the matrix $\hat \bchi_0$ is lifted to a finite value, as one can see that
\beq
\det \big(\hat \bchi_0 + \hat \bchi_{fluct}\big) = N \, \frac{\chi_{s.p.}^3}{2} >0 \; ,
\eeq
and is of order $N$ (instead of $N^2$ as could be expected on the basis of the scaling of the leading term $\chi_0$).
Summarizing, data with small connected correlations (i.e., $c-m^2 \sim 1/N$) are described by a fully connected model with finite $h$.
Conversely, it must hold that the whole space $\mathcal G (\bphi)$ stripped of the quasi-one dimensional region $(Nm,\frac{2}{N-1} m^2 + \delta c)$ is mapped on the region of the $(J,h)$ plane in which $J>1$ and $h \sim 1/N$. To show this, we consider the approximation in which the only relevant terms of the free energy $F(h,J)$ are $F(h,J) = F_0(h,J) + F_{trans}(h,J)$.
\prop{
The inverse problem for the fully connected ferromagnet described by the terms $F_0  + F_{trans}$ has solution for any point $(m,c) \in \mathcal G(\bphi)$ excluding the region $c-m^2 \sim 1/N$. The points $(h^\star, J^\star)$ satisfy the equations
\beqa
m &=& m_{s.p.} - (h \chi_{s.p.} + m_{s.p.} ) [1-\tanh (N h m_{s.p.})]  \\
c &=& m_{s.p.}^2 + h m_{s.p.} \chi_{s.p.} [1-\tanh (N h m_{s.p.})] \\
m_{s.p.} &=& \tanh (J m_{s.p.} + h) \; .
\eeqa
}
Also in this case one can show that in the limit $h \ll N$, the null mode of $\hat \bchi_0$ is lifted due to
\beq
\det (\hat \bchi_0 + \hat \bchi_{trans}) \xrightarrow[N \to \infty]{} N^3 \chi_{s.p.} m_{s.p.}^4 \textrm{sech}(h N m_{s.p.}) \; .
\eeq
Finally, one can draw the following conclusion, which despite being a trivial consequence of what shown above, shows that the $N \to \infty$ limit can lead to counter-intuitive results.
\rmk{
Consider the solution of the inverse problem for a fully connected ferromagnet and a point $\bar \bphi = (N m, \frac{N-1}{2} c)$ drawn from the space of empirical averages $\mathcal G (\bphi)$ with uniform measure. Then for any $\epsilon > 0$, $J^\star(\bar \bphi) > 1$ and $h^\star (\bar \bphi) \in [-\epsilon, \epsilon]$ with probability $P \xrightarrow[N\to\infty]{} 1$.
}
This simple example shows some of the features discussed above concerning the limit of large $N$, namely:
\begin{enumerate}
\item{
The free energy loses (strict) concavity, as one has $\det \hat \bchi \xrightarrow[N\to\infty]{}\det \hat \bchi_0 = 0$. This indicates that some directions in the coupling space cannot be discriminated. In this example, when $N$ is large, interactions are no longer distinguishable from external fields due to the presence of an eigenvector $(-m,1)$ associated with the null eigenvalue.
}
\item{
Model condensation takes place, as all the region $\mathcal G (\bphi)$ but a set of null measure is mapped on a one-dimensional strip. This will be better elucidated in chapter \ref{ch:Geometry}, where we will be able to quantify the density of models contained in a finite region of the space $(h,J)$.
}
\end{enumerate} 

\section{Saddle-point approach to mean-field systems\label{sec:ErgBreak}}
In this section we generalize the procedure employed in the case of the fully connected ferromagnet to the case in which a saddle-point approach is used to solve the direct problem for a generic system.
In particular, we consider a statistical model $(\bphi,\bg)$ with partition function
\beq
Z(\bg) = \sum_{s\in \Omega} \exp \left( \sum_{\mu=1}^M g_\mu \phi_{\mu,s} \right) \; ,
\eeq
and suppose that the operators $\phi_{\mu,s}$ can be written as functions of a small set of parameters $\bpsi (s)= (\psi_1(s),\dots,\psi_A(s))$, so that for any $\mu$ one has $\phi_\mu (s) = \phi_\mu [\bpsi(s)]$. Then it is possible to write
\beqa
Z(\bg) &=& \int d\bpsi \sum_s \exp \left(  \sum_{\mu=1}^M g_\mu \phi_\mu (\bpsi) \right) \delta \left( \bpsi - \bpsi (s) \right) \nonumber \\
&=& \int d\bpsi \exp \left(  \sum_{\mu=1}^M g_\mu \phi_\mu (\bpsi) + \Sigma (\bpsi) \right)  \; ,
\eeqa
where $e^{\Sigma (\bpsi)} = \sum_s \delta \left( \bpsi - \bpsi (s) \right)$, and $\Sigma(\bpsi)$ is often referred as \emph{entropy} for the value of the order parameter $\bpsi$. For many statistical models, one has that the limit
\beq
f(\bpsi) = \lim_{N\to \infty} \frac{1}{N} \left( - \sum_{\mu=1}^M g_\mu \phi_\mu (\bpsi)  - \Sigma(\bpsi) \right)
\eeq
is finite, and $f(\bpsi)$ is often called (intensive) \emph{free-energy} for the value of the order parameter $\bpsi$. In this case, one can exploit a saddle-point approximation to evaluate the partition function $Z(\bg)$ at large $N$. It results
\beq
Z(\bg) \xrightarrow[N\to\infty]{} e^{-N f (\bpsi^\star)} \sqrt{\frac{2 \pi}{N \det \left( f^{(2)}(\bpsi^\star) \right) }} \; . \label{eq:SPPartFunc}
\eeq 
where we use the notation $f^{(n)}(\bpsi)$ for the tensor with components $f^{(n)}_{a_1,\dots, a_n} = \partial_{\psi_{a_1}} \dots \partial_{\psi_{a_n}} f (\bpsi)$, and $\bpsi^\star$ is the global minimum of the function $f(\bpsi)$, which in particular satisfies
\beq
\frac{\partial}{\partial \psi_a} f(\bpsi) = 0 \; . \label{eq:SPEquations}
\eeq
Besides providing us with a mean to calculate the free energy $F(\bg)$, the ensemble averages $\langle\bphi\rangle$ and the susceptibilities $\hat \bchi$, the notions defined above allow us to introduce the concept of \emph{state}, which we will use to characterize the phenomenon of ergodicity breaking.

\defin{
Consider a statistical model $(\bphi,\bg)$ which can be described by a set of order parameters $\bpsi$, and such that at large $N$ its partition function can be approximated by (\ref{eq:SPPartFunc}). Then we call a \emph{state} any local minima of the saddle-point equations (\ref{eq:SPEquations}).
}
We will label any of those minima as $\bpsi^{(\alpha)}$ with $\alpha=1,\dots,P$, and use the a superscript $\alpha$ to identify quantities associated with the state $\alpha$, as for example
\beq
F^{(\alpha)}(\bg) = - \log Z^{(\alpha)} (\bg) \; .
\eeq
In principle just the state with smallest free energy $F^{(\alpha)}$ should be relevant for the computation of the partition function (\ref{eq:SPPartFunc}). Indeed all the other states have an interpretation according to the dynamics which governs the system. Such states are relevant in order to model the phenomenon of ergodicity breaking, which occurs whenever the configurations of a large system $s \in \Omega$ cannot be sampled according to the probability distribution $p(s)$ in experiments of finite length $T$.\footnote{
We won't explicitly refer to the dynamics leading to the loss of ergodicity, even though this phenomenon is naturally associated with the stochastic process leading to the stationary distribution (\ref{eq:ProbDensity}) and is more naturally discussed in the framework of a Markov chain \cite{Feller:1950zr}.
}
\\
In particular we informally remind that for large statistical models $(\bphi,\bg)$ endowed with a realistic dynamics (e.g., Metropolis-Hastings \cite{Teller:1953kx,Hastings:1970vn}) leading in the limit of exponentially large $T$ to the stationary distribution $\bp$ associated with $(\bphi,\bg)$, states naturally emerge when observing a finite amount of configurations. In fact, the iteration of a dynamics for $T \ll 2^N$ time steps typically produces configurations belonging to the same state as the initial one, while in the opposite limit of large $T$ the probability of observing a state belonging to a configuration $\alpha$ is proportional to $e^{N f(\bpsi^{(\alpha)})}$.
%The relevance of the notion of state is linked with the dynamical properties of the statistical model $(\bphi,\bg)$, which we will informally remind in the following in order to justify our approach.
%\rmk{
%Given a realistic dynamic process (e.g., Metropolis-Hastings \cite{Teller:1953kx,Hastings:1970vn}) leading in the limit of exponentially large $T$ to the stationary distribution $\bp$ associated with $(\bphi,\bg)$:
%\begin{itemize}
%\item{
%Given an initial configuration $s$ belonging to a state $\alpha$ (i.e., $\bpsi^{(\alpha)} \approx \bpsi(s)$), the configurations obtained by iterating the dynamics for a time $T$ polynomial in $N$ belong to the state $\alpha$ with probability exponentially close to 1. 
%}
%\item{
%Configurations obtained by iterating the dynamics for a time $T$ exponentially large in $N$ on any initial state belong to the state $\alpha$ with probability proportional to $e^{N f(\bpsi^{(\alpha)})}$.
%}
%\end{itemize}
%}
Hence, unless data obtained from an experiment are exponentially large in the size of the system (which isn't typically the case in real world applications of the inverse problem), one expects empirical averages to concentrate around averages which are in principle different from the ensemble ones, and that are associated with a specific state $\alpha$.
Accordingly, we define the notion of state average $\langle \phi^{(\alpha)}\rangle$, which is expected in the regime of $T \ll 2^N$ to model the averages obtained by experiments of finite length as follows:
\defin{Given a system $(\bphi,\bg)$ whose partition function can be approximated by the partition function (\ref{eq:SPPartFunc}), we define the  \emph{state averages}
\beq
\langle \phi_\mu^{(\alpha)} \rangle = - \frac{\partial F^{(\alpha)}}{\partial g_\mu}
\eeq
and the state susceptibilities
\beq
\chi_{\mu,\nu}^{(\alpha)} = - \frac{\partial^2 F^{(\alpha)}}{\partial g_\mu \partial g_\nu} \; .
\eeq
}
The correctness of above construction has been verified for several statistical models subject to different dynamics \cite{Monasson:2010hc,Zamponi:2010fk}, nevertheless to the best of our knowledge no fully general, rigorous result concerning this phenomenon is available yet. In particular, in order to rigorously motivate the notion of state average, it would be necessary to show that for a generic, local dynamics a decomposition property of the form $p_s \xrightarrow[N\to\infty]{} \sum_{\alpha=1}^P q_\alpha p^{(\alpha)}_s $ where $\sum_{\alpha=1}^P q_\alpha = 1$ and $p^{(\alpha)} \in \mathcal M (\Omega)$ holds for the Gibbs measure, which again is known to be correct just for specific models.

%Then one can define the notion of state average $\bar \bphi^{(\alpha)}$ in order to model this idea.
%\defin{
%We assume that the probability density $\bp$ associated with $(\bphi,\bg)$ can be decomposed in the large $N$ limit as
%\beq
%p_s \xrightarrow[N\to\infty]{} \sum_{\alpha=1}^P q_\alpha p^{(\alpha)}_s \; ,
%\eeq
%where $\sum_{\alpha=1}^P q_\alpha = 1$ and $p^{(\alpha)} \in \mathcal M (\Omega)$, with
%\beq
%\langle \phi_\mu^{(\alpha)} \rangle = \sum_s p^{(\alpha)}_s \phi_{\mu,s} = - \frac{\partial F^{(\alpha)}}{\partial g_\mu}
%\eeq
%and
%\beq
%\chi_{\mu,\nu}^{(\alpha)} = \sum_s p^{(\alpha)}_s  \phi_{\mu,s} \phi_{\nu,s} - \langle \phi_\mu^{(\alpha)} \rangle \langle \phi_\nu^{(\alpha)} \rangle = - \frac{\partial^2 F^{(\alpha)}}{\partial g_\mu \partial g_\nu} \; .
%\eeq
%We also assume that averages obtained in any experiment of finite length $T$ are i.i.d.\ samples distributed according to $p^{(\alpha)}_s$, where $\alpha$ is the state associated with the initial configuration of the system, and will call \emph{empirical state average} the averages produced by such measure, and denote them with $\bar \phi^{(\alpha)}$.
%}
%This approach models the phenomenon of ergodicity breaking by assuming the decomposability of the probability distribution (\ref{eq:ProbDensity}), which can be shown for several models to hold \cite{Zamponi:2010fk}. We don't prove this result in general (TODO: can be done?), but rather present its consequences in the context of the inverse problem.
In that case, the state averages and the susceptibilities can be explicitly computed by explicitly deriving the above free-energy, allowing to prove the following result.
\prop{
The direct problem for a statistical model $(\bphi,\bg)$ which can be described with order parameters $\bpsi$ and an order parameter free-energy $f(\bpsi)$ can be solved in saddle-point approximation in any state $\alpha$, leading to
\beqa
F^{(\alpha)}(\bg) &=& N f(\bpsi^{(\alpha)}) + \frac{1}{2} \log \det f^{(2)}(\bpsi^{(\alpha)}) -\frac{1}{2} \log \frac{2\pi}{N} \\
\langle \phi_\mu^{(\alpha)} \rangle &=& \phi_\mu (\bpsi^{(\alpha)}) + \frac{1}{N} \left[ (f^{(2)})^{-1}_{a,b} \phi_{\mu ; a,b}^{(2)} - (f^{(2)})^{-1}_{a,b} f^{(3)}_{b,a,d} (f^{(2)})^{-1}_{d,e} \phi_{\mu ;e}^{(1)} \right]  \label{eq:SPAvg} \\
\chi_{\mu,\nu}^{(\alpha)} &=& [(f^{(2)})^{-1}_{a,b}]_\nu^{(1)} \phi^{(2)}_{\mu ; b,a}
- [(f^{(2)})^{-1}_{a,b}]_\nu^{(1)} f_{b,a,d}^{(3)} (f^{(2)})^{-1}_{d,e} \phi^{(1)}_{\mu ; e}  \\
&-& (f^{(2)})^{-1}_{a,b} f_{b,a,d;\nu}^{(3,1)} (f^{(2)})^{-1}_{d,e} \phi^{(1)}_{\mu ; e} - (f^{(2)})^{-1}_{a,b} f^{(3)}_{b,a,d} [(f^{(2)})^{-1}_{d,e}]_\nu^{(1)} \phi^{(1)}_{\mu ; e} \nonumber \\
&+& \frac{1}{N} \Bigg[[(f^{(2)})^{-1}_{a,b}]_c^{(1)} (f^{(2)})^{-1}_{c,d} \phi_{\nu ; d}^{(1)} \phi^{(2)}_{\mu ; b,a} 
+ (f^{(2)})^{-1}_{a,b} \phi^{(3)}_{\mu ; a,b,c} (f^{(2)})^{-1}_{c,d} \phi^{(1)}_{\nu ; d}  \nonumber \\
&-& [(f^{(2)})^{-1}_{a,b}]_f^{(1)} f^{(3)}_{b,a,d} (f^{(2)})^{-1}_{d,e} \phi_{\mu; e}^{(1)}f_{f,g}^{(2)} \phi_{\nu ; g}^{(1)}
- (f^{(2)})^{-1}_{a,b} f^{(4)}_{b,a,d,f} (f^{(2)})^{-1}_{d,e} \phi_{\mu; e}^{(1)} f_{f,g}^{(2)} \phi_{\nu ; g}^{(1)} \nonumber \\
&-&  (f^{(2)})^{-1}_{a,b} f^{(3)}_{b,a,d} (f^{(2)})^{-1}_{d,e} \phi_{\mu; e}^{(1)} f_{f,g}^{(2)} \phi_{\nu ; g}^{(1)}
- (f^{(2)})^{-1}_{a,b} f^{(3)}_{b,a,d} [(f^{(2)})^{-1}_{d,e}]^{(1)}_f \phi_{\mu; e,f}^{(2)} f_{f,g}^{(2)} \phi_{\nu ; g}^{(1)} \Bigg] \nonumber
\eeqa
where $\phi_{\mu; a_1,\dots,  a_n}^{(n)}$ indicates the tensor $\frac{\partial}{\partial \psi_{a_1}} \dots \frac{\partial}{\partial \psi_{a_n}} \phi_\mu(\bpsi^{(\alpha)})$, \\$f^{(m,n)}_{a_1,\dots, a_m ; \mu_1 \dots, \mu_N} = \frac{\partial}{\partial \psi_{a_1}} \dots \frac{\partial}{\partial \psi_{a_m}} \frac{\partial}{\partial g_{\mu_1}} \dots \frac{\partial}{\partial g_{\mu_n}}  f(\bpsi^{(\alpha)})$ and by convention repeated index are summed.
}
This result allows us to characterize the behavior of the inverse problem in the large $N$ limit. In fact one can see that at leading order in $N$, the momentum matching condition (\ref{eq:MomentumMatching}) becomes
\beq
\langle \phi_\mu^{(\alpha)} \rangle \xrightarrow[N\to\infty]{} \phi_\mu (\bpsi^{(\alpha)}) = \bar \phi \; ,
\eeq
where we remark that the averages in the state $\alpha$ do not depend explicitly on $\bg$, being their dependence contained in the order parameter $\bpsi^{(\alpha)}$. 
This implies that, given two statistical models $(\bphi, \bg)$ and  $(\bphi, \bg^\prime)$ with $\bg \neq \bg^\prime$ such that there exist a couple of states (respectively $\alpha$ and $\alpha^\prime$) solving the saddle point equations with $\bpsi^{(\alpha)} = \bpsi^{\prime (\alpha^\prime)}$, in the large $N$ limit those models cannot be discriminated.
\rmk{Consider an empirical dataset $\bar \phi^{(\alpha)}$ generated by a system in state $\alpha$. Unless one doesn't consider a matching condition in which the state average contains the corrections of order $1/N$ indicated in the right term of formula (\ref{eq:SPAvg}), it is not generally guaranteed that it is possible to reconstruct the state $\alpha$ which generated the empirical averages.}
A rough criteria which can be used in order to check the expected number of solutions for the inverse problem is provided by the comparison of the number of solutions of the saddle-point equations $P$, the number of order parameters $A$ and the number of couplings $M$. If in particular $M > A$, then the saddle-point equations are expected to have a continuous number of solutions $\bg^\star$ specifying the same value of the order parameters for any of the $P$ states $\bpsi^{(\alpha)}$. If $M < A$ a unique set of couplings is expected to be associated with a value of an order parameters. Finally, if $M=A$, then a finite number of solutions for the couplings has to be expected.
\subsection{Ergodicity breaking for a fully connected pairwise model}
\begin{figure}[h] %  figure placement: here, top, bottom, or page
   \centering
   \begin{tabular}{cc}
   \includegraphics[width=2.9in]{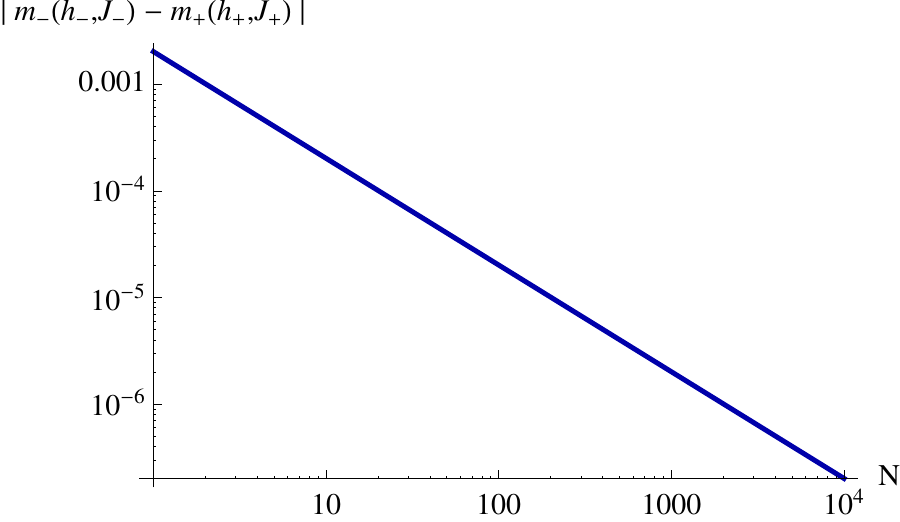}
   \includegraphics[width=2.9in]{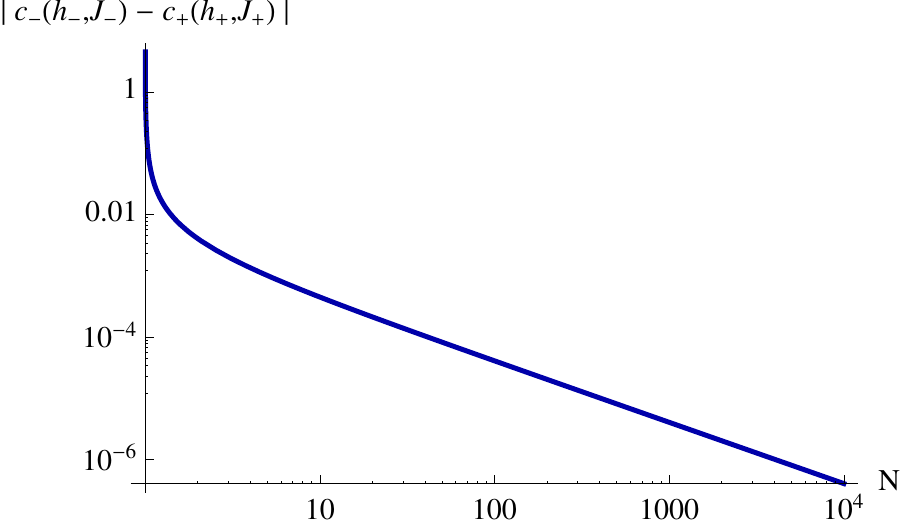}
   \end{tabular}
   \caption{Absolute difference among the ensemble averages of models describing two different states of a fully connected model, as a function of the system size $N$.}
   \label{fig:ErgBreak}
\end{figure}

Consider the fully-connected pairwise model of section \ref{sec:FCFerromagnet}. In that case the construction above can be trivially applied by considering the only order parameter $ \psi (s) = \sum_{i=1}^N s_i$ (so that $A=1$). The saddle-point equation for this model
\beq
m = \tanh \left( J m + h\right)
\eeq
can have either one solution $m^\star$ (thus, $P=1$) or two stable solutions $m_+^\star$ and $m_-^\star$ ($P=2$) according to the values of $h$ and $J$.
We consider as an illustrative example the case in which $J=J_-=4$ and $h=h_-=0.1$, hence $P=2$ solutions are present. For this model the metastable state is characterized by $m_-^\star \approx -0.9991754$, and it is easy to show that any pair $(J,h)$ satisfying
\beq
m^\star_- = \tanh   \left( J m_-^\star + h \right)
\eeq
has the same saddle point magnetization. In particular, it is possible to find $h<0$ solutions corresponding to the stable $\alpha = +$ state characterized by the same value of the magnetization. For example, the stable state of the model $(J_+ \approx 3.39950 , h_+ = -0.5)$ has magnetization $m_-(J_-,h_-) = m_+ (J_+,h_+)$. In figure \ref{fig:ErgBreak} we show how the models $(J_-,h_-)$ and $(J_+, h_+)$ lead to the same value of the state averages $m$ and $c$ in the thermodynamic limit $N \to \infty$: not even the state of a large fully connected ferromagnet can be reconstructed on the basis of a finite length experiment, unless the state averages are known with large precision.
The difference of this result with respect to what found in section (\ref{sec:FCFerromagnet}) lies in the fact that state averages can be matched by any solution of the form
\beqa
h^\star &=& \arctanh  m - \delta J \, m \\
J^\star &=&  \delta J \; ,
\eeqa
regardless of the sign of $h^\star$ (while in that case it had to be taken $\textrm{sign }(h^\star) = \textrm{sign } (\bar m )$). In both cases a continuous number of solutions for the inverse problem is present.

\section{Disorder and heterogeneity: the regime of large $N$ and large $M$ \label{sec:LargeNLargeM}}
The results presented in section \ref{sec:FCFerromagnet} for the Curie-Weiss model refer to a specific statistical model whose associated inverse problem shows interesting features in the limit of large $N$. Despite the fact that such properties generally hold for similar kind of models (section \ref{sec:ErgBreak}) one could wonder whether this behavior is retained in the more relevant case in which a large number of inferred parameters is present.
Consider for example a general pairwise model (\ref{eq:IsingModel}), characterized by a set of $N$ external fields and $\frac{N(N-1)}{2}$ pairwise couplings. In this case one may have several problems in studying the features introduced in section \ref{sec:LargeNFiniteM} as we did above. In particular:
\begin{itemize}
\item{
The averages $\bmg$, $\hat \bc$  and the generalized susceptibility $\hat \bchi$ are hard to compute for a generic value of $\hat \bJ$ and $\bh$. Therefore, it is not possible to understand which points of marginal polytope $\mathcal G(\bphi)$ are associated with zero modes in $\hat \bchi$. Moreover, the limit $N\to\infty$ is ambiguously defined if no prescription is provided for how should the empirical averages scale with $N$.
}
\item{
For the same reason, it is not possible to find in which points one expects model concentration to occur, as this would require knowing which eigenvalues of $\hat \bchi$ are divergent in the thermodynamic limit $N \to \infty$ for generic points  $(\bmg, \hat \bc) \in \mathcal G (\bphi)$.
}
\item{
No saddle-point approach is justified for generic empirical averages $\bmg$, $\hat \bc$. Thus, an approach analogous to the one in \ref{sec:ErgBreak} cannot be considered, and the notion of state cannot be described in such terms.
}
\end{itemize}
These difficulties could be overcome by resorting to the notion of \emph{disorder}, which is commonly used in the field of statistical mechanics of heterogeneous systems. In particular we want to show, as a possible outlook of this work, an approach to the analysis of the large $M$ limit borrowed from that field \cite{Mezard:1987fk} which could be applied to this problem.

\subsection{Self-averaging properties and inverse problem}
Given an operator set $\bphi$, consider a set of statistical models $\mathcal M(\bphi)$ and a prior $P_0(\bg)$ on this space. Then, suppose that a statistical model $(\bphi, \bg)$ is sampled according to $P_0(\bg)$, and successively a set of empirical data of length $T$ is drawn by such distribution. Several functions of the estimator $\bg^\star(\bar \bphi)$ can be built in order to analyze the properties of an instance of the inverse problem, such as the quantities
\beq
\Delta (\bar \bphi, \bg) =  \sqrt{\sum_{\mu=1}^M \frac{1}{M} \left(g^\star_\mu(\bar \bphi) - g_\mu \right)^2} \; , \label{eq:AvgErrorCoup}
\eeq
which quantifies the average error in the inferred coupling and
\beq
\frac{1}{M} \log \det \hat \bchi (\bg^\star(\bar \bphi) ) = \frac{1}{M} \textrm{tr} \log \hat \bchi (\bg^\star(\bar \bphi) ) \; , \label{eq:AvgLogFish}
\eeq
whose divergence signals critical properties of the generalized susceptibility matrix $\hat \bchi$.
If these of quantities are self-averaging for large $N$ and $T$ (i.e., they concentrate around an average value determined by $P_0(\bg)$), then one expects that specific instances of inverse problems drawn by the same prior $P_0(\bg)$ to share the same collective features.
As an example, if one considers a Gaussian prior for the ferromagnetic model of the type $P_0(\bh, \hat \bJ) \propto \exp \left( - N \sum_{i<j}  \frac{(J_{ij}-J_0/N)^2}{2 \, \delta J^2} \right) \exp \left( - \sum_i \frac{(h_i - h_0)^2}{2 \, \delta h^2} \right)$ with $J_0 \neq 0$, then it is known that the macroscopic behavior of the model approaches in the large $N$ limit the one of a fully-connected ferromagnet (\ref{eq:FCFerromagnet}) defined by the only parameters $(h_0,J_0)$ \cite{Mezard:1987fk}. In section \ref{sec:GeomApplications} we will support this claim through a specific example, showing a case in which the properties of a homogeneous model allow to describe very accurately the collective features of the inverse problem for an heterogeneous one. Nevertheless, it would be interesting to repeat the calculations shown in the previous sections in this more general scenario in which disorder is present, and prove through the so-called \emph{replica} formalism \cite{Mezard:1987fk} the correctness of these expectations.
\rmk{
The idea of disorder in the context of the inverse problem is obviously linked to the existence of a prior $P_0(\bg)$ on the space $\mathcal M(\bphi)$, so that in principle the case of a flat prior cannot be treated with these techniques. Nevertheless, fixing implicitly a specific class of models through $P_0(\bg)$ is the price to pay to answer to very interesting questions, which wouldn't otherwise be well-posed namely: (i) can a specific model be learnt with high probability according to a given inference prescription? (ii) Are the global properties of an heterogeneous system equivalent the the ones of an homogeneous one? (iii) Is it possible to understand the generic properties of $\hat \bchi$?}

\chapter{Complete representations \label{ch:Completeness}}
In this chapter we will introduce the notion of complete family of operators, which can be used to gain some insight about the inverse problem. Although in general this approach may introduce a high degree of over fitting, dealing with complete families allows to discuss very transparently some features of inference which are related to algorithmic complexity (section \ref{sec:CompleteInverseProblem}). Moreover, completeness allows for an explicit reparametrization of the probability distribution (\ref{eq:ProbDensity}) in terms of state probabilities, allowing for a complete understanding of properties of the inverse problem which are less clear by using the Gibbs form for the probability density. More interestingly, in this language we will be able to differentiate local features of the direct and of the inverse problem, which in turn rely on the locality of the marginals. In this chapter the inverse problem for some models will be exactly and explicitly solved, while some ideas will be presented in order to generalize this methods to more relevant problems (sections \ref{sec:InvPairTree} and \ref{sec:Inv1DChain}). In section \ref{sec:Applications} we will present some specific examples illustrating these ideas.
\section{Orthogonality and completeness}
We define in this section the notion of \emph{orthogonality} and \emph{completeness} for families of operators. While the orthogonality condition is related to the one of minimality, the one of completeness will allow to formally invert the relation among ensemble averages $\langle \bphi \rangle$ and couplings $\bg$.

\defin{
Given a family of operators $\bphi$, we call it \emph{orthogonal} if it satisfies
\beq
\frac{1}{|\Omega|} \sum_s \phi_{\mu, s} \phi_{\nu, s\phantom{^\prime}} = \delta_{\mu,\nu} \label{eq:Orthogonality} \; ,
\eeq
while it will be called \emph{complete} if it holds
\beq
\frac{1}{|\Omega|} \sum_\mu \phi_{\mu, s} \phi_{\mu, s^\prime} = \delta_{s,s^\prime} \label{eq:Completeness} \; .
\eeq
}
Property $(\ref{eq:Orthogonality})$ can be seen as expressing the fact that in an orthogonal family any pair of operators decorrelate when averaged with respect to a uniform probability density (at infinite temperature in the language of statistical mechanics). Additionally, if $\phi_0 \in \bphi$, one can see that in an orthogonal family, for $\mu \neq 0$
\beq
\frac{1}{|\Omega|} \sum_s \phi_{\mu, s} = 0 \; ,
\eeq
i.e., $\phi_\mu$ has zero mean at infinite temperature for any $\mu \neq 0$. Finally, if $\bphi$ is an orthogonal family, then it is easy to see that $\bphi \backslash \{ \phi_0\}$ is minimal.
The main result that derives instead from equation (\ref{eq:Completeness}) is the explicit one-to-one mapping between couplings $\bg$, state probabilities $\bp$ and averages $\langle \bphi \rangle$, as clarified by the next proposition.

\prop{Given a family $\bphi$ satisfying (\ref{eq:Orthogonality}) and (\ref{eq:Completeness}), the statistical model $(\bphi \backslash \{ \phi_0 \}, \bg)$ associated with the probability density $\bp$ satisfies
\beqa
\langle \phi_\mu \rangle &=& \sum_s \phi_{\mu,s } \exp \left( \sum_\nu g_\nu \phi_{\nu,s} \right) \label{eq:FormalAvg} \\
g_\mu &=& \frac{1}{|\Omega|} \sum_s \phi_{\mu, s} \log \left( \frac{1}{|\Omega|} \sum_\nu \langle \phi_\nu \rangle \phi_{\nu,s}  \right) \label{eq:FormalCoup} \; .
\eeqa
 Additionally, state probabilities can be expressed as
\beq
p_s = \frac{1}{|\Omega|}\sum_\mu \langle \phi_{\mu} \rangle \phi_{\mu,s} \label{eq:ProbReparam} \; .
\eeq
}
\prf{
These relations are a direct consequence of the axioms (\ref{eq:Orthogonality}) and (\ref{eq:Completeness}) and can be checked by direct substitution.
}
\subsubsection{Monomials}
Throughout most of the following discussion, we will focus on families of operators $\bphi$ formed by  \emph{monomials}, for which axiom (\ref{eq:Orthogonality}) trivially applies. More precisely, given a cluster of spins $\Gamma$, we define the monomial $\phi_\Gamma (s)$  as
\beqa
\phi_\Gamma (s) = \prod_{i \in \Gamma} s_i \; ,
\eeqa
while the identity is associated with the empty cluster $\phi_0 (s) = \phi_{\emptyset}(s) = 1$. It is easy to show the following:

\prop{
Given a collections of clusters $(\Gamma_0,\dots, \Gamma_M)$ with $\Gamma_i \neq \Gamma_j \; \forall \; (i,j)$ it holds for the family $\bphi = \{ \phi_{\Gamma_0},\dots , \phi_{\Gamma_M}\}$ that
\begin{itemize}
\item{$\bphi$ is an orthogonal family;}
\item{$\bphi$ is complete if and only if it contains all possible monomials, whose number is $|\Omega| = 2^N$.}
\end{itemize}
}
Moreover, monomials satisfy a very important relation which will be used extensively in the following.
\prop{
Consider a complete family of monomials $\bphi$. Then the marginals of the probability density $\bp$ associated with the model $(\bphi \backslash \{ \phi_0 \}, \bg)$ can be expressed as
\beq
p^\Gamma(s^\Gamma )  = \frac{1}{2^{|\Gamma|}}  \sum_{\Gamma^\prime \subseteq \Gamma} \langle \phi_{\Gamma^\prime} \rangle \, \phi_{\Gamma^\prime,s}
\eeq
}
\prf{This can be checked by using equation (\ref{eq:ProbReparam}) and showing that for each monomial it holds
\beq
\frac{1}{2}\sum_{s_i} \phi_{\Gamma,s} = \delta_{i \not\in \Gamma} \, \phi_{\Gamma,s} \; . \label{eq:LocalMarginal}
\eeq
}
This  property expresses the locality of marginals once they are expressed in terms of ensemble averages. This should be compared with the expression of a marginal written as a function of the couplings (\ref{eq:MarginalDef}), in whose form the locality properties are hidden by the interaction structure.

\section{Inference on complete models}
\subsection{The complete inverse problem \label{sec:CompleteInverseProblem}}
The techniques shown in the above section can be used to write a formal solution of the inverse problem in full generality. The main drawback of this procedure is the overfitting issue which has to be associated with the presence of an exponential number of couplings, which in practical cases makes this approach unfeasible unless the system has small size (typically $N \sim 10^{1}$). Indeed, as the solution of the complete inverse problem illustrates with simplicity some very general features of many inverse problem, we choose to present its solution.
\defin{
The \emph{complete inverse problem} is the inverse problem associated with the statistical model defined by the complete family of monomials $\{\phi_{\Gamma}(s) \}_{\Gamma \subseteq V}$. Its probability density can be written as
\beq
p(s) =  \exp \left( \sum_{\Gamma \subseteq V} g_\Gamma \phi_{\Gamma}(s) \right) \; . \label{eq:CompleteInverseProblem}
\eeq
}
It is easy to write the formal solution for the entropy by using the relation (\ref{eq:Completeness}), while its differentiation (or the direct use of the relation (\ref{eq:FormalCoup})) leads to an exact expression for the couplings and the susceptibility matrix.
\prop{
The expression for the entropy of the complete inverse problem reads
\beq
S(\bar \bphi) = - \frac{1}{|\Omega|} \sum_s \left( \sum_{\Gamma \subseteq V}  \bar \phi_\Gamma \, \phi_{\Gamma,s} \right)
\log  \left( \frac{1}{|\Omega|} \sum_{\Gamma^\prime \subseteq V}  \bar \phi_{\Gamma^\prime} \, \phi_{\Gamma^\prime,s} \right) \; ,
\eeq
while the inferred couplings $\bg^\star$ and the inverse susceptibility matrix $\hat \bchi^{-1}$ result
\beqa
g^\star_\Gamma &=& \frac{1}{|\Omega|} \sum_s \phi_{\Gamma,s} \log  \left( \frac{1}{|\Omega|} \sum_{\Gamma^\prime \subseteq V}  \bar \phi_{\Gamma^\prime} \, \phi_{\Gamma^\prime,s} \right)  \\
\chi_{\Gamma,\Gamma^\prime}^{-1} &=& \frac{1}{|\Omega|^2} \sum_s \frac{\phi_{\Gamma,s}  \phi_{\Gamma^\prime,s} }{\frac{1}{|\Omega|} \sum_{\Gamma^\prime \subseteq V}  \bar \phi_{\Gamma^\prime} \, \phi_{\Gamma^\prime,s}} \label{eq:FormalSusc} \; .
\eeqa
}
This solution has a simple interpretation in terms of empirical frequencies, once one rewrites above expression using the relation $\bar p_s = |\Omega|^{-1}\sum_{\Gamma^\prime \subseteq V}  \bar \phi_{\Gamma^\prime} \, \phi_{\Gamma^\prime,s}$ as
\beqa
g_\Gamma^\star &=& \frac{1}{|\Omega|} \sum_s \phi_{\Gamma,s} \log  \bar p_s \label{eq:FormalCoupling} \\
\chi_{\Gamma,\Gamma^\prime}^{-1} &=& \frac{1}{|\Omega|^2} \sum_s \frac{\phi_{\Gamma,s}  \phi_{\Gamma^\prime,s} }{\bar p_s} \label{eq:FormalInvFisher} \; .
\eeqa
In this form it is possible to appreciate that the solution simply corresponds to a matching of state probabilities with empirical probabilities.
\rmk{This last observation can be made more precise by exploiting the identity
\beq
\log P_T(\hat \bs | \bg) = T \sum_\Gamma g_\Gamma \bar \phi_\Gamma = T \sum_s \log p_s \, \bar p_s \; ,
\eeq
which can be used to express the log-likelihood function as a function of the probabilities $\bp$ instead of the coupling vector $\bg$. Its maximization can be seen equivalently as performed over the state probabilities $\bp$. In this case, the obvious solution is $\bp^\star = \bar \bp$, so that the expression (\ref{eq:FormalCoupling}) is describing an approach in which the state probabilities are matched with the empirical ones one-by-one. In particular, if a configuration is not observed, the inferred probability for that configuration is strictly zero.
}
\subsubsection{Divergencies}
The formal solution (\ref{eq:FormalCoupling}) shows that the inferred couplings can be infinite if there are states which are never sampled in the data $\hat \bs$. In particular if data are generated by an actual probability distribution $\bp$ assigning zero weight to some configuration, the $\Omega$ space splits into an accessible and a non-accessible sector, and divergencies can be seen as are required to implement an hard constraint on the set of accessible configurations. Couplings obtained by using this scheme are finite either when all states are measured or when divergencies cancel out for a given region of the coupling space. Indeed, the presence of an unaccessible sector has to be considered a spurious result unless $p_s \approx \bar p_s$ , which is expected to hold just in the large $T$ limit. In particular for $T < |\Omega|$ , $\bar p_s = 0$ for at least $ |\Omega| - T> 0$ configurations, regardless of the presence or absence of a forbidden sector. Therefore it is not possible to distinguish if divergencies are due to the presence of an unaccessible sector or to poor sampling. In this case, regularization schemes such as the use of Laplacian smoothing or an L-2 norm can be used to obtain finite results. This basically corresponds to lift the probability for non-measured configurations from zero to some finite value. For example, Laplacian smoothing procedure \cite{Russell:2010ve} corresponds to the choice:
\beq
p_s^\star = \frac{ \bar p_s  + \lambda}{1+ |\Omega| \lambda}
\eeq
Finally, we remark that the same type of divergence arises in all the cases that will be analyzed (see sections \ref{sec:InvPairTree} and \ref{sec:Inv1DChain}), and is a very general characteristic of inverse problems, which typically relates to under sampling. This is the simplest setting in which this problem can be analyzed in full generality.
\subsubsection{Observed sector}
The expression for the inferred couplings (\ref{eq:FormalCoupling}) involves a summation over all the configuration space $\Omega$, so that a summation over $|\Omega|=2^N$ terms seems to be required to calculate any of them. Indeed, those expressions may be rewritten exploiting the orthogonality relation (\ref{eq:Orthogonality}), which implies that
\beq
\frac{1}{|\Omega|}\sum_{s \in \bar {\mathcal I}} \phi_{\Gamma,s} = \delta_{\Gamma,0} - \frac{1}{|\Omega|}\sum_{s \not\in \bar {\mathcal I}} \phi_{\Gamma,s}
\eeq
where $\bar {\mathcal I}= \{ s \in \Omega \; | \; \bar p_s > 0 \}$ is the set of observed configuration. Then, one can rewrite (\ref{eq:FormalCoupling}) as
\beq
g_\Gamma^\star = \frac{1}{|\Omega|} \sum_{s \in \bar {\mathcal I}} \phi_{\Gamma,s} \log  \bar p_s + \log \bar p_0 \left( \delta_{\Gamma,0} -  \frac{1}{|\Omega|} \sum_{s \in \bar {\mathcal I}} \phi_{\Gamma,s} \right) \; ,
\eeq
where the term proportional to $\bar p_0 = 0$ account for the divergencies, and the sum over states runs over a number $|\bar {\mathcal I} | \leq T$ terms. In the case of the regularized complete inverse problem (section \ref{sec:RegCompInvProb}), we will see that it will be possible to write an analogous expression for the couplings, in which the weight assigned to non-observed configuration will be finite.

\subsubsection{Rate of convergence}
Given an underlying statistical model $\bp$ for the complete inverse problem, large deviation theory (as described in section \ref{sec:InvSLDev}) states that for large $T$ the variance of the inferred couplings $\bg$ with respect to the measure given by $P_T(\bp | \bar \bp) \propto P_T(\bar \bp | \bp) $ is
\beq
\textrm{Var}(g_\Gamma^\star) = \frac{\chi^{-1}_{\Gamma,\Gamma}}{T} = \frac{1}{T}  \left( \frac{1}{|\Omega|^2 } \sum_s \frac{1}{p_s} \label{eq:ConvRateCompProb} \right) \; .
\eeq
Incidentally, the same quantity can also be obtained by averaging with respect to the $\langle \dots \rangle_T$ measure, a result which allows to express the rate of convergence for the complete inverse problem (appendix \ref{app:ConvCompInvProb}).
While the $1/T$ pre factor expresses the expected scaling for the error on the inferred coupling, the $\chi^{-1}_{\Gamma,\Gamma}$ term is non trivial. In particular we observe that:
\begin{enumerate}
\item{The fluctuations of the inferred couplings are identical for all the operators.}
\item{The value of the fluctuations is bound by the inequality:
\beq
\frac{1}{T} \leq \textrm{Var}(g_\Gamma^\star)  \leq \frac{1} { T |\Omega| p_{min}} \label{eq:BoundConvRateCompProb}
\eeq
where $p_{min} = \min_s p_s$.
}
\item{The speed of convergence is limited by the presence of rare configurations. In particular if $p_{min} = 0$, the variance diverges.}
\end{enumerate}
The generalization to the case in which the sector of observable states $\mathcal{I} = \{s \in \Omega \; | \; p_s > 0 \}$ is smaller than the entire phase space $\mathcal{I} \subset \Omega$ is straightforward (appendix \ref{app:ConvCompInvProb}). Indeed, it is necessary to define a set of \emph{regular} operators $\bphi^{reg}$ such that $\bphi^{reg} = \{ \phi_\Gamma \in \phi \;|\; \sum_{s \in \mathcal{I}} \phi_{\Gamma,s} = 0 \}$. For couplings associated with regular operators it holds the asymptotic property
\beq
\textrm{Var} (g_\Gamma^{\star reg }) = \frac{1}{T |\Omega|^2} \sum_{s \in \mathcal{I}} \frac{1}{p_s} \quad . \label{eq:ConvRateCompProbHidSector}
\eeq
If the sector of observable states has cardinality $|\mathcal{I}| = \alpha |\Omega|$, then the fluctuations on the regular couplings satisfy the bound
\beq
\frac{\alpha^2}{T} \leq {\rm Var} (g_\Gamma^{\star reg})  \leq \frac{\alpha}{T |\Omega| p_{min}}
\eeq
where $p_{min} = \min_{s \in \mathcal{I}} p_s$. Even in the cases analyzed in section \ref{sec:InvPairTree} and \ref{sec:Inv1DChain} the presence of rare configurations will limit the speed of convergence of the inferred couplings to their actual value.

\subsection{Regularization of the complete inverse problem \label{sec:RegCompInvProb}}
The generality of the complete inverse problem renders its regularization relevant for a strong theoretical reason. In fact, the complete inverse problem is totally non-parametric in the sense that the probability distribution (\ref{eq:CompleteInverseProblem}) contains all possible statistical models describing a set of $N$ binary variables. Then one could think of selecting the most appropriate statistical model to describe a dataset of binary data simply by applying a suitable regularizer to this general problem, and let the regularization term itself perform the task of model selection (an approach successfully adopted in~\cite{Ravikumar:2010ys,Wainwright:2007zr} in a less general scenario). We present in the following the results obtained by using different regularization terms, and comment about the interpretation of the solutions of the regularized inverse problem. Finally, we will characterize a symmetry property of regularizers which can be used to study their suitability in the field of high-dimensional inference (i.e., for large values of $N$).
\subsubsection{L-2 regularization}
The simplest regularized version of the complete inverse problem is the one defined by the function
\beq
H(\bg|\hat \bs) = - T \sum_{\mu=0}^M g_\mu \bar \phi_\mu + \frac{\beta}{2} \sum_{\mu = 1}^M g_\mu^2 \; \label{eq:CompleteL2Reg}
\eeq
which implements the Gaussian prior over the L-2 norm of the coupling vector described in section \ref{sec:L2Regular}. In terms of state probabilities, equation (\ref{eq:CompleteL2Reg}) can be written as
\beq
H(\bp|\hat \bs) = -T \sum_s \log p_s \, \bar p_s + \frac{\beta}{2} \left[ \left( \frac{1}{|\Omega|} \sum_s \log^2 p_s \right) - \left(  \frac{1}{|\Omega|} \sum_s \log p_s \right)^2 \right]
\eeq
and its minimization with respect to $p_s$ (constrained to $\sum_s p_s = 1$) leads to the set of implicit equations
\beq
p_s^\star = \bar p_s - \frac{\beta}{T |\Omega|} \left( \log p_s^\star - \frac{1}{|\Omega|} \sum_{s^\prime} \log p_{s^\prime}^\star \right) \; . \label{eq:RegulProbDens}
\eeq
Its solution determines the value of the couplings $g$ through the relation
\beq
g_\mu^\star = \frac{1}{|\Omega|} \sum_s \phi_{\mu,s} \log p_s^\star \; .
\eeq
We observe that:
\begin{enumerate}
\item{The summation over the configuration space requires considering in principle an exponential number of terms, but this issue can be avoided as
explained in section \ref{sec:CompleteInverseProblem}.}
\item{The expression for $g_\mu^\star$ is always finite, as the presence of infinite couplings is suppressed by the cost associated with the L-2 norm.}
\item{The parameter $\beta$ controls the total value of the L-2 norm of the coupling vector $\bg^\star$ and the entropy of the inferred distribution. In particular the total L-2 norm can be expressed as
\beq
\sum_\mu g_\mu^{\star 2} = \frac{1}{|\Omega|} \sum_s \log^2 p_s^\star \; ,
\eeq
where the statistical weights $\bp^\star$ are fixed by equation (\ref{eq:RegulProbDens}).
}
\item{The additional problem of solving the system of equations for $p_s$ requires in principle the numerical solution of $|\Omega|=2^N$ equations. Indeed, all equations
linked with unobserved configurations are equal, and defining as above the probability $p_0$ for non-measured configurations, the number of independent
equations that have to be solved is $| \bar {\mathcal I} | +1 \leq | \mathcal I | +1 \leq T + 1$.}
\end{enumerate}
This considered, the expression for the couplings obtained using this regularization scheme is
\beq
g_\mu^\star = \frac{1}{|\Omega|} \sum_{s \in \mathcal I} \phi_{\mu, s} \log \left( \frac{p_s^\star}{p_0^\star} \right) +\delta_{\mu 0}  \log p_0^\star \; ,
\eeq
where the $p_s$ and the $p_0$ satisfy the set of implicit equations:
\beq
\left\{
\begin{array}{ccl}
p_s^\star &=& - \frac{\beta}{|\Omega| T} \left( \log p_s^\star - \frac{1}{|\Omega|} \sum_{s^\prime \in \mathcal I} \log p_{s^\prime}^\star - \frac{|\Omega| - |\mathcal I |}{|\Omega|} \log p_0^\star \right) + \bar p_s \\
&& \\
p_0^\star &=& - \frac{\beta}{|\Omega| T} \left( \log p_0^\star - \frac{1}{|\Omega|} \sum_{s^\prime \in \mathcal I} \log p_{s^\prime}^\star - \frac{|\Omega| - |\mathcal I |}{|\Omega|} \log p_0^\star  \right)  \; .
\end{array} \right. \label{eq:L2RegCompInvProb}
\eeq
We remark that the calculation of the regularized couplings can be performed in polynomial time in $T$.
\subsubsection{Entropy regularization}
Another choice for the regularization is motivated by the following argument. If a dataset $\hat \bs$ of length $T$ is associated with an entropy $S(\bar \bp) \sim \log T$, with $\log T \ll N $, it is likely for the model to be in the under sampled regime, as the entropy per variable is expected to be finite (i.e.,  $S(\bp) \sim N $) for well-behaved models. Then, it is possible to consider a regularizing term which penalizes low entropy distribution, so that
\beq
H(\bg|\hat \bs) = - T \sum_{\mu=0}^M g_\mu \bar \phi_\mu - \beta S(p) \;, \label{eq:CompleteEntrReg}
\eeq
where as usual $\bp$ is the density associated with the statistical model $(\bphi,\bg)$, so that $S(\bp) = - \sum_{\mu=0}^M g_\mu \langle \phi_\mu \rangle$.
The minimization of above expression with respect to $g_\mu$ leads to
\beq
\bar \phi_\mu =   \langle \phi_\mu \rangle  + \frac{\beta}{T} \sum_{\nu = 1}^M g_\nu \frac{\partial \langle \phi_\nu \rangle}{\partial g_\mu} \; .
\eeq
After some manipulation and after using the completeness relation (\ref{eq:Completeness}) one finds that
\beq
\bar p_s = p_s + \frac{\beta}{T} p_s \log p_s - \frac{\beta}{T} p_s \left( \sum_{s^\prime} p_{s^\prime} \log p_{s^\prime} \right) \quad .
\eeq
Finally, by writing $s_s = - p_s \log p_s$, one is led to a set of implicit equations
\beq
p_s = \frac{\bar p_s + \frac{\beta}{T} s_s}{1 + \frac{\beta}{T} \sum_{s^\prime} s_{s^\prime}} \label{eq:EntrRegCompInvProb} \; ,
\eeq
which is analogous to the one described in the L-2 case. Also in this case the system has to be solved numerically, by exploiting the fact that the probabilities $p_s$ depend on the $s$ index through the empirical frequency $\bar p_s$ (i.e., states visited the same number of times are associated with the same inferred probability). Equation (\ref{eq:Completeness}) can finally be used to extract the inferred couplings from the probability density of $p_s$.
\subsubsection{Susceptibility regularization}
The inverse generalized susceptibility of a model $\hat \bchi^{-1}$ provides an indication of the generalizability of an inference procedure through equation (\ref{eq:GenFuncInvFisher}), which implies that the response of the inferred couplings $\bg^\star$ to a shift of the empirical averages $\bar \bphi$ is
\beq
\chi^{-1}_{\mu,\nu} = \frac{\partial g^\star_\mu}{\partial \bar \phi_\nu} \; .
\eeq
Then one could think to favor generalizability in an inference procedure by introducing a regularization term of the form
\beq
H(\bg | \hat \bs) = -T \sum_{\mu=0}^M g_\mu \bar \phi_\mu + \beta \, \textrm{tr}  \left( \hat \bchi^{-1} \right) \; .
\eeq
By employing equation (\ref{eq:FormalSusc}) it is easy to see that the inverse susceptibility matrix can be written as a function of the coupling vector $g$ as
\beq
\chi^{-1}_{\mu,\nu} = \frac{1}{|\Omega|^2}\sum_s \phi_{\mu,s} \phi_{\nu,s} \exp \left( - \sum_{\rho=0}^M g_\rho \phi_{\rho,s} \right) \; ,
\eeq
so that the total energy can be written as
\beq
H(\bg | \hat \bs) = -T \sum_{\mu=0}^M g_\mu \bar \phi_\mu + \beta  \left(  \frac{ |\Omega| - 1}{|\Omega|^2} \right)\sum_s p_s^{-1} \; .
\eeq
Its minimization leads to
\beq
\bar \phi_\mu = \langle \phi_\mu \rangle + \frac{\beta}{T}  \left(   \frac{ |\Omega| - 1}{|\Omega|^2} \right) \sum_s  \big[\langle \phi_\mu \rangle - \bar \phi_\mu \big] p_s^{-1}
\eeq
whose solution requires solving a set of implicit equations analogous to (\ref{eq:L2RegCompInvProb}) and (\ref{eq:EntrRegCompInvProb}) of the form
\beq
p_s = \frac{\bar p_s +  \frac{\beta}{T} \left(   \frac{ |\Omega| - 1}{|\Omega|^2} \right)  p_s^{-1}}{1 +  \frac{\beta}{T} \left(   \frac{ |\Omega| - 1}{|\Omega|^2} \right) \sum_{s^\prime} p_{s^\prime}^{-1}} \; .
\eeq
By using equation (\ref{eq:ProbReparam}) the solution $\bp^\star$ can be used to explicitly express $\bg^\star$.
\rmk{
Notice that this regularization scheme artificially pushes the inferred couplings $\bg^\star$ towards regions of the space $\mathcal M(\bphi)$ in which fluctuations are high. This is a very general feature of inference procedures which favor the stability of the inferred model: requiring a model to be stable forces the generalizes susceptibility to be large, or equivalently, ensemble averages to have strong fluctuations. 
}
\subsubsection{L-1 regularization}
We will write the L-1 regularized problem for the complete inverse problem described as in section \ref{sec:L1Regular}, with the idea that its solution it would be equivalent to a complete, non-parametric solution of the problem of binary inference. In the more optimistic scenario, the problem of model selection would be implicitly solved by the L-1 norm, without the need of explicitly breaking the symmetry among the operators by choosing (\emph{a priori}) the more relevant ones, as it is usually done by means of the maximum entropy principle (appendix \ref{app:MaxEntPr}). Relevant operators should arise as conjugated to non-zero couplings in a regularized problem of the form
\beq
H(\bg | \hat \bs) = - T \sum_{\mu=0}^M g_\mu \bar \phi_\mu + \beta \sum_{\mu=1}^M |g_\mu | \; .
\eeq
The minimization of above expression with respect to $\bg$ leads to
\beq
p_s \in \bar p_s - \frac{\beta}{T |\Omega|} \sum_{\mu=1}^M \textrm{sgn} (g_\mu) \phi_{\mu,s} \; ,
\eeq
where we define the set valued function $\textrm{sgn} (x) $ as in appendix \ref{app:ConvexOpt}.
We remark several issues concerning this regularizer:
\begin{enumerate}
\item{Unlike the L-2 case, the completeness relation does not allow to switch from a summation on operators to a summation over configurations, hence algebraic properties cannot be fully exploited to manipulate the above equation.}
\item{The minimization condition is a system of $|\Omega|$ implicit equations in which the inferred values of $p_s$ on non-observed configurations are generally different.
This is due to the term $\sum_\mu ({\rm sgn }\,  g_\mu) \phi_{\mu,s}$, which is different even for $s \notin \mathcal I$ (see example \ref{sec:L1SymmBreak}).}
\item{L-1 norm is associated with a compact description of the probability distribution (it is used to enforce sparsity in the number of non-zero couplings), while in the case $T \ll 2^N$ one deals with few observations of the system (sparsity in the number of observed configurations). As the change of parametrization (\ref{eq:ProbReparam}) from $\bp$ to $\bg$ is strongly non-local (i.e., what is sparse in a parametrization is not sparse in the other one), the problem becomes hard to solve due to \emph{frustration}, alias the simultaneous request of incompatible conditions in a constraint satisfaction problem.
\item{
Even if a fast (i.e., polynomial in $N$) algorithm to find a solution for a single coupling $g_\mu$ was available, a preliminary selection of the couplings to focus on would nevertheless be needed. In fact, even in the scenario in which the calculation of a single $g_\mu$ can be achieved in polynomial time, a constrained optimization problem should be formulated in order to select \emph{which} subset of couplings is non-zero given a specific value of $\beta$.}
}
\end{enumerate}
\subsubsection{Explicit selection of couplings}
An interesting case is the one in which a specific inverse problem -- such as the inverse pairwise model -- is seen as a regularized version of the complete inverse problem. This implicitly implies that unlike with the previous regularizers, in this particular example we are not interested in the problem of model selection, but we mean to offer a different perspective on a problem which is known to be hard, in order to characterize it from a different point of view. In particular, we consider the regularized minus-log-likelihood
\beq
H(\bg|\hat \bs) =  - T \sum_{\mu=0}^M g_\mu \bar \phi_\mu + \frac{\beta}{2} \sum_{\mu = 1}^M \theta_\mu \, g_\mu^2 \; \label{eq:CompletePairwiseReg}
\eeq
in which $\theta \in \{ 0,1 \}^M$ determines the couplings that are penalized by the L-2 norm, and we consider the limit of large, positive $\beta$, so that $g_\mu^\star \approx 0$ if $\theta = 1$.
The minimization of (\ref{eq:CompletePairwiseReg}) leads to the set of equations
\beq
\bar \phi_\mu = \sum_s p_s^\star \phi_{\mu,s} + \frac{\beta}{T} \theta_\mu g_\mu^\star \; ,
\eeq
which in the parametrization of states becomes
\begin{equation}
\bar p_s = p_s^\star + \frac{\beta}{T |\Omega|} \sum_\mu \phi_{\mu, s} \theta_\mu g_\mu^\star \; . \label{eq:ExpCoupSel}
\end{equation}
The last term in (\ref{eq:ExpCoupSel}) is finite in the limit of large $\beta$, and encodes the constraint specified by $\theta$.
Within this formulation the intrinsic difficulty of an inverse, non-complete problem emerges as the fact that the probabilities $p_s^\star$ can be different for states visited with the same frequency. This is associated with the dependence of the second term of equation (\ref{eq:ExpCoupSel}) upon the index $s$ associated with the operators $\phi_{\mu,s}$, and is analogous to the case of the L-1 norm described above.\\
\rmk{A formal solution for this problem can be written by studying the limit $\beta \rightarrow \infty$, which is associated with couplings $g_\mu^\star = 0$ for $\theta_\mu \neq 0$. The equation $g_\mu^\star=0$ can be expressed in term of operator averages by using equation (\ref{eq:FormalCoup}) as follows:
\begin{equation}
1 = \prod_s  \left( \frac{1}{|\Omega|} \sum_\nu \langle \phi_\nu^\star \rangle \phi_{\nu,s}  \right)^{\phi_{\mu, s}} \; ,\label{eq:GeneralWick}
\end{equation}
where $\langle \phi_\nu^\star \rangle$ indicates the ensemble average of the operator $\phi_\mu$ under the distribution $p_s^\star$. This result expresses a relation among observables which must hold whenever couplings are zero, which is typically used to express higher order correlations in terms of low order ones, which can be expressed as roots of polynomial equations.
Then equation (\ref{eq:FormalCoup}) can be used to write the remaining couplings, and the roots of equation (\ref{eq:GeneralWick}) can in principle be used to obtain an expression for the non-zero components of $\bg^\star$.
}

\subsubsection{Symmetry properties of the regularizers}
The limit of large $N$ of the regularized complete inverse problem provides an insight on the structure of the regularizers which have been examined in the previous sections. In particular, we can consider the regime in which $N$ is large, while $T$ scales polynomially in $N$ ($T \sim N^\alpha$) so that $T \ll |\Omega|= 2^N $, and provide an argument about the behavior of the regularized inverse problem. Indeed, we will first need to define the notion of symmetric regularizer.
\defin{
Consider the complete inverse problem defined by the model $(\{ \phi_\Gamma\}_{\Gamma \subseteq V} \backslash \phi_\emptyset,\bg)$ and a regularizer $H_0(\bg)$. Then, we call $H_0(\bg)$ a \emph{symmetric} regularizer if for any pair of states $s$ and $s^\prime$ it holds
\beq
\bar p_s = \bar p_{s^\prime} \Rightarrow p_s^\star = p_{s^\prime}^\star
\eeq
}
For example, the L-2 regularizer, the entropy regularizer and the susceptibility regularizer analyzed above are symmetric regularizers. Obviously, the non-regularized problem $H_0(\bg) =0$ is also symmetric.
The following proposition holds for symmetric regularizers.
\prop{Consider the complete inverse problem defined by the model $(\{ \phi_\Gamma\}_{\Gamma \in V}\backslash \phi_\emptyset,\bg)$ and a symmetric regularizer $H_0(\bg)$. Suppose additionally that the empirical probability vector $\bar \bp$ has elements only in $\bar \bp \in \{ 0,1/T \}^{|\Omega|}$. Then the solution of the regularized inverse problem is given by
\beq
g_\mu^\star \propto \bar \phi_\mu \; .
\eeq
}
This result intuitively indicates that symmetric regularizers are unable to distinguish correlations and interactions unless states are sampled more than once. As in the large $N$ regime described above one expects (for well-behaved probability distributions) single states to appear either one or zero times, then this indicates that non-parametric inference procedures should be performed with non-symmetric regularizers in order to extract informative results about interactions. From another perspective, this shows that in the extremely under sampled limit $T \ll |\Omega|$, the more biased couplings are the ones associated with biased empirical averages.
Notice that while in the case of the explicit coupling selection the regularizer is expected not to be symmetric by construction (states are biased according to their overlaps with the explicitly selected operators), it is interesting to see that the L-1 norm breaks the state symmetry without the need of biasing specific operators (example \ref{sec:L1SymmBreak}).
\prf{To prove the above proposition it is sufficient to notice that by symmetry the coupling vector $\bg^\star$ depends on the two values $p_0^\star$ and $p_{1/T}^\star$ associated with the states sampled zero ($\bar p_s = 0$) and once ($\bar p_s = 1/T$). Then equation (\ref{eq:ProbReparam}) implies that
\beqa
g_\mu^\star &=& \frac{1}{|\Omega|} \log \left( \frac{p_{1/T}^\star}{p_0^\star} \right) \sum_{s \in \bar{\mathcal I}} \phi_{\mu, s} +\delta_{\mu 0}  \log p_0^\star \nonumber \\
&=& \frac{T}{|\Omega|} \log \left( \frac{p_{1/T}^\star}{p_0^\star} \right) \bar \phi_\mu +\delta_{\mu 0}  \log p_0^\star \; .
\eeqa
}
\rmk{
The symmetry broken by the L-1 regularizer and by the explicit coupling selection is associated with the following consideration: in principle, unless there is an explicit information that allows to distinguish between states $s$ and $s^\prime$ that are observed the same number of times, then inference should assign the same weight to those states. In the first case such symmetry is spontaneously broken (the information injected by the prior doesn't specifically favor any state), while in the second it is explicitly broken.
}

\subsection{Pairwise model on trees \label{sec:InvPairTree}}
One of the simplest cases in which the pairwise model defined by equation (\ref{eq:IsingModel}) can be explicitly solved is when the topology of the interaction matrix $\bJ$ is the one of a tree. In that case it is well known that \emph{message passing} algorithms \cite{Mezard:2009ko} can find the solution to the direct problem in a time linear in $N$. Indeed, there are several reasons which make the inverse problem worth studying. The first one is the observation that the factorization property (\ref{eq:TreeFactorization}) allows to write an explicit, closed form solution of the inverse problem. The second one is the exceptional stability of the inverse problem with respect to the direct one. Finally, the a full analogy with the complete case can be discussed, and a general scheme for the structure of solutions for inverse problems can be sketched speculating on this simple example.

\defin{
Consider the pairwise model described in section \ref{sec:IsingModel}, defined by the probability density
\beq
p( s) = \frac{1}{Z(\bh,\hat \bJ)} \exp \left( \sum_{i\in V} h_i s_i + \sum_{(i,j) \in E} J_{ij} s_i s_j \right) \; ,
\eeq
in the case in which the set of edges $E$ does not contain any cycle. Then this model is called a \emph{tree} (see appendix \ref{app:TreeFactorization} for a more precise definition).
}
For such models the inverse problem is easy to solve due to the factorization property shown in appendix \ref{app:TreeFactorization}, which allows to write the probability density as
\beq
p(s) = \prod_{(i,j) \in E} p^{\{i,j\}} (s_i,s_j) \prod_{i \in V}  \,[ p^{\{i\}}(s_i) ]^{1-|\partial i |}\; , \label{eq:TreeFactorization}
\eeq
where $\partial i = \{ \phi_{\{i,j\}} \in \bphi  \; | \; (i,j) \in E  \}$. Hence, the entropy can be written as
\beq
S(\bmg,\hat \bc) = \sum_{(i,j) \in E} S^{\{i,j\}}(m_i,m_j,c_{ij}) + \sum_{i \in V} (1-|\partial i|)S^{\{ i\}}(m_i)
\eeq
and the inverse problem can be solved, as shown in the next proposition.
\prop{
For a the pairwise model of the form (\ref{eq:IsingModel}) with a tree topology, the entropy $S(\bmg,\hat \bc)$ can be written as
\beqa
S(\bmg,\hat \bc) &=& \sum_{(i,j) \in E} \sum_{s_i,s_j}\Bigg[  \frac{1}{4}  (1 + m_i s_i + m_j s_j + c_{ij} s_i s_j) \Bigg] \log \Bigg[ \frac{1}{4}(1 + m_i s_i + m_j s_j + c_{ij} s_i s_j) \Bigg] \nonumber \\
&+& \sum_{i \in V} (1-|\partial i |) \sum_{s_i} \Bigg[ \frac{1}{2}  (1 + m_i s_i) \Bigg] \log \Bigg[ \frac{1}{2}(1 + m_i s_i ) \Bigg] \; ,
\eeqa
while the fields $\bh^\star$ and the couplings $\bJ^\star$ result
\beqa
h_i^\star &=& \frac{1}{4} \sum_{j \in \partial i}  \sum_{s_i,s_j}  s_i \log \Bigg[ \frac{1}{4}(1 + m_i s_i + m_j s_j + c_{ij} s_i s_j) \Bigg]  \nonumber \\
&+& \frac{1}{2} (1-|\partial i |) \sum_{s_i}  s_i \log \Bigg[ \frac{1}{2}(1 + m_i s_i ) \Bigg]  \label{eq:InfCoupPairTree} \\
J_{ij}^\star &=& \frac{1}{4} \sum_{s_i,s_j}  s_i s_j \log \Bigg[ \frac{1}{4}(1 + m_i s_i + m_j s_j + c_{ij} s_i s_j) \Bigg]  \; , \nonumber
\eeqa
and the inverse susceptibility matrix $\hat \bchi^{-1}$ is given by
\beqa
\chi^{-1}_{\{i,j\},\{k,l\}} &=& \frac{1}{16} \sum_{s_i,s_j} \frac{ \delta_{i,k} \delta_{j,l} + \delta_{i,l} \delta_{j,k}}
{\bar p^{\{i,j\}}(s_i,s_j)}  \nonumber \\
\chi^{-1}_{\{i,j\},\{k\}} &=&  \frac{1}{16} \sum_{s_i,s_j} \frac{\delta_{i,k} s_j + \delta_{j,k} s_i}{\bar p^{\{i,j\}}(s_i,s_j)} \label{eq:FisherInfoPairTree} \\
\chi^{-1}_{\{i\},\{j\}} &=& \frac{1}{16} \sum_{k \in \partial i} \sum_{s_i s_k} \frac{\delta_{i,j} + s_i s_k \delta_{k,j} }{\bar p^{\{i,k\}}(s_i,s_k)} +
\frac{1}{4} (1-|\partial i|) \sum_{s_i} \frac{\delta_{i,j}}{\bar p^{\{i\}}(s_i)} \nonumber
\eeqa
}
The structure of this solution is reminiscent of the one shown in the case of the complete inverse problem described in section \ref{sec:CompleteInverseProblem}, and can intuitively be understood as follows. To solve an inverse problem it is necessary to find the clusters which allow to express the entropy (in that case all clusters had to be included, while in this case single spin and two spins clusters alone are sufficient to write the full entropy). Couplings are obtained as sums over cluster contributions, in which each of them contributes with a value proportional to the average of the conjugated operator, weighted by the log-probability of each cluster configuration. Inverse generalized susceptibilities quantify the amount of cluster fluctuations, and are large if local fluctuations are rare. \\
The presence of a large number of delta functions is due to the fact that the entropy is built by a small number of cluster contributions, so that the response of the couplings to a shift in the value of the conjugated average is strongly localized: either the perturbation is applied to a neighbor, in whose case the response is finite, or it is zero. This has to be compared with the direct problem, in which a perturbation in the couplings changes the average of a finite number of operators in general. In that case, one roughly expects that
\beq
\chi_{\{i\},\{j\}} \propto e^{|i-j| / \xi} \; ,
\eeq
where $\xi$ is the correlation length of the system. This was first noted in \cite{Cocco:2011fk,Cocco:2012uq}, where it is shown that for a large number of statistical models that the structure of $\hat \bchi$ is dense, while the one of $\hat \bchi^{-1}$ tends to be sparse.

\subsection{One-dimensional periodic chain with arbitrary range couplings \label{sec:Inv1DChain}}
An interesting application of the inference scheme presented in this chapter concerns the solution of the inverse problem for one-dimensional chains. Despite the fact that an exact solution of this problem has been first presented in \cite{Gori:2011ly}, we will be interested in providing a rigorous proof relying on completeness properties. Also in this case a complete analogy with the previous example can be drawn.
Consider a set of  binary spins $s \in \Omega$ and a family of operators of range $R$ (i.e.\ acting on the first $R$ spins) $\bphi (s_1,\dots , s_R) = (\phi_1(s_1,\dots, s_R), \dots, \phi_{M}(s_1,\dots, s_R))$ subject to the periodic boundary conditions $s_i = s_{i+N}$. Then the notion of one-dimensional chain can be introduced through the action of \emph{translation operators} $\bT=\{T_n\}_{n=0}^{N/\rho-1}$, defined through their action on the $\bphi$
\beq
T_n \phi_\mu (s_1, \dots s_R) = \phi_\mu(s_{1+n \rho}, \dots , s_{R+n \rho}) \; , \label{eq:OneDimChain}
\eeq
which corresponds to a shift of the argument of $\bphi$ on the next set of $n \rho$ spins, so that $\rho < R$ is characterized as the periodicity of the chain.

\defin{
A one-dimensional chain is defined as the probability distribution on the space $s \in \Omega$
\beq
p(s) = \frac{1}{Z(\bg)} \exp \left( \sum_{\mu=1}^{M} g_\mu \sum_{n=0}^{N/\rho-1} T_n \phi_\mu (s) \right) \; ,
\eeq
where $\bT$ is a set of translation operators characterized by a periodicity parameter $\rho$ and $\bphi$ is a set of $M$ operators of range $R$.
}
We are interested in solving the inverse problem for this type of system, which means to calculate the entropy $S(\bT \bar \bphi)$ as a function of the empirical averages of the operators $\bT\bphi = \sum_{n=0}^{N/\rho} T_n \bphi$. In order for the entropy to be well-behaved, and in order to exploit the property of completeness (\ref{eq:Completeness}), we need to require a specific choice for the set $\bphi$.
\defin{
A one-dimensional chain defined by a family of operators $\bphi$ and translation operators $\bT$ is \emph{orthogonal} and \emph{complete} if
\begin{itemize}
\item{
For any $m,n \in (0,\dots , N/\rho-1)$, $\sum_s T_n \phi_{\mu,s} T_m \phi_{\nu,s} = \delta_{m,n} \delta_{\mu,\nu}$
}
\item{
For any generic operator $\phi \neq 1$ of range $R$, and any $m \in (0,\dots , N/\rho-1)$, there exist $n$ and $\mu$ such that $T_m \phi = T_n \phi_\mu$.
}
\end{itemize}
}
A possible choice for a family $\bphi$ satisfying those requirements is provided by a suitable choice of monomials. More precisely, one can to define a set $\Gamma_0 = \{1,\dots, R\}$ and a set $\gamma_0 = \{ \rho+1,\dots , R\}$, so that the family of operators $\bphi = \{ \phi_{\Gamma} \}_{\Gamma \subseteq \Gamma_0} \backslash \{ \phi_{\gamma} \}_{\gamma \subseteq \gamma_0}$ describes the $|\bphi| = 2^R(1 - 2^{-\rho})$ monomials belonging to $\Gamma_0$ which are not contained in $\gamma_0$ (appendix \ref{app:1DChain}). Intuitively, this corresponds to define the problem through all operators located inside the unit cell, so that any other operator of range $R$ can be generated in a unique way by using the translation operators $\bT$.

For a one-dimensional chain, it is possible to prove (appendix \ref{app:1DChain}) that the probability density $\bp$ can be factorized as
\beq
p(s) = \prod_{n=0}^{N/\rho-1} \frac{p^{\Gamma_n}(s^{\Gamma_n})}{p^{\gamma_n}(s^{\gamma_n})} \; ,
\eeq
where $\Gamma_n = T_n  \Gamma_0 =\{ 1+n\rho,\dots,R+n\rho\}$ while $\gamma_n =T_n \gamma_0 =\{ 1+(n+1) \rho,\dots,R+n\rho\}$.
Consequently the entropy can be written as
\beq
S(\bT \bar \bphi) = \sum_{n=0}^{N/\rho -1} \left[ S^{\Gamma_n}(\bp^{\Gamma_n}) - S^{\gamma_n}(\bp^{\gamma_n}) \right] \; .
\eeq
This relation, together with equation (\ref{eq:LocalMarginal}) which expresses the locality of marginals,  allows to explicitly find the expression of the entropy of a one-dimensional chain.
\prop{
The inverse problem for an orthogonal, complete one-dimensional chain of monomials has the following solution. The entropy can be expressed as\footnote{
Notice that with abuse of notation we are writing $\bar \phi_\mu$ instead of $\frac{\rho}{N} \sum_n T_n \bar \phi_\mu$.
}
\beqa
S(\bT \bar \bphi) &=& \frac{N}{\rho} \Bigg\{  \sum_{s^{\Gamma_0}} \left[ \frac{1}{2^R} \sum_{\Gamma \in \Gamma_0} \sum_{\mu \in \phi} c_{\mu,\Gamma} \, \bar \phi_\mu \, \phi_{\Gamma,s^{\Gamma_0}}\right]  \log
\left[ \frac{1}{2^R} \sum_{\Gamma \in \Gamma_0} \sum_{\mu \in \phi} c_{\mu,\Gamma} \, \bar \phi_\mu \, \phi_{\Gamma,s^{\Gamma_0}}\right]  \nonumber  \\
&-& \sum_{s^{\gamma_0}}\left[ \frac{1}{2^{R-\rho}} \sum_{\gamma \in \gamma_0} \sum_{\mu \in \phi} c_{\mu,\gamma} \, \bar \phi_\mu \, \phi_{\gamma,s^{\gamma_0}}\right]  \log
\left[ \frac{1}{2^{R-\rho}} \sum_{\gamma \in \gamma_0} \sum_{\mu \in \phi} c_{\mu,\gamma} \, \bar \phi_\mu \, \phi_{\gamma,s^{\gamma_0}}\right] \Bigg\} \; , \nonumber \\
&&
\eeqa
where $c_{\mu,\Gamma} = 1$ if $\exists n $ such that $T_n \phi_\mu = \phi_\Gamma$ and $c_{\mu,\Gamma} = 0$ otherwise. The couplings result
\beqa
g_\mu^\star &= & \sum_{s^{\Gamma_0}} \left[ \frac{1}{2^R} \sum_{\Gamma \in \Gamma_0} c_{\mu,\Gamma}  \, \phi_{\Gamma,s^{\Gamma_0}}\right]  \log
\left[ \frac{1}{2^R} \sum_{\Gamma \in \Gamma_0} \sum_{\nu \in \phi} c_{\nu,\Gamma} \, \bar \phi_\nu \, \phi_{\Gamma,s^{\Gamma_0}}\right]  \nonumber  \\
&-& \sum_{s^{\gamma_0} }\left[ \frac{1}{2^{R-\rho}} \sum_{\gamma \in \gamma_0} c_{\mu,\gamma} \, \phi_{\gamma,s^{\gamma_0}}\right]  \log
\left[ \frac{1}{2^{R-\rho}} \sum_{\gamma \in \gamma_0} \sum_{\nu \in \phi} c_{\nu,\gamma} \, \bar \phi_\nu \, \phi_{\gamma,s^{\gamma_0}}\right] \label{eq:Coupl1DChain}
\eeqa
while the inverse susceptibilities are given by
\beqa
\chi^{-1}_{\mu,\nu} &=&   \frac{\rho}{N} \Bigg\{  \sum_{s^{\Gamma_0}} \frac{
\left[ \frac{1}{2^R} \sum_{\Gamma \in \Gamma_0} c_{\mu,\Gamma}  \, \phi_{\Gamma,s^{\Gamma_0}}\right] 
\left[ \frac{1}{2^R} \sum_{\Gamma \in \Gamma_0} c_{\nu,\Gamma}  \, \phi_{\Gamma,s^{\Gamma_0}}\right] 
}{
\bar p^{\Gamma_0} (\bT \bar \bphi)
} \nonumber \\
&-&
\sum_{s^{\gamma_0}} \frac{
\left[ \frac{1}{2^R} \sum_{\gamma \in \gamma_0} c_{\mu,\gamma}  \, \phi_{\gamma,s^{\gamma_0}}\right] 
\left[ \frac{1}{2^R} \sum_{\gamma \in \gamma_0} c_{\nu,\gamma}  \, \phi_{\gamma,s^{\gamma_0}}\right] 
}{
\bar p^{\gamma_0} (\bT \bar \bphi)
}\Bigg\} \; . \label{eq:Fisher1DChain}
\eeqa
}
Also in this case the structure of the solution is analogous to the one found in section \ref{sec:CompleteInverseProblem} for the complete inverse problem and in section \ref{sec:InvPairTree} for the inverse pairwise tree. The expression of the entropy is a sum of cluster contributions associated with unit cells. Such contributions are all equal due to periodicity, so that two clusters only ($\Gamma^0$ and $\gamma^0$) are sufficient to write the exact expression for the full entropy. The stability of the problem is instead determined by the fluctuations inside $\Gamma^0$ and $\gamma^0$, and divergencies occur whenever any state in $\Gamma^0$ is not observed.
\rmk{
In the case of a one-dimensional chain, the role which in the previous examples was played by of the number of observations $T$ is played by the quantity $T N / \rho$, which measures the number of sampled unit cells. For this type of system even in the case of one single observation ($\,T=1$) the noise on the inferred couplings can be small if the system is large enough. 
}
\rmk{
In order to apply these ideas to empirical data, the information about the one-dimensional nature of the problem should be \emph{a priori} known. Indeed, the exact nature of the interactions needs not to be known, provided that the $R$ parameter is larger than the actual range of the interactions.
}

\section{Applications \label{sec:Applications}}
\subsection{Complete inverse problem}
\begin{figure}[h] %  figure placement: here, top, bottom, or page
   \centering
   \includegraphics[width=4in]{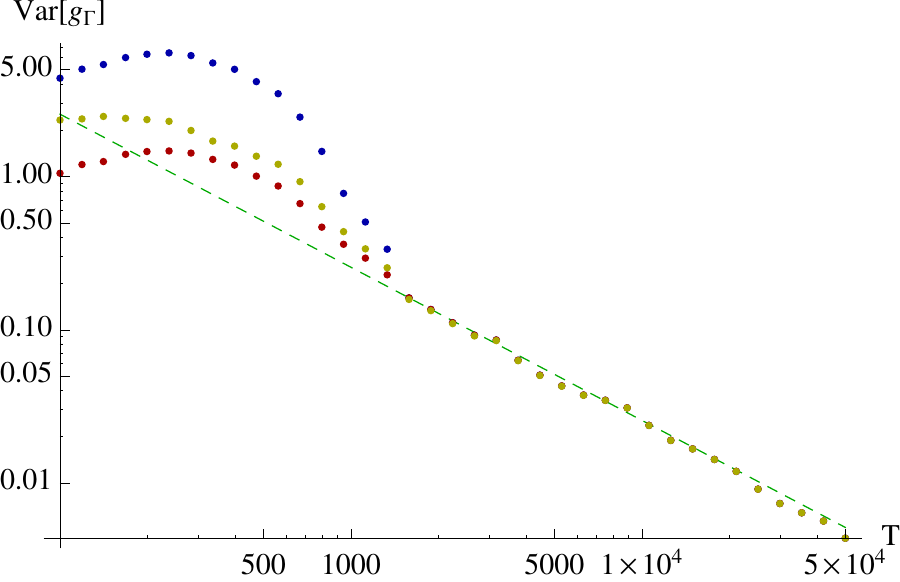} 
   \caption{Variance of the inferred coupling vector as a function of the number of samples $T$ for a flat probability distribution. Un-regularized inference procedure (\ref{eq:FormalCoupling}) corresponds to the blue line, the yellow one indicates an L-2 regularization scheme with $\beta=10$ while the red one is obtained by using a cutoff in the divergencies of the form $p_0 \propto \log \epsilon = - \frac{1}{2} \log T$. The green line corresponds to the expected scaling for the error (\ref{eq:ConvRateCompProb}) in the case of a flat distribution.}
   \label{fig:PureNoiseError}
\end{figure}

\begin{figure}[h] %  figure placement: here, top, bottom, or page
   \centering
   \begin{tabular}{cc}
   \includegraphics[width=2.9in]{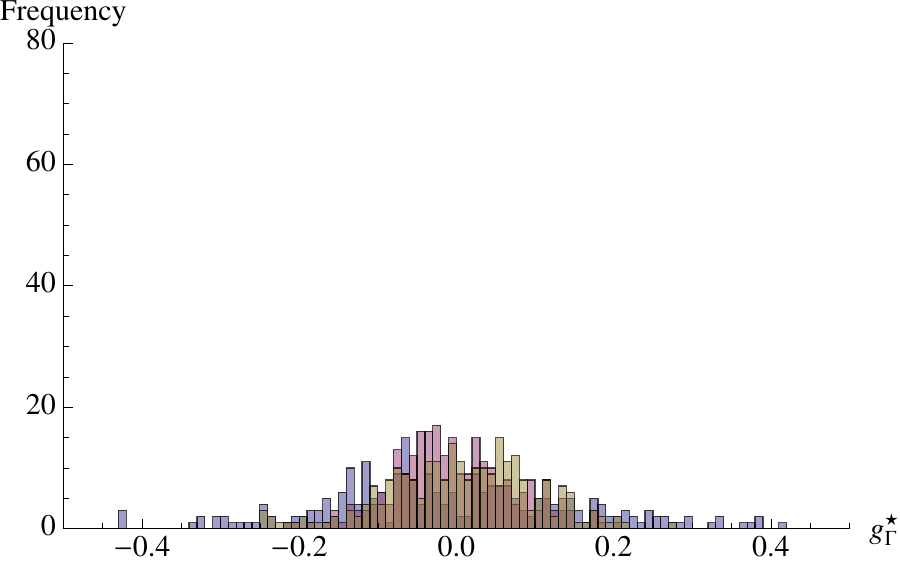}
   \includegraphics[width=2.9in]{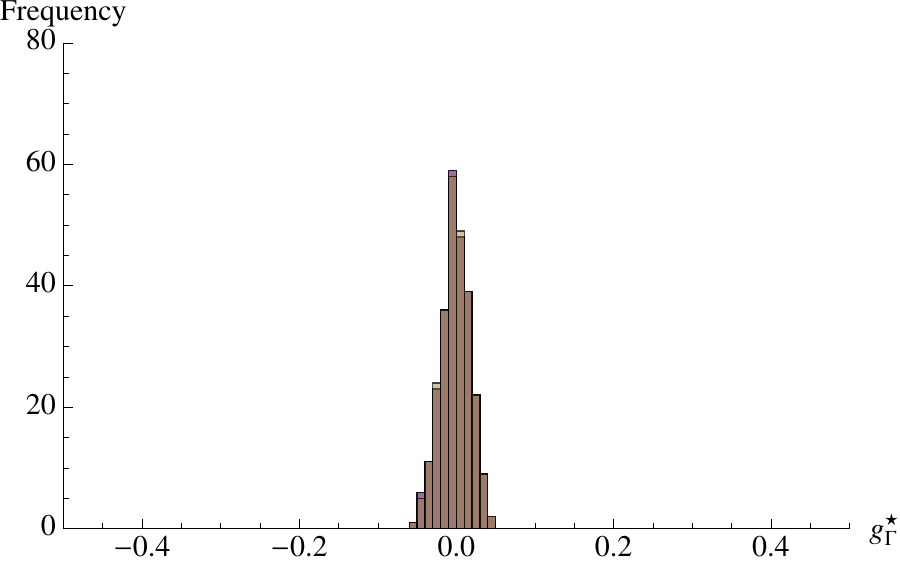}
   \end{tabular}
   \caption{We plot the histogram of the inferred couplings for a complete model with $N=8$ and $g_\Gamma = 0$, hence describing the posterior probability $P_T(\bg | \hat \bs)$ for $T=237$ (left panel) and $T=2657$ (right panel). We employed the same type of regularizers as in figure \ref{fig:PureNoiseError}.}
   \label{fig:PureNoiseHistogram}
\end{figure}

\begin{figure}[h] %  figure placement: here, top, bottom, or page
   \centering
   \begin{tabular}{cc}
   \includegraphics[width=2.9in]{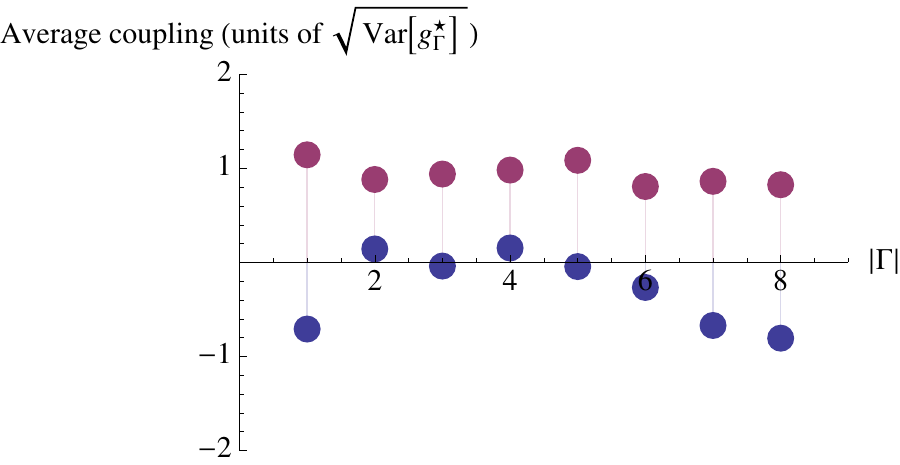}
   \includegraphics[width=2.9in]{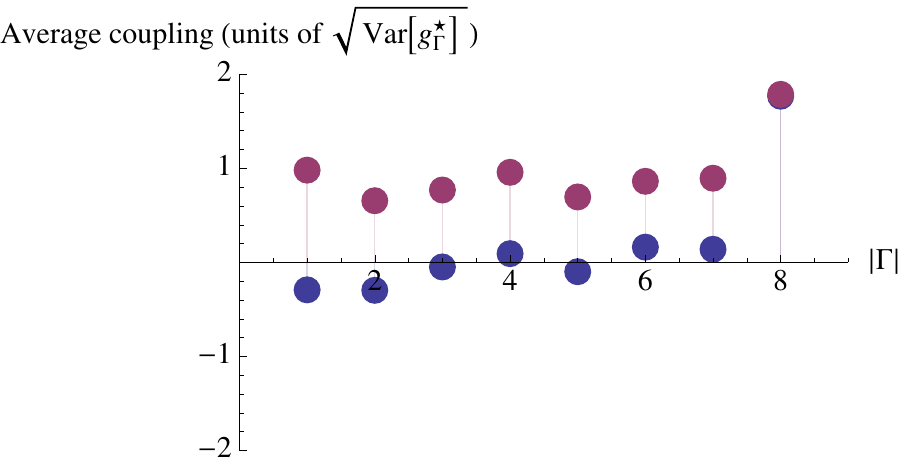}
   \end{tabular}
   \caption{We plot the average mean inferred coupling $\binom{N}{k}^{-1} \sum_{|\Gamma| = k}g_\Gamma^\star$ (blue line) and the average mean absolute coupling (red dots) $\binom{N}{k}^{-1} \sum_{|\Gamma| = k} |g^\star_\Gamma|$ for $T=237$ (left panel) and $T=2657$ (right dots) in units of the error $\sqrt{\textrm{Var}(g_\Gamma^\star)}$. The figure indicates that no specific size for the cluster $\Gamma$ is preferred, as predicted by the expression for the error (\ref{eq:ConvRateCompProb}).}
   \label{fig:PureNoisekBodyDist}
\end{figure}

\begin{figure}[h] %  figure placement: here, top, bottom, or page
   \centering
   \includegraphics[width=4in]{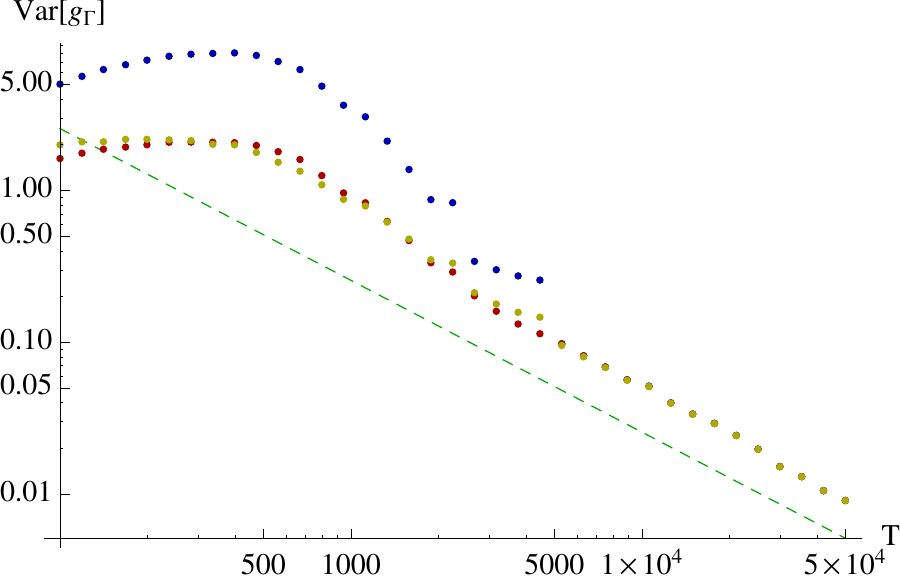} 
   \caption{Variance of the inferred coupling vector as a function of the length of the number of samples $T$, for a pairwise model with $N=8$, $h_i = 0$ and $J_{ij} = 1/N$. See figure \ref{fig:PureNoiseError} for the color convention and the type of regularizers adopted. The green line shows the expected scaling of the variance for a flat distribution, indicating that the reconstruction of a pairwise model is affected by a higher error than the one of a flat distribution.}
   \label{fig:PairwiseError}
\end{figure}

\begin{figure}[h] %  figure placement: here, top, bottom, or page
   \centering
   \includegraphics[width=4in]{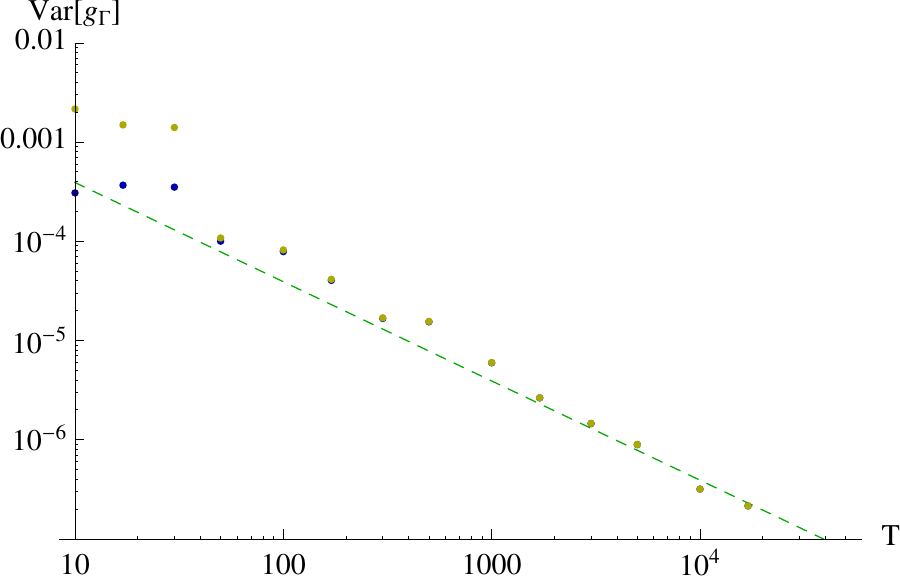} 
   \caption{Variance of the regular (i.e., non-divergent) couplings as a function of the length of the number of samples $T$, for a model with $N=8$ a set of $|\mathcal I | = 16$ observable states. Blue and yellow line respectively denote the non-regularized and the L-2 regularized value of the couplings (with $\beta=5$). The green line shows the expected scaling of the variance for a flat distribution over the set of observable states.}
   \label{fig:HiddenSectorError}
\end{figure}

\begin{figure}[h] %  figure placement: here, top, bottom, or page
   \centering
   \includegraphics[width=4in]{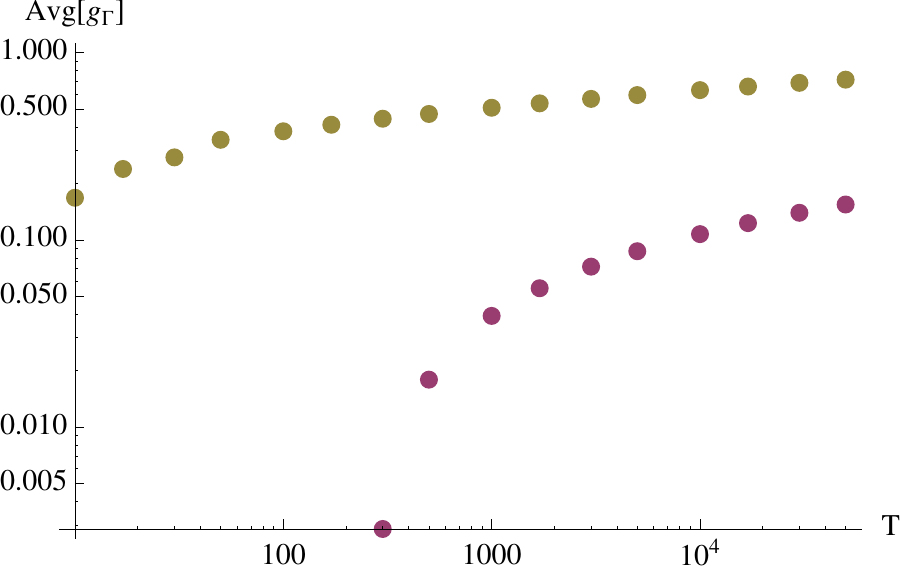} 
   \caption{Divergence with $T$ of the non-regular couplings, for the model described in previous plot. Red and yellow line respectively denote the values obtained putting $\log \epsilon = -\frac{1}{2} \log T$ and using an L-2 regularization ($\beta = 5$). Notice that the divergence is very slow, as it is expected to be logarithmic in $T$.}
   \label{fig:HiddenSectorDiv}
\end{figure}

The techniques shown in section \ref{sec:CompleteInverseProblem} have been tested on synthetic datasets in order to check their performance. As expected, they are suitable for systems in the small $N$ regime, due to the slow convergence in $T$ of the inferred coupling vector $\bg^\star$ to the true coupling vector $\bg$, which can be seen as a consequence of the over fitting problem associated with the presence of an exponential number of couplings. We have considered for simplicity a system of $N=8$ spins, with couplings corresponding to several models, namely:
\begin{enumerate}
\item{
{\bf Pure Noise:} A model with $g_\Gamma=0$ describing the flat distribution $p_s = 1/ |\Omega|$.
}
\item{
{\bf Pairwise model:} A model with two body interactions (i.e., $g_\Gamma = 0$ if $\phi_{\Gamma}$ is such that $|\Gamma| \neq 2$), and couplings equal to $g_\Gamma = 1/N$.
}
\item{
{\bf Arbitrary couplings and hidden sector:} A model with infinite couplings associated to four random operators, in order to test the behavior of the algorithm in presence of divergent couplings.
}
\end{enumerate}
In all those cases, we were able to compute by enumeration the partition function of the model, and to sample from the exact probability distribution a set of $T \in \{100,\dots,50000\}$ states which have been used to construct the vectors of empirical frequencies $\bar \bp$ and empirical averages $\bar \bphi$. Formulas derived in the above sections have been used to solve the inverse problem for those sets of sampled states.
For the case 1.\ of a flat probability distribution, we were able to check formula  (\ref{eq:ConvRateCompProb}) describing the concentration of the couplings towards their expected value $g_\mu = 0$, as shown in figure \ref{fig:PureNoiseError}.
Beyond the naive inference scheme described in section \ref{sec:CompleteInverseProblem}, we have employed an L-2 regularization scheme (yellow line) and a simple cutoff for divergencies of the form $\log \epsilon = - \frac{1}{2} \log T$ (red line). This last prescription is motivated by the simple consideration that for a multinomial distribution the variance on the empirical probabilities scales as $T^{-1}$, so that the error on the sampled probability $p_0$ is expected to be of the order of $T^{-1/2}$. In figure \ref{fig:PureNoiseHistogram}, we plot an histogram of the couplings obtained for various values of $T$ in order to show the shape of the posterior $P_T(\bg | \hat \bs)$. Finally, we show in figure \ref{fig:PureNoisekBodyDist} that there is no cluster size $|\Gamma|$ which dominates the coupling vector for any value of $T$, implying that no model is favored by this inference scheme. 
For the case 2.\ of a pairwise model (section \ref{sec:IsingModel}), we have considered a model with $h_i = 0 \; \forall i$ and $J_{ij} = 1/N \; \forall i < j$. We performed the same analysis and collected the same statistics as in the previous case. In figure \ref{fig:PairwiseError} we plot the variance of the inferred coupling distribution against the number of samples $T$, finding that as indicated by the inequality (\ref{eq:BoundConvRateCompProb}), the pre factor $\frac{1}{|\Omega|^2} \sum_s \frac{1}{p_s}$ controlling the convergence to zero of the errors is higher than for a flat probability distribution.
Also in this case we plot the histogram of the inferred coupling for various values of $T$, comparing the unimodal distribution of couplings in the noise-dominated regime ($T \lesssim 10^3 $) with the bimodal distribution emerging for large sample size ($T \gtrsim 10^3$), in which the shrinking noise peak leaves room for the genuine signal concentrated in $g_\Gamma \approx 1$. The plot of the mean value and the mean absolute value of the couplings with fixed cluster size shows that even in this case no particular cluster size is biased except for $|\Gamma| = 2$.
\begin{figure}[h] %  figure placement: here, top, bottom, or page
   \centering
   \begin{tabular}{cc}
   \includegraphics[width=2.9in]{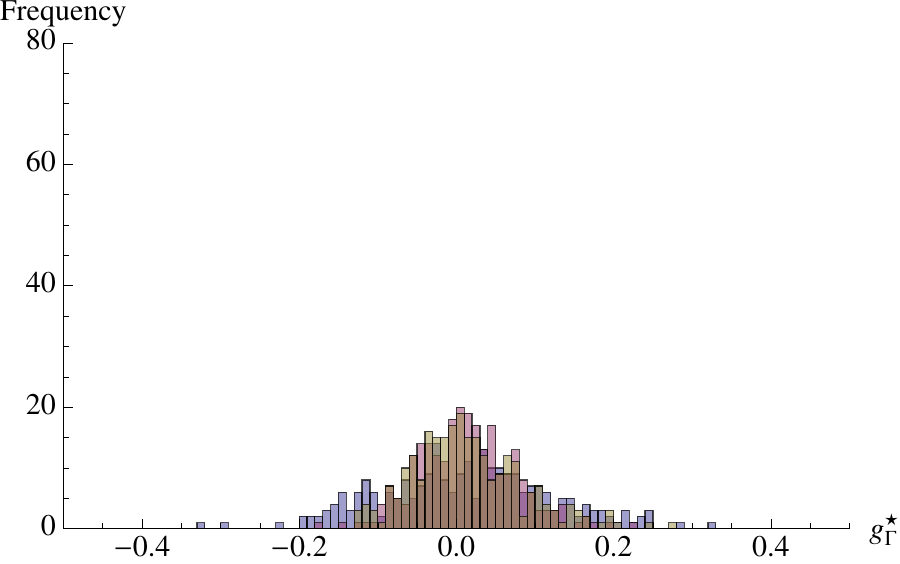}
   \includegraphics[width=2.9in]{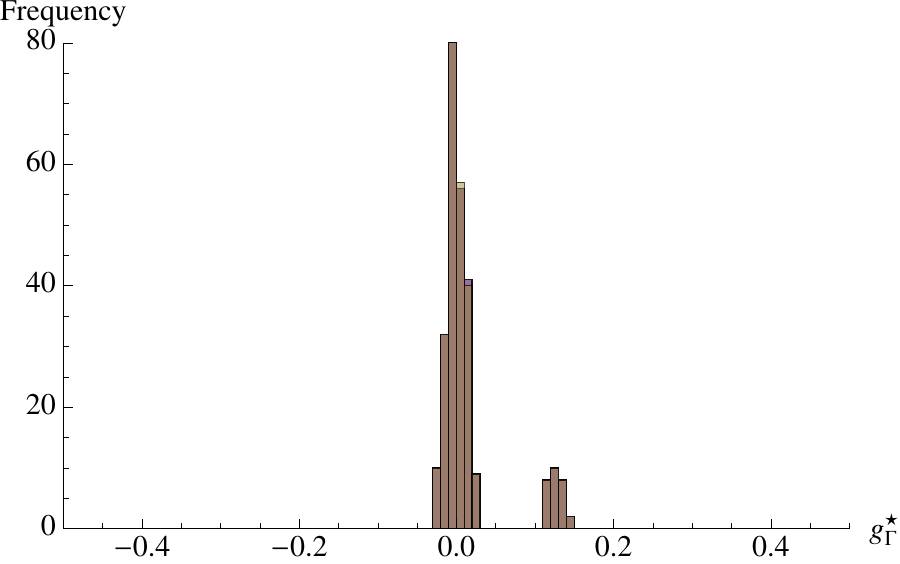}
   \end{tabular}
   \caption{Histogram of the inferred couplings for the pairwise model described in figure \ref{fig:PairwiseError} for $T=1121$ (left panel) and $T=14934$ (right panel), where the color convention is also described. Notice the transition from a unimodal distribution in the noise-dominated regime to the bimodal distribution obtained for large $T$.}
   \label{fig:PairwiseHistogram}
\end{figure}
\begin{figure}[h] %  figure placement: here, top, bottom, or page
   \centering
   \begin{tabular}{cc}
   \includegraphics[width=2.9in]{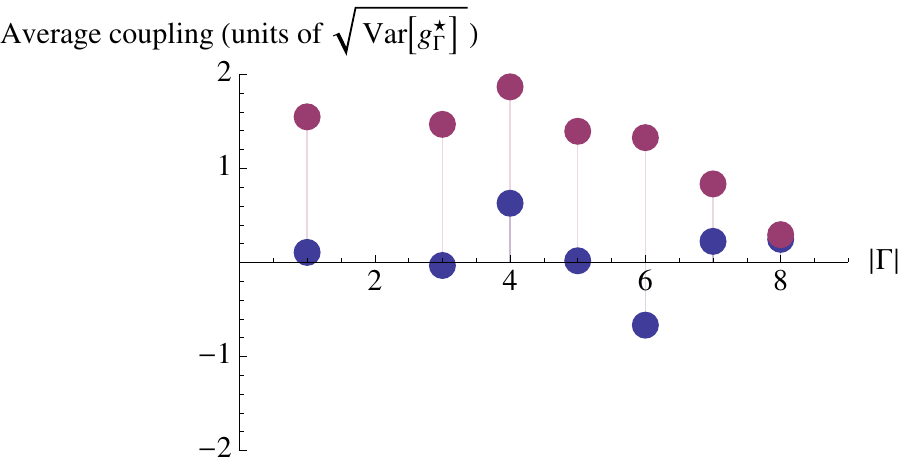}
   \includegraphics[width=2.9in]{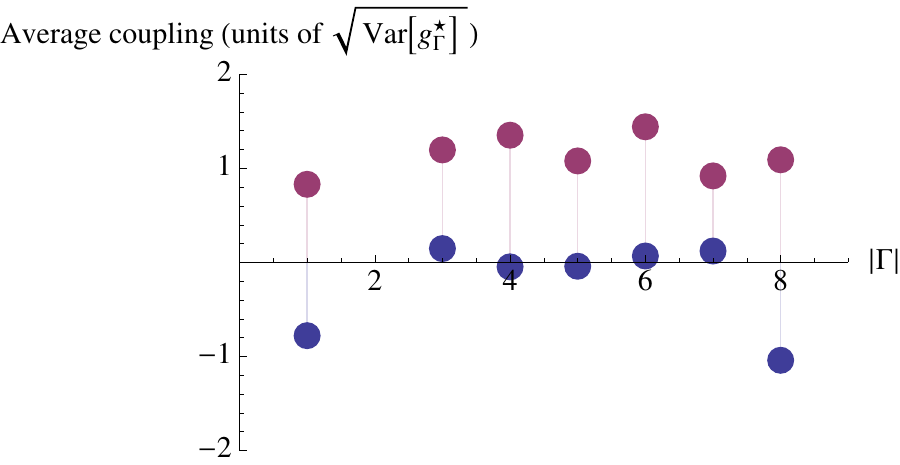}
   \end{tabular}
   \caption{Average mean inferred coupling (blue points) and average mean absolute coupling (red points) for $T=1121$ (left panel) and $T=14934$ (right panel) in units of the error $\sqrt{\textrm{Var}(g_\Gamma^\star)}$. Just clusters with $|\Gamma| = 2$ are favored (and hence out of scale in this plot).}
   \label{fig:PairwisekBodyDist}
\end{figure}
Finally, we show how this procedure might be employed in the case in which one or more couplings are infinite. We consider complete models in which all couplings $g_\Gamma$ are put to zero, but a random set which are set to $g_\Gamma = \infty$. As an illustrative example, we consider the case $g_{\{1\}} = g_{\{7\}} = {\{3,6\}} = {\{1,4,5,7\}} = \infty$, which lead to a set of observable states $\mathcal I$ with $|\mathcal I | = 2^4$, and a set of regular (i.e., non divergent) couplings $\bg^{reg}$ of size $|\bg^{reg}| = 240$. We plot in figure \ref{fig:HiddenSectorError} the variance of the regular inferred couplings to their exact value against the length of the dataset $T$, while in figure \ref{fig:HiddenSectorDiv} we show how non-regular couplings approach infinity.

\subsection{L-1 norm vs L-2 norm: emergence of state symmetry breaking \label{sec:L1SymmBreak}}
\begin{figure}[h] %  figure placement: here, top, bottom, or page
   \centering
   \begin{tabular}{cc}
   \includegraphics[width=2.9in]{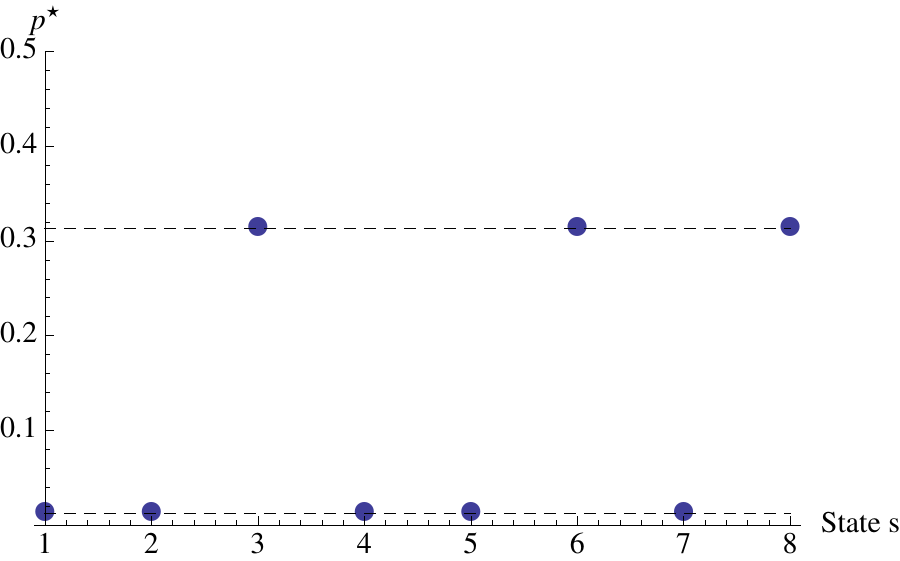}
   \includegraphics[width=2.9in]{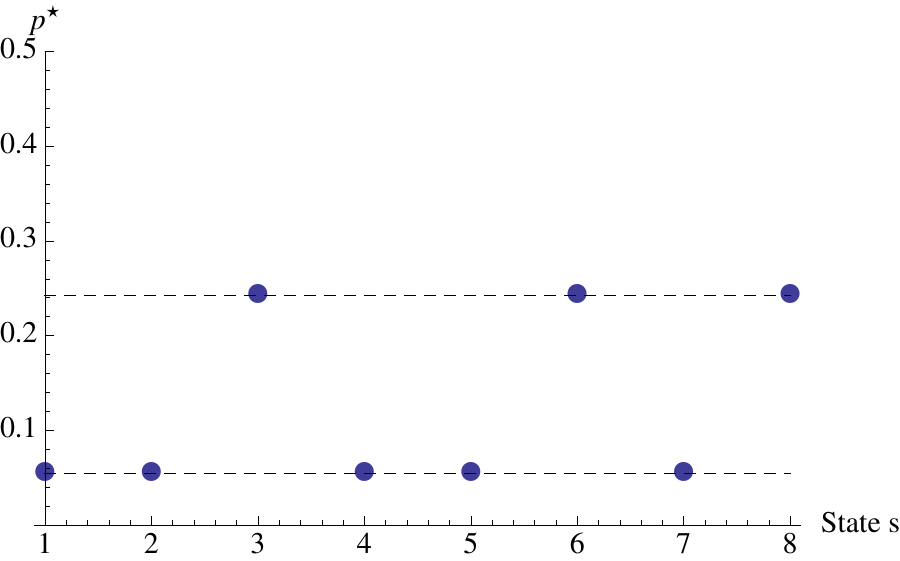}
   \end{tabular}
   \caption{Inferred probability $\bp^\star$ for the L-2 regularized complete inverse problem, in the case $\beta = 0.1$ (left panel) and $\beta = 0.8$ (right panel) in the highly under sampled limit $\bar p_s \in \{ 0,1/T\}$. Equal empirical frequencies $\bar p_s$ are mapped to equal inferred probabilities $p_s^\star$.}
   \label{fig:SymmRegL2Norm}
\end{figure}

\begin{figure}[h] %  figure placement: here, top, bottom, or page
   \centering
   \begin{tabular}{cc}
   \includegraphics[width=2.9in]{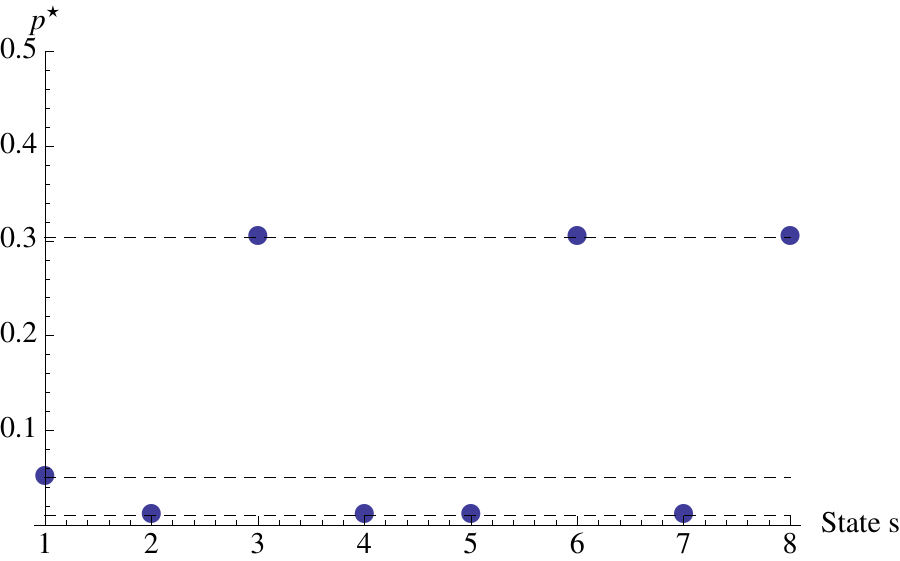}
   \includegraphics[width=2.9in]{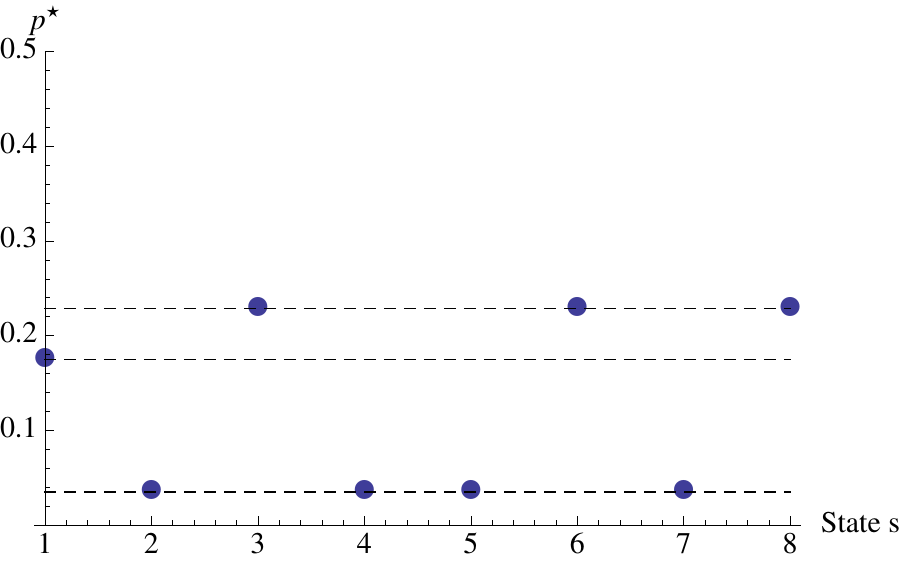}
   \end{tabular}
   \caption{Inferred probability $\bp^\star$ for the L-1 regularized complete inverse problem with $N=3$, in the case $\beta = 0.1$ (left panel) and $\beta = 0.3$ (right panel) in the highly under sampled limit $\bar p_s \in \{ 0,1/T\}$. The state symmetry which associates the same weight to configurations sampled the same number of times is spontaneously broken.}
   \label{fig:SymmRegL1Norm}
\end{figure}

In section \ref{sec:RegCompInvProb} we have defined a notion of symmetry for the regularizers of the complete inverse problem, by saying that a regularizer is \emph{symmetric} if it holds for any pair of states $s,s^\prime$ that $\bar p_s = \bar p_{s^\prime} \Rightarrow p_s^\star = p^\star_{s^\prime}$. We want to show through a very simple example that the L-1 norm is non-symmetric and hence, according to the argument presented in section \ref{sec:RegCompInvProb}, it is not expected to have a trivial limit in the high-dimensional inference regime $T \sim N^\alpha \ll |\Omega|$. To show this, we consider a system of $N=3$ spins, described by a complete model consisting of $|\phi| = 7$ operators and compare the inferred probability $\bp^\star$ obtained by using an L-1 regularization with the one obtained by using an L-2 regularization. To do this, we numerically minimized (see appendix \ref{app:ConvexOpt} for the details) the function
\beq
H(\bg | \hat \bs)  = -T \left( F(\bg) + \sum_{\Gamma \subseteq V \neq \emptyset} g_\Gamma \bar \phi_\Gamma \right) + H_0(\bg)
\eeq
with either $H_0(\bg) = \beta \sum_{\Gamma \subseteq V, \neq \emptyset } |g_\Gamma|$ or $H_0 = \frac{\beta}{2}\sum_{\Gamma \subseteq V \neq \emptyset} g^2_\Gamma$ for respectively the L-1 and the L-2 norm. We assumed the sampled configuration vector to be $\bar p = \frac{1}{3} (\delta_{s,- - -} + \delta_{s,+-+} + \delta_{s,+++})$, in order to deal with only two different values for the empirical probability vector $\bar \bp$. The results obtained in the case of the L-2 norm for the inferred probabilities are shown in figure \ref{fig:SymmRegL2Norm}, where it is possible to appreciate the uniform lifting of non-observed configurations, while probabilities associated with observed states are uniformly decreased as predicted by equation (\ref{eq:L2RegCompInvProb}).
In the case of the L-1 norm (figure \ref{fig:SymmRegL1Norm}) we found that the vector of inferred probabilities can assign three different weights to the inferred state probability vector $\bp^\star$. In particular the configuration corresponding to the non-observed state $(-1,-1,-1)$ is lifted to a non-trivial value which breaks the state symmetry.

\subsection{Pairwise model on a tree}
\begin{figure}[h] %  figure placement: here, top, bottom, or page
   \centering
   \includegraphics[width=4in]{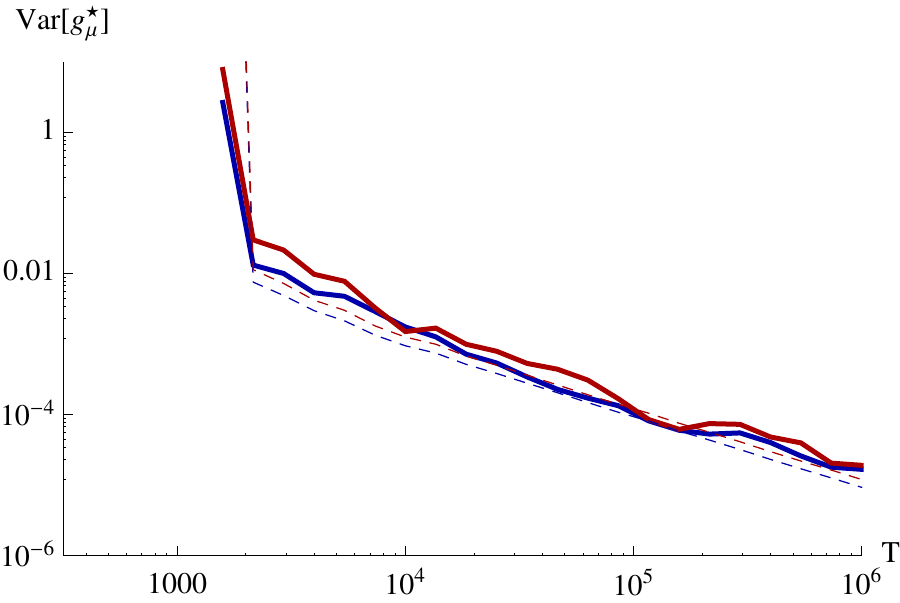} 
   \caption{Variance of the inferred couplings $h$ (red line) and $J$ (blue line) against the number of samples $T$ for a pairwise tree. The dashed lines plotted for reference indicates the error predicted by equation \eqref{eq:FisherInfoPairTree}.}
   \label{fig:TreeError}
\end{figure}

\begin{figure}[h] %  figure placement: here, top, bottom, or page
   \centering
   \begin{tabular}{cc}
   \includegraphics[width=2.9in]{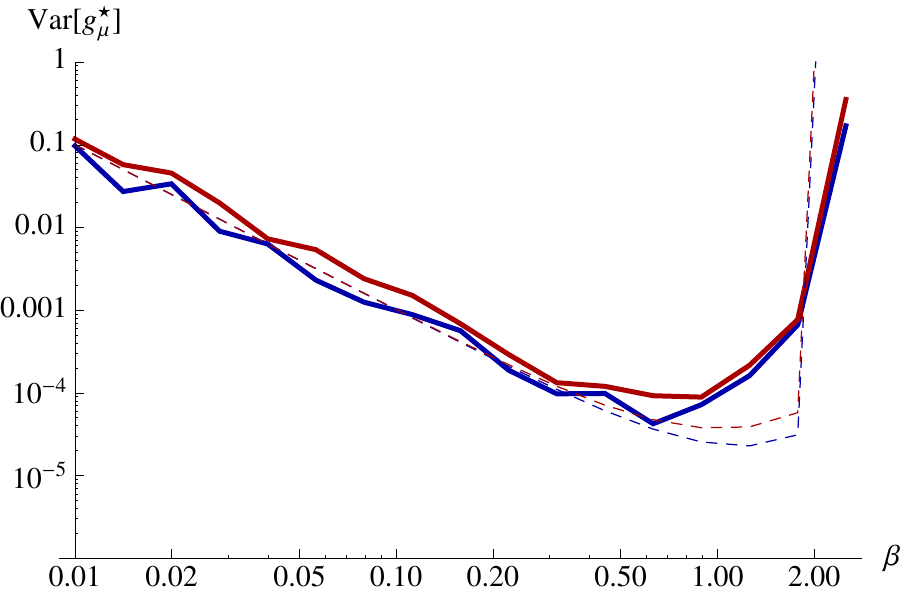}
   \includegraphics[width=2.9in]{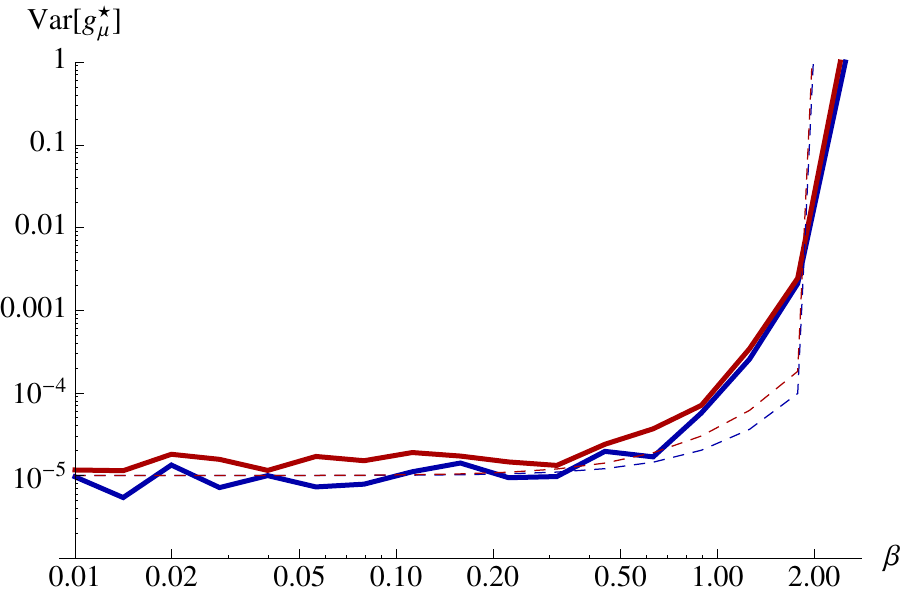} 
   \end{tabular}
   \caption{Variance of the inferred couplings $\bh^\star$ (red line) and $\bJ^\star$ (blue) against the inverse temperature $\beta$ for a pairwise tree, obtained by using $T=10^5$ MonteCarlo samples. We plot both the variance of $\bh^\star$ and $\bJ^\star$ (left panel) and the one of the products $\bh^\star\beta$ and $\bJ^\star \beta$ (right panel), in order to show that this inference procedure cannot discriminate an overall interaction strength from an inverse temperature. The dashed lines indicate the value of the error estimated through equation \eqref{eq:FisherInfoPairTree}.}
   \label{fig:TreeVarBeta}
\end{figure}

We tested the results shown in section \ref{sec:InvPairTree} providing a solution for the inverse problem for pairwise models with tree-like structure. We considered trees of size $N=50$, and studied the behavior of the solution of the inverse problem for samples of length $T$ up to $10^6$. The model which we considered was defined by the couplings $J$ and $h$ randomly and uniformly drawn in the interval $[0,1]$. Datasets that we used did not consist of i.i.d.\ configurations sampled from the exact probability distribution, rather we sampled the states by using a Monte-Carlo simulation of $T$ sweeps with a Metropolis-Hastings algorithm \cite{MacKay:2003vn,Krauth:1998fk}. We selected an initial condition of the form $\{1,\dots, 1\}$ in order to enforce a solution of positive $\bmg$ in case of ergodicity breaking.
Figure \ref{fig:TreeError} shows the variance of the inferred couplings as a function of the length of the time series $T$, comparing it against a reference scaling $1/T$ for a random instance of a problem (i.e., a specific choice of $\bh$ and $\bJ$). We find that formula (\ref{eq:InfCoupPairTree}) correctly predicts the inferred couplings and their scaling to the actual ones. We remark that in this case, errors arise not only due to the finite number of samples, but is also introduced from an imperfect sampling of the empirical averages $m$ and $c$. Indeed, as long as $\langle \bphi \rangle - \bar \bphi \sim T^{-1/2}$, the results obtained display the correct scaling of the variance.
We also considered the case in which we produce a random instance of the problem, and consider all the models obtained by multiplying the couplings with an \emph{inverse temperature} $\beta$ controlling the width of the fluctuations, in order to model the cases in which the noise is enhanced ($\beta$ large) and the one in which it is suppressed ($\beta \to 0$). In particular, we considered a random instance of the model defined by couplings $g$ randomly extracted in $[0,1]$, and multiplied by a parameter $\beta \in [1/2N,1]$, from which we extracted via MonteCarlo a set of $T=10^5$ samples. In figure \ref{fig:TreeVarBeta} we plotted the variance of the inferred coupling against the inverse temperature $\beta$.
This plot shows that it is not possible to discriminate an overall strength of a couplings from a temperature parameter modulating the fluctuations. This implies that the maximum accuracy in inferring the products $\beta \bh$ and $\beta \bJ$ is obtained when fluctuations are maximum ($h_i=J_{ij}=0$), while the maximum accuracy for the inferred vector $(h,J)$ is achieved by finding a compromise between maximum signal (favoring high couplings) and minimum noise (favoring high temperature, or equivalently low $\beta$).
We also studied how the quality of the reconstruction of the couplings degrades  by raising the $\beta$ parameter. We find that within this inference scheme it is possible to reconstruct accurately the couplings as long as local fluctuations are sampled. More precisely, expression (\ref{eq:FisherInfoPairTree}) states that couplings can be accurately reconstructed as long as all the four possible states belonging to clusters of interacting spins $(i,j)$ are well-sampled. This indicates that pushing $\beta$ to large values, the configuration $(s_i,s_j) = (1,1)$ gets more biased, eventually leading to the absence of other states if $T$ is finite. Then, error can be large or divergent as shown in section \ref{sec:CompleteInverseProblem} for the case of the complete inverse problem.
\rmk{
Notice that an accurate reconstruction of the couplings is obtained when local fluctuations (i.e., fluctuations relative to clusters of two spins) are sampled. It is not necessary to probe \emph{global} fluctuations, which indicates that even in a phase in which ergodicity is broken, it is possible to accurately reconstruct the couplings, although no global fluctuations of the empirical average $m = \frac{1}{N} \sum_i m_i$ are observed. This indicates that it is not crossing the \emph{critical point} what degrades the quality of the inference procedure, rather it is the lack of local fluctuations in the empirical samples.
}

\subsection{One-dimensional periodic chain}
\begin{figure}[h] %  figure placement: here, top, bottom, or page
   \centering
   \includegraphics[width=4in]{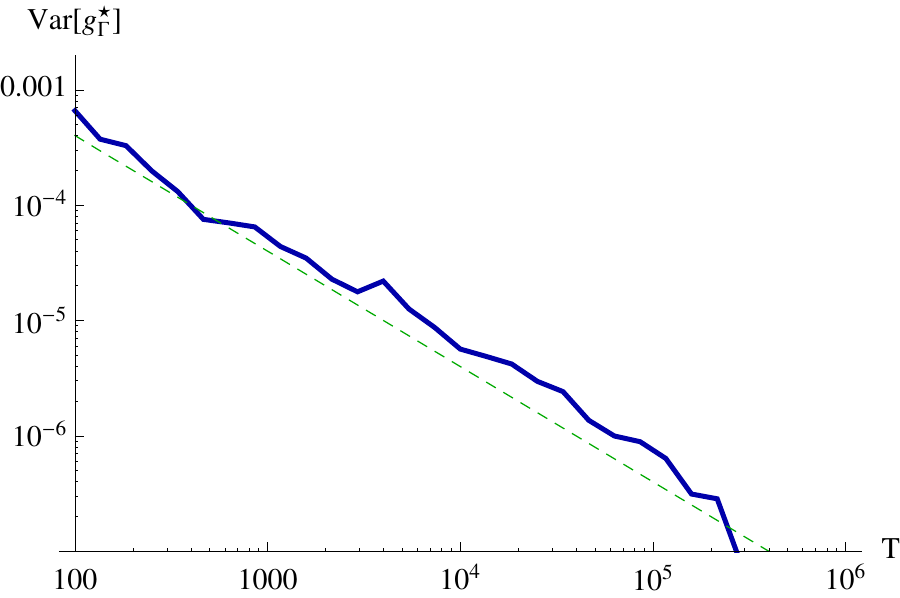}
   \caption{Variance of the inferred coupling vector $\bg^\star$ (blue line) plotted against the number of sampled unit cells $N T/\rho$, obtained by MonteCarlo sampling of a model describing a complete one-dimensional periodic chain of size $N=50$, range $R=4$ and periodicity $\rho=2$. The green dashed line shows the error predicted by equation \eqref{eq:Fisher1DChain}.}
   \label{fig:1DimChainError}
\end{figure}

\begin{figure}[h] %  figure placement: here, top, bottom, or page
   \centering
   \begin{tabular}{cc}
   \includegraphics[width=2.9in]{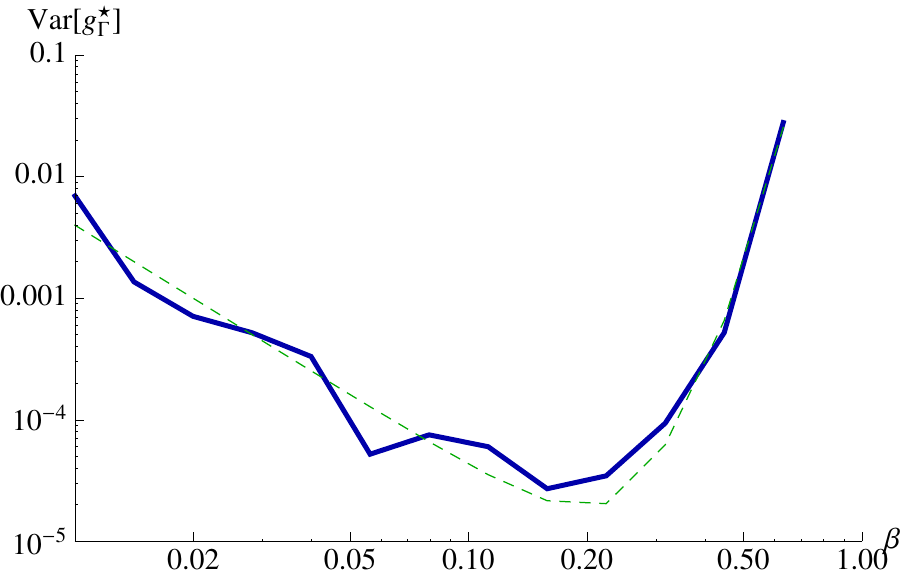}
   \includegraphics[width=2.9in]{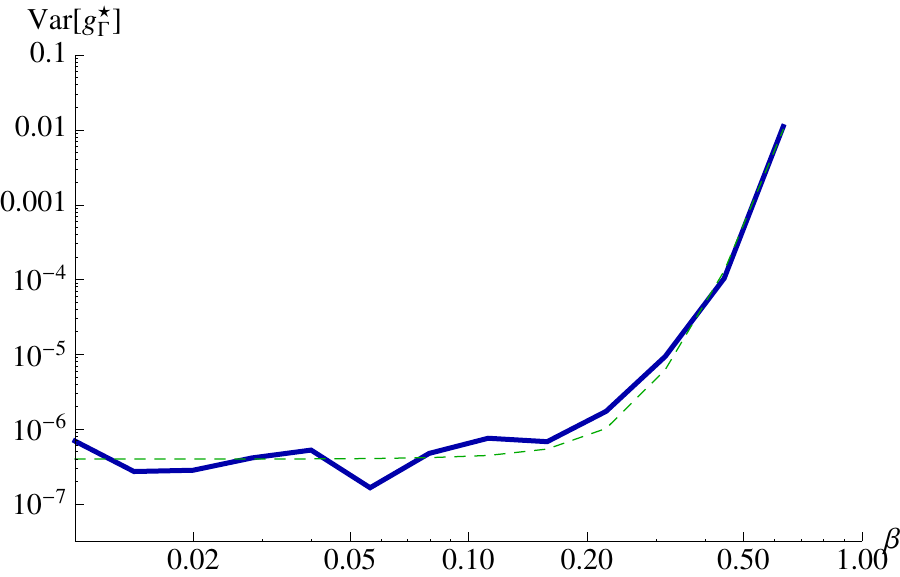} 
   \end{tabular}
   \caption{Variance of the inferred couplings $\bg^\star$ (blue line) against the inverse temperature $\beta$ for a one dimensional periodic chain. We have sampled $10^5$ configurations via MonteCarlo to construct the empirical averages $\bar \bphi$. The left panel shows the results for the inferred couplings $\bg^\star$, while the right one displays the results for the product $\beta \bg^\star$. The dashed lines indicate the estimation of the error obtained through equation \eqref{eq:Fisher1DChain}.}
   \label{fig:1DimChainVarBeta}
\end{figure}

We studied the performance of the inference procedure described in section \ref{sec:Inv1DChain} in inferring the couplings of a one-dimensional periodic chain with arbitrary range interactions. The analysis confirms the validity of the expression (\ref{eq:Coupl1DChain}) for the couplings and (\ref{eq:Fisher1DChain}) for the inverse susceptibilities. As an illustrative example, we consider the case of a periodic complete chain of size $N=50$ with interactions of range $R=4$ and periodicity parameter $\rho =2$. We sampled via MonteCarlo a set of up to $10^6$ configurations for a model in which the couplings $g_\Gamma$ have been randomly and uniformly extracted from the interval $[0,1/2N]$ (see above section for the details about the sampling procedure). The results for the variance of the inferred couplings (a set of $|\bg^\star| = 2^R(1-2^{-\rho})$ values) are represented in figure \ref{fig:1DimChainError}, where we study their dependence on the number of sampled unit cells $N T/\rho$.
As we did above, we studied the behavior of this inference procedure after modulating the interaction strength with an overall inverse temperature parameter $\beta$ controlling the intensity of the fluctuations for a random instance of the model . The results are shown in figure \ref{fig:1DimChainVarBeta}, where we show both the variance for the parameters $\bg^\star$ and the one for the product $\beta \bg^\star$.
Also in this case it is apparent that for a flat distribution ($\beta = 0$) the error on $\beta \bg^\star$ is minimum, while for the parameters $\bg^\star$ the reconstruction error is minimal for a finite value of $\beta$ which optimize the signal-to-noise ratio. We remark that also in this case the quality of the reconstruction of the couplings is determined by the sampling of the configurations belonging to clusters of $R$ spins. If local fluctuations are not sampled well-enough, the error on the inferred couplings is large as predicted by equation (\ref{eq:Fisher1DChain}). As observed above, it is not necessary to probe global fluctuations of the system in order to accurately reconstruct the couplings.

\chapter{Information geometry and criticality \label{ch:Geometry}}
In this chapter we will be interested in studying the natural structure of Riemannian manifold which characterizes the space of probability distributions \cite{Amari:1985ay}. This structure provides a mean to rigorously define a distance between statistical models, which can be used to characterize the consistency of the solution of the inverse problem through the notion of distinguishable distribution \cite{Myung:2000jy}. The metric structure of the coupling space becomes especially interesting in the case of models displaying a critical behavior at large $N$, as it allows for a characterization of (second-order) criticality from the point of view of information theory. In this scenario critical points can be seen as regions of the space of statistical models which are infinitely descriptive, in the sense that any finite region of the coupling space around a critical point can encode an anomalously high number of distinguishable statistical models. We call this phenomenon \emph{model condensation}. An illustrative example is presented by discussing the thermodynamic limit of a fully connected ferromagnet. Finally, we will introduce a model of a stochastic point-process known as Hawkes process which we will use as a toy model to study the features of the inverse problem when applied to a realistic dataset, and compare the results to the ones obtained by studying real data describing financial transactions in a stock market. This will allow to distinguish among spurious and the genuine collective features which emerge from the analysis of empirical data similar to the one considered in \cite{Schneidman:2006vg,Shlens:2006uq,Cocco:2009mb} in the context of neurobiology.

\section{Metric structure of the probability space}

\subsection{Fisher information as a metric \label{sec:FisherMetric}}
Any statistical model $(\bphi, \bg)$ of the form  (\ref{eq:ProbDensity}) defines a probability density $p(s)$ on the configuration space $\Omega$ which is parametrically specified by a coupling vector $\bg$. As such, one can see the space $\mathcal M (\bphi)$ of all the probability densities obtained by varying the coupling vector $\bg \in \mathbb{R}^M$ as an $M$-dimensional, smooth manifold, in which the role of the coordinates is played by the coupling vector $\bg$. The advantage gained by taking this point of view is that the space $\mathcal M(\bphi)$ is no longer associated with any particular parametrization of the probability space, rather it is characterized in term of the densities $\bp$ independently of their functional form.
This is the point of view taken in the field of \emph{information geometry}, in which the geometric properties of the space of probability distributions are inquired by using methods of differential geometry (see \cite{Amari:1985ay,Amari:1987lj} and \cite{Amari:2007fk} for a pedagogical review), which we will briefly present in the following sections. We will be interested in using these methods to answer several questions, namely: (i) is it possible to define a meaningful measure of distance in the space $\mathcal M (\bphi)$?(ii) Is it possible to define a notion of \emph{volume} in such $\mathcal M (\bphi)$? (iii) Can a measure of \emph{complexity} be defined?
We will see that a positive answer to those points can be given by means of the Fisher information matrix.
\defin{
Consider a minimal family $\bphi$ and its corresponding manifold $\mathcal M (\bphi)$. Then its tangent space $\mathcal T (\bphi)$ is equipped by a canonical basis $ (\partial_1, \dots \partial_M)= (\frac{\partial}{\partial g_1}, \dots, \frac{\partial}{\partial g_M})$, and given two tangent vectors\footnote{
It is customary in literature to use superscripts for contravariant tensors and superscripts for covariant ones. We will disregard for simplicity this distinction and use lower indices for any tensor or vector field, as their use will be unambiguous.
} $X = \sum_{\mu=1}^M X_\mu \partial_\mu $ and $Y = \sum_{\mu=1}^M Y_\mu \partial_\mu $ and a point $\bp \in \mathcal T(\bphi)$ one can define the scalar product $\langle  \cdot, \cdot \rangle_\bp : \mathcal T (\bphi) \times \mathcal T (\bphi) \to \mathbb{R}$ as:
\beq
\langle X , Y \rangle_\bp = \sum_{\mu,\nu} \chi_{\mu,\nu} \, X_\mu Y_\nu  \; . \label{eq:FisherMetric}
\eeq
It can be shown (appendix \ref{app:FreeEnConc}) that for any $X,Y \in \mathcal T(\bphi)$ one has
\beqa
\langle X, Y \rangle_\bp &>& 0 \\
\langle X, Y \rangle_\bp &=& \langle Y, X \rangle_\bp
\eeqa
Hence, $\langle  \cdot, \cdot \rangle_\bp$ is a metrics which we define the \emph{Fisher metrics} associated with $\mathcal M(\bphi)$.
}
Notice that the scalar product $\langle X, Y \rangle_\bp$ is independent of the parametrization used to describe the distribution $\bp$ due to the transformation law of $\chi_{\mu,\nu} = \left< \partial_\mu \log p(s) \partial_\nu \log p(s) \right>$. This fact, and the choice of this metric itself, will be intuitively justified in the next section where the notion of distinguishable distribution will be introduced. 
The Fisher metrics allows to define the length of a curve in the space $\mathcal M(\bphi)$.

\defin{
Given a curve $\gamma$, i.e., a one-to-one function $\gamma : [a,b] \subset \mathbb{R} \to \mathcal M(\bphi)$ with components $\bGam = (\gamma_1,\dots \gamma_M)$, we define its \emph{length} as
\beq
\ell (\gamma) = \int_a^b dt \sqrt{ \sum_{\mu,\nu} \frac{d \gamma_\mu}{dt} \frac{d \gamma_\nu}{dt} \chi_{\mu,\nu}} 
\eeq
}
It is easy to show that the length of a curve (i) is independent of the parametrization of $\gamma$, (ii) is independent of the parametrization of $\mathcal M(\bphi)$ (iii) is additive, i.e., given $a<b<c$, $\gamma_1 : [a,b] \to \mathcal M (\bphi)$, $\gamma_2 : [b,c] \to \mathcal M (\bphi)$ and  $\gamma : [a,c] \to \mathcal M(\bphi) $ such that $\gamma(t) = \gamma_1$ if $t < b$ and $=\gamma_2$ if $t \geq b$, one has $\ell (\gamma) = \ell(\gamma_1) + \ell(\gamma_2)$. Finally, a notion of distance $d(\cdot, \cdot) : \mathcal M(\bphi) \times \mathcal M(\bphi) \to \mathbb{R}$ between points in $\mathcal M(\bphi)$ can be defined through
\beq
d(\bp, \bq) = \min_{\gamma \in \gamma(\bp,\bq)} \ell(\gamma)
\eeq
where $\gamma(\bp,\bq)$ denotes the set of curves in $\mathcal M(\bphi)$ starting in $\bp$ and ending in $\bq$.

\defin{The curve
\beq
\gamma^\star = \arg\min_{\gamma \in \gamma(\bp,\bq)} \ell(\gamma)
\eeq
is called a \emph{geodesics}, and its coordinates $\bGam^\star = (\gamma^\star_1, \dots, \gamma^\star_M)$ satisfy the linear differential equation
\beq
\frac{\partial^2 \gamma_\mu}{\partial t^2} + \sum_{\nu,\rho} \Gamma^\mu_{\nu,\rho} \frac{\partial \gamma_\nu}{\partial t}\frac{\partial \gamma_\rho}{\partial t} = 0\; , \label{eq:Geodesics}
\eeq
where the  \emph{Christoffel symbols} $\Gamma^\mu_{\nu,\rho}$ are given by
\beq
\Gamma^\mu_{\nu,\rho} = \frac{1}{2} \chi^{-1}_{\mu,\sigma} \left( \frac{\partial \chi_{\sigma,\nu}}{\partial g_\rho} + \frac{\partial \chi_{\sigma,\rho}}{\partial g_\nu} -  \frac{\partial \chi_{\nu,\rho}}{\partial g_\sigma} \right) 
\eeq
}
In appendix (\ref{app:Geodesics}) we prove this well-known result by explicitly varying the length functional $\ell(\gamma)$.
\prop{
The function $d(\cdot, \cdot): \mathcal M(\bphi) \times \mathcal M(\bphi) \to \mathbb{R}$ satisfies for any $\bp,\bp^\prime,\bp^{\prime \prime} \in \mathcal M (\bphi)$ the following relations: (i) $d(\bp,\bp^\prime) \geq 0$, (ii) $d(\bp,\bp^\prime) = 0$ if and only if $\bp = \bp^\prime$, (iii) $d(\bp,\bp^\prime) = d(\bp^\prime, \bp)$, (iv) $d(\bp,\bp^\prime) \leq d(\bp,\bp^{\prime\prime}) + d(\bp^{\prime\prime},\bp^\prime)$. Hence, it is a proper measure of distance.
}
We will show in the next section that this distance relates to the inverse problem by intuitively counting how many error bars away are two distributions away one from the other, given a fixed experiment length $T$. A related concept is the one of volume, which can be used to quantify the number distributions that cannot be distinguished one from the other on the basis of an experiment of finite length.
\defin{
Given a sub-manifold $\mathcal M \subseteq \mathcal M (\bphi)$, we define the \emph{volume} of $\mathcal M$ as the value
\beq
\mathcal N ( \mathcal M) = \int_{\mathcal M} d\bg \sqrt{\det \hat \bchi} \; ,
\eeq
which can trivially be shown to be invariant under reparametrization of $\bp$. 
}
Finally, we define along the lines of \cite{Myung:2000jy} the \emph{complexity} of a manifold $\mathcal M (\bphi)$ as the integral
\beq
\mathcal N (\mathcal M (\bphi)) = \int_{\mathcal M (\bphi)} d\bg \sqrt{\det \hat \bchi} \; . \label{eq:GeomCompl}
\eeq
The relevance of this measure will be elucidated in section \ref{sec:Complexity}.

\subsection{Sanov theorem and distinguishable distributions \label{sec:SanovDistDist}}
\begin{figure}[h] %  figure placement: here, top, bottom, or page
   \centering
   \includegraphics[width=4in]{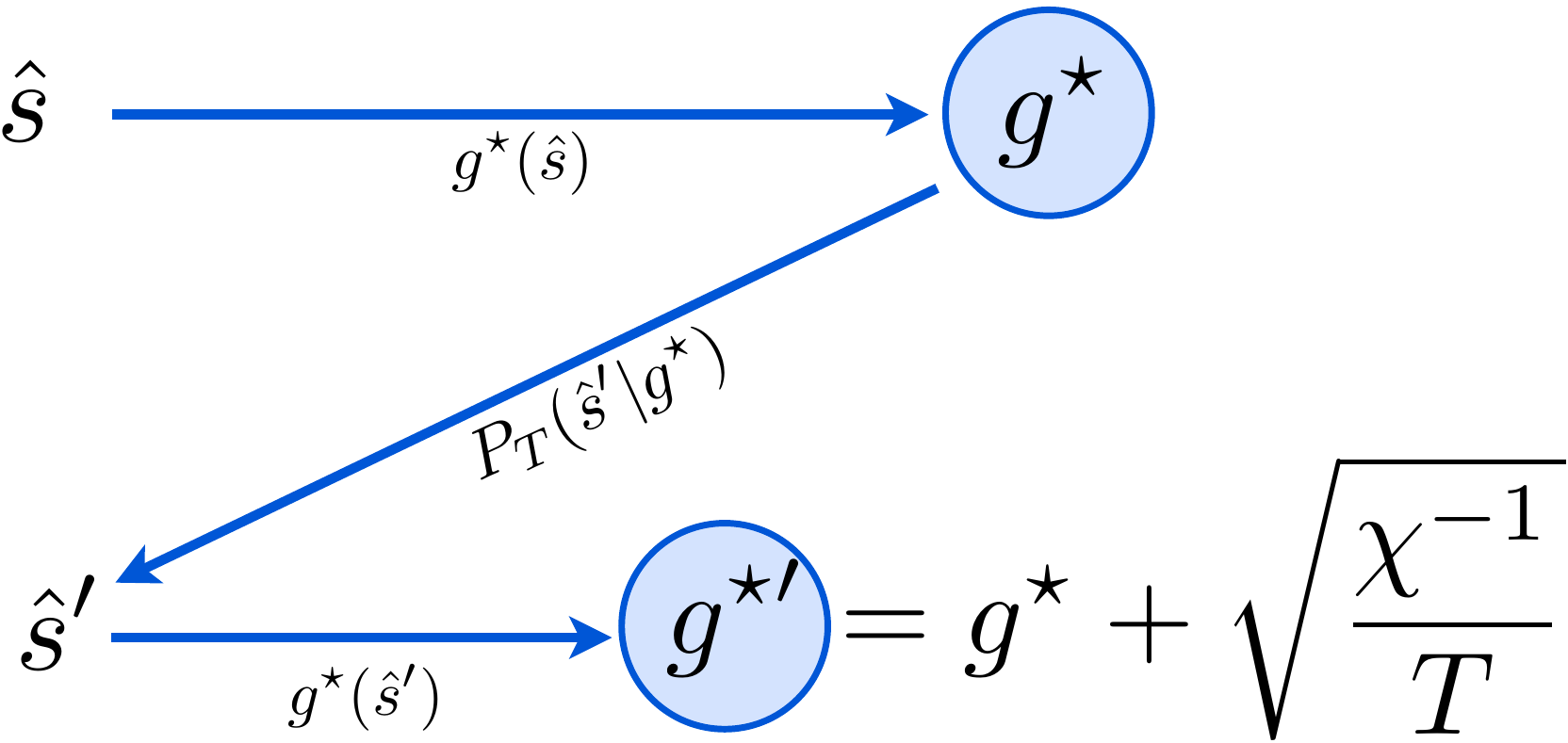} 
   \caption{Cartoon illustrating the notion of indistinguishable distributions.}
   \label{fig:IndDistrib}
\end{figure}

The metric introduced in section \ref{sec:FisherMetric} can be justified by providing an intuitive interpretation in terms of distinguishable distribution, a concept which we will present starting from a simple consistency requirement. Suppose to be given a dataset $\hat \bs$ of length $T$ generated by an underlying (unknown) distribution. Then, given an operator set $\bphi$ it is possible to construct the empirical averages $\bar \bphi$ and to infer the maximum likelihood estimate of the couplings $\bg=\bg^\star (\bar \bphi)$ describing the data, and to use them to generate a \emph{different} dataset $\hat \bs^\prime$ of the same length as $\hat \bs$. The maximum likelihood estimator $\bg^\prime = \bg^\star(\bar{\bphi^\prime})$ of $\hat \bs^\prime$ will, in general, be different from $\bg$. Thus, distributions labeled by $\bg$ and $\bg^{\prime}$ cannot be distinguished on the basis of a dataset of length $T$, as sketched in figure \ref{fig:IndDistrib}.
What one expects is that by increasing $T$, the model $\bg^\prime$ gets closer and closer to $\bg$. This idea can be rigorously formulated by means of Sanov theorem (presented in section \ref{sec:InvSLDev}), which allows to prove the following corollary.
\corol{Consider a statistical model $(\bphi,\bg)$ associated with a probability density $\bp$. Then, given a set of empirical averages $\bar \bphi$ generated by $\bp$ and a maximal likelihood estimator $\bg^\star$, the probability that the maximum likelihood estimator $\bg^\star$ takes a value close to $\bg^\prime$ on the dataset associated with $\bar \bphi$ is given by
\beq
\lim_{\delta \to \bm{0}} \lim_{T\to\infty} -\frac{1}{T} \log \textrm{Prob}(\bg^\star(\bar \bphi) - \bg^\prime \in \delta \bg) = D_{KL}( \bp^\prime || \bp ) \; , \label{eq:SanovCorol}
\eeq
where $\bp^\prime$ is defined by the statistical model $(\bphi,\bg^\prime)$ and $\delta \bg = [-\delta, \delta]^M$.
}
The proof of this corollary is presented in appendix (\ref{app:ExpansionKLIndist}). What it implies is that the Kullback-Leibler divergence controls the probability that after the resampling procedure explained above one ends in a model very different from the starting one. As expected, such probability is exponentially small in $T$.
We will informally rewrite above corollary in the form
\beq
-\frac{1}{T} \log \textrm{Prob}(\bg^\star(\bar \bphi) = \bg^\prime) \xrightarrow[T\to \infty]{} D_{KL}( \bp^\prime || \bp )  \; ,
\eeq
implying a choice of $\delta$ enforcing $\delta \bg$ very close to $\bm{0}$. 
This will allow us to characterize the concept of indistinguishable distribution.
 
\defin{
Consider two models $\bg$ and $\bg^\prime$ within the same family of operators $\bphi$. Then, given a dataset of length $T$ and empirical averages $\bar \bphi$ sampled by the model $(\bphi,\bg)$ and an \emph{accuracy} $\epsilon > 0 $, we say that $\bg$ and $\bg^\prime$ are \emph{indistinguishable} if the maximum likelihood estimator $\bg^\star$ satisfies
\beq
- \log \textrm{Prob}[\bg^\star (\bar \bphi) = \bg^\prime] \leq \epsilon \label{eq:FormalIndistinguishability}
\eeq
}
Given corollary (\ref{eq:SanovCorol}), it is easy to prove (appendix \ref{app:ExpansionKLIndist}) that for large $T$ the distinguishability of two distributions is determined by the generalized susceptibility, as stated in the next proposition.
\prop{Given two models $(\bphi,\bg)$ and $(\bphi,\bg^\prime)$, in the limit of large $T$ they are indistinguishable if
\beq
\frac{1}{2} \left[ (\bg^\prime-\bg)^T \hat \bchi \, (\bg^\prime-\bg) \right] \leq \frac{ \epsilon}{T} \; . \label{eq:Indistinguishability}
\eeq
}
\rmk{Although the notion of indistinguishability inherits asymmetry in $\bg$ and $\bg^\prime$ from the Kullback-Leibler divergence $D_{KL}(\bp || \bp^\prime)$, above proposition shows that for large $T$ the definition  symmetrizes.}
\rmk{
This proposition clarifies the role of the Fisher metric (\ref{eq:FisherMetric}): it shows that the distance among two close-by distributions is proportional to the log-probability that the maximum likelihood estimator of a statistical model $(\bphi, \bg)$ takes value $\bg^\star = \bg^\prime$. From this perspective, it is non-trivial to notice that this result is invariant after reparametrization of the probability densities.
}
This last property identifies an approximatively elliptical region of indistinguishability in the space $\mathcal M (\bphi)$ around each statistical model $(\bphi,\bg)$, whose volume $\mathcal{V}_{T,\epsilon}(\bg)$ can be easily calculated in the large $T$ limit, and is given by
\beq
\mathcal{V}_{T,\epsilon}(\bg) = \frac{1}{\sqrt{\det \hat \bchi}} \left[ \frac{1}{ \Gamma ( \frac{M}{2} + 1)} \left( \frac{2 \pi \epsilon}{T}\right)^{\frac{M}{2}} \right] \, \label{eq:VolIndDist}
\eeq
as shown in appendix \ref{app:VolumeIndist}. Besides displaying the scaling of the volume with $T$ expected by dimensional analysis, equation (\ref{eq:VolIndDist}) shows that the Fisher information controls how wide is each region of indistinguishability inside the space $\mathcal M (\bphi)$. In particular, the more the fluctuations are relevant in a given region  $\mathcal M \subseteq \mathcal M (\bphi)$, the better models in $\mathcal M$ can be discriminated on the basis of a finite length experiment. Finally, the volume $\mathcal V_{T,\epsilon}(\bg)$ allows to define the concept of density of models, and to link it to the metrics described in section \ref{sec:FisherMetric}.
\defin{
Consider the statistical model $(\bphi,\bg)$ and the space of models $\mathcal M(\bphi) $. Then for any fixed $T$ and $\epsilon > 0$ we define the \emph{density of states} $\rho_{T,\epsilon} (\bg)$ as
\beq
\rho_{T,\epsilon} (\bg) = \frac{1}{\mathcal V_{T,\epsilon}(\bg)} \propto \sqrt{\det \hat \bchi} \; .
\eeq
}
For large enough values of $T$, the density of models can be used to count the number of distinguishable models $\mathcal N_{T,\epsilon} (\mathcal M) = \int_{\mathcal M} dg \, \rho_{T,\epsilon}(\bg) \propto \mathcal N(\mathcal M)$ in a region of the space $\mathcal M(\bphi)$. Then the Fisher metrics (\ref{eq:FisherMetric}) has a natural interpretation through the notion of indistinguishable distributions, and the integration measure $\sqrt{\det \hat \bchi}$ induced by the metric $\hat \bchi$ is proportional to the density of models $\rho_{T,\epsilon}(\bg)$.
The notion of distance defined in the previous section also has a simple interpretation in this setting. Consider in fact the discretization of the manifold $\mathcal M (\bphi)$ induced by a sample size $T$ and an accuracy $\epsilon$, in which a curve $\gamma : [a,b] \in \mathbb{R} \to \mathcal M(\bphi)$ is given. Suppose that one is interested in counting the number of ellipsoids (i.e., regions of indistinguishability) crossed by $\gamma$. Then one can see using equation (\ref{eq:Indistinguishability}) that the number of such regions $\ell_{T,\epsilon}(\gamma)$ tends in the large $T$ limit to
\beq
\ell_{T,\epsilon}(\gamma) \left( \frac{2 \epsilon}{T} \right)^{1/2} \xrightarrow[T\to\infty]{} \int_a^b dt \sqrt{\chi_{\mu,\nu} \dot \gamma_\mu(t) \dot \gamma_\nu (t)} = \ell(\gamma) \; .
\eeq
A geodesic is interpreted in this setting as measuring the minimum number of models which have to be crossed to link two probability densities $\bp$ and $\bq$ with a curve $\gamma$, and the corresponding distance $d(\bp, \bq)$ is proportional to such number through the trivial pre factor $(T / 2 \epsilon)^{1/2}$. Summarizing, the link among the notions of length and volume defined in \ref{sec:FisherMetric} and the corresponding notions in the field of statistical learning is provided by the relations
\beqa
\ell_{T,\epsilon} (\gamma) &=& \ell(\gamma) \left( \frac{T}{2 \epsilon} \right)^{1/2} \\
d_{T,\epsilon}(\bp,\bq) &=& d(\bp,\bq) \left( \frac{T}{2 \epsilon} \right)^{1/2} \label{eq:DistanceDistrib} \\
\mathcal N_{T,\epsilon} (\mathcal M) &=& \mathcal N (\mathcal M)  \left[ \Gamma \left( \frac{M}{2} + 1\right) \left( \frac{T}{2 \pi \epsilon}\right)^{\frac{M}{2}} \right]
\eeqa

\subsection{Complexity measures and criticality \label{sec:Complexity}}
One of the most relevant problems in the field of statistical learning is the one of choosing the most appropriate model in order to fit an empirical dataset $\hat \bs$ generated by an unknown distribution. In particular it is well-known that models containing a large number of parameters typically lead to large values for the likelihood function $P_T(\hat \bs | \bg)$, while parsimonious models tend to produce worst in-sample values. Conversely, parsimonious models tend to generalize better, while complex models tend to fit noisy components of data leading to a poor out-of-sample performance. Using a prior function $P_0(\bphi, \bg)$ which keeps into account the complexity of the model itself is a practical strategy which can be used to find an optimal compromise between faithfulness to the data and generalizability of the model. Popular priors used to achieve those goals are:
\begin{itemize}
\item{
{\bf Akaike informetion criterion:} The Akaike information criterion (AIC) can be associated with the choice of a prior which penalizes the number of inferred parameters $M$ through \cite{Akaike:1973uq}
\beq
P_0(\bphi,\bg) = e^{-M} \; ,
\eeq
which leads to the score
\beq
AIC = 2H(\bphi,\bg | \hat \bs) = 2M + 2H_0(\bphi,\bg | \hat \bs) \; .
\eeq
}
\item{
{\bf Bayesian informetion criterion:}  The Bayesian information criterion (BIC) considers a prior of the type \cite{Schwarz:1978kx}
\beq
P_0(\bphi,\bg) = e^{-\frac{M}{2} \log T} \; ,
\eeq
leading to a score of the form
\beq
BIC = 2H(\bphi,\bg | \hat \bs) = M \log T + 2 H_0(\bphi,\bg | \hat \bs) \; ,
\eeq
in which both the number of parameters and the sample size are taken into account. The BIC is closely related to the so-called Minimal Description Length criterion (MDL), in the sense that the score function $H(\bg | \hat \bs)$ is proportional to the one obtained in \cite{Rissanen:1986qz,Rissanen:1984ix} by favoring models which lead to compressible data descriptions. In this sense, the notion of \emph{simplicity} for a statistical model is related to the one compressibility and algorithmic complexity.
}
\end{itemize}
We will show in the following that the above results of information geometry allow to construct a measure of complexity which generalizes the BIC stated above, retaining the main feature of being completely invariant under reparametrization of the model \cite{Myung:2000jy,Balasubramanian:2005oz}. In order to do this, we consider the prior:
\beq
P_0(\bphi,\bg) =  \frac{\sqrt{\det \hat \bchi}}{\mathcal N(\bphi)} \; , \label{eq:JeffreysPrior}
\eeq
where the term $\mathcal N(\bphi)$ is the volume of $\mathcal M(\bphi)$ defined in (\ref{eq:GeomCompl}).
\prop{
Consider the probability for an unknown dataset $\hat \bs$ of length $T$ to belong to a given class of statistical models $\bphi$. Under the prior (\ref{eq:JeffreysPrior}) this is given by
\beq
P(\bphi | \hat \bs) \propto \int_{\mathcal M(\bphi)} d\bg P_{T}(\hat \bs | \bg) \left( \frac{\sqrt{\det \hat \bchi}}{\mathcal N(\bphi)} \right) \; .
\eeq
In the limit $T \to \infty$, this quantity concentrates according to:
\beq
P(\bphi | \hat \bs) \xrightarrow[T\to\infty]{} \left( \frac{P_T(\hat \bs | \bg^\star)}{\mathcal N(\bphi)} \right) \left( \frac{2\pi}{T} \right)^{M/2} \; ,
\eeq
where $\bg^\star$ is the maximum likelihood estimator of $\bg$.
}
The proof of this result is completely analogous to the one shown in appendix \ref{app:ConvInfCoup}, and is obtained through a saddle-point expansion of the likelihood function $P_T(\hat \bs | \bg)$. This result implies that the score assigned to the model $\bphi$ converges to (up to an irrelevant  constant in $\bphi$)
\beq
-\log P(\bphi | \hat \bs)  \xrightarrow[T\to\infty]{} - \log P_T(\hat \bs | \bg^\star) + \frac{M}{2} \log T + \log \mathcal N(\bphi) \; . \label{eq:ScoreGeomCompl}
\eeq
\rmk{The first two terms of the score (\ref{eq:ScoreGeomCompl}) match the ones obtained by considering the BIC. The extra term $\log \mathcal N(\bphi)$ quantifies a geometric contribution to the complexity of the model, which takes into account not only the number of parameters $M$, but also the detailed shape of the manifold $\mathcal M (\bphi)$.
}
Our interest lies in the fact that, assuming that on the basis of dimensional analysis the complexity measure $\log \mathcal N(\bphi)$ scales  like
\beq
\log \mathcal N(\bphi) \sim M \log \ell \; ,
\eeq
where $\ell$ is a characteristic length scale, when high-dimensional models are considered, the scaling of the complexity might be \emph{anomalous}, in the sense that $\ell$ can scale in the limit $N \to \infty $ as a power of $N$. %(we will show for example that for a fully-connected ferromagnet, one has that $\ell \sim 2 \log N$ (TOCHECK))
This argument additionally suggests that models $\bphi$ containing critical points should be penalized by the prior (\ref{eq:JeffreysPrior}), which assigns low scores to complex models. Intuitively, it has to become very costly to describe critical points even if the number of parameters of the model $M$ is not large. \\
More specifically, if one assumes the scaling $\ell \sim N^\alpha$, then it is
\beq
H(\bphi | \hat \bs) = -\log P(\bphi | \hat \bs) \sim H_0(\bphi, \bg^\star | \hat \bs) + M \left( \frac{1}{2} \log T + \alpha \log N \right) \; , \label{eq:ComplexityTotal}
\eeq
where one has the scaling $H_0(\bphi, \bg^\star | \hat \bs) \sim T$.
Then one can intuitively expect that a fixed scaling of $T,N$ and $M$ is required in order for the inverse problem to be meaningful (i.e., the left term side of (\ref{eq:ComplexityTotal}) to dominate the score). Hence, when dealing with high-dimensional inference, avoiding overfitting requires not only to study how $M$ scales with $N$, but also to consider that the geometric properties of the model themselves can play a role through the logarithmic correction in the last term of (\ref{eq:ComplexityTotal}).

\subsection{Examples \label{sec:ExampleGeometry}}
\subsubsection{The independent spin case}
Consider the independent spin model described in section \ref{sec:IndepSpins}. By using equation (\ref{eq:GenSuscIndepSpin}) it is possible to find that
\beq
\det \hat \bchi = \prod_{i\in V} \cosh^{-2} h_i \; .
\eeq
Hence, the number of distinguishable independent spin models which can be described in an experiment of final length $T$ with accuracy $\epsilon$ is
\beq
\mathcal N_{T,\epsilon} = \int d \bh \, \rho_{T,\epsilon}(\bh) =  \left[ \Gamma \left( \frac{N}{2} +1 \right) \left( \frac{\pi \, T}{- 2 \log \epsilon}\right)^{\frac{N}{2}} \right]  \; ,
\eeq
so that for example, just $\mathcal N_{T,\epsilon} \approx 5$ distinguishable models can be described by means of $T=100$ observations of $N=1$ spin with an accuracy of $e^{-\epsilon} =1 \%$, while for $T=1000$ and $e^{-\epsilon} = 10 \% $ one gets $\mathcal N_{T,\epsilon} \approx 23 $.
The finiteness of $\mathcal N_{T,\epsilon}$ also implies that infinite regions of the $\bh \in \mathbb{R}^N$ space belong to the same distinguishable distribution. This can easily be checked, and one can see for example that for $N=1$ the condition
\beq
1 = \int_{-\infty}^{h_{min}} d h \rho_{T,\epsilon}(h)
\eeq
implies that for $T=100$ and $e^{-\epsilon} = 1 \%$ all models with $h$ smaller than $h_{min} \approx - 1.16$ (or $h$ larger than $h_{max} = - h_{min}$) belong to the same region of indistinguishability.

\subsubsection{Fully connected ferromagnet}
\begin{figure}[h] %  figure placement: here, top, bottom, or page
   \centering
   \includegraphics[width=4in]{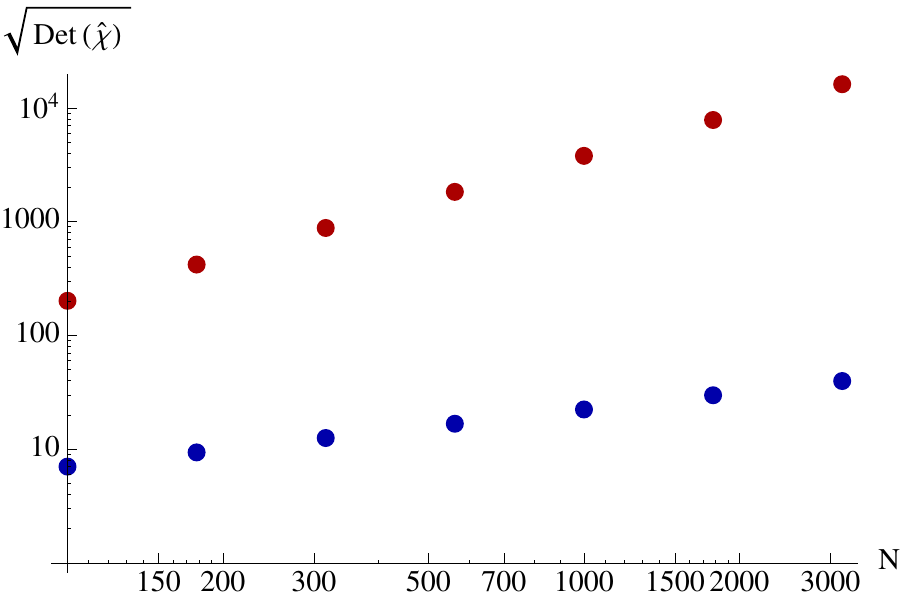}
   \caption{Finite size scaling of the measure $\sqrt{\det \hat \bchi}$ for a fully connected ferromagnet computed via exact enumeration. The value obtained for the models $(h,J) = (0,1)$ (red points) and $(h,J) = (0,0)$ (blue points) are plotted.}
   \label{fig:FinSizeScalVol}
\end{figure}

\begin{figure}[h] %  figure placement: here, top, bottom, or page
   \centering
   \begin{tabular}{cc}
   \includegraphics[width=2.5in]{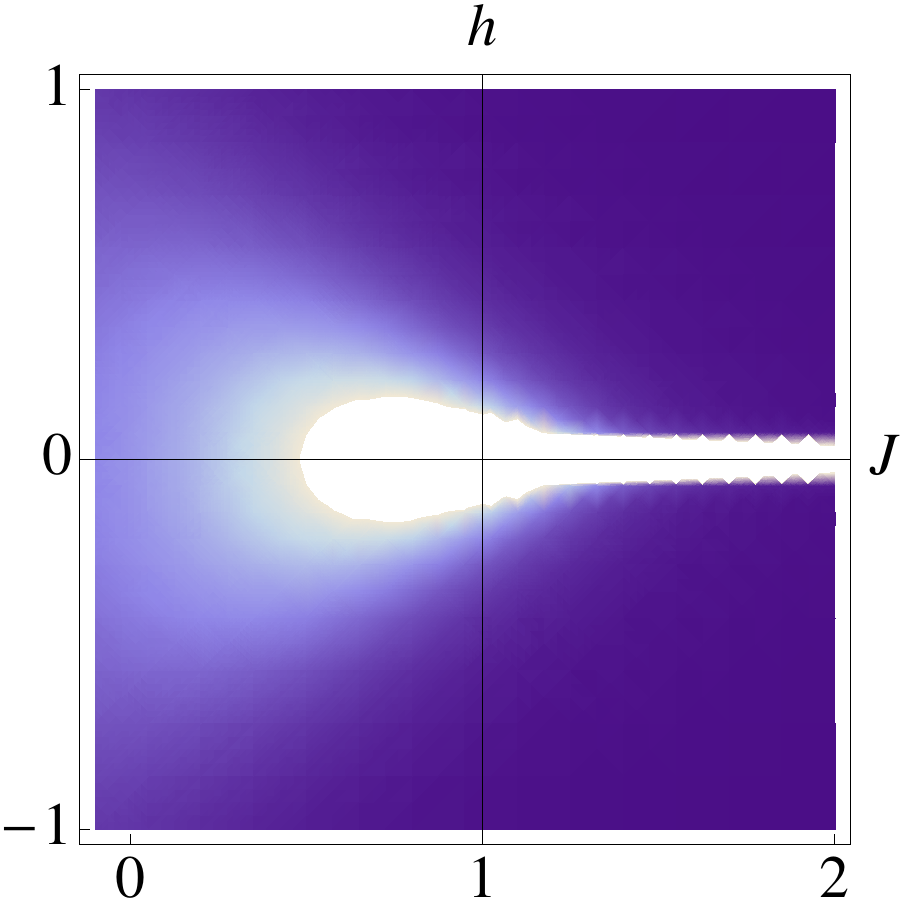}
   \includegraphics[width=2.5in]{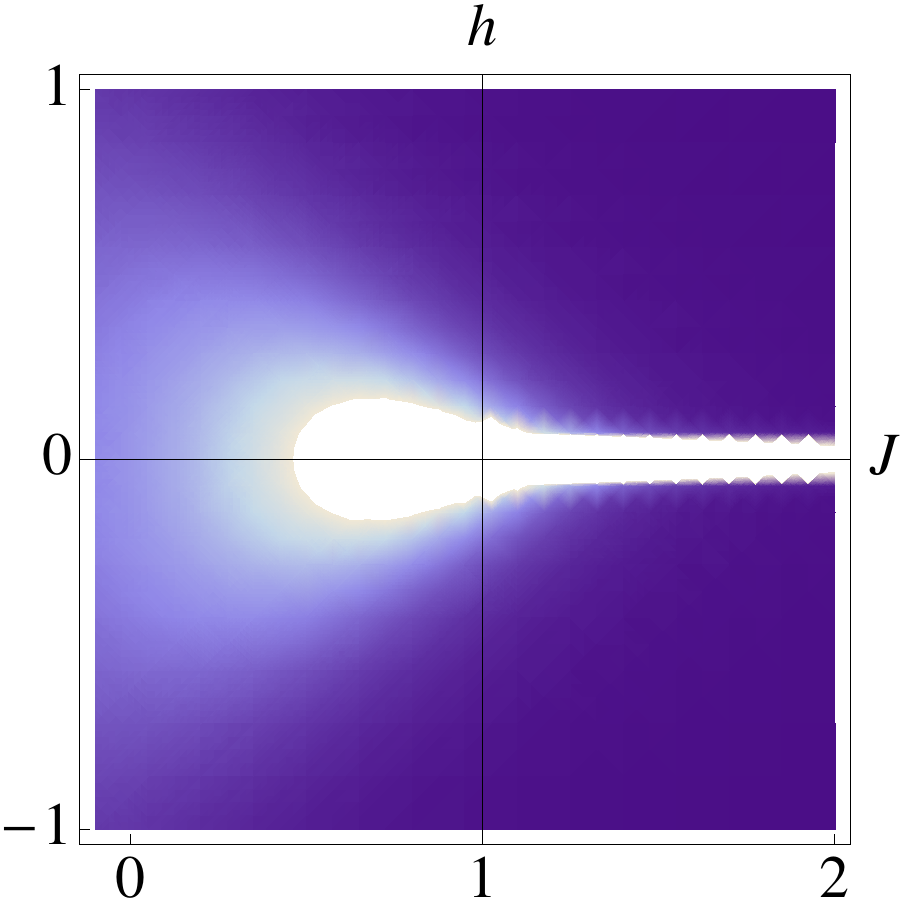}
   \end{tabular}
   \caption{Density of models $\rho(h,J) \propto \det \hat \bchi$ for the fully connected ferromagnet. The left panel shows the exact value calculated for $N=100$, while the right panel displays the saddle-point approximation described in appendix \ref{app:SPFerromagnet}.}
   \label{fig:FCDensityModels}
\end{figure}

\begin{figure}[h] %  figure placement: here, top, bottom, or page
   \centering
   \includegraphics[width=3in]{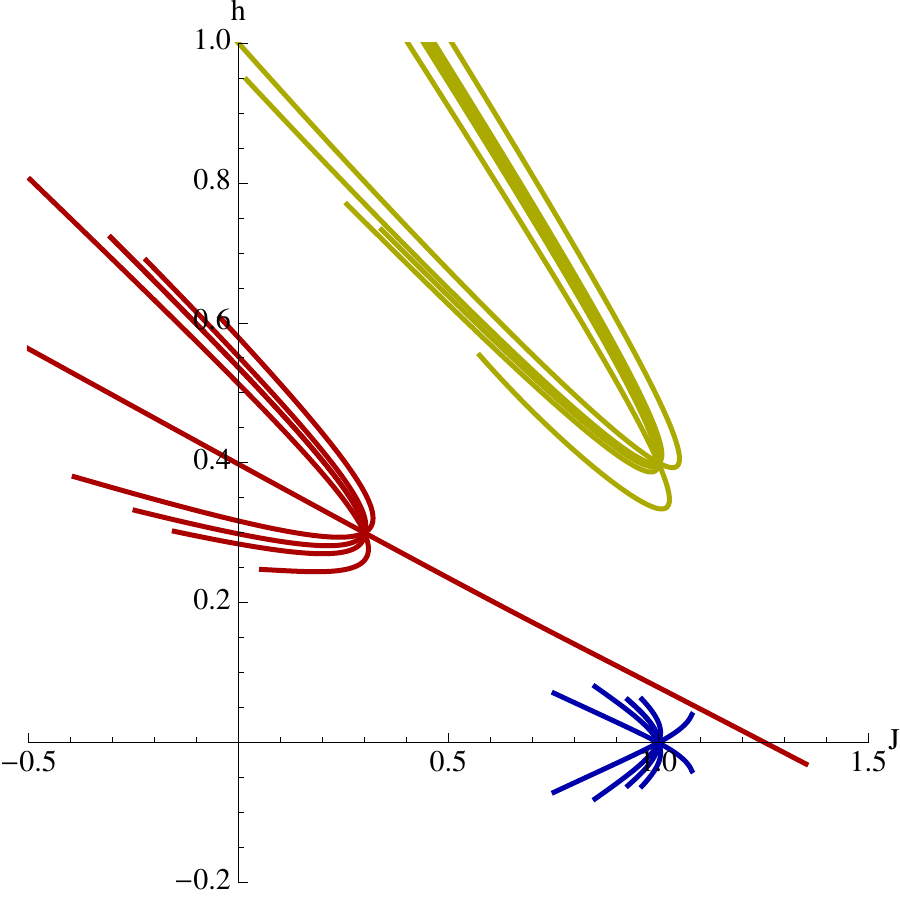}
   \caption{Three sets of geodesics of length $\ell(\gamma) = 1$ plotted for a system of size $N=50$. Blue (respectively, red, yellow) lines describe curves passing through the points $(h,J) = (0,1), (0.3,0.3), (0.4,1)$. It is possible to notice how volume shrinks around the critical point (0,1) and the presence of a quasi-null mode of $\hat \bchi$ along the direction $(-m,1)$.}
   \label{fig:FCGeodesics}
\end{figure}

Let's consider the fully connected ferromagnet described in section \ref{sec:FCFerromagnet}. In that case the calculation of $\det \hat \bchi$ is non-trivial, and requires an analysis of the finite $N$ corrections to the saddle-point solution of the model presented in appendix \ref{app:SPFerromagnet}, where it is shown that to leading order in $N$ one has
\beq
\sqrt{\det \hat \bchi} = \sqrt{\frac{N}{2}} \left( \chi_{s.p.}^{3/2} + \delta(h) \theta (J-1) \sqrt{2 \pi^2 m_{s.p.}^2 \chi_{s.p.}} \right) \label{eq:MeasureFC}
\eeq
Also in this case it is possible to count the number of distinguishable models in a given region of space by explicitly integrating this measure. For example, we can calculate $\mathcal N_{T,\epsilon}$ in the semiplane $J \geq J_{max} \gg 1$ stripped of the $h=0$ line. In that case it results that
\beq
\det \hat \bchi \approx \sqrt{N} \,\left( 4\, \sqrt{2} \, e^{-3 (J + |h|)} \right) \; ,
\eeq
which implies that in such region $\mathcal N_{T,\epsilon} \approx T \sqrt{N} \left( \frac{4 \sqrt{2}}{-9 \pi \log \epsilon} \right) e^{-3 J_{max}}$. This indicates that no $J \gtrsim J_{max} \sim \frac{1}{3} \log T + \frac{1}{6} \log N$ can be discriminated by $J_{max}$ unless $h \approx 0$. Interestingly, the number of models contained in the critical line $h \approx 1/N$ dominates $\mathcal N_{T,\epsilon}$ in the semiplane $J > J_{max}$. In fact the term of (\ref{eq:MeasureFC}) proportional to $\delta(h)$ contributes with
\beq
\int_{-\infty}^{+\infty} d h \det \hat \bchi \approx  \sqrt{N}  \left( 2 \pi e^{-J}\right) \; ,
\eeq
so that keeping into account the transition line one gets $\mathcal N_{T,\epsilon} \approx T \sqrt{N} \left( \frac{1}{- \log \epsilon} \right) e^{- J_{max}} $, and values of $J$ which cannot be discriminated by $J_{max}$ are the ones for which  $J \gtrsim \log T + \frac{1}{2} \log N$.
Finally, one can notice that $\chi_{s.p.}$ is divergent for $(h,J)=(0,1)$. In particular the analysis of $\sqrt{\det \hat \bchi}$ shows that along the line $J=1$, the divergence is of the type $\sqrt{\det \hat \bchi} \propto |h|^{-1}$, while for $h = 0$ and $J < 1$, one has $\sqrt{\det{\hat \bchi}} \propto |1-J|^{-3/2}$. Both divergencies are non-integrable, implying that the number of distinguishable models contained in a finite region around the point $(0,1)$ dominates the total volume of the coupling space. This singularity is smeared out by finite-size effects when $N < \infty$, indeed those characteristics emerge by studying the scaling for finite $N$ of the volume $\mathcal N$, as shown in figure \ref{fig:FinSizeScalVol}.
We plot in figure \ref{fig:FCDensityModels} the density of distinguishable models for this model in the case $N=100$, computed both by exact enumeration and via saddle point approximation.
The geodesics for this model can also be numerically computed by solving the differential equation (\ref{eq:Geodesics}) explicitly. As an example, we plot in figure \ref{fig:FCGeodesics} a set of geodesics of length $\ell(\gamma) = 1$ calculated for a system of size $N=50$.

\section{Inference of a non-equilibrium model}
Many recent works in the field of neurobiology focus on neuronal ensembles which are described by means of strings of binary variables encoding the activity pattern of a set of $N \sim 10^1$ or $N\sim10^2$ neurons \cite{Schneidman:2006vg,Shlens:2006uq,Cocco:2009mb}. Such compact description of the fundamental units of those system has been argued to be meaningful, triggering the expectation that techniques such as the ones described in chapter \ref{ch:Inference} might be applied on empirical data in order to extract relevant information about the interaction patterns of networks of real neurons. As a result of those expectations, striking features of neural ensembles started to emerge from the solution of the inverse problems applied to experimental data \cite{Tkacik2006:mm,Mora:2011kq,Stephens:2008cg}. These findings posed a challenging question, whose answer has yet to be fully clarified in order to assess their validity, namely: \emph{how much of those emerging features depends on the inference procedure which has been applied, and how much is intrinsically associated with structural properties of the system}? The implications of the answer go well beyond the field of neurobiology, and apply more generally to the field of statistical learning.
In this section we want to provide a partial answer to this point, and show that procedures similar to the ones used to study such neural networks may generate spurious features in the inferred models, as well as genuine ones. We address in particular the issue of \emph{criticality}, which we identify from the point of view of statistical mechanics with the presence of long-range correlations in a system as a result of strong collective interactions among its constituents. We apply those ideas to two datasets whose nature is similar to the one considered in \cite{Schneidman:2006vg,Shlens:2006uq,Cocco:2009mb}: a set of simulated realizations of a Hawkes point-process \cite{Hawkes:1971nq,Hawkes:1971lc} and a dataset describing transactions in a financial market.

\subsection{The Hawkes process}
We will introduce the Hawkes point-process as a null-model to describe a system consisting of $N$ interacting units which are able produce \emph{events} in time and cross-influence each other in \emph{absence} of remarkable collective behaviors (i.e., the emergence of long-range correlations in time or space). The study of the discretized version of this model will allow an analysis of the genuine and the spurious features of the inferred model under the procedure described in chapter \ref{ch:Inference}.
\subsubsection{Definition and basic properties}
We will briefly remind the notion of point process, which we will use to construct the Hawkes process, addressing the reader to \cite{Bauwens:2009ys,Bowsher:2002jw} for a more detailed description.
\defin{
We consider an $N$-variate point-process described by a non-decreasing, right-continuous \emph{counting function} $\bX = (X_1,\dots, X_N): [0,\infty)\subset \mathbb{R} \to \mathbb{N}^N$, such that
\beqa
 \frac{\textrm{Prob} (dX_i (\tau) = 1| \bX (\tau^\prime) (\tau^\prime < \tau))}{d\tau} &\xrightarrow[d\tau \to 0]{} & \lambda_i(\tau | \bX (\tau^\prime) (\tau^\prime < \tau)) \\
 \frac{\textrm{Prob} (dX_i (\tau) > 1| \bX (\tau^\prime) (\tau^\prime < \tau))}{d\tau} &\xrightarrow[d\tau \to 0]{} & 0  \; ,
\eeqa
where $dX_i (\tau) = X_i (\tau+ d\tau) - X_i (\tau)$, and the (possibly stochastic) value $\bLam (\tau) = (\lambda_1(\tau), \dots , \lambda_N( \tau))$ is referred as the \emph{conditional intensity} (or more simply, \emph{intensity}) function.
}
Intuitively, $X_i(\tau)$ counts the number of \emph{events} of type $i$ falling in the interval $[0, \tau]$, being the probability of the occurrence of an event in $(\tau, \tau+d\tau]$ equal to $ \lambda_i(\tau) d\tau$, and being the one associated with the outcome of two events of order smaller than $d\tau$. A well-known example is provided by the Poisson process, which is a point-processes specified by a constant, deterministic value for the intensity $\lambda_i(\tau) = \mu_i$. Finally, we will say that a point-process is (asymptotically, weak-sense) \emph{stationary} if the mean $\mathbb{E}[d \bX(\tau)] \xrightarrow[d\tau\to 0]{\tau \to \infty} \bLam(\tau) \, d\tau$ is independent of $\tau$ and the covariance $\textrm{Cov}(dX_i(\tau),dX_j(\tau^\prime)) \xrightarrow[d\tau\to 0]{\tau \to \infty} \sigma_{ij}(\tau,\tau^\prime) \, d\tau^2$ depends just upon the difference $\tau - \tau^\prime$.
\defin{
We will call a \emph{Hawkes point-process} the stationary, $N$-variate point-process $\bX(\tau) = (X_1(\tau), \dots X_N(\tau))$  defined by a stochastic intensity vector $\bLam(\tau) = (\lambda_1(\tau), \dots,\lambda_N(\tau))$ of the form
\beq
\lambda_i(\tau) = \mu_i + \sum_{j=1}^N \int_{-\infty}^{\tau} dX_j (\tau^\prime) K_{ij}(\tau - \tau^\prime) \; , \label{eq:HawkesDynamics}
\eeq
such that $\hat \bK(\tau)$ is a positive matrix kernel satisfying
\beqa
K_{ij}(\tau) &\geq& 0 \qquad \textrm{if } \tau \geq 0 \\
K_{ij}(\tau) &=& 0 \qquad \textrm{if } \tau < 0 \\
\max_{n} |K_n| &<& 1 \label{eq:HawkesStationarity} \; ,
\eeqa
where  $\{K_n\}_{n=1}^N$ are the eigenvalues of the Fourier transform $\hat \bK(\omega) = \int d \omega e^{i \omega \tau} \hat \bK(\tau)$ calculated in the point $\omega=0$, so that condition (\ref{eq:HawkesStationarity}) ensures the stationarity of the process (\ref{eq:HawkesDynamics}).
}
This model describes a self-excitatory process (i.e., $\textrm{Cov}(dX_i(\tau),dX_j(\tau^\prime)) \geq 0 $) due to the positive, linear coupling of the stochastic intensities $\lambda_i(\tau)$ with the process itself. The interest in this model resides in the fact that it can  describe \emph{clustering} of events: just as non-interacting (i.e., Poisson) point-processes describe events which occur at times uniformly drawn from the time axis, Hawkes point-processes model events which tend to take place in close-by regions in time due to an attractive interaction modeled by the kernel $\hat \bK(\tau)$.

We focus on the properties of this model in the stationary regime, which is guaranteed to exist for the choice of the spectral radius of the kernel $\hat \bK(\tau)$ that we specified through (\ref{eq:HawkesStationarity}). Despite the fact that both averages and two point correlations of $\bX(\tau)$ can be analytically computed for a large class of functions $\hat \bK(\tau)$ \cite{Hawkes:1971nq,Hawkes:1971lc}, in the following discussion we will just require the knowledge of the average intensity $\bLam = \expect{\bLam(\tau)}$.
\prop{
Given a stationary Hawkes point-process, the average intensity vector $\bLam$ is given by
\beq
\bLam = \left( \hat \bDl - \hat \bK(\omega=0) \right)^{-1} \bMu \; , \label{eq:AvgHwk}
\eeq
as one can easily see by taking the expectation value of equation (\ref{eq:HawkesDynamics}) and imposing the stationarity condition $\bLam(\tau) = \bLam$.
}
We employ the notation $\expect{\dots}$ to indicate an average taken in the stationary state of the model, and $\hat \bDl$ denotes the identity matrix in dimension $N$. \\
We want to highlight some of the features of the Hawkes process which differentiate it from statistical models such as the ones described in section \ref{sec:DirectProblem}.
\begin{itemize}
\item{
{\bf Dynamics:} The Hawkes process describes a stochastic process characterized by the dynamics (\ref{eq:HawkesDynamics}), while a statistical model $(\bphi, \bg)$ of the form (\ref{eq:ProbDensity}) describes a stationary probability density. This implies that any information  concerning the directionality in time (e.g., causality) of the interactions is lost when passing to a description in terms of i.i.d.\ binary strings.\footnote{This can be understood by noting that any dataset $\pi [\hat \bs] = \{ s^{(\pi_t)}\}_{t=1}^T$ obtained by applying any permutation $\pi_t$ to a raw dataset   $\hat \bs = \{ s^{(t)}\}_{t=1}^T$ leads to the same inverse problem.}
}
\item{
{\bf Non-stationarity:} For any non-stationary generalization of the Hawkes process in which the kernel changes in time (i.e., it is of the form $\hat \bK(\tau,\tau^\prime)$), or the exogenous intensity is a function $\bMu(\tau)$, it is likely that inferring a stationary model may lead to errors in the interpretation of the results. In particular, what is described as an interacting, stationary system in the language of the inferred model $p(s) \propto \exp \left( J \sum_{i<j} s_i s_j + h \sum_i s_i  \right)$ may correspond to a non-interacting, non-stationary real system \cite{Tyrcha:2012fk}.
}
\item{
{\bf Criticality:} The divergence of the mean intensity $\expect{\bLam(\tau)}$ doesn't indicate criticality of the statistical model describing the stationary state of the Hawkes process. In particular, the divergence of $\bLam(\tau)$ is not linked to collective effects, as it is present even for finite $N$, while a proper phase transition in the statistical mechanics sense can arise just in the large $N$ limit. 
}
\end{itemize}
These considerations also apply when considering the binary encoding of the stochastic process describing spiking neurons, or more generally when considering any point-process which is binned and discretized in order to perform an inference procedure such as the one described in chapter \ref{ch:Inference}.

\subsubsection{The fully-connected Hawkes process}
We introduce here the notion of fully-connected Hawkes process, which we will relate to the fully-connected pairwise model in the following part of the discussion.
\defin{
Consider an $N$-dimensional Hawkes point-process, whose intensity vector $\bLam(\tau)$ is defined by
\beq
\lambda_i (\tau) = \mu_i + \sum_j \int_{-\infty}^t dX_j (\tau^\prime) \; \alpha_{ij} e^{- \beta (\tau-\tau^\prime)} \; ,
\eeq
which corresponds to the choice of an exponentially decaying influence kernel $K_{ij}(\tau-\tau^\prime) = \alpha_{ij} e^{-\beta (\tau-\tau^\prime)} \theta(\tau-\tau^\prime)$. Let $\hat \bAl$ be a matrix of the form
\beq
\alpha_{ij} = \frac{\alpha}{N-1} (1 - \delta_{ij})
\eeq
and the vector $\bMu$ to be equal to $\mu_i = \mu$ for each $i$. Then such process will be called a \emph{fully-connected Hawkes process}.
}
For a fully-connected Hawkes process, it is easy to see by employing formula (\ref{eq:AvgHwk}) that
\beq
\expect{\lambda_i(\tau)} = \mu \left( 1 - \frac{\alpha}{\beta} \right)^{-1} \; ,
\eeq
while the stationarity condition (\ref{eq:HawkesStationarity}) reduces to $\alpha < \beta$.

\subsubsection{Binning and discretization}
In order to establish a connection between a spin system and a Hawkes process, we consider a discretization in time and a binarization of the signal dealt according to the following procedure.

\defin{
Given a realization of an $N$-dimensional Hawkes process described by a counting function $\bX(\tau)$ with $\tau \in [0,\tau_{max}]$ and a \emph{bin size} $\delta \tau$, we define for any $i \in \{1,\dots, N\}$ and $t \in \{1 , \dots, \tau_{max} / \delta \tau = T \}$ the \emph{binning functions}
\beq
b_i^{(t)}(\bX,\delta \tau) = \min \left\{ 1, X_i(t \, \delta \tau ) - X_i (t \, \delta \tau -\delta \tau)\right\}  \label{eq:BinningB}
\eeq
which is $1$ is any event of type $i$ occurred in the interval $\tau \in \delta \tau \, [t-1,t]$ and zero otherwise.
We analogously define the functions
\beq
s_i^{(t)}(\bX,\delta \tau) =  2 b_i^{(t)}(\bX,\delta \tau) - 1 \; , \label{eq:BinningS}
\eeq
which evaluate to $1$ if an event of type $i$ occurred in the interval $\tau \in \delta \tau \, [t-1,t]$ and to $-1$ otherwise.
}
In order to shorten the notation, we will often write $b^{(t)}_i = b_i^{(t)}(\bX,\delta \tau)$ and $s_i^{(t)}(\bX,\delta \tau) = s_i^{(t)}$. Those functions provide a mean to map an Hawkes process to an empirical dataset $\hat \bs$ through $( \bX,\delta \tau) \to \hat \bs = \{ s_i^{(t)}(\bX,\delta \tau) \}_{t=1}^T$. Notice that an empirical dataset $\hat \bs$ constructed according to this procedure does not consist in general of i.i.d.\ observations.

\subsection{Trades in a financial market}
Financial markets are complex systems in which a large number of individuals interacts by buying and selling contracts at variable prices according to an unknown, dynamically varying set of criteria (e.g., their specific needs, their past experience, their future expectations). In this sense, markets can be seen as intermediary entities implicitly defined by a set of trading rules which mediate the interactions of individuals. Those rules should be such that efficient allocation of resources is achieved, so that price of traded goods reflects correct information about their fundamental value \cite{Fama:1970ly}. Evidence that this is not always the case has dramatically emerged in recent times \cite{Bouchaud:2009ct,Kirman:2010fk,Trichet:2010la}. Part of the responsibility has been attributed to the instability of the microscopic mechanism by means of which financial markets process information, producing prices and providing liquidity for investors \cite{Bouchaud:2008pb}. Hence, it makes sense to characterize empirically how such mechanism operates, and to identify its weaknesses, its sources of inefficiencies and potential causes of its instability. With this ideas in mind, we want to characterize from the empirical point of view a part of the complex process leading to price formation.
\subsubsection{Types of market data}
In modern financial markets the action of the participants is constantly recorded, and most of the events taking place during its activity are electronically stored. In some cases, part of this data is available for investigation.
In particular some main categories of datasets describing market activity which can be identified and classified on the basis of the timescale they are associated with. The most detailed level of description (timescales ranging from tens of milliseconds to the second) is achieved when informations about single market events triggered by individual agents are available \cite{Lillo:2008ht,Moro:2009wy,Lachapelle:2010kv}. A more coarse-grained description of the market is obtained by focusing on the price process and its variations. More precisely, it is possible to define an instantaneous price for any contract, and to keep track of all its variations (tick-by-tick data) throughout the duration of the market activity. Data describing all events changing the price (called either \emph{trade} or \emph{quote} events) are necessary to achieve this level of description (being the typical time resolution required the one of the second) \cite{Dacorogna:2001uf}. Finally, data corresponding to market behavior at lower frequencies are often publicly available, and involve, beyond the daily opening and closing price, the volume traded and the highest and lowest daily price for all traded goods (e.g., they can be found in \cite{:kx}).
In this discussion, we focus on data describing \emph{trade} events, which belong to the intermediate regime in which the price process is monitored with the resolution of around one second. Any of those trade events corresponds to the transfer of a contract from a seller to a buyer at a given price, for a given quantity (volume) of a good.

\subsubsection{Cross-correlation of trade events}
It has been observed in empirical data across several markets that trade events of single securities are not independent one from another, rather they influence each other leading to interesting clustering phenomena. Moreover, by considering multiple securities traded in the same market venue, it is possible to check that even event times associated with the to trade of different instruments are strongly correlated among each other. Then, one can be interested in answering the following question: \emph{do correlations in trading times arise from correlated exogenous phenomena driving market activity, or do they form due to an endogenous contagion process spreading across the market?} While the former scenario would correspond to a picture in which market activity reflects fundamental exogenous information, the latter would be associated with the scenario of a (potentially unstable) market which self-interacts without necessarily assimilating external information. Those scenarios can in principle coexist, although it is not easy to construct a quantitative, empirically measurable notion distinguishing the two regimes \cite{Joulin:2008bv}. It should also be added that part of the explanation for long-range correlation in trading times has been identified in the mechanism of \emph{order-splitting}: the finite amount of liquidity available in the market forces traders to split large orders (\emph{meta-orders}, \emph{care-orders} or \emph{hidden-orders}) in smaller lots which are traded incrementally, leading to long-range correlation of trading times (from hours to days, sometimes even up to weeks).
Indeed a relevant role could also be played by collective interactions across different securities, which could lead to correlated order flow. This possibility is empirically inquired in section (\ref{sec:PairwiseTradeEv}), where we apply the techniques described in chapter \ref{ch:Inference} to this type of financial system, and try to understand the results on the basis of what we presented in the early part of this chapter.

\section{Applications \label{sec:GeomApplications}}
With these ideas in mind, we consider two sets of realizations of a point-processes $\bX(\tau)$.
\begin{itemize}
\item{{\bf Hawkes processes:} We considered simulated data corresponding to several realizations of a multivariate ($N=100$) fully-connected Hawkes processes with parameters in a variable range.
}
\item{{\bf Financial data:}
We studied trade events corresponding to one year of activity (2003) in a specific stock market, the New York Stock Exchange (NYSE), for the $N=100$ most traded assets.
}
\end{itemize}
The counting functions $\bX(\tau)$  have been discretized in both cases by using a sliding window of size $\delta t $ in order to build the datasets $\hat \bs (\bX,\delta t)$ by using the binning function (\ref{eq:BinningS}).
Datasets $\hat \bs$ have been used to construct the empirical magnetizations $\bmg = (m_i)_{i \in V}$ and the correlation matrix $ \hat \bc = \{ c_{ij} \}_{i<j \in V}$, together with the average magnetization $m = \sum_{i=1}^N m_i$ and the average correlation $c=\frac{2}{N(N-1)} \sum_{i<j \in V} c_{ij}$. Then we solved the inverse problem for this sets of data by considering two type of models:
\begin{itemize}
\item{
{\bf Fully-connected ferromagnet:} We considered the operator set $\bphi = \{ \sum_i s_i , \frac{1}{N} \sum_{i<j} s_i s_j\}$ defining the model (\ref{eq:FCFerromagnet}) and extracted the conjugated parameters $\bg^\star = (h^\star,J^\star)$ given the empirical averages $\bar \bphi = (N m, \frac{N-1}{2} c)$ as shown in section \ref{sec:FCFerromagnet}.
}
\item{
{\bf Disordered fully-connected ferromagnet: } We considered the operator set $\bphi = \{ s_i \}_{i \in V} \cup \{ \frac{1}{N}  s_i s_j\}_{i<j \in V}$ defining the model (\ref{eq:IsingModel}) and extracted the conjugated parameters $\bg^\star = \{ \bh^\star,\hat \bJ^\star)$ given the empirical averages $\bar \bphi = (\bmg,\hat \bc)$ by using the algorithms described in section \ref{sec:MFApproach}
}
\end{itemize}

\subsection{Pairwise fully-connected model for Hawkes processes}
\begin{figure}[htbp] %  figure placement: here, top, bottom, or page
   \centering
   \begin{tabular}{cc}
   \includegraphics[width=2.9in]{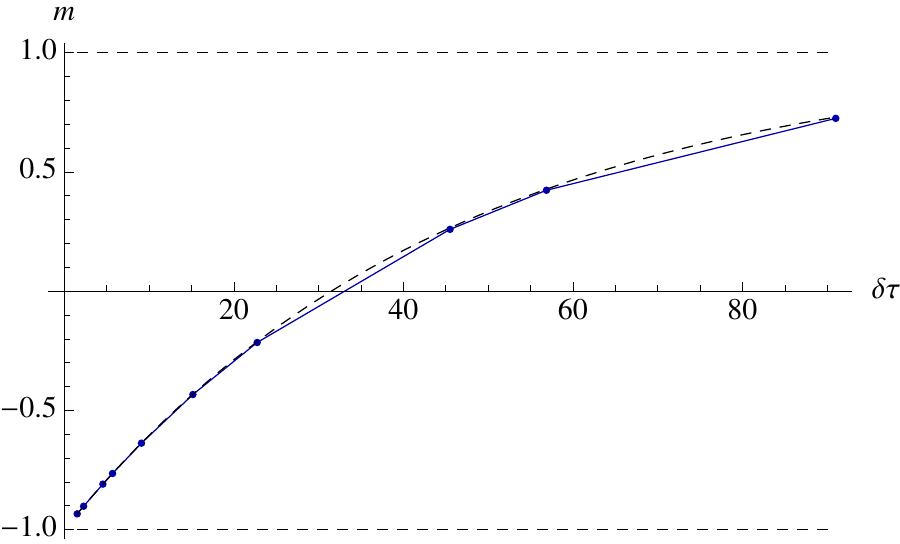} 
   \includegraphics[width=2.9in]{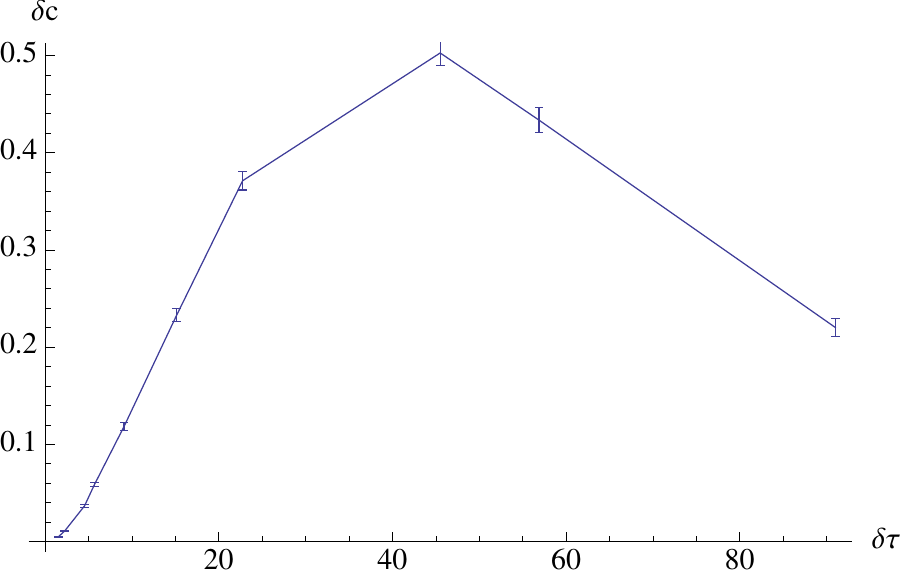} 
   \end{tabular}
   \caption{Average magnetization (left panel) and average correlation (right panel) as functions of the bin size $\delta \tau$ (in units of seconds) for simulated data corresponding to a fully-connected Hawkes process defined by parameters $\mu = 0.011\,\textrm{s}^{-1}, \alpha = 0.015 \,\textrm{s}^{-1},\beta = 0.03 \,\textrm{s}^{-1}$. $\delta c$ indicates the normalized connected correlation $\delta c = N (c - m^2)$ defined in section \ref{sec:FCFerromagnet}. The dashed line in the left panel indicates the reference value $m = 1 - 2 e^{- \mu \delta t \, / \,(1-\alpha/\beta)}$.}
   \label{fig:CorrMagnHawkes}
\end{figure}

In the case of the fully-connected Hawkes process, we considered $N$-variate models with $N=100$ for various set of parameters $(\mu, \alpha, \beta)$. We fixed without loss of generality $\mu=0.011 \,\textrm{s}^{-1}$ (as a common factor in the choice of the parameters can be reabsorbed into a suitable definition of the time coordinate $\tau$) and simulated datasets consisting of $5 \times 10^3$ events with $\alpha$ in the range $[0,\beta]$. We first studied the behavior of the average magnetization and correlations, finding the results summarized in figure \ref{fig:CorrMagnHawkes} for the generic case $\mu = 0.011\,\textrm{s}^{-1}, \alpha = 0.015 \,\textrm{s}^{-1},\beta = 0.03 \,\textrm{s}^{-1}$ and described in the following.
\subsubsection{Relations among bin size and empirical observables}
\begin{itemize}
\item{
The average magnetization ranges from -1 to 1 depending on $\delta \tau$, being the crossover determined from the value of $\mathbb{E}[\lambda(\tau)]$. We plot for reference the curve $1- 2 e^{- \delta \tau \mu /(1-\alpha/\beta)}$ corresponding to the average value  of the magnetization in the stationary state.
}
\item{
Correlations drop to zero as the window $\delta \tau$ is made smaller (a phenomenon known in the field of finance as Epps effect \cite{Epps:1979rp}). In particular if the $\delta \tau$ is smaller than the natural scale for the dynamics of the system $\beta^{-1}$, one expects correlations not to be fully developed. Conversely, when the bin size includes on average multiple events ($\delta \tau \sim \mathbb{E}[\lambda(\tau)]^{-1}$) correlations start to drop due to the binarization of the data.
}
\end{itemize}
This leads to a general consideration involving the optimal bin size required to perform inference: while a large $\delta \tau$ implies less statistics (due to $T=\tau_{\max}/\delta \tau$) and leads to multiple events thus decreasing correlations, it generates less correlated samples (as the auto-correlation decays exponentially in $\beta \delta \tau$). Conversely, small values of $\delta \tau$ imply more statistics, at the price of decreasing the independence of the samples. Eventually, for $\delta \tau$ very small no dynamics is observed due to Epps effect.
All those features can be qualitatively motivated in a simple approximation which allows to compute the averages $\mathbb{E}[s^{(t)}_i]$ and $\mathbb{E}[s^{(t)}_i s^{(t)}_j]$ (appendix \ref{app:ApproxHawkes}).

\subsubsection{Features of the inferred models}
\begin{figure}[h] %  figure placement: here, top, bottom, or page
   \centering
   \includegraphics[width=4in]{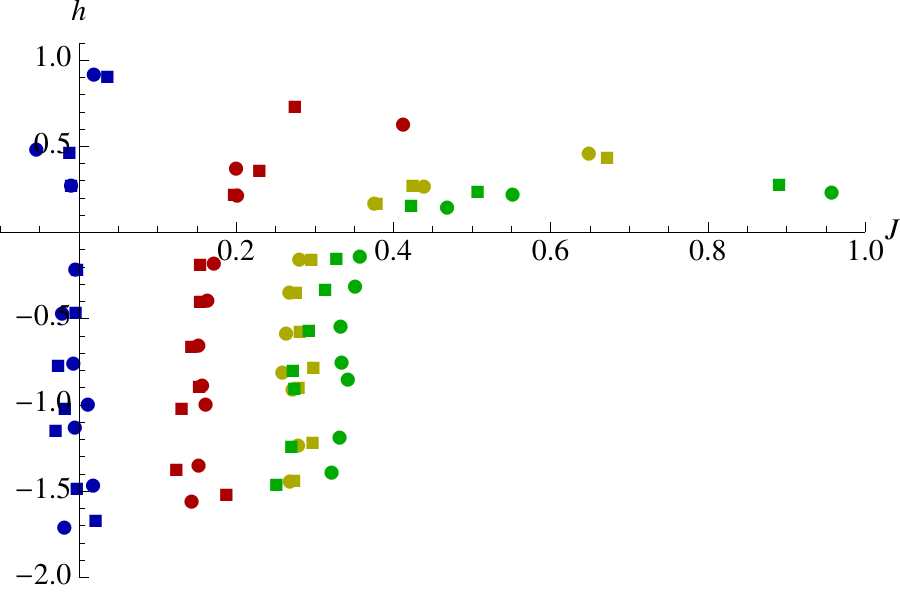} 
   \caption{Inferred couplings obtained for several choices of Hawkes processes, for various choices of the bin size $\delta \tau$. We considered models with $\mu = 0.01\,\textrm{s}^{-1}, \alpha = 0,0.0075,0.015,0.0225 \,\textrm{s}^{-1}, \beta = 0.03\,\textrm{s}^{-1}$ (respectively, blue, red , yellow, green line), and bin sizes ranging from 20 to 80 s. Circles correspond to average couplings inferred from a heterogeneous model, while squares indicate couplings obtained by fitting a homogeneous model.}
   \label{fig:AvgInterHawkes}
\end{figure}

\begin{figure}[h] %  figure placement: here, top, bottom, or page
   \centering
   \begin{tabular}{cc}
   \includegraphics[width=2.9in]{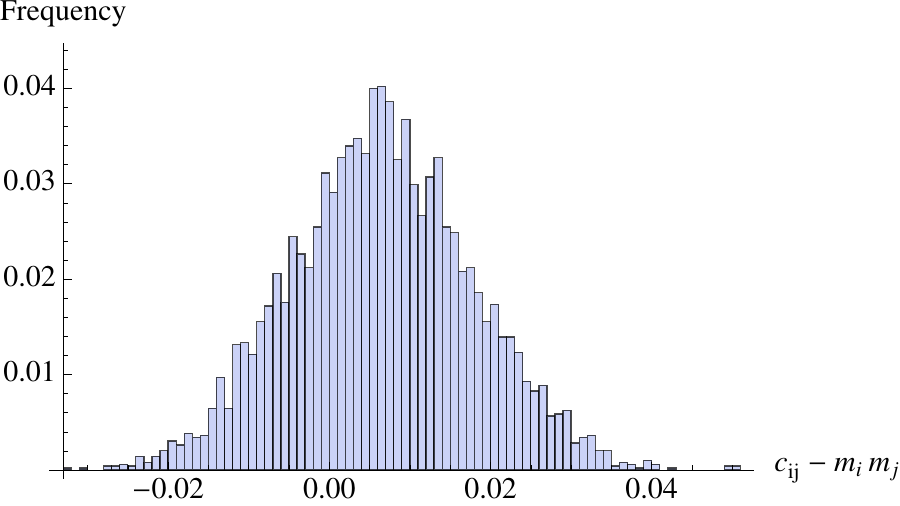} 
   \includegraphics[width=2.9in]{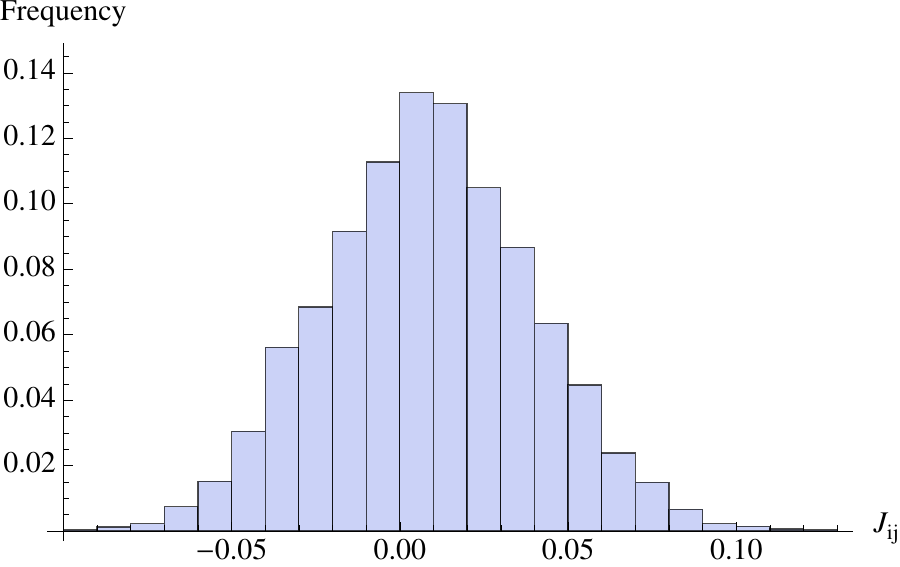} 
   \end{tabular}
   \caption{Histogram of the off-diagonal values of the correlation matrix $\hat \bc - \bmg \bmg^T$ (left panel) and the inferred interaction matrix $\hat \bJ^\star$ (right panel) for an Hawkes process defined by $\mu = 0.01 \, \textrm{s}^{-1}$, $\alpha = 0.025 \, \textrm{s}^{-1}$, $\beta = 0.03 \, \textrm{s}^{-1}$, binned with a resolution of $\delta \tau \approx 30 \, \textrm{s}$. Data corresponds to $5000$ events, corresponding to approximatively $T = \tau_{max} / \delta \tau \approx 4167$.}
   \label{fig:HistCorrelInteracHawkes}
\end{figure}

\begin{figure}[h] %  figure placement: here, top, bottom, or page
   \centering
   \includegraphics[width=4in]{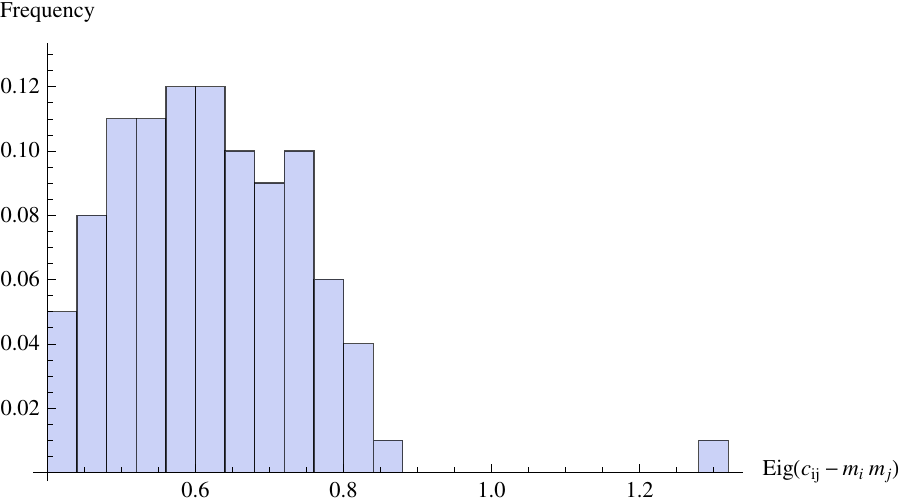} 
   \caption{Histogram of the eigenvalues of the connected correlation matrix $\hat \bc - \bmg \bmg^T$ for the Hawkes process described in figure \ref{fig:HistCorrelInteracHawkes}. Notice that due to symmetry, one would expect for large $T$ to have $N-1$ degenerate eigenvalues of size $1-m^2- \delta c/N \approx 0.62$ and a larger eigenvalue of size $1-m^2 + \delta c \, (N-1) /N \approx 1.26$, whose associated eigenvector is of the form $ (1,\dots, 1)/\sqrt{N}$.}
   \label{fig:HistEigenHawkes}
\end{figure}

Extracting the couplings of a fully connected model from the values of magnetization and correlation described above leads to the results depicted in figure \ref{fig:AvgInterHawkes}, where we consider both the disordered case $\bg = (\bh , \hat \bJ)$ and the two-parameter model $\bg =(h,J)$.
We stress in the following some of the main features.
\begin{itemize}
\item{The Poisson point-process is mapped on the line $J=0$, while models with increasing interaction parameter $\alpha$ for a fixed $\beta$ are mapped on monotonically increasing values of $J$. In this sense, interactions in the original model are genuinely mapped in couplings $J$ within the inferred model. Moreover, increasing interaction parameters lead to curves which are closer to the critical point.}
\item{The inferred fields do not increase monotonically in $\delta \tau$, and the asymptotic behavior when $\delta \tau \to + \infty$ may be either $h\to +\infty$ (for $\alpha > \beta/2$) or $h \to -\infty$ ($\alpha < \beta / 2$). This indicates that the inference procedure that we use can generate \emph{metastable} states (see section \ref{sec:ErgBreak}) as legitimate solutions of the inverse problem. The metastability can be understood as a spurious result of the inference procedure, as it doesn't correspond to any instability of the underlying Hawkes point-process.}
\item{
Adopting a criterium of maximum information efficiency in order to select $\delta \tau$ would lead to a choice of an inferred model which is maximally close to the critical point, where the stability of the model is infinite (section \ref{sec:ExampleGeometry}). Equivalently, adding to the criteria required to choose $\delta \tau$ listed above also the stability would poise the inferred model artificially close to the critical point, where statistical models generalize better.}
\item{
Interestingly, the inferred model doesn't lie on the line $h=0$ where most models concentrate (section \ref{sec:FCFerromagnet}). This is because the scaling $\sim 1/N$ of the kernel $\hat \bK (\tau)$ leads to correlations proportional to $1/N$ (see appendix \ref{app:ApproxHawkes} for a qualitative understanding of this behavior).
}
\end{itemize}
These results have been obtained both for the disordered (using naive mean-field and TAP equations, which lead to similar results) and the non-disordered model (using formulae (\ref{eq:FCFerroInfField}) and (\ref{eq:FCFerroInfCoup})) in order to check the artificial degree of heterogeneity which the inference procedure would have induced if the permutational symmetry among the $N$ spins wouldn't have been known in advance. In figure \ref{fig:HistCorrelInteracHawkes} we plot an histogram of the off-diagonal elements of the connected correlation matrix $c_{ij} - m_i m_j$ and of the inferred couplings $J_{ij}$ for a specific case. In figure \ref{fig:HistEigenHawkes} we plot the histogram of the eigenvalues obtained in the same case. 
\subsection{Pairwise fully-connected model for NYSE trade events \label{sec:PairwiseTradeEv}}
\begin{figure}[htbp] %  figure placement: here, top, bottom, or page
   \centering
   \begin{tabular}{cc}
   \includegraphics[width=2.9in]{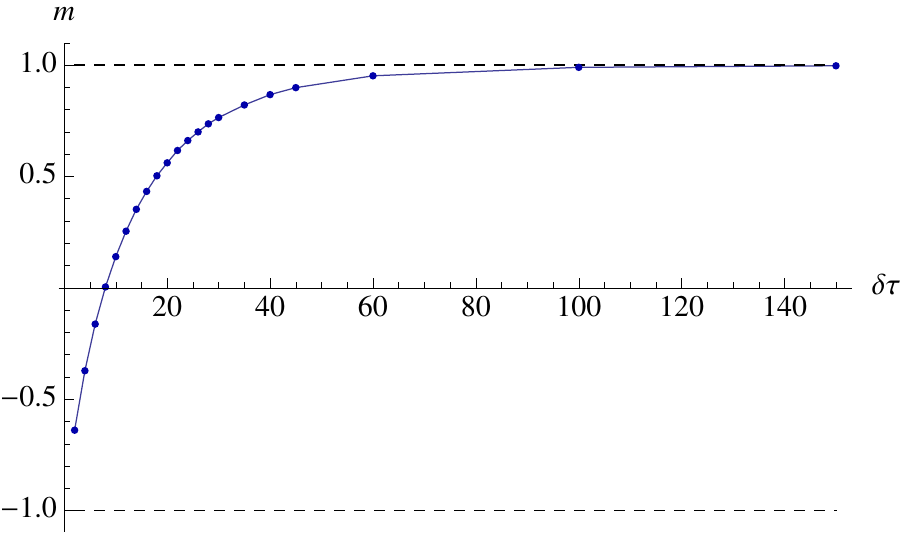} 
   \includegraphics[width=2.9in]{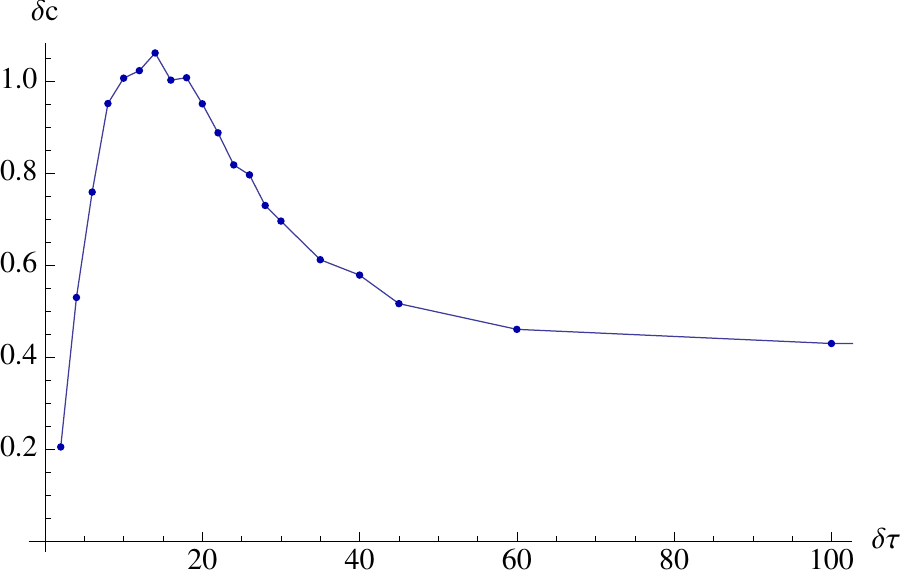} 
   \end{tabular}
   \caption{Average magnetization (left panel) and average correlation (right panel) for data corresponding to 100 days of financial transactions in the NYSE. $\delta \tau$ indicates the bin size in seconds, $\delta c$ is the normalized correlation coefficient. The plot refers to a representative stock of the ensemble, specifically it is associated with the asset Analog Devices Inc. (ADI).}
   \label{fig:CorrMagnFinData}
\end{figure}

We now focus on a dataset describing 100 days of trading activity (from 02.01.2003 to 05.30.2003) in the NYSE for the 100 most traded stocks. We consider only on the central part of each trading day ($\tau_{max} = 10^4$ s), in order to avoid non-stationary effects linked with the opening and the closing hours of the market \cite{Bowsher:2002jw}. Any financial transaction in this period has been defined as an event, independently on the buy or sell direction of the trade. The total data available allowed us to study $10^6$ s of market activity corresponding to $\sim 10^5$ trade events, which have been binned by using sliding windows of size $\delta t \in \{ 2,\dots, 100\} $ s. The results obtained for the average magnetization and the average correlations as functions $\delta t$ are reported in figure (\ref{fig:CorrMagnFinData}), in which it is possible to appreciate at which scale the magnetization changes from -1 to 1 (around $10$ s), the one at which correlations form ($\sim 10$ s) and decrease due to the presence of multiple events ($\sim 30$ s).
\subsubsection{Features of the inferred model}
\begin{figure}[h] %  figure placement: here, top, bottom, or page
   \centering
   \includegraphics[width=3in]{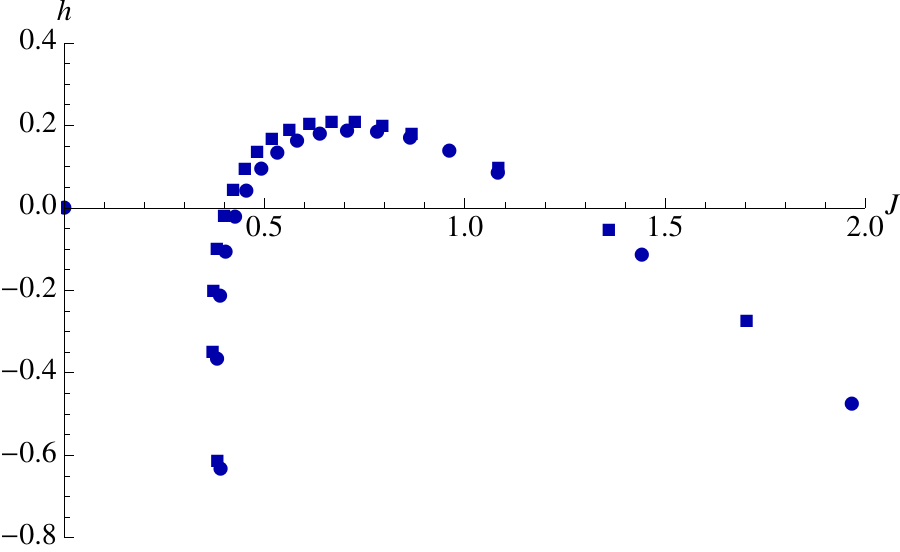} 
   \caption{Inferred couplings $h^\star$ and $J^\star$ obtained with financial data, for various choices of the bin size $\delta \tau$. Squares indicate the result of inferring a homogeneous model, while circles indicate the averages of the vector $\bh$ and of the matrix $\hat \bJ$ obtained by inferring a disordered model.}
   \label{fig:AvgInterFinData}
\end{figure}

\begin{figure}[htbp] %  figure placement: here, top, bottom, or page
   \centering
   \begin{tabular}{cc}
   \includegraphics[width=2.9in]{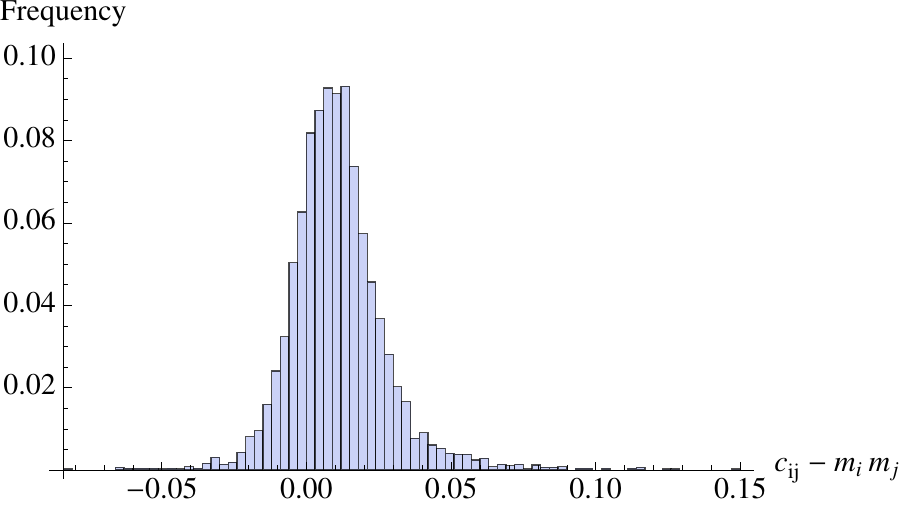} 
   \includegraphics[width=2.9in]{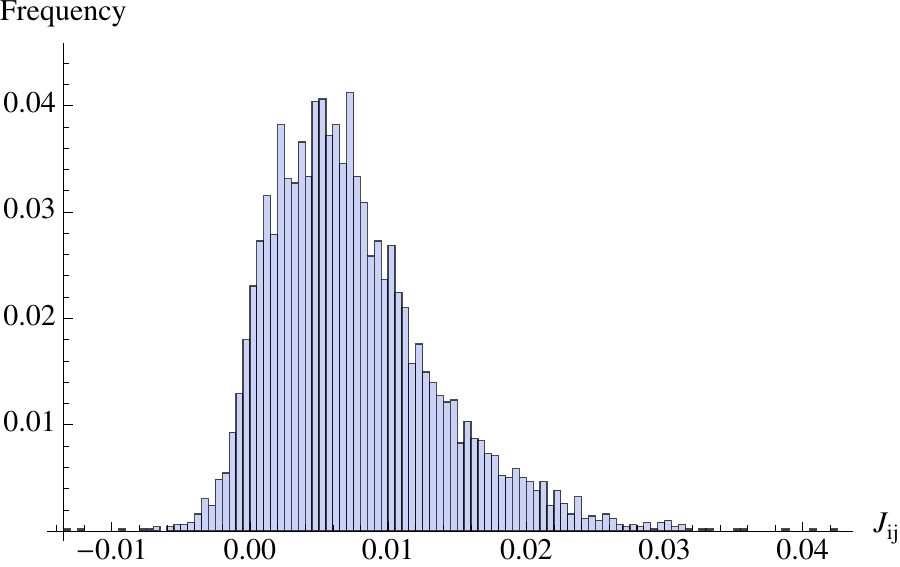} 
   \end{tabular}
   \caption{Histogram of the off-diagonal values of the correlation matrix $\hat \bc - \bmg \bmg^T$ (left panel) and the inferred interaction matrix $\hat \bJ^\star$ (right panel) for financial data, binned with a resolution of $30 \, \textrm{s}$.}
   \label{fig:HistCorrelInteracFinData}
\end{figure}

\begin{figure}[htbp] %  figure placement: here, top, bottom, or page
   \centering
   \includegraphics[width=4in]{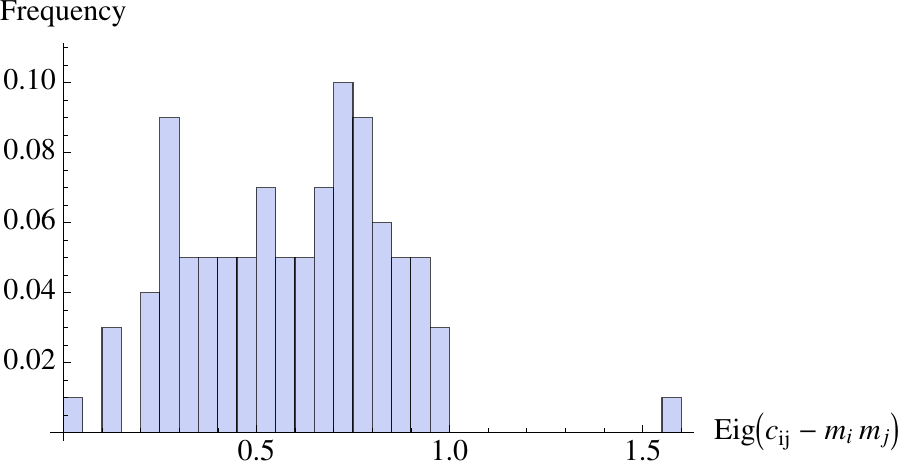} 
   \caption{Histogram of the eigenvalues of the correlation matrix $\hat \bc - \bmg \bmg^T$ for financial data, binned with a resolution of $30 \, \textrm{s}$.}
   \label{fig:HistFinData}
\end{figure}

By considering a fully-connected ferromagnet, such $\bar \bphi = (m,c)$ data has been inverted in order to obtain the interactions $\bg = (h,J)$, as shown in figure \ref{fig:AvgInterFinData}, where we also plotted the quantities $(\frac{1}{N} \sum_{i=1}^N h_i, \frac{2}{N (N-1)} \sum_{i<j} J_{ij})$ obtained by considering a disordered fully-connected ferromagnet.
While for the non-disordered model we used formula (\ref{eq:FCFerroInfCoup}) to invert the averages for the couplings, in the case of the disordered model we used mean-field equations -- both naive Mean-Field (equations (\ref{eq:MFInversionJ}) and (\ref{eq:MFInversionH})) and TAP equations (equations (\ref{eq:TAPInversionJ}) and (\ref{eq:TAPInversionH})) --  which produced consistent results. We stress some features of the results we obtained:
\begin{itemize}
\item{
The ratio $h/J$ changes according to $\delta t$, so that it is not possible to interpret $h$ as measuring exogenous driving factors and $J$ as a genuine interaction. Moreover, as explained in section \ref{sec:FCFerromagnet}), this inference procedure may mix interactions with external fields due to the approximate symmetry $\bg^\star \to \bg^\star + \delta J (-m,1)$. What is possible to say is that the Hawkes process which best describes this $(h,J)$ curve is defined by parameters $\mu \approx 0.011 \, \textrm{s}^{-1}$, $\alpha \approx 0.022 \, \textrm{s}^{-1}$, $\beta \approx 0.03 \, \textrm{s}^{-1}$, so that the exogenous intensity $\mu$ corresponds to approximatively one fourth of the average intensity $\mathbb{E}[\lambda(\tau)]$.
}
\item{
Results describing the Hawkes process allow to understand the proximity of the inferred parameters $(h,J)$ to the critical point  as related to the divergence of the average intensity $\mathbb{E} [\lambda(\tau)]$, rather than arising from a collective effect.
}
\item{
As in the previous case, the inferred model doesn't lie on the critical line $h=0$. This is due to the fact (section \ref{sec:FCFerromagnet}) that correlations are of the order of $1/N$, so that a description in terms of fully-connected ferromagnet  leads to a non-degenerate description of the data.
}
\end{itemize}
\rmk{
A procedure which has been proposed to estimate the distance of an inferred model from the critical point consists in rescaling of all the couplings by a common factor $\beta$ (i.e., performing a shift $\bg^\star \to \beta \bg^\star$) which is interpreted as a fictitious inverse temperature. Studying how the elements of the susceptibility matrix $\hat \bchi$ vary with respect to $\beta$ should allow to identify criticality in the inferred model by the presence of peaks close to $\beta = 1$ in specific components of the matrix. We perform this procedure with our data and plot the results in figure \ref{fig:FictInvTemp}, finding that:
\begin{itemize}
\item{
This procedure is not isotropic, in the sense that the shift $\bg \to \beta \bg$ implicitly indicates that the direction $(1,\dots, 1)$ should be the preferred one in order to evaluate distances in the coupling space.
}
\item{
This type of measure does not describe the distance of the inferred model from the critical point in term of distinguishable models (equivalently, this measure of distance is not invariant under reparametrization of the statistical model $(\bphi, \bg)$). 
}
\end{itemize}
\begin{figure}[htbp] %  figure placement: here, top, bottom, or page
   \centering
   \begin{tabular}{cc}
     \includegraphics[width=2.9in]{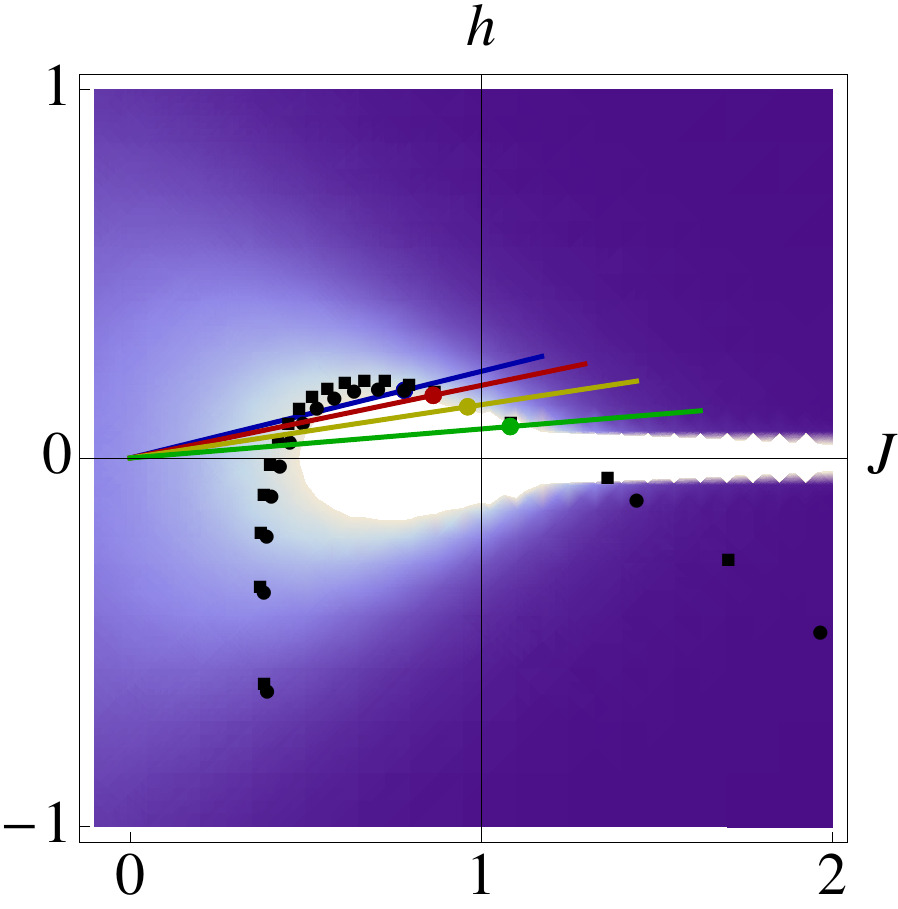} 
     \includegraphics[width=2.9in]{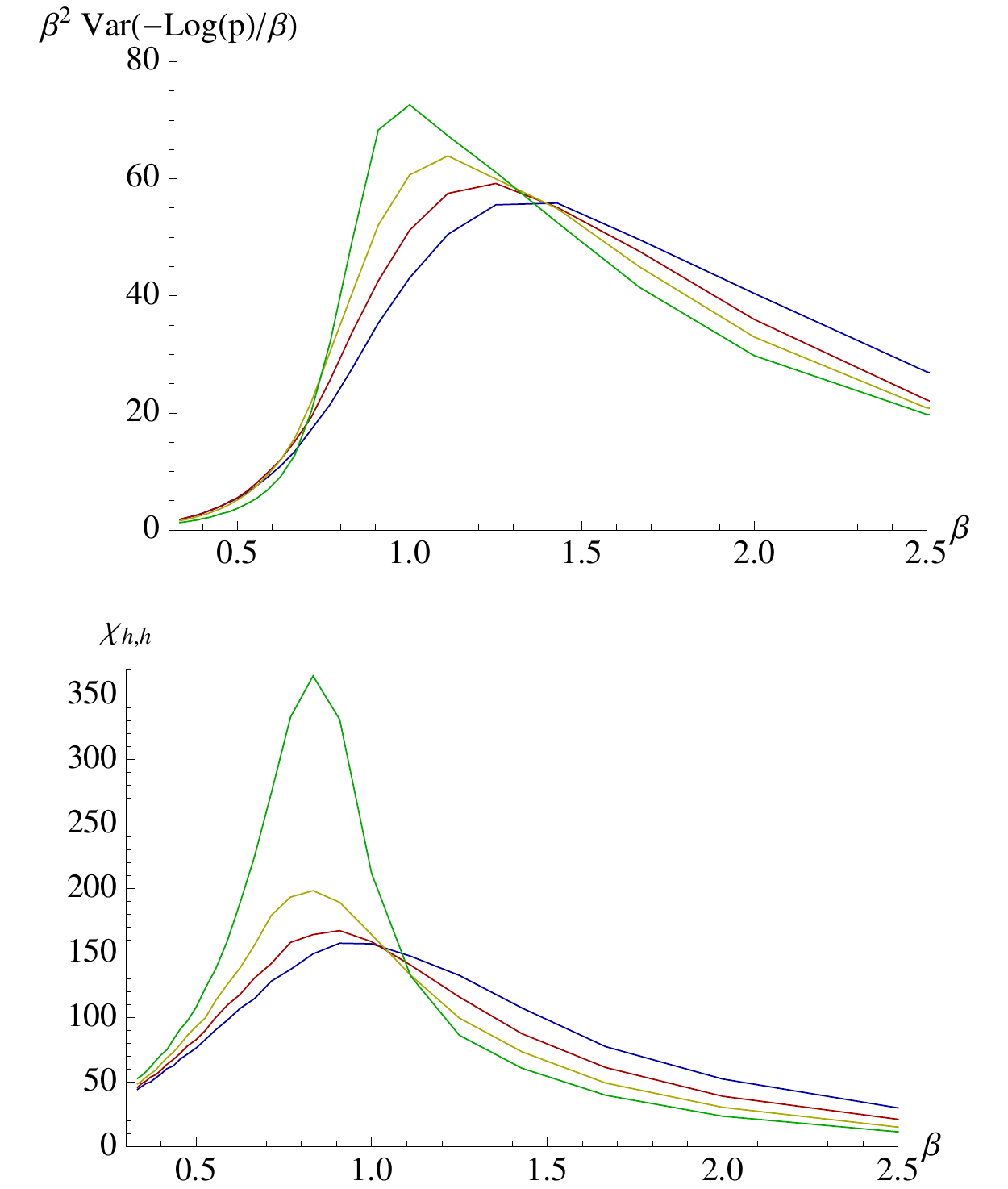}

   \end{tabular}
   \caption{In the left panel we plot the regions of the phase space which are probed by shifting the inferred couplings for financial data $\bg^\star$ by a fictitious inverse temperature $\beta$ for various bin sizes (blue, red yellow and green correspond, respectively, to $\delta \tau = 24,26,28,30$ s), while the background shows the model density $\rho(\bg) \propto \det \hat \bchi$. In the right plots we show the \emph{specific heat} $\beta^2 \textrm{Var}[-\log p/\beta]$ and the susceptibility $\chi_{h,h}$ as a function of the inverse temperature for the same bin sizes as in the left plot.}
   \label{fig:FictInvTemp}
\end{figure}
These points can lead to problems when model condensation is present: very different models may be described by slightly shifting $\beta$. Moreover an inferred model may lie close to the critical point due not only due to model condensation, but also due to the choice of a stable inference procedure, so that it is likely that $\hat \bchi$ attains large value in the point $\bg^\star$, and that by moving from that point one can expect  fluctuations to strongly decrease.
A better measure of distance would be provided by considering \emph{geodesics} in the coupling space under the Fisher metrics $\hat \bchi$, as shown in section \ref{sec:SanovDistDist}. We remind that properties (\ref{eq:InvMeanMaxLik}) and (\ref{eq:InvCovMaxLik}) allows to informally identify this measure as counting \emph{how many error bars is one away from the critical point}. This approach doesn't specify any privileged direction in the coupling space (as the geodesic distance is associated with whatever path in the coupling space is minimizing the number of such error bars), nor varies according to the reparametrization (as in that case also error bars are reparametrized accordingly). As an example, one finds that the distance $d_{T,\epsilon}(\bp_{crit},\bp^\star)$ defined by equation \ref{eq:DistanceDistrib} between the critical point and the inferred parameters for $\delta \tau = 28 \, \textrm{s}$ ($\,h^\star \approx 0.14 , J^\star \approx 0.96$) is $d_{T,\epsilon}(\bp_{crit}, \bp^\star) \gtrsim 10^2$ for $\epsilon = -\log 1\%$ and $T \sim 10^6 \, \textrm{s} / 28 \, \textrm{s}$.
}
We also performed an analysis of the empirical connected correlation matrix $\hat \bc - \bmg \bmg^T$ and of the inferred interaction matrix $\hat \bJ^\star$ in order to check the compatibility of data with an homogeneous model. The corresponding histograms have been plotted in figure \ref{fig:HistCorrelInteracFinData}, showing that data is qualitatively similar to the one which would have been obtained with an homogeneous model.
The principal component analysis of the matrices $\hat \bc - \bmg \bmg^T$ and $\hat \bJ$ indicates in both cases the presence of a large eigenvalue, whose associated eigenvector is roughly of the form $\frac{1}{\sqrt{N}} (1,\dots, 1)$ as shown in the histogram (\ref{fig:HistFinData}).
This findings can be interpreted as indicating that a significant part of the structure of the cross-excitatory network can be captured by an homogeneous model.\footnote{This is somewhat similar to what one finds for the statistics of stock price variations \cite{Mantegna:1999ay,Bouchaud:2003fy}. In that case the correlation matrix has a large eigenvalue of size proportional to $N$ (also called \emph{market mode}), together with a small number of isolated eigenvalues, whose associated eigenvectors usually identify \emph{financial sectors}. Interestingly, the inspection of the eigenvalues of $\hat \bc$ and $\hat \bJ$ beyond the largest one evidences different sectors with respect to the ones found by studying stock price variations.}

\chapter{Conclusion\label{ch:Conclusion}}

In this work we have presented a general approach to the field of statistical learning, in which the problem of estimating parameters describing a complex system is seen as an inverse problem in the field of statistical mechanics. This perspective has been proven to be especially relevant in order to study extended systems, which in this language are associated to physical systems in the thermodynamic limit. This regime is well-known in physics, and several techniques (mean-field approximations) are known to solve the inverse problem in this framework with high accuracy. Interesting collective features emerge in this regime: models can \emph{condensate} leading to regions of the space of parameters which are able to describe anomalously well very diverse datasets, and \emph{null-modes} can develop leading to degenerate representations of a dataset. All of these factors have to be kept into account when studying inverse problem for empirical datasets, in order to disentangle the genuine features of a system from the spurious ones depending on the inference procedure which is applied.
We have also shown that \emph{complete representations} of the inverse problem lead to the exact solution of several systems (complete systems, one-dimensional systems, tree-like interaction networks), and allow a general understanding of the locality and stability features of the inverse problems, which are \emph{easier} and more \emph{resilient} to noise than the direct ones.
The notion of regularizer has also been discussed and its use has been clarified by specific, solvable examples, in order to show general features of non-parametric inference. We find that a symmetry property characterizes the regularizers, and a tradeoff between computational complexity and relevance of the inference procedure has to be sought on the basis of such symmetry in order to perform model selection.
Finally, we have shown how differential geometry can be used to understand the consistency of the inverse problem, and how the special features associated with the large $N$ limit have a clear geometric interpretation in terms of \emph{distance} and \emph{volume}. In this language, criticality of the inferred model is related to the strong divergence of the number of datasets which can be described through a small shift of the inferred parameters. Finally, we have presented the application of these ideas to two datasets, a synthetic one describing a self-excitatory point process and an empirical one describing transactions in a financial market. We used those datasets in order to illustrate our ideas by separating genuine features of the inferred model and spurious ones, finding that the dataset describing financial transactions can be well-described by a fully-connected ferromagnet in which interactions play a prominent role with respect to external driving factors.
\appendix
\chapter{Binary Inference}
\section{Maximum entropy principle \label{app:MaxEntPr}}
Consider a set of data $\hat \bs=\{ s^{(t)}\}_{t=1}^T$, and a family of operators $\bphi$. The maximum entropy principle states that among all probability distributions $\bp$ such that $\langle \bphi \rangle = \bar \bphi$ (where as usual $\bar \bphi= \frac{1}{T} \sum_{t=1}^T \phi_{\mu,s^{(t)}}$), the one which maximizes the Shannon entropy $S(\bp)$ is given by the statistical model (\ref{eq:ProbDensity})
\beq
p_s^\star = \frac{1}{Z(\bg)} \exp \left( \sum_{\mu=1}^M g_\mu^\star \phi_{\mu,s}\right)
\eeq
in which each of the $g_\mu^\star$ is seen as a Lagrange multiplier enforcing the condition $\langle \phi_\mu \rangle = \bar \phi_\mu$.\\
This principle is often invoked in order to justify the model (\ref{eq:ProbDensity}) as the simplest (i.e., with higher entropy) one which is able to explain a given set of empirical averages \cite{Wainwright:2008kx}. Indeed it should be observed that this principle doesn't completely solve the problem of selecting the most appropriate model in order to explain data $\hat \bs$, rather it converts it into the problem of selecting the best set of observables $\bar \bphi$. In both cases a family $\bphi$ has to be specified, and this has to be done on the basis of some \emph{a priori} information (e.g., which operators are likely to be contained in the model), or according to the specific goal of the inference problem which one is trying to solve (e.g., which observables are considered relevant for a particular application).
\prf{
The proof of this result amounts to solve the constrained optimization problem
\beq
\bp^\star = \arg\max_\bp \left[ S(\bp) + (g_0 + 1) +\sum_{\mu=1}^M g_\mu \left( \langle \phi_\mu \rangle - \bar \phi_\mu \right) \right] \; ,
\eeq
in which the Lagrange multipliers $\{ g_\mu \}_{\mu=1}^M$ constrain the averages $\langle \phi_\mu \rangle$ to their empirical values, while $g_0$ enforces the normalization. By differentiation with respect to $p_s$, one can easily obtain equation (\ref{eq:ProbDensity}). The conditions for the existence and the uniqueness of such solution are the same ones required in order to solve the inverse problem, and are described in section \ref{sec:InvProblem}.
}

\section{Concavity of the free energy \label{app:FreeEnConc}}
Consider the free energy $F(\bg)$ defined as in section \ref{sec:DirectProblem}. We want to prove that it is a concave function by showing that the susceptibility matrix $\hat \bchi$ defined in equation (\ref{eq:GenSusc}) is positive semidefinite.
\prf{
First one can show that
\beq
\chi_{\mu,\nu} = - \frac{\partial^2 F}{\partial g_\mu \partial g_\nu} = \sum_s (\phi_{\mu,s} - \langle \phi_\mu \rangle) \, (\phi_{\nu,s} - \langle \phi_\nu \rangle )\,  p_s \;
\eeq
which allows to proof that for any vector $\bx$, the quadratic form $\bx^T \hat \bchi \bx$ is greater or equal than zero. In fact one has that
\beqa
\sum_{\mu,\nu > 0} x_\mu \chi_{\mu,\nu} x_\nu &=& \sum_s p_s  \Bigg[ \sum_{\mu > 0} x_\mu (\phi_{\mu,s} - \langle \phi_\mu \rangle) \Bigg]  \Bigg[ \sum_{\nu > 0} x_\nu (\phi_{\nu,s} - \langle \phi_\nu \rangle ) \Bigg] \\
&=& \left< \Bigg[ \sum_{\mu > 0} x_\mu (\phi_{\mu,s} - \langle \phi_\mu \rangle)\Bigg]^2\right> \geq 0 \; .
\eeqa
Additionally, if the operators $\phi_{\mu,s}$ are minimal in the sense defined in section \ref{sec:DirectProblem} above expression has to be strictly larger than zero for $\bx \neq \bm{0}$. In fact if $\bx^T \hat \bchi \bx = 0$, then it must hold for each state $s$ that
\beq
\sum_{\mu>0} (\phi_{\mu,s} - \langle \phi_\mu \rangle) x_\mu = 0 \; ,
\eeq
which by minimality of $\bphi$ implies that $x_\mu = 0$ for each $\mu$.
}

\section{Small deviations of the empirical averages \label{app:SmallDevEmpAv}}
We want to show that given a statistical model $(\bphi,\bg)$, equations (\ref{eq:MeanEmpAvg}) and (\ref{eq:CovEmpAvg}) hold for the empirical averages $\bar \bphi$.
\prf{
For the averages, it is sufficient to show that due to the factorization property of $P_T(\hat \bs | \bg)$ one has
\beq
\langle \bar \phi_\mu \rangle_T = \frac{1}{T} \sum_{t=1}^T \langle \phi_{\mu,s^{(t)}} \rangle_T = \frac{1}{T} \sum_{t=1}^T \langle \phi_\mu \rangle = \langle \phi_\mu \rangle \; ,
\eeq
while for the covariances one can write
\beq
\langle \bar \phi_\mu \bar \phi_\nu\rangle_T - \langle \bar \phi_\mu \rangle_T \langle \bar \phi_\nu \rangle_T
=
\frac{1}{T^2} \sum_{t,t^\prime=1}^T\left[  \langle \phi_{\mu,s^{(t)}} \phi_{\nu,s^{(t^\prime)}} \rangle_T - \langle \phi_{\mu,s^{(t)}} \rangle_T \langle \phi_{\nu,s^{(t^\prime)}} \rangle_T \right] \; .
\eeq
By noting that due to independence all terms with $t \neq t^\prime$ vanish from previous expression, one recovers equation (\ref{eq:CovEmpAvg}).
}

\section{Sanov theorem \label{app:SanovTheor}}
We want to prove Sanov theorem (\ref{eq:Sanov}), which states that given a probability distribution $\bp$ and a compact set of probability densities $\mathcal M \subseteq \mathcal M(\Omega)$, one has that the empirical frequencies $\bar \bp$ sampled from $P_T (\hat \bs | \bp)$ obey the large deviation principle
\beq
\lim_{\delta \to 0} \lim_{T\to\infty} - \frac{1}{T} \log \textrm{Prob} (\bar \bp \in \mathcal M^\prime) =  D_{KL} (\bq^\star || \bp) \; .
\eeq
where $\bq^\star = \arg\min_{\bq \in \mathcal M} D_{KL}(\bq || \bp)$ and $\mathcal M^\prime$ is the compact set $\mathcal M^\prime = \{ \bp^\prime = \bp + \delta \bp \; \in \mathcal M(\Omega) \; |\;  \bp \in \mathcal M, \delta \bp \in [-\delta,\delta]^{|\Omega|} \; \} $
\prf{
We will provide a simple combinatorial proof of Sanov theorem along the lines of \cite{Morters:2008fv}, which requires some preliminary definitions. Given an empirical frequency $\bar \bq$, we denote with $\hat \bs (\bar \bq)$ the set $\hat \bs (\bar \bq) = \{ \hat \bs \in \Omega^T \; | \; \bar q_s = \frac{1}{T} \sum_{t=1}^T \delta_{s,s^{(t)}}\}$ of empirical datasets compatible with $\bar \bq$. We also define the set of all possible empirical frequencies as $\overline{\mathcal M_T} (\Omega)$.
For those sets it holds that:
\begin{itemize}
\item{
The cardinality of $\hat \bs(\bar \bq)$ is bound by
\beq
\frac{1}{\mathcal P_1(T)}  e^{T S(\bar \bq) } \leq |\hat \bs(\bar \bq)| \leq \mathcal P_2(T) e^{T S(\bar \bq)} \; ,
\eeq
where $\mathcal P_1(T),\mathcal P_2(T)$ are polynomials in $T$ with positive coefficients.
This descends from applying Stirling bounds on the factorial to the exact relation
\beq
| \hat \bs (\bar \bq) | = \frac{T ! }{\prod_s \left( T \bar q_s \right)!} \; .
\eeq
and plugging the definition of Shannon entropy (\ref{eq:ShannonEntropy}) in the resulting expression.
}
\item{
The cardinality of $\overline{\mathcal M_T} (\Omega)$ is bound by
\beq
|\overline{\mathcal M_T} (\Omega) | \leq (T+1)^{|\Omega|} \; .
\eeq
because each configuration $s$ is visited a number of times between $0$ and $T$.
}
\item{
Due to compactness of $\mathcal M$ and continuity of $D_{KL}(\bq || \bp)$, one has that $\min_{\bq \in \mathcal M} D_{KL}(\bq || \bp)$ exists and is attained in the (unique, due to convexity) point $\bq^\star \in \mathcal M$.
}
\end{itemize}
By using those properties, we can find an upper bound for the large deviation function as follows:
\beqa
 \textrm{Prob}(\bar \bp \in \mathcal M^\prime) &=& \sum_{\bar \bq \in \overline{\mathcal M_T} (\Omega) \cap \mathcal M^\prime}  \textrm{Prob}(\bar \bp = \bar \bq )  \nonumber \\
 &=& \sum_{\bar \bq \in \overline{\mathcal M_T} (\Omega) \cap \mathcal M^\prime}  \sum_{\hat \bs \in \hat \bs(\bar \bq)} P_T(\hat \bs | \bp) \nonumber \\
 &\leq& \sum_{\bar \bq \in \overline{\mathcal M_T} (\Omega) \cap \mathcal M^\prime} \mathcal P_2(T) e^{T S(\bar \bq)} e^{-T [ S(\bar \bq) + D_{KL} (\bar \bq || \bp)] } \nonumber \\
 &\leq& (T+1)^{|\Omega|} \mathcal P_2(T) e^{-T  D_{KL} ( \bq^{\star \prime} || \bp)} \; .
\eeqa
where $\bq^{\star \prime} = \arg\min_{\bq \in \mathcal M^\prime} D_{KL}(\bq || \bp)$.
This trivially implies
\beq
\lim_{T\to\infty} \frac{1}{T} \log \textrm{Prob}(\bar \bp \in \mathcal M) \leq - D_{KL}(\bq^{\star \prime} || \bp) \; .
\eeq
By taking the limit $\delta \to 0$, one recovers $\bq^{\star \prime} \to \bq^\star $.
For the lower bound, one needs to notice that for any $\delta$ it is possible to find a sufficiently large $T$ and a $\delta \bp \in [-\delta, \delta]^{|\Omega|}$ such that $\bar \bq^\star \in \overline{\mathcal M_T}(\Omega) \cap \mathcal M^\prime$ is close enough to $\bq^\star$ (due to density of rational numbers into real numbers), so that $| D_{KL} (\bar \bq^\star || \bp) - D_{KL}(\bq^\star || \bp) | < \epsilon$ with $\epsilon$ arbitrary.
Then one can write that
\beq
\textrm{Prob} (\bar \bp \in \mathcal M^\prime) \geq \textrm{Prob}(\bar \bp = \bar \bq^\star) = \sum_{\hat \bs \in \hat \bs(\bar \bq^\star)} P_T(\hat \bs | \bp) \geq \frac{1}{\mathcal P_1(T)} e^{-T D_{KL} (\bar \bq^\star || \bp)} \; ,
\eeq
which due to the arbitrarity of $\epsilon$ allows to prove the lower bound
\beq
\lim_{T\to\infty} \frac{1}{T} \log \textrm{Prob}(\bar \bp \in \mathcal M)  \geq  -D_{KL} (\bar \bq^\star || \bp) \geq -D_{KL} (\bq^\star || \bp) - \epsilon \; .
\eeq
}

\section{Cram\'er-Rao bound \label{app:CRBound}}
Cram\'er-Rao bound states that given a statistical model $(\bphi,\bg)$ with $F(\bg)$ strictly convex and an unbiased estimator of $\bg$ denoted as $\bg^\star$, the covariance matrix of $\bg^\star$ under the measure $\langle \dots \rangle_T$ is bound according to equation (\ref{eq:CRBound}).
\prf{First, it is necessary to prove that to prove that, after defining $V_\mu = \frac{\partial \log P_T(\hat \bs | g)}{\partial g_\mu} $, and using equation (\ref{eq:Likelihood}) one has
\beq
\left< V_\mu \right>_T = \left< T \left[ \bar \phi_\mu - \langle \phi_\mu \rangle \right] \right>_T = 0 \; , \label{eq:AvgMomMatching}
\eeq
where we also used equation (\ref{eq:MeanEmpAvg}) (i.e., $\langle \phi_\mu \rangle = \langle \phi_\mu \rangle _T$). Then, it is possible to compute the covariance
\beqa
\textrm{Cov} (V_\mu, g^\star_\nu - g_\nu) &=& \left< V_\mu [g^\star_\nu - g_\nu] \right>_T - \left< V_\mu \right>_T \left< g^\star_\nu - g_\nu \right>_T \nonumber \\
&=&\left< V_\mu g^\star_\nu \right>_T - \left< V_\mu \right>_T g_\nu- \left< V_\mu \right>_T \left< g^\star_\nu - g_\nu \right>_T \nonumber \\
&=& \sum_{\hat \bs} \left[ \frac{1}{P_T(\hat \bs | g)} \frac{\partial P_T(\hat \bs | g)}{\partial g_\mu} g^\star_\nu \right] P_T(\hat \bs | g) = \frac{\partial g_\nu}{\partial g_\mu} \nonumber \\
&=& \delta_{\mu,\nu}
\eeqa
and exploit Cauchy-Schwartz inequality, which implies that for any pair of vectors $\bx, \by$ it holds that
\beq
\left(  \bx^T \left< \bV [\bg-\bg^\star]^T \right>_T \by  \right)^2 \leq  \left< (\bx^T \bV)^2 \right>_T \left< ([\bg^\star-\bg]^T \by)^2 \right>_T \label{eq:CauchySchwartz}
\eeq
Equation (\ref{eq:AvgMomMatching}) fixes the value of the left-hand side term of equation (\ref{eq:CauchySchwartz}) to be $\bx^T \by$, while the right-hand side can be expanded into
\beq
 \left< (\bx^T \bV)^2 \right>_T \left< ([\bg^\star-\bg]^T \by)^2 \right>_T = T \left( \bx^T \, \hat \bchi \, \bx \right) \left( \by^T \, \left< [\bg^\star - \bg][\bg^\star - \bg]^T \right>_T \, \by \right),
\eeq
where we used that $\langle \bV \bV^T\rangle_T = T \hat \bchi$ due to equation  (\ref{eq:CovEmpAvg}). Finally, by choosing the arbitrary vector $\bx$ to be $\bx = \hat \bchi^{-1} \by / T$ ($\hat \bchi$ is invertible due to strict concavity of $F(\bg)$), it holds for any $\by$ that
\beq
\frac{1}{T} \left( \by^T \, \hat \bchi^{-1} \, \by  \right)  \leq \left( \by^T \left< [\bg^\star - \bg][\bg^\star - \bg]^T \right>_T  \by \right)
\eeq
which proves the thesis (\ref{eq:CRBound}).
}
\section{Convergence of the inferred couplings \label{app:ConvInfCoup}}
Given a set of empirical frequencies $\bar \bp$, we want to prove that for a generic set of models described by an operator set $\bphi$ the mean and the covariances of couplings $\bg$ defining a probability distribution $\bp$, weighted by the measure provided by the posterior $P_T(\bg | \bar \bp)$ are given in the limit of large $T$ by equations (\ref{eq:InvMeanMaxLik}) and (\ref{eq:InvCovMaxLik}).
\prf{
To calculate them, we first notice that, by defining
\beq
\mathcal Z(\bar \bphi) = \int d \bg \; e^{-T D_{KL}(\bar \bp || \bp)}  = \int d \bg \; e^{T \sum_{\mu=0}^M g_\mu \bar \phi_\mu}
\eeq
it is possible to write
\beqa
\frac{\partial \mathcal Z(\bar \bphi)}{\partial \bar \phi_\mu} &=& T \int d \bg \; g_\mu \, e^{-T D_{KL}(\bar \bp || \bp)} \\
\frac{\partial^2 \mathcal Z(\bar \bphi)}{\partial \bar \phi_\mu \partial \bar \phi_\nu}  &=& T^2 \int d \bg \; g_\mu g_\nu \, e^{-T D_{KL}(\bar \bp || \bp)} \; ,
\eeqa
so that the calculation of the generating function $\log \mathcal Z(\bar \bphi)$ allows to find the required momenta of $\bg$. In the limit of large $T$ it is possible to perform a saddle-point estimation of the function $\mathcal Z(\bar \bphi)$ around the minimum of the convex function $D_{KL}(\bar \bp || \bp)$ (or, equivalently, the maximum of the concave free energy $F(\bg)$), which requires the expansion of the Kullback-Leibler divergence. This procedure yields
\beqa
\mathcal Z(\bar \bphi) &=& \int d \bg \; \exp \left[ T \left( F(\bg^\star) + \sum_{\mu=1}^M g_\mu^\star \bar \phi_\mu +
\frac{1}{2} \sum_{\mu,\nu} \frac{\partial^2 F(\bg)}{\partial g_\mu \partial g_\nu} (g_\mu - g_\mu^\star)(g_\nu - g_\nu^\star) + \dots \right) \right] \nonumber \\
&\xrightarrow[T\to \infty]{}& e^{-T S(\bar \bphi)}  \sqrt{ \frac{2 \pi}{T \det \hat \bchi} } \; ,
\eeqa
where -- as shown in section \ref{sec:InvProblem}  -- the maximum likelihood estimator $\bg^\star$ can be defined as the minimizer of the Kullback Leibler divergence. The differentiation of $\log \mathcal Z(\bar \bphi)$ finally leads to
\beqa
\frac{1}{T} \frac{\partial \log \mathcal Z(\bar \bphi)}{\partial \bar \phi_\mu} &\xrightarrow[T\to \infty]{}& -\frac{\partial S(\bar \bphi)}{\partial \bar \phi_\mu} = g_\mu^\star \\
\frac{1}{T^2} \frac{\partial^2 \log \mathcal Z(\bar \bphi)}{\partial \bar \phi_\mu \partial \bar \phi_\nu}  &\xrightarrow[T\to \infty]{}& - \frac{1}{T} \frac{\partial^2 S(\bar \bphi)}{\partial \bar \phi_\mu \partial \bar \phi_\nu} = \frac{\chi^{-1}_{\mu,\nu}}{T} \; ,
\eeqa
where we used equation (\ref{eq:GenFuncCoupling}) and (\ref{eq:GenFuncInvFisher}) to express the derivatives of the entropy $S(\bar \bphi)$.
}

\chapter{High-dimensional inference}
\section{The fully-connected ferromagnet: saddle-point calculation \label{app:SPFerromagnet}}
We want to prove that the free energy $F(h,J)$ of a fully connected ferromagnet described by the probability density (\ref{eq:FCFerromagnet}) can be written as in (\ref{eq:SaddlePointFCFerro}).
\prf{
This can be shown by noting that by using Stirling formula and approximating the sum with an integral one can write
\beqa
Z(h,J) &=& e^{- J/2} \sum_{N_+ = 0}^N \delta \left[ Nm - \left( \frac{N+N_+}{2} \right) \right] \binom{N}{N_+} \exp \left[ N \left( \frac{J  m^2}{2}  + h m \right) \right]  \nonumber \\
&\xrightarrow[N\to \infty]{}& e^{-J/2} \int_{-1}^1 dm \exp \left[ N \left( \frac{J  m^2}{2}  + h m + s(m) \right) \right] \; , \label{eq:SPIntegralFerro}
\eeqa
with $s(m) = -\frac{1+m}{2} \log \frac{1+m}{2} -  \frac{1-m}{2} \log \frac{1-m}{2}$. For $(h,J)$ independent of  $N$, above integral can be evaluated by saddle-point, and is dominated by the (absolute) minimum $m_{s.p.}(h,J)$ of the function $f_{h,J}(m) = -\frac{J  m^2}{2}  - h m - s(m)$. By substituting $F(h,J) = -\log Z(h,J)$ one finds
\beq
F(h,J) \xrightarrow[N\to \infty]{}  \frac{J}{2} + N f_{h,J}(m_{s.p. (h,J)}) - \frac{1}{2} \log \frac{2\pi}{N \partial_m^2 f_{h,J}(m_{s.p.}(h,J))} \, ,
\eeq
where $m_{s.p.}(h,J)$ satisfies the saddle-point equation
\beq
m = \tanh \left( J m + h \right) \label{eq:SPeqFCFerro} \; .
\eeq
Instead for large, finite $N$, $J>1$ independent of $N$ and $0 \leq h \propto 1/N$ equation (\ref{eq:SPeqFCFerro}) has two minima $m_+$ and $m_-$, whose contribution can be kept into account through
\beqa
Z_+ + Z_- &=& Z_+ \left( 1+ \frac{Z_-}{Z_+}\right) = Z_+ \, e^{\log (1+ Z_-  / Z_+)} \nonumber \\
&=&   Z_+\,  e^{- F_{trans}} \,
\eeqa
which yields the last term of equation (\ref{eq:SaddlePointFCFerro}).
}

\subsection{The leading contribution $F_{0}$.}
The main features of the model can be described by keeping into account the term $F_0(h,J)$, which is the only one in equation (\ref{eq:SaddlePointFCFerro}) proportional to $N$. It is given by
\beq
F_{0}(h,J) = N f_{h,J}(m_{s.p.}(h,J)) \; ,
\eeq
 where
\beq
f_{h,J}(m) = - h m - \frac{J m^2}{2} + \left( \frac{1+m}{2} \log \frac{1+m}{2} + \frac{1-m}{2} \log \frac{1-m}{2} \right) \; ,
\eeq
and $m_{s.p.}(h,J)$ is defined as the absolute minimum of the function $f_{h,J}(m)$, hence it satisfies the transcendental equation
\beq
m = \tanh \left( J m + h \right)  \; .
\eeq
The contribution of $F_0(h,J)$ to the ensemble averages is
\beqa
\left< \sum_i s_i \right>_{0} &=& - \frac{\partial F_{0}}{\partial h} = N \, m_{s.p.}\\
\left< \frac{1}{N} \sum_{i<j} s_i s_j\right>_{0} &=& - \frac{\partial F_{0}}{\partial J} = N \, \frac{m_{s.p.}^2}{2}\; ,
\eeqa
while the one to the susceptibility matrix $\hat \bchi$ is given by
\beq
\hat \bchi_{0} = N \, \chi_{s.p.} \left( \begin{array}{cc} 1 & m_{s.p.} \\ m_{s.p.} & m^2_{s.p.} \end{array} \right) \; ,
\eeq
where $\chi_{s.p.} =\partial m_{s.p.} / \partial h $. Its eigenvalues are given by $N \chi_{s.p.} (0, 1 + m^2_{s.p.})$.
\subsubsection{The role of Gaussian fluctuations}
The term $F_{fluct}(h,J)$ allows to compute the eigenvalue decomposition for the matrix $\hat \bchi$, whose smallest eigenvalue receives a contribution which grows in $N$, and is related to the Gaussian integral (\ref{eq:SPIntegralFerro}). It results
\beq
F_{fluct}(h,J) = - \frac{1}{2} \log \left( \frac{2\pi}{N \partial_m^2 f_{h,J}(m_{s.p.}(h,J)) } \right) \; .
\eeq
The contribution of $F_{fluct}(h,J) $ to the solution of the direct problem is
\beqa
\left< \sum_i s_i \right>_{fluct} &=&  -\chi_{s.p.}^2 \frac{m}{(1-m^2)^2}\\
\left< \frac{1}{N} \sum_{i<j} s_i s_j\right>_{fluct} &=& - \frac{\chi_{s.p.}^2}{2} \left( J - \frac{1-3m^2}{(1-m^2)^2} \right)\;
\eeqa
and
\beq
\hat \bchi_{fluct} = \chi_{s.p.}^4 \; (1-m^2_{s.p.})^{-3} \; \hat{ \bm{a}} (J,m)\; ,
\eeq
with
\beqa
a_{11}(J,m) &=& - (1-J-3Jm^2) \\
a_{12}(J,m) = a_{2,1} (J,m) &=& -(3-3J - 3Jm^2) \\
a_{22}(J,m) &=& 1-2J +J^2 -11m^2 + 14 Jm^2 \\
&-& 3 J^2 m^2 - 4 J m^4 + 3 J^2 m^4 -J^2 m^6 \nonumber
\eeqa

\subsection{Transition line and metastability}
The function $f_{h,J}(m)$ may display either one or two local minima according to the value of the couplings $h$ and $J$. In the case $h\geq 0$ that we are considering, whenever two local minima $m_+$ and $m_-$ are present, one has $m_{s.p.} = m_+$ with $\delta f_{h,J} = f_{h,J}(m_-) -  f_{h,J}(m_+) \geq 0$. The contribution of the \emph{state} $m_-$ to the saddle point integral vanishes in the large $N$ limit as long as $\delta f_{h,J} $ is finite, but for $\delta f_{h,J} \approx 1/N$, the contribution of the $m_-$ cannot be neglected, and requires the introduction of a term in the free energy of the form
\beq
F_{trans} (h,J) = - \log \left( 1 + e^{-N \delta f_{h,J}} \sqrt{ \frac{J- \frac{1}{1-m_+^2}}{J- \frac{1}{1-m_-^2}} } \right)  \; .
\eeq
For small enough values of $h$, the values of the minima become $m_+ = - m_-$, and above term can be written as
\beq
F_{trans} (h,J) = - \log \left( 1 + e^{-2 N h m_{s.p.}} \right) \; .
\eeq
Hence, this term describes the region of the coupling space which we call \emph{transition line}, where $h m_{s.p.} \ll 1/N$. The contribution to the averages and to the generalized susceptibility of this term is given by
\beqa
\left< \sum_i s_i \right>_{trans} &=& - N\,  [1-\tanh(N h m_{s.p.}) ] (h \chi_{s.p.} + m_{s.p.})\\
\left< \frac{1}{N} \sum_{i<j} s_i s_j\right>_{trans} &=& N \, h \, m_{s.p.} \, \chi_{s.p.}[1-\tanh(N h m_{s.p.})]
\eeqa
and
\beq
\hat \bchi_{trans} = N^2 \, \hat {\bm{b}} (h,J,m)  \; .
\eeq
The matrix $\hat{\bm{b}}(h,J,m)$ (whose explicit form is not particularly illuminating) can be obtained by deriving above averages with respect to $h$ and $J$.
\subsubsection{Determinant of the generalized susceptibility}
The term $\sqrt{\det \hat \bchi}$ is shown in chapter \ref{ch:Geometry} to be relevant in order to count the number of distinguishable statistical models inside a given region of the space $(h,J)$. It can be calculated at leading order in $N$ by keeping into account the different contributions to the free energy $F(h,J)$. The region in which $|h| \gg 1/N$ is described by $F \xrightarrow[N\to\infty]{} F_0 + F_{fluct}$, and it results
\beqa
\det \hat \bchi &\xrightarrow[N\to\infty]{}& \det (\hat \bchi_0 + \hat \bchi_{fluct})  \xrightarrow[N\to\infty]{}  \det \hat \bchi_0 +  \frac{N}{2} \chi_{s.p.}^3 \nonumber \\
&=& \frac{N}{2} \chi_{s.p.}^3 \; ,
\eeqa
while the region $h \ll 1/N$ is dominated by the contribution $F_0 + F_{trans}$, implying
\beq
\det \hat \bchi \xrightarrow[N\to\infty]{} \det (\hat \bchi_0 + \hat \bchi_{trans})  \xrightarrow[N\to\infty]{} N^3 \left( \frac{m_{s.p.}^4 \chi_{s.p.}}{\cosh^2 (N h m_{s.p.})} \right) + O(N^2) \; .
\eeq
\subsection{Marginal polytope for a fully connected ferromagnet \label{app:FCMargPoly}}
We want to characterize the marginal polytope $\mathcal G (\bphi)$ for the fully connected ferromagnet (\ref{eq:FCFerromagnet}), that is, the set of empirical averages $(m,c) \in \mathbb{R}^2$ compatible with at least one probability density $p \in \mathcal M(\Omega)$.
\prf{Due to density of the empirical frequency $\bar p$ in the space $\mathcal M(\Omega)$, we will consider the large $T$ limit of a sequence of observations $\{ m^{(t)}\}_{t=1}^T$. Fixed any $m \in [-1,1]$, one needs to require
\beq
m = \frac{1}{T} \sum_{t=1}^T m^{(t)} 
\eeq
and ask for a possible arrangement of the sequence $\{ m^{(t)}\}_{t=1}^T$ compatible with a correlation $c$, that is,
\beq
c = \frac{1}{T} \sum_{t=1}^T \frac{(m^{(t)})^2 - 1/N}{1-1/N} \, ,
\eeq
where, after easy combinatorics, we used the fact that the correlation $c^{(t)}$ measured in the observation number $t$ depends just upon the total magnetization $m^{(t)}$. Finding a solution to this problem is easy due to convexity of $\sum_t (m^{(t)})^2$. In particular by taking the limit $T \to \infty$ a solution can be found for any $m$, while then the minimum and the maximum value of $c$ are given respectively by $\frac{m^2 - 1/N}{1-1/N}$ and $1$. Interestingly, the same result can be obtained with more simplicity by exploiting the necessary condition $\textrm{Var}[\sum_i s_i] \geq 0$. Notice also that for large $N$ the connected correlation coefficient $c-m^2$ is bound from below by $(m^2 -1)/N$: equivalently no large system, subject to whatever type of interaction, can be globally anti-correlated.
}

\chapter{Convex optimization \label{app:ConvexOpt}}
In this appendix we will briefly remind part of the theory which has been developed in order solve unconstrained minimization problems of convex functions of the form $H(\bg) : \mathbb{R}^M \to \mathbb{R}$, addressing the interested reader to \cite{Boyd:2004oq} for a more complete analysis.
\section{Differentiable target}
Consider a convex, differentiable function $H(\bg) : \mathbb{R}^M \to \mathbb{R}$. Then for each point $\bg$ it exists a \emph{gradient} $\nabla H(\bg) = \left( \frac{\partial }{\partial g_1}, \dots , \frac{\partial}{\partial g_M} \right) H(\bg)$ and a positive semi definite \emph{Hessian} matrix $\hat \bchi (\bg)$ with elements $\chi_{\mu,\nu} = \partial_\mu \partial_\nu H(\bg)$. Then the following properties hold:
\begin{enumerate}
\item{
The gradient is a global under estimator of $H(\bg)$, namely for any $\bg^\prime$ one has that
\beq
H(\bg) \geq H(\bg^\prime) + \nabla H(\bg^\prime)^T (\bg - \bg^\prime) \; .
\eeq
}
\item{
The gradient defines a \emph{descent direction} $\bm{v} = - \nabla H(\bg)$, which means that for all $\bg$ it exists an $\epsilon$ such that
\beq
H(\bg - \epsilon \nabla H(\bg)) \leq H(\bg)
\eeq
}
\item{
The Hessian defines the descent direction $\bm{v} = - \hat{\bchi}^{-1} (\bg)  \nabla H(\bg)$. Algorithms exploiting this property usually go under the name of \emph{Newton's methods}.
}
\end{enumerate}
These properties are simples consequences of differentiability and convexity of $H(\bg)$, and allow to solve the problem the problem of its minimization. The first property implies that given a $\bg$ such that $\nabla H(\bg) = \bm{0}$, $\bg$ is a global minimum of $H(\bg)$. If this equation can be explicitly solved, the minimum can be found. Indeed if, as it often is the case, the condition $\nabla H(\bg) = \bm{0}$ is non-analytically solvable, it is possible to exploit properties 2.\ and 3.\ in order to build iterative algorithms which decrease the target function $H(\bg)$ at each step. In particular, iterative algorithms exploiting property 2.\ are expected to achieve linear convergence to the minimum, while more sophisticated algorithms (Newton methods) constructed by using the Hessian can achieve quadratic convergence. More efficient schemes (quasi-Newton methods) such as the L-BFGS approximation \cite{Liu:1989uq,Byrd:1995fk} exploit an approximation for the Hessian in order to save memory and computational power by exploiting successive updates of the gradient. We present in the following an example of a simple algorithm which can be used to minimize a convex differentiable $H(\bg)$, which we use mainly as a proof of principle for the solvability of this type of problem. Secondly, the efficiency of the Boltzmann learning algorithm presented in section \ref{sec:LimitationsHighN} is rooted in the gradient descent method.
\subsection{Gradient descent algorithm \label{app:GradDesc}}
Given a convex, differentiable $H(\bg)$ and starting point $\bg^{(0)}$, we consider a sequence $\{ \bg^{(k)}\}_{k=1}^K$ built according to the iterative scheme
\beq
\bg^{(k+1)} = \bg^{(k)} - \epsilon_k \nabla H(\bg^{(k)}) \; , \label{eq:GradDesc}
\eeq
where we introduced the \emph{schedule} $\{ \epsilon_k \}_{k=1}^K$. Suppose that each of the $\epsilon_k$ is chosen in order to satisfy the (Armijo) condition
\beq
H(\bg^{(k+1)}) \leq H(\bg^{(k)}) - \epsilon_k \beta || \nabla H(\bg) ||^2 \; , \label{eq:Armijo}
\eeq
for a given $0 < \beta < 1$, by considering the initial value $\epsilon_k = 1$ and iterating the map $\epsilon_k \leftarrow \epsilon_k / \tau$ for $\tau < 1$ until (\ref{eq:Armijo}) is satisfied.
Then it holds that either $\min_{\bg \in \mathbb{R}^M} H(\bg) = -\infty$ or $\lim_{k\to\infty} || \nabla H(\bg) ||^2 = 0$, that is, if a minimum exists, the sequence $\{ g^{(k)}\}_{k=1}^K$ can approximate it with arbitrary precision.
\rmk{
Searching the optimal $\epsilon_k$ is usually called a \emph{line search}, and the procedure that we introduce to find it is guaranteed to find an $\epsilon_k$ satisfying (\ref{eq:Armijo}) if the maximum eigenvalue of $\hat \bchi$ is bounded by a given $\chi_{max}$. In particular the convexity of $H(\bg)$ and a straightforward application of Taylor theorem allow to prove that any $\epsilon_k$ in the interval
\beq
0 \leq \epsilon_k \leq \frac{2 (1-\beta)}{\chi_{\max}} 
\eeq
satisfies the Armijo condition (\ref{eq:Armijo}).
}
\prf{
In order to prove that the convergence of the algorithm, one can use (\ref{eq:Armijo}) to iteratively build the inequality
\beq
H(\bg^{(K)}) \leq H(\bg^{(K-1)}) - \beta \epsilon_k || \nabla H(\bg^{(K-1)}) ||^2 \leq  H(\bg^{(0)}) - \beta \sum_{k=0}^K \epsilon_k || \nabla H(\bg^{(k)}) ||^2 \; .
\eeq
Then, as the succession $H(\bg^{(K)}) - H(\bg^{(0)})$ is strictly decreasing in $K$, it has a limit. Such limit can be either $-\infty$ (in which case $H(\bg)$ has no minimum) or can be finite. The finiteness of the limit implies that
\beq
H(\bg^{(\infty)}) - H(\bg^{(0)}) = \lim_{K\to\infty} \sum_{k=0}^K \epsilon_k || \nabla H(\bg^{(k)}) ||^2 
\eeq
which leads to
\beq
\lim_{K\to\infty} \epsilon_K || \nabla H(\bg^{(K)}) ||^2 = 0 \; .
\eeq
} 
Notice that the rate of convergence in $K$ of this algorithm can be rather slow, which is the reason why more sophisticated algorithms are commonly used to perform this task (see \cite{Boyd:2004oq}).

\section{Non-differentiable target}
If a convex function is not differentiable in all of its domain the solution of the minimization problem is technically more complicated, but it is still possible to take advantage of the convexity property in order to build efficient minimization algorithms (see \cite{Boyd:2010tg,Boyd:2004oq}).
Consider a convex function $H(\bg) : \mathbb{R}^M \to \mathbb{R}$. Then one can define a \emph{sub-gradient} as any global underestimator of $H(\bg)$, namely $\bm{v} \in \mathbb{R}^M$ is a subgradient of $H(\bg)$ in $\bg^\prime$ if for any $\bg$ one has
\beq
H(\bg) \geq H(\bg^\prime) + \bm{v}^T (\bg - \bg^\prime) \; . \label{eq:SubGrProp}
\eeq
The set of all the sub-gradients of $H(\bg)$ in $\bg^\prime$ is called the \emph{sub-differential} of $H(\bg)$, and is denoted with $\tilde \nabla H(\bg)$. One can show that
\begin{itemize}
\item{
$\tilde \nabla H(\bg)$ is non-empty if $H(\bg)$ is locally convex and bounded around $\bg$.
}
\item{
$\tilde \nabla H(\bg)$ is closed and convex.
}
\item{
The sub-differential is \emph{additive}, so that $\tilde \nabla [ H_1(\bg) + H_2 (\bg) ] = \tilde \nabla  H_1(\bg) + \tilde \nabla H_2 (\bg) $.
}
\item{
The sub-differential has the \emph{scaling} property $\tilde \nabla \lambda H(\bg)  = \lambda \tilde \nabla H(\bg) $ for $\lambda>0$.
}
\item{
If $H(\bg)$ is differentiable, then $\tilde \nabla H(\bg)  =\{  \nabla H(\bg) \}$.
}
\end{itemize}
This properties characterize the sub-differential as a notion generalizing the ordinary differential, which is suitable to solve problems involving non-differentiable functions. In particular the properties shown above for differentiable functions generalize to:
\begin{enumerate}
\item{
If $\bm{0} \in \tilde \nabla H(\bg)$ then $\bg$ is a global minimum of $H(\bg)$.
}
\item{
The direction $\bm{v} = - \epsilon \tilde \nabla H(\bg)$ is not in general a descent direction.
}
\end{enumerate}
This implies that in order to minimize a non-differentiable function it is still possible to find the points whose sub-differential is equal to zero, but that a naive sub-gradients descent similar to (\ref{eq:GradDesc}) is not guaranteed to find a solution.
\subsubsection{An example: the absolute value}
Consider the function $H(g): \mathbb{R} \to \mathbb{R}$ defined as $H(g) = H_d(g) + |g|$, with $H_d(g)$ convex and differentiable. Then the sub-differential of $H(\vec g)$ is given by
\beq
\tilde \nabla H(g) = \nabla H_d(g) + \textrm{sgn} (g)
\eeq
where
\beq
\textrm{sgn} (g) =  \left( \begin{array}{ccc}   \textrm{sign} (g) & \textrm{if} & g\neq0 \\ \left[-1,1\right] &  \textrm{if} & g= 0\end{array} \right. \; . \label{eq:SgnFunct}
\eeq
which is minimum for
\beqa
x = 0 &\textrm{if} & | \nabla H_d(0) | \leq  1 \nonumber \\
x \gtrless 0 &\textrm{if} & \nabla H_d(x) = \mp 1 \; .
\eeqa

The notion of sub-gradient also allow us to generalize the gradient descent algorithm to non-differentiable functions, as shown in the following.
\subsection{Sub-gradient descent algorithm}
Consider a convex $H(\bg)$, a starting point $\bg^{(0)}$, and a sequence $\{ \bg^{(k)}\}_{k=1}^K$ built according to the iterative scheme
\beq
\bg^{(k+1)} = \bg^{(k)} - \epsilon_k \bm{v}^{(k)} \; , \label{eq:SubGradDesc}
\eeq
where $\bm{v}^{(k)} \in \tilde \nabla H(\bg^{(k)})$ is a sub-gradient in $\bg^{(k)}$, and where we introduced the schedule $\{\epsilon_k \}_{k=1}^K$. Then, one can show that if $H(\bg)$ has a minimum $\bg^\star$, then
\beq
H(\bg^{best}) - H(\bg^\star) \leq \frac{R^2 + G^2 \sum_{k=1}^K \epsilon_k^2}{2 \sum_{k=1}^K \epsilon_k } \label{eq:SubGrBound}
\eeq
where $\bg^{best} = \arg\min_{\bg \in \bg^{(k)}} H(\bg)$, while $R$ and $G$ enforce respectively a bound of the initial distance from the minimum $||\bg^{(1)} - \bg^\star ||^2 \leq R$ and the Lipschitz bound $\frac{|H(\bg) - H(\bg^\prime)|}{|| \bg - \bg^\prime||}\leq G$\footnote{Although the hypothesis of Lipschitz bounded $H(\bg)$ is not strictly required, for the sake of clarity we have chosen choose to present the algorithm in this simpler form.}. In particular, by choosing $\epsilon_k \propto 1 / k$, one can show that in that case
\beq
\lim_{k\to\infty} H(\bg^{best}) - H(\bg^\star) = 0
\eeq
\prf{
To prove this result, it is necessary to consider the Euclidean distance to the minimum $\bg^\star$, which due to the property (\ref{eq:SubGrProp}) satisfies
\beqa
|| \bg^{(K)} - \bg^\star||^2 &=& || \bg^{(K-1)} - \epsilon_{K-1} \bm{v}^{(K-1)} - \bg^\star ||^2  \\
&=& || \bg^{(K-1)} - \bg^\star ||^2   - 2 \epsilon_{K-1} (\bg^{(K-1)} - \bg^\star)^T \bm{v}^{(K-1)} + \epsilon_k^2 ||\bm{v}^{(K-1)} ||^2 \nonumber \\
&\leq& || \bg^{(K-1)} - \bg^\star ||^2   - 2 \epsilon_{K-1} (H(\bg^{(K-1)}) - H(\bg^\star)) + \epsilon_k^2 ||\bm{v}^{(K-1)} ||^2 \nonumber \; ,
\eeqa
so that one can recursively build the inequality
\beqa
|| \bg^{(K)} - \bg^\star||^2 &\leq& || \bg^{(0)} - \bg^\star ||^2  - 2 \sum_{k=0}^K \epsilon_k (H(\bg^{(k)}) - H(\bg^\star)) + \sum_{k=0}^K \epsilon_k^2 ||\bm{v}^{(k)} ||^2 \nonumber \\
&\leq & R^2 - 2 [H(\bg^{best}) - H(\bg^\star)] \sum_{k=0}^K \epsilon_k + G^2 \sum_{k=0}^K \epsilon_k^2 \; .
\eeqa
Finally, by using $|| \bg^{(K)} - \bg^\star||^2 \geq 0$, one can rearrange the terms and obtain the bound (\ref{eq:SubGrBound}).
}
Notice that in this case the sequence $\epsilon_k$ is not optimized on-line, rather it is fixed at the beginning of the algorithm. This is because the sub-gradient doesn't specify necessarily a descent direction, hence the sub-gradient descent may increase the function $H(\bg)$, requiring the values $\bg^{best}$ and $H(\bg^{best})$ to be stored at each iteration step.

\chapter{Complete families}
\section{Rate of convergence for the complete inverse problem \label{app:ConvCompInvProb}}
Consider a statistical model $\bp$ in which all states have strictly positive probability (i.e.\ it exists a $p_{min} \neq 0$ such that $ \forall s \quad p_{min} \leq p_s$). We want to show how within inference scheme (\ref{eq:FormalCoupling})
the inferred couplings concentrate around their actual values at fixed $N$ in the limit $T \rightarrow \infty$.
\prf{
The expression for $g_\mu^\star$ is:
\beq
g_\mu^\star = \frac{1}{|\Omega|} \sum_s \phi_{\mu,s} \log \bar  p_s \; ,
\eeq
while the probability to observe a given set empirical frequencies $\bar \bp$ out of the measure of $T$ samples is given by the multinomial distribution described in section \ref{sec:DirSLDev}.
Its mean and correlations are sufficient to completely determine the convergence for large enough values of $T$. In particular one finds that
\beq
\langle g_\mu^\star \rangle_T \xrightarrow[T\to\infty]{}  \frac{1}{|\Omega|} \sum_s \phi_{\mu,s} \log  \langle  \bar p_s \rangle = g_\mu \, ,
\eeq
while the fluctuations of the inferred couplings are equal to
\beqa
{\rm Var} (g_\mu^\star) = \langle (g_\mu^\star)^2 \rangle_T - \langle g_\mu^\star \rangle^2_T & \xrightarrow[T\to\infty]{}& \frac{1}{|\Omega|^2} \sum_{s,s^\prime} \phi_{\mu,s}\phi_{\mu,s^\prime} \frac{{\rm Cov} (\bar p_s,\bar p_{s^\prime})}{p_s p_{s^\prime}}  \\
&=& \frac{1}{T} \left[ \left( \frac{1}{|\Omega|^2} \sum_s \frac{1}{p_s} \right) - \delta_{\mu 0} \right] \, ,
\eeqa
which is the result shown in equation (\ref{eq:FormalInvFisher}).
This can be generalized to the case in which the set of states with strictly positive probabilities is a subset $\mathcal I \subset \Omega$, so that one can define the set of \emph{regular} operators $\bphi^{reg} = \{ \phi_\mu \in \phi \; | \; \sum_{s \in \mathcal I} \phi_{\mu,s} = 0\}$. The same proof as above can be performed for regular operators on the estimator
\beq
g_\mu^{\star reg} = \frac{1}{|\Omega|} \sum_{s \in \mathcal I} \phi_{\mu, s} \log \bar p_s \; ,
\eeq
finding the result described in equation (\ref{eq:FormalInvFisherHidSector}).
}

\section{Factorization property for tree-like models \label{app:TreeFactorization}}
In this section we prove a fundamental property of statistical models whose the interaction structure is loop-less, which we call \emph{trees} and rigorously define as follows.\footnote{This definition corresponds to what is often referred in literature as a \emph{forest}, while the word tree is typically reserved to each connected component of a forest. For simplicity we will disregard such difference, and make no distinction among trees and forests.}

\defin{
Consider a statistical model $(\bphi,\bg)$ of the form (\ref{eq:ProbDensity}), with $g_\mu \neq 0$ for all $g_\mu \in \bg$. Then the set $\bphi$ is called a \emph{tree} if it is not possible to find a \emph{cycle} connecting any set of vertices, i.e., it doesn't exist a closed path $\{i_1,\dots,i_{L-1}, i_{L} = i_1\} \in V^L$ such that for each couple $\{ i_n, i_{n+1} \} $ there exist an operator $\phi_{i_n, i_{n+1}} \in \bphi$ depending on both $s_{i_n}$ and $s_{i_{n+1}}$, with $\phi_{i_n, i_{n+1}} \neq \phi_{i_m, i_{m+1}} $ for all $n \neq m \in \{ 1, \dots,L-1\}$.
}
For trees we will show along the lines of \cite{Mezard:2009ko} that the following factorization property holds.
\theor{
Consider a tree-like statistical model $(\bphi,\bg)$. Then its associated probability density $\bp$ can be written as
\beq
p(s) = \prod_{\mu = 1}^M p^{\partial \phi_\mu} (s^{\partial \phi_\mu}) \prod_{i \in V} p^{\{ i\}} (s^{i})^{1- |\partial i|} \; , \label{eq:FactorizationTree}
\eeq
where $\partial i = \{ \phi_\mu \in \bphi \; | \;  \phi_\mu (s) \textrm{ depends upon } s_i \}$ while $\partial \phi = \{ i \in V \; | \; \phi (s) \textrm{ depends upon } s_i \}$.
}
\prf{
The theorem can be proved by induction on the number of operators $M$. Consider the case $M=1$ in which just one operator is present ($\bphi = \{ \phi \}$). Then, it is trivial to see that equation (\ref{eq:FactorizationTree}) holds due to
\beq
p(s) \propto \exp \left[ g \phi(s) \right] \propto p^{\partial \phi}(s^{\partial \phi}) \prod_{i \in V \backslash \partial \phi_\mu} p^{\{ i \}}(s_i) \; .
\eeq
Let then property (\ref{eq:FactorizationTree}) hold for the case of $M$ operators, and consider a statistical model in which $|\bphi|=M+1$. Then, as $(\bphi,\bg)$ is a tree, it is possible to consider without loss of generality an operator $\phi_\mu \in \bphi$ such that $|\partial j | = 1$ for all $j \in \partial \phi_\mu$ but at most a single variable. Suppose that such variable exists, and label it as $i$. Then by defining the cluster $\Gamma = \{ j \in V | j \not\in \partial \phi_\mu \} \cup \{ i \}$, a straightforward application of Bayes rule yields
\beqa
p(s) &=& p^\Gamma(s^\Gamma) p^{V \backslash \Gamma} (s^{V \backslash \Gamma} | s^{\Gamma}) \nonumber \\
&=&  p^\Gamma (s^\Gamma) \frac{p^{\partial \phi_\mu} (s^{\partial \phi_\mu})}{p^{\{i\}}(s^{\{i\}})} \label{eq:BayesFactorTree} \; .
\eeqa
Additionally, the marginal $p^\Gamma (s^\Gamma)$ can be written in the form
\beq
p^\Gamma(s^\Gamma) \propto \exp \left( \sum_{\phi_\nu \in \bphi \backslash \partial \phi_\mu} g_\nu \phi_\nu (s^\Gamma) \right) \underbrace{ \sum_{s_j \; |\; j \not\in \Gamma} \exp \left( g_\mu \phi_\mu(s^{\partial \phi_\mu}) \right)}_{ \equiv \; \psi(s_i)} \; .
\eeq
Then it is possible to reabsorb the $\psi(s_i)$ factor inside a new operator obtained by the following change on a generic $\phi_\rho \in \partial i \backslash \phi_\mu$:
\beq
\phi_\rho (s^{\partial \phi_\rho} ) \to \phi^\prime_\rho (s^{\partial \phi_\rho}) = \phi_\rho (s^{\partial \phi_\rho} ) + \frac{\log \psi (s_i)}{g_\rho} \; .
\eeq
The statistical model describing the reduced problem for $\Gamma$ spins can thus be described by using $M$ operators, so that it is possible to use the inductive hypothesis to show that
\beq
p^\Gamma  (s^\Gamma) = p^{\{ i\}} (s_i)^{1 - (| \partial i |-1)} \prod_{\phi_\nu \in \bphi \backslash \{\phi_\mu\}} p^{\partial \phi_\nu } (s^{\partial \phi_\nu}) \prod_{j \in V \backslash \{i\}} p^{\{ j\}} (s_j)^{1- | \partial j|}\; .
\eeq
Above expression can finally be plugged into equation (\ref{eq:BayesFactorTree}) so to obtain equation (\ref{eq:FactorizationTree}). In order to prove the thesis (\ref{eq:FactorizationTree}) in full generality it is nevertheless necessary to perform an analogous derivation in the case in which no such $i$ variable exist, an exercise which for the sake of conciseness we leave to the reader.
}
 
\section{Factorization property of the one-dimensional periodic chain \label{app:1DChain}}
Consider a one-dimensional periodic chain of size $N$, range $R$ and periodicity $\rho$ defined by a complete, orthogonal set of operators $\bphi$ and a set of translation operators $\bT$. We want to show that for such chain it holds the factorization property
\beq
 p(s) = \prod_{n=0}^{N/\rho-1} \frac{p^{\Gamma_n}(s^{\Gamma_n})}{p^{\gamma_n}(s^{\gamma_n})} \; .
\eeq
where the sets $\Gamma_n$ and $\gamma_n$ are defined as in section \ref{sec:Inv1DChain}.
\prf{
To obtain this result, one needs to define a two-dimensional model defined by the log-probability
\beqa
\log p_\lambda (s,t) &=& - \log Z(\bg) + \sum_{n=0}^{N/\rho-1} \sum_{\mu \in \phi}  g_\mu \phi_\mu(s_{1+n\rho}^n,\dots,s_{R+n\rho}^n) \nonumber \\
&+& \lambda \sum_{n=0}^{N/\rho-1} \sum_{i=(n+1)  \rho + 1}^{n \rho + R} \left[ (t_i^n - s_i^n)^2+ (t_i^n - s_i^{n+1})^2 \right] \; ,
\eeqa
in which the configuration space contains the degrees of freedom are $s_i^n \in \{ -1,1\}$ (with $n=0,\dots,N/\rho-1$ and $i=1+n\rho,\dots, R+n\rho$) and $t_i^n \in \{ -1,1\}$ (with $n=0,\dots,N/\rho-1$ and $i=1+(n+1)\rho,\dots, R+n\rho$). The model is sketched in figure \ref{fig:2DMap}, in which it is possible to appreciate the connection with the original one-dimensional chain. In particular, the interaction mediated by $\lambda$ controls the strength of the bonds in the auxiliary dimension (labeled by $n$), so that in the limit $\lambda \to \infty$ the model describes the original chain, with the obvious identification $s^n_i \to s_i$ and $t^n_i \to s_i$.
\begin{figure}[h] %  figure placement: here, top, bottom, or page
   \centering
   \includegraphics[width=4in]{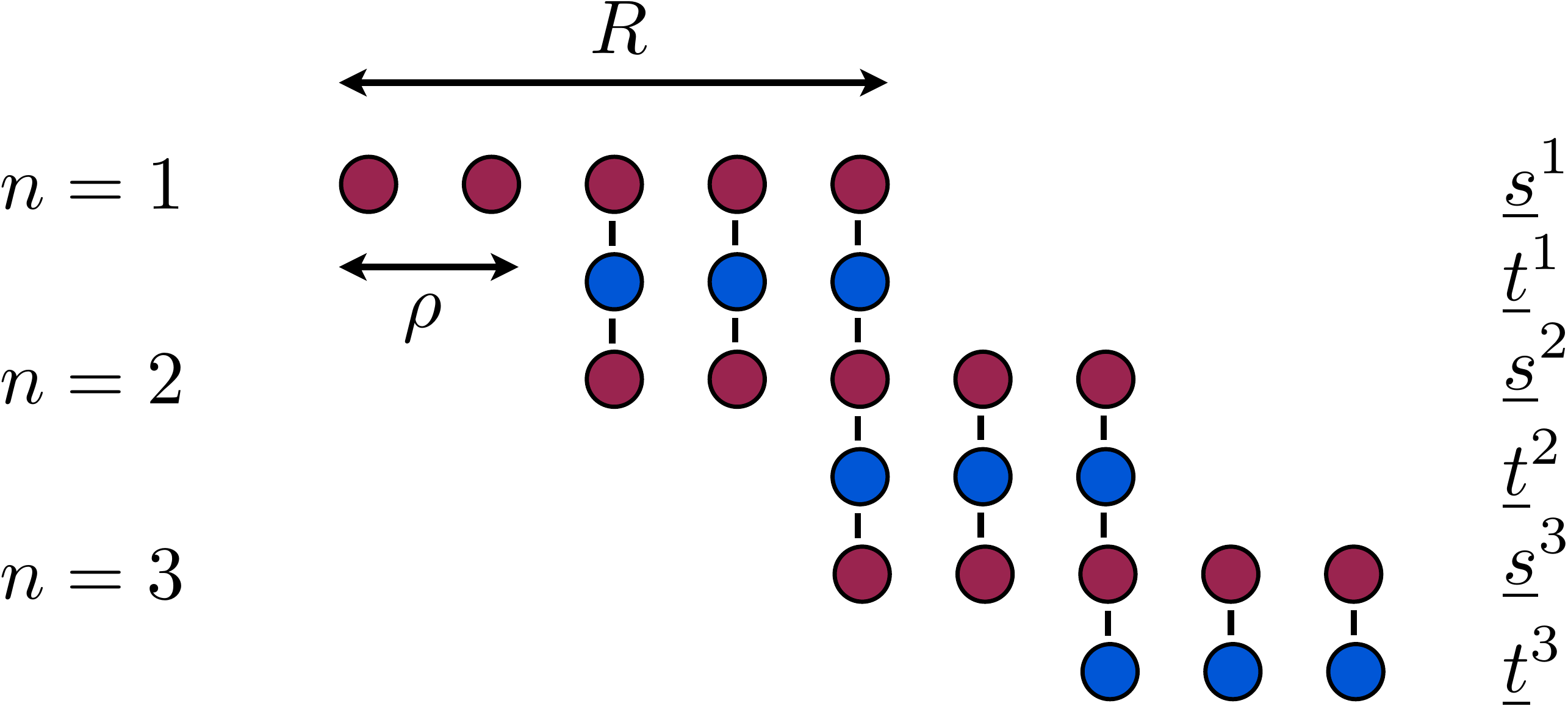} 
   \caption{Two dimensional auxiliary model $p_\lambda(s)$ associated with the original distribution $p(s)$ describing a one-dimensional periodic chain.}
   \label{fig:2DMap}
\end{figure}
By defining the row variables $\underline s^n = \{ s_i^n\}_{i=1+n\rho}^{i=R+n\rho}$ and $\underline t^n = \{ t_i^n\}_{i=1+(n+1)\rho}^{i=R+n\rho}$, one can see that the log-probability for the two dimensional model can be written as
\beq
\log p_\lambda (s,t) = -\log Z_\lambda(\bg) - \sum_{n=0}^{N/\rho-1} \bigg[ \mathcal H_\lambda^n(\underline s^n)  + \mathcal H^{n,n}_\lambda(\underline s^n, \underline t^n) + \mathcal H^{n,n+1}_\lambda (\underline t^n, \underline s^{n+1})\bigg]   \; , \label{eq:2DMapTree}
\eeq
hence the distribution over the degrees of freedom $\underline s^n$ and $\underline t^n$ and whose log-probability is given by (\ref{eq:2DMapTree}) defines a \emph{tree}, because only successive row of variables interact\footnote{Periodic boundary conditions enforce the presence of a single loop of length $N$, so that the model is not exactly a tree. Nevertheless, for $N$ large enough and for $\bg$ sufficiently distant from critical points of the model, if any, the presence of such loop can be neglected.}. For such a model, one can straightforwardly generalize the result of appendix \ref{app:TreeFactorization} to the case of the non-binary variables $\underline s^n$ and $\underline t^n$ to show that the full measure $p_\lambda(s,t)$ can be decomposed into the product of the marginals
\beq
p_\lambda(s,t) = \frac{
\prod_n p_\lambda^{\Gamma_n \cup  \gamma_n} (\underline s^{n} ,\underline t^{n} ) p_\lambda^{\gamma_n \cup \Gamma_{n+1}} (\underline t^n ,\underline s^{n+1} )
}{
\prod_n p_\lambda^{\Gamma_n} (\underline s^n) p_\lambda^{\gamma_n} (\underline t^n)
} \; ,
\eeq
where $\Gamma_n$ and $\gamma_n$ are analogously defined in the case of the two-dimensional model. By taking the $\lambda \to \infty$ limit, the identification
\beqa
p_\lambda^{\Gamma_n \cup  \gamma_n} (\underline s^{n} ,\underline t^{n} )&\xrightarrow[\lambda \to \infty]{}& p^{\Gamma_n}(s_{n\rho +1} ,\dots , s_{n \rho +R}) \\
p_\lambda^{\Gamma_n} (\underline s^n) &\xrightarrow[\lambda \to \infty]{}&  p^{\Gamma_n}(s_{n\rho +1} ,\dots , s_{n \rho +R}) \\
p_\lambda^{\gamma_n} (\underline t^n) &\xrightarrow[\lambda \to \infty]{}&  p^{\gamma_n} (s_{(n+1)\rho +1} ,\dots , s_{n \rho +R}) \; .
\eeqa
allows to recover the factorization property which had to be proven.
}

\chapter{Geometry}
\section{Geodesics \label{app:Geodesics}}
We want to find that the condition which a curve $\gamma : [a,b] \in \mathbb{R} \to \mathcal M(\bphi)$ has to satisfy order to minimize a functional $\ell(\gamma)$ of the form
\beq
\ell(\gamma) = \int_a^b dt \sqrt{\chi_{\mu,\nu} \frac{d\gamma_\mu}{dt} \frac{d \gamma_\nu}{dt}} \; ,
\eeq
(in which summations on repeated indices are implicit) is given by equation (\ref{eq:Geodesics}).
\prf{In order for $\gamma$ to be a minimum, it needs to extremize the functional $\ell(\gamma)$, so that by constructing the variation $\gamma \to \gamma + \delta \gamma$ we can impose $\delta \ell(\gamma+\delta \gamma) - \ell(\gamma) = \delta \ell (\gamma) = 0$. This implies
\beq
\delta \ell(\gamma) = \int_a^b dt \left(\chi_{\mu,\nu} \frac{d\gamma_\mu}{dt} \frac{d \gamma_\nu}{dt} \right)^{-1/2} \left( \frac{1}{2} \partial_\rho \chi_{\mu,\nu} \frac{d\gamma_\mu}{dt} \frac{d \gamma_\nu}{dt} \delta \gamma_\rho
+ \chi_{\mu,\nu} \frac{d\gamma_\mu}{dt} \frac{d }{dt}\delta \gamma_\nu   \right) = 0\; .
\eeq
By changing variable to
\beq
dt = \left(\chi_{\mu,\nu} \frac{d\gamma_\mu}{dt} \frac{d \gamma_\nu}{dt} \right)^{-1/2} du \; ,
\eeq
one obtains
\beq
\delta \ell(\gamma) = \int_{u_a}^{u_b} du  \left( \frac{1}{2} \partial_\rho \chi_{\mu,\nu} \frac{d\gamma_\mu}{du} \frac{d \gamma_\nu}{du} \delta \gamma_\rho
+ \chi_{\mu,\nu} \frac{d\gamma_\mu}{du} \frac{d }{du}\delta \gamma_\nu   \right) = 0\; ,
\eeq
which after integration by parts and some manipulation reads
\beq
\delta \ell(\gamma) = - \int_{u_a}^{u_b} du  \left[  \chi_{\mu,\rho} \frac{d^2 \gamma_\mu}{du^2} + \frac{1}{2} \left( \partial_\nu \chi_{\mu,\rho} + \partial_\mu \chi_{\nu,\rho} - \partial_\rho \chi_{\mu,\nu} \right) \frac{d \gamma_\mu}{du} \frac{d \gamma_\nu}{du} \right] \delta \gamma_\rho = 0 \; .
\eeq
Imposing the integrand of above expression to be equal to zero and composing with the inverse Fisher information $\hat \bchi^{-1}$ yields equation (\ref{eq:Geodesics}).
}

\section{Property of the maximum likelihood estimator}
We want to prove that, given a probability density $\bp$ defined by a statistical model $(\bphi,\bg)$, for any empirical dataset of length $T$ generated by $\bp$ producing empirical averages $\bar \bphi$, the probability of the maximum likelihood estimator $\bg^\star(\bar \bphi)$ taking a given value $\bg^\prime$ satisfies
\beq
\lim_{\delta \bg \to \bm{0}} \lim_{T\to \infty} -\frac{1}{T} \log \textrm{Prob}(\bg^\star(\bar \bphi) = \bg^\prime + \delta \bg) = D_{KL}( \bp^\prime || \bp ) \; ,
\eeq
being $\bp^\prime$ the density associated with the statistical model $(\bphi, \bg)$.
\prf{
To prove this relation, we first need to define the set $\mathcal M(\bphi,\bg^\prime)$ of probability distributions \emph{compatible} with $\bg^\prime$, defined by
\beq
\mathcal M(\bphi,\bg^\prime) = \left\{ \bq \in \mathcal M(\Omega) \; \Bigg| \; \forall \phi_\mu \in \bphi , \;  \sum_s q_s \phi_{\mu,s} = \sum_s \phi_{\mu,s} \exp \left( \sum_{\mu=0}^M g^\prime_\mu \phi_{\mu,s}\right) = \langle \phi_\mu \rangle_{\bg^\prime} \right\}
\eeq
It can be shown that:
\begin{enumerate}
\item{
$\mathcal M (\bphi, \bg^\prime)$ is compact.
}
\item{
$\bar \bp \in \mathcal M (\bphi, \bg^\prime)$, if and only if $\bg^\star (\bar \bphi) = \bg^\prime$.
}
\item{Due to continuity of the functions $\bg^\star(\bar \bphi)$ and $\bar \bphi (\bar \bp)$ it holds
\beq
\lim_{\bar \bq \to \bar \bp} \bg^\star (\bar \bphi(\bar \bq)) = \bg^\star (\bar \bphi(\bar \bp))
\eeq
}
\end{enumerate}
Then, Sanov theorem (section \ref{sec:InvSLDev}) applied to the set $\mathcal M(\bphi,\bg)$ implies that
\beq
\lim_{\delta \to 0} \lim_{T\to \infty} -\frac{1}{T} \log  \textrm{Prob} [\bar \bp \in \mathcal M^\prime (\bphi, \bg^\prime)] = D_{KL} (\bq^\star || \bp) 
\eeq
where $\bq^\star = \arg\min_{\bq \in \mathcal M(\bphi,\bg^\prime)} \left[ D_{KL} (\bq || \bp)\right]$. In order to find the minimum, one can show that for $\bq \in \mathcal M(\bphi,\bg^\prime)$ it holds
\beq
D_{KL}(\bq || \bp ) = - S(\bq) + F(\bg) + \sum_{\mu=1}^M g_\mu \langle \phi \rangle_{\bg^\prime} \; ,
\eeq
where only the term $-S(\bq)$ depends on the distribution $\bq$. Then the maximum entropy principle (appendix \ref{app:MaxEntPr}) states that the density $\bq \in \mathcal M(\bphi, \bg^\prime)$ maximizing $S(\bq)$ is the statistical model described $(\bphi, \bg^\prime)$, whose associated probability density has been called $\bp^\prime$. Then one has
\beq
\lim_{\delta \to 0} \lim_{T\to \infty} -\frac{1}{T} \log \textrm{Prob} [\bar \bp \in \mathcal M^\prime (\bphi, \bg^\prime)] =  D_{KL} (\bp^\prime || \bp) \label{eq:AppSanovCorol}
\eeq
finally, by using property 2.\ of $\mathcal M (\bphi, \bg^\prime)$ and the continuity property 3.\ one has that for $\delta$ sufficiently small, $\textrm{Prob} [\bar \bp \in \mathcal M^\prime (\bphi, \bg^\prime)]$ is arbitrarily close to $\textrm{Prob} [\bg^\star (\bar \bphi) - \bg^\prime \in \delta \bg]$, which together with (\ref{eq:AppSanovCorol}) proves the thesis (\ref{eq:SanovCorol}).
}

\section{Expansion of the Kullback-Leibler divergence \label{app:ExpansionKLIndist}}
We want to prove that, given a pair of statistical models $(\bphi,\bg)$ and $(\bphi,\bg^\prime)$ and an accuracy parameter $\epsilon$, for large $T$ they are indistinguishable if condition (\ref{eq:Indistinguishability}) holds.
\prf{
If $\bg$ and $\bg^\prime$ are indistinguishable, then corollary (\ref{eq:SanovCorol}) implies that, for large $T$, $D_{KL} (\bp^\prime || \bp) \leq \epsilon/ T$. As $D_{KL} (\bp^\prime || \bp) = 0 \Leftrightarrow \bp = \bp^\prime$, one can expand $D_{KL}(\bp^\prime || \bp)$ around the point $\bg^\prime = \bg$, obtaining
\beqa
D_{KL} (\bp^\prime || \bp) &\approx& D_{KL} (\bp || \bp) + \sum_{\mu=1}^M \frac{\partial D_{KL} (\bp^\prime || \bp)}{\partial g_\mu^\prime} \Bigg|_{\bg^\prime =\bg} (g^\prime_\mu - g_\mu)  \\
&+& \frac{1}{2} \sum_{\mu,\nu=1}^M \frac{\partial^2 D_{KL} (\bp^\prime || \bp)}{\partial g_\mu^\prime \partial g_\nu^\prime}\Bigg|_{\bg^\prime =\bg} (g^\prime_\mu - g_\mu) (g^\prime_\nu - g_\nu) \; .
\eeqa
It is easy to see that for the probability distributions $\bp$ and $\bp^\prime$ associated respectively to $\bg$ and $\bg^\prime$ it holds equation (\ref{eq:RelDKLFreeEn}), which reads
\beq
D_{KL} (\bp^\prime || \bp) = F(\bg^\prime)  - F(\bg) + \sum_{\mu =1}^M (g_\mu^\prime - g_\mu) \langle \phi_\mu \rangle_{\bp^\prime} \; . \label{eq:KLFreeEnRel}
\eeq
As equation (\ref{eq:KLFreeEnRel}) implies $D_{KL} (\bp || \bp) = 0$, $\partial_\mu D_{KL} (\bp^\prime || \bp) |_{\bg^\prime = \bg}= 0$ and \\
$\partial_\nu \partial_\mu D_{KL} (\bp^\prime || \bp) |_{\bg^\prime = \bg}= \chi_{\mu,\nu}$, equation (\ref{eq:Indistinguishability}) is proven.
}

\section{Volume of indistinguishability \label{app:VolumeIndist}}
Given the space $\mathcal M(\bphi)$ identified by the minimal operator set $\bphi$, we want to show that the volume of the space of indistinguishable distributions around a point $\bg$ is given by equation (\ref{eq:VolIndDist}), where $T$ is the length of the dataset and $\epsilon > 0$ is the accuracy parameter.
\prf{
The volume $\mathcal V_{T,\epsilon}(\bg)$ is given by
\beq
\mathcal V_{T,\epsilon}(\bg) = \int_{\mathcal M_{ind}} d\bg \; ,
\eeq
while property (\ref{eq:Indistinguishability}) characterizes the region of indistinguishability $\mathcal M_{ind}$ around $\bg$ as $\mathcal M_{ind} \xrightarrow[T\to\infty]{} \left\{ \bp^\prime \in \mathcal M(\Omega) \; | \frac{1}{2}(\bg^\prime-\bg)^T \hat \bchi (\bg^\prime-\bg) \leq \frac{ \epsilon}{T} \right\} \subseteq \mathcal M (\bphi)$.
We also need to require that $T$ large enough in order to neglect the variations of $\bchi$ in $\mathcal M_{ind}$, so that we can treat it as constant in $\bg^\prime$.
Due to symmetry  of $\hat \bchi$, the components of Fisher information matrix can be decomposed as $\chi_{\mu,\nu} = \sum_{\lambda=1}^M u_{\mu,\lambda} \, \chi_\lambda \, u_{\nu,\lambda}$, while due to minimality of $\bphi$ the eigenvalues $\chi_\lambda$ are strictly positive, suggesting the change of coordinates
\beq
\eta_\lambda = \sum_{\mu=1}^M (g_\mu^\prime-g_\mu) u_{\mu,\lambda} \sqrt{\chi_\lambda} \; . \label{eq:JacobVolInd}
\eeq
Then the region $\mathcal M_{ind}$ is mapped into the spherical region \\
$\mathcal M_{ind} = \left\{ \bp^\prime \in \mathcal M(\Omega) \; | \frac{1}{2}\bm{\eta}^T  \bm{\eta} \leq  \frac{ \epsilon}{T} \right\}$ so that the volume becomes
\beq
\mathcal V_{T,\epsilon}(\bg) = \frac{1}{\sqrt{\det \hat \bchi}} \int_{\mathcal M_{ind}} d\bm{\eta}  \; ,
\eeq
where $1 / \sqrt{\det \hat \bchi} \neq 0$ is the Jacobian of transformation (\ref{eq:JacobVolInd}). It is then sufficient to remind that the volume of a sphere of radius $\sqrt{\frac{2 \epsilon}{T}}$ in $M$ dimensions is given by
\beq
\int_{\mathcal M_{ind}} d \bm{\eta} = \left( \frac{\pi^{\frac{M}{2}}}{\Gamma (\frac{M}{2} + 1)} \right) \left( \frac{2 \epsilon}{T} \right)^{\frac{M}{2}}
\eeq
to prove equation (\ref{eq:VolIndDist}).
}

\section{Estimation of the empirical observables for an Hawkes point process \label{app:ApproxHawkes}}
\begin{SCfigure} %  figure placement: here, top, bottom, or page
   \centering
   \begin{tabular}{c}
   \includegraphics[width=2.5in]{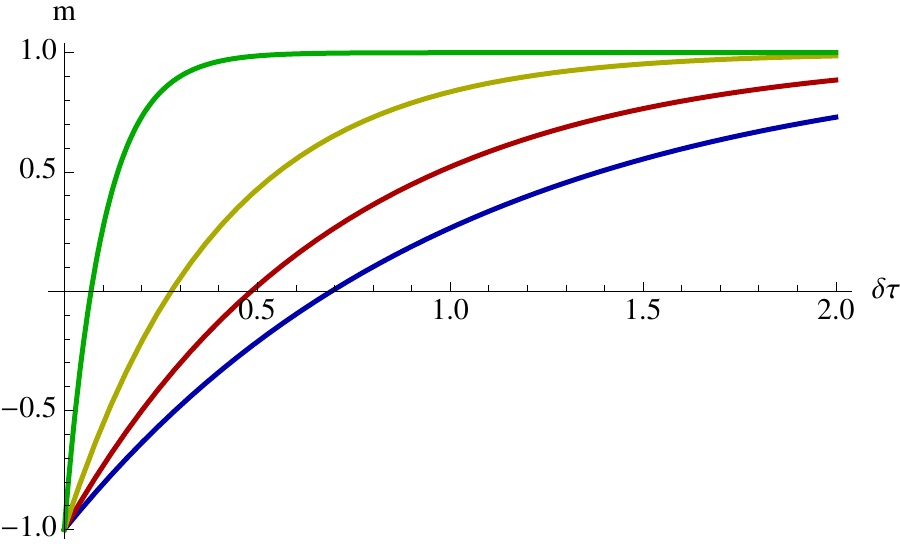} \\
   \; \includegraphics[width=2.5in]{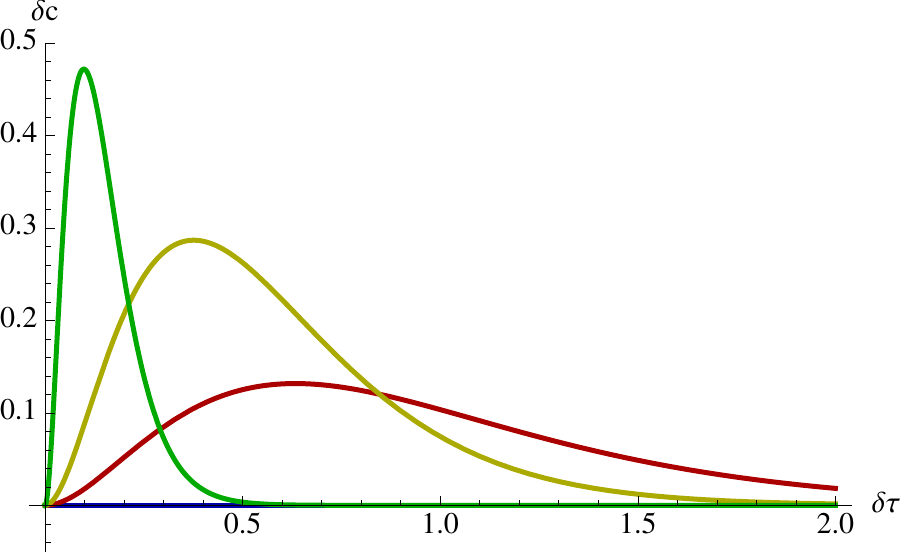} 
   \end{tabular}
   \caption{Approximate values of average magnetization $m$ and rescaled connected correlation $\delta c$ associated with a fully-connected pairwise model used to describe a fully-connected Hawkes process. We consider in particular models for which $\mu=1$, $\beta = 2$ and $\alpha = 0, 0.3, 0.6, 0.9$ (respectively blue, red, yellow and green line). The corrected qualitative features of the model are captured in this approximate scheme.}
   \label{fig:HawkesApprox}
\end{SCfigure}

\begin{figure} %  figure placement: here, top, bottom, or page
   \centering
   \begin{tabular}{cc}
   \includegraphics[width=2.5in]{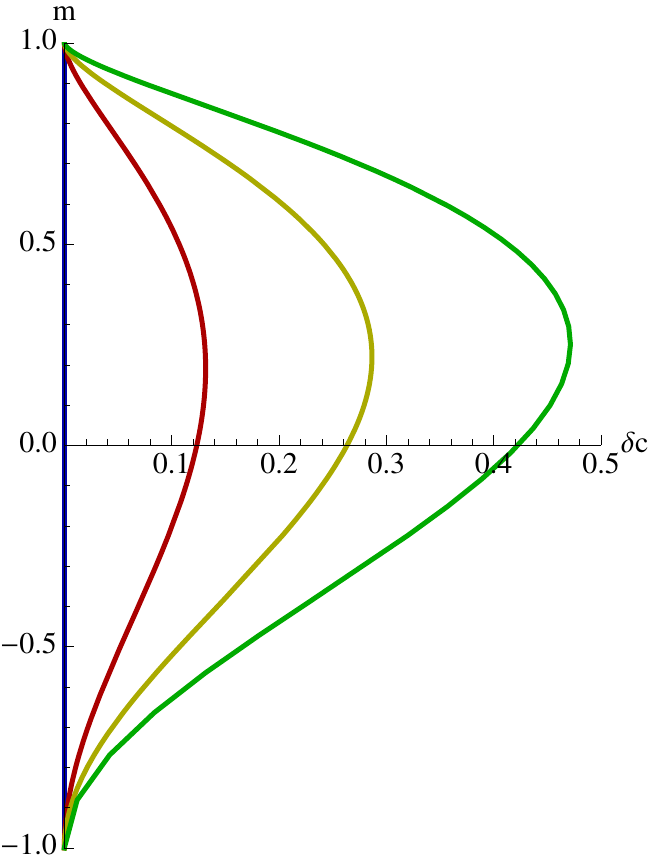}
   \includegraphics[width=2.5in]{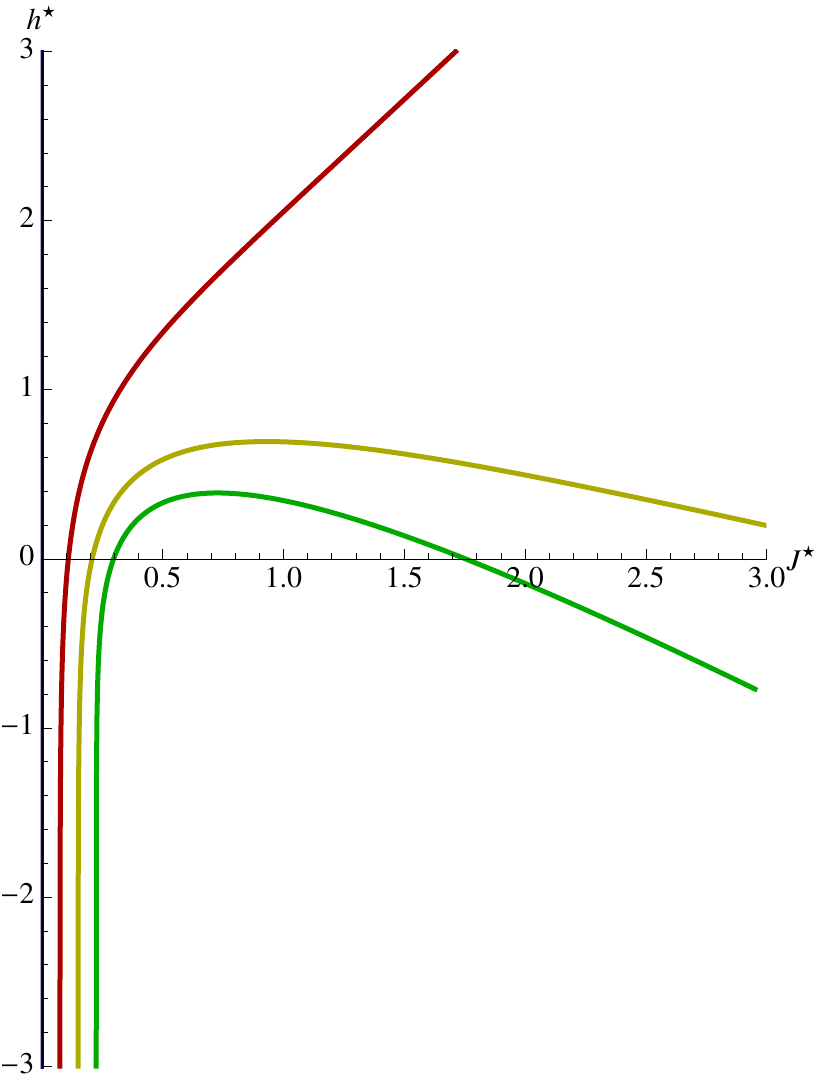} 
   \end{tabular}
   \caption{Approximate values of the empirical averages $(m,c)$ and of the inferred couplings $(h^\star, J^\star)$ obtained by using a fully connected pairwise model to fit a set of Hawkes point process, for the same choice of models and color conventions as in the previous plot, parametrically plotted as a function of the bin size $\delta \tau$.}
   \label{fig:HawkesPhaseDiag}
\end{figure}

Consider a fully connected Hawkes process defined as in (\ref{eq:HawkesDynamics}), characterized by exogenous intensity $\mu$ and kernel parameters parameters $\alpha$ and $\beta$. We will show that the qualitative features of the  fully connected pairwise model associated through the binning functions (\ref{eq:BinningB}) and (\ref{eq:BinningS}) can be obtained by using an approximate scheme.
More precisely, given a realization of a fully connected Hawkes point-process $\bX$ and a bin size $\delta \tau$ we will calculate the quantities
\beqa
m_i &=& \frac{1}{T} \sum_{t=1}^T s^{(t)}_i (\bX,\delta \tau) \label{eq:MagnFCHawkes} \\
\delta c_{ij} &=& \frac{N}{T} \left[ \left( \sum_{t=1}^T s^{(t)}_i  (\bX,\delta \tau) \, s^{(t)}_j (\bX,\delta \tau) \right) - m_i m_j \right] \label{eq:CorrFCHawkes} \; .
\eeqa
First, one can easily notice (expanding the minimum inside the binning functions) that any correlation function of the quantities $b_i^{(t)}$ and $s_i^{(t)}$ can be linked to the properties the Hawkes processes under convolution. In particular, one has for the first two momenta
\beqa
\mathbb{E}[b^{(t)}_i(\bX,\delta \tau) ] &=& f_i(\delta \tau) \\
\mathbb{E}[b^{(t)}_i (\bX,\delta \tau) \, b^{(t)}_j(\bX,\delta \tau) ] &=& f_i(\delta \tau) + f_j (\delta \tau) - f_{i+j}(\delta \tau) \; ,
\eeqa  
where $f_i(\delta \tau)$ is the average number of events of type $i$ during time $\delta \tau$ in the stationary state, while $f_{i+j}(\delta \tau)$ is the average number of events of type $i$ or $j$, which is associated with the convolution $X_{i+j} = X_i + X_j$.
Thus, to calculate the quantities (\ref{eq:MagnFCHawkes}) and (\ref{eq:CorrFCHawkes}) one needs to calculate
\beqa
\mathbb{E} [b^{(t)}_{c}] &=& \sum_{K=0}^{\infty} \textrm{Prob}[\delta X_{c}(t\, \delta \tau  ) \geq 1 , \delta X_{\backslash c}(t \, \delta \tau ) = K ] \nonumber \\
&=& 1 - \sum_{K=0}^{\infty} \textrm{Prob}[\delta X_{c}(t\, \delta \tau  ) = 0 , \delta X_{\backslash c}(t \, \delta \tau ) = K  ]
\label{eq:ProbBinnedHawkes}\; ,
\eeqa
where $\delta X_{c}(t\, \delta \tau  ) = X_{c}(\delta \tau(t+1)) - X_{c}(\delta \tau t)$, while $c \in \{ i, j, i+j \}$ and $\backslash c \in \{ V \backslash \{ i\},V \backslash \{ j\}, V \backslash \{ i+ j \} \} $ refer to the channels which one needs to take into account to calculate magnetizations and correlations.
Above probability can be computed by taking into account that:
\begin{itemize}
\item{
The convolution of a set of Hawkes processes is a Hawkes process.
}
\item{
Probability (\ref{eq:ProbBinnedHawkes}) can be reduced via convolution to the probability of a 2-variate Hawkes processes describing channel $c$ and the environment $\backslash c$.
}
\end{itemize}
The parameter set describing such convolution is given for $c=i$ by $\bMu = \mu (1,N-1)$, $\beta$ unchanged and
\beq
\hat \bAl =  \alpha \left( \begin{array}{cc} 0 & 1 \\ N-1 & N-2\end{array} \right)  \; .
\eeq
The one describing the case $c=i+j$ has $\bMu = \mu (2,N-2)$, $\beta$ unchanged and
\beq
\hat \bAl = \alpha \left( \begin{array}{cc} 1 & 2 \\ N-2 & N-3\end{array} \right)  \; .
\eeq
With this in mind, one can expand probability (\ref{eq:ProbBinnedHawkes}) in term of the intensities and obtain
\beq
\mathbb{E} [b^{(t)}_{c}]  = 1 - \sum_{K=0}^\infty \frac{1}{K!} \int_0^{\delta \tau} d\tau_K \dots \int_0^{\delta \tau} d\tau_1 \; e^{-\int_0^{\delta \tau} du \, \lambda_c (u) +\lambda_c (u) } \,  \lambda_{\backslash c}(\tau_k) \prod_{k=0}^K \lambda_{\tau_k}^{\backslash c} \label{eq:FullIntHawkes} \; .
\eeq
In principle, one should plug the initial conditions in the stochastic intensities $\bLam(\tau)$ inside previous formula and compute the integral. For example, if one supposes the initial intensities to correspond to the stationary state intensities, one should insert into (\ref{eq:FullIntHawkes}) the following expression
\beqa
\lambda_c (\tau) &=&  \mu_c + e^{-\beta \tau} (\bar \lambda_c - \mu_c) + \sum_{k=1}^K \alpha_{c,\backslash c} \; e^{-\beta(\tau-\tau_k)} \theta(\tau-\tau_k) \\
\lambda_{\backslash c} (\tau) &=& \mu_{\backslash c} + e^{-\beta \tau} (\bar \lambda_{\backslash c} - \mu_{\backslash c}) + \sum_{k=1}^K \alpha_{\backslash c, \backslash c} \; e^{-\beta(\tau-\tau _k)} \theta(\tau-\tau_k)
\eeqa
and perform explicitly the integral. This is very hard to do analytically, so that we consider an approximate scheme in which it is possible to obtain a qualitatively correct result for the averages, motivated by the fact that in both the cases that we consider ($c=i$ and $c=i+j$) we have that $\alpha_{\backslash c,c}, \alpha_{\backslash c,c} \gg \alpha_{c,c},\alpha_{c, \backslash c}$ and $\mu_{\backslash c} \gg \mu_c$. This regime justifies the approximation in which the trajectory $\lambda_{\backslash c}(\tau)$ is described by the \emph{deterministic} function
\beq
\lambda_{\backslash c} (\tau) = L_{\backslash c}^0 \, \psi (\delta t) + L_{\backslash c} \, [1- \psi (\delta t)] \; ,
\eeq
where
\beq
L_{\backslash c} (\tau)^0 = \left[ \left( \bDl - \frac{\bAl}{\beta} \right)^{-1}\right]_{\backslash c,c} \mu_c + \left[ \left( \bDl - \frac{\bAl}{\beta} \right)^{-1}\right]_{\backslash c, \backslash c} \mu_{\backslash c}
\eeq
is the average intensity of channel $\backslash c$ in the stationary state in which channel $c$ is free to produce events, while
\beq
L_{\backslash c} = \left( 1 - \frac{\alpha_{\backslash c,\backslash c} }{\beta}  \right)^{-1} \mu_{\backslash c}
\eeq
is the average in the stationary state in which channel $c$ is conditioned in order not to produce events. Finally $\psi (\delta \tau)$ is a generic function such that $\phi(0) = 1$ and $\phi (\infty) = 0$.
Then, one can insert this approximation into equation (\ref{eq:FullIntHawkes}), supposing that the number of events $K$ is deterministic and concentrated around its average number.
Then we have
\beqa
\mathbb{E} [ b_c(\delta t) ]  &=& 1 - \sum_k \textrm{Prob} [ \delta X_c (\delta \tau) = 0, \delta X_{\backslash c} (\delta \tau = K )]  \\
&\approx& 1 - e^{- \int du [ L_c^0 \, \psi(u) + L_{\backslash c} \, (1-\psi (u) ] }  \; .
\eeqa
If for example we suppose that $\psi(\tau) = e^{-\beta \tau}$, so that the relaxation dynamics for the intensity is ruled by the same parameter $\beta$ controlling the dynamics, we get
\beqa
\mathbb{E}[ b_i^{(t)} ]&\xrightarrow[N\to\infty]{}& 1 - \exp \left(- \frac{\mu \, \delta t}{1-\alpha / \beta} \right) \\
\mathbb{E}[ b_i^{(t)} ]&\xrightarrow[N\to\infty]{}& 1 - \exp \left(-  \frac{2 \mu \, \delta t}{1-\alpha / \beta} \right) \nonumber \\
N \left( \mathbb{E}[  b_i^{(t)} \, b_j^{(t)} ] - \mathbb{E}[   b_i^{(t)}  ] \mathbb{E}[  b_j^{(t)} ]  \right) &\xrightarrow[N\to\infty]{}&
\frac{2 \, \alpha \, \mu \, e^{-2 \mu \delta t /(1-\alpha/\beta)} [e^{-\beta \delta t} -1 + \beta \delta t ]}{(\alpha-\beta)^2}  \nonumber \; .
\eeqa
This information can be exploited to compute $m$ and $\delta c$, which after using the rule $s_c^{(t)} = 2 b_c^{(t)} - 1$ result
\beqa
m &=& 1 - 2 e^{-\mu \, \delta t / ({1- \alpha/\beta})} \\
\delta c &=& \left( \frac{8 \, \alpha \, \mu \, e^{-2 \mu \delta t /(1-\alpha/\beta)} [e^{-\beta \delta t} -1 + \beta \delta t ]}{(\alpha-\beta)^2} \right) \; .
\eeqa
This result provides a simple qualitative picture, whose degree of inaccuracy lies in the choice of the function $\psi(\tau)$, and in the hypothesis that the trajectory of the stochastic intensity concentrates around a deterministic function. Nevertheless, this approximation captures some of the features that we find by computing magnetization and correlations for various realizations of Hawkes processes for various bin sizes, as shown in figure \ref{fig:HawkesApprox}.
Notice in particular the qualitative features of the model correctly reproduced in this scheme, namely (i) correlations drop to zero for small bin sizes (Epps effect) or values $\delta \tau$ larger then the average inter-event time, (ii) correlations increase with the interaction parameter $\alpha$ and are zero for the Poisson case $\alpha= 0$. The magnetizations calculated in this way correspond instead to the exact value. In figure \ref{fig:HawkesPhaseDiag} we plot the ensemble averages of the model and the average inferred couplings ($h^\star,J^\star$) in the case of a fully connected pairwise model for various choices of the bin size and of the interaction parameter $\alpha$.
Finally, notice that this approximation is able to capture the finiteness of $\delta c$, which implies that the description of data in term of a fully connected ferromagnet doesn't lead to a degenerate representation of the model (see section \ref{sec:FCFerromagnet}).

% Make the bibliography single spaced
\singlespacing
\bibliographystyle{amc}

% add the Bibliography to the Table of Contents
\cleardoublepage
\ifdefined\phantomsection
  \phantomsection  % makes hyperref recognize this section properly for pdf link
\else
\fi
\addcontentsline{toc}{chapter}{Bibliography}

% include your .bib file
\bibliography{thesis}

\begin{thebibliography}{10}

\bibitem{SEC2010:dq}
Findings regarding the market events of may 6, 2010.
\newblock Tech. rep., U.S. Commodity Futures Trading Commission and the U.S.
  Securities and Exchange Commission.

\bibitem{:zr}
Google translate.
\newblock \url{http://translate.google.com}.

\bibitem{:ly}
Kaggle official website.
\newblock \url{http://www.kaggle.com}.

\bibitem{:uq}
Netflix official website.
\newblock \url{http://www.netflixprize.com/}.

\bibitem{:kx}
Yahoo finance.
\newblock \url{http://yahoo.finance.com}.

\bibitem{Ackley:1985hc}
{\sc Ackley, D., Hinton, G., and Sejnowski, T.}
\newblock A learning algorithm for boltzmann machines.
\newblock {\em Cogn. Sci. 9\/} (1985), 147.

\bibitem{Akaike:1973uq}
{\sc Akaike, H.}
\newblock Information theory and an extension of the maximum likelihood
  principle.
\newblock In {\em Second international symposium on information theory\/}
  (1973), vol.~1, Springer Verlag, pp.~267--281.

\bibitem{Almeida:1978uq}
{\sc Almeida, J., and Thouless, D.}
\newblock Stability of the sherrington-kirkpatrick solution of a spin glass
  model.
\newblock {\em J. Phys. A-Math. Gen. 11\/} (1978), 983--990.

\bibitem{Amari:1985ay}
{\sc Amari, S.}
\newblock {\em Differential Geometrical Methods In Statistics}.
\newblock Springer, 1985.

\bibitem{Amari:1987lj}
{\sc Amari, S., Barndorff-Nielsen, O., Kass, R., Lauritzen, S., and Rao, C.}
\newblock {\em Differential Geometry in Statistical Inference}.
\newblock Institute of Mathematical Statistics, 1987.

\bibitem{Amari:2007fk}
{\sc Amari, S., and Nagaoka, H.}
\newblock {\em Methods of information geometry}, vol.~191.
\newblock American Mathematical Society, 2007.

\bibitem{Aurell:2012ys}
{\sc Aurell, E., and Ekeberg, M.}
\newblock Inverse ising inference using all the data.
\newblock {\em Phys. Rev. Lett. 108}, 9 (2012), 090201.

\bibitem{Balasubramanian:2005oz}
{\sc Balasubramanian, V.}
\newblock {\em Advances in Minimum Description Length: Theory and
  Applications}.
\newblock MIT Press, 2005, ch.~MDL, Bayesian inference, and the geometry of the
  space of probability distributions.

\bibitem{Bauwens:2009ys}
{\sc Bauwens, L., and Hautsch, N.}
\newblock {\em Handbook of Financial Time Series}.
\newblock Springer, 2009, ch.~Modelling financial high frequency data using
  point processes, pp.~953--979.

\bibitem{Baxter:1982vn}
{\sc Baxter, R.}
\newblock {\em Exactly solved models in statistical mechanics}.
\newblock Academic Press, 1982.

\bibitem{Bouchaud:2009ct}
{\sc Bouchaud, J.}
\newblock Economics needs a scientific revolution.
\newblock {\em Nature 455}, 7217 (2008), 1181.

\bibitem{Bouchaud:2008pb}
{\sc Bouchaud, J., Farmer, J.~D., and Lillo, F.}
\newblock {\em Handbook of Financial Markets: Dynamics and Evolution}.
\newblock Elsevier, 2008, ch.~How Markets Slowly Digest Changes in Supply and
  Demand, pp.~57--156.

\bibitem{Bouchaud:2003fy}
{\sc Bouchaud, J., and Potters, M.}
\newblock {\em Theory of financial risk and derivative pricing: from
  statistical physics to risk management}.
\newblock Cambridge University Press, 2003.

\bibitem{Bowsher:2002jw}
{\sc Bowsher, C.}
\newblock Modelling security market events in continuous time: Intensity based,
  multivariate point process models.
\newblock Tech. Rep. 2002-W22, Nuffield College, Oxford, 2002.

\bibitem{Boyd:2010tg}
{\sc Boyd, S.}
\newblock Subgradient methods.
\newblock {\em Lecture notes, \url{http://www.stanford.edu/
  class/ee364b/lectures.html}\/} (2010).

\bibitem{Boyd:2004oq}
{\sc Boyd, S., and Vandenberghe, L.}
\newblock {\em Convex optimization}.
\newblock Cambridge University Press, 2004.

\bibitem{Braunstein:2008sa}
{\sc Braunstein, A., Pagnani, A., Weigt, M., and Zecchina, R.}
\newblock Inference algorithms for gene networks: a statistical mechanics
  analysis.
\newblock {\em J. Stat. Mech. 2008}, 12 (2008), P12001.

\bibitem{Byrd:1995fk}
{\sc Byrd, R., Lu, P., Nocedal, J., and Zhu, C.}
\newblock A limited memory algorithm for bound constrained optimization.
\newblock {\em SIAM J. Sci. Comp. 16}, 5 (1995), 1190--1208.

\bibitem{Cocco:2009mb}
{\sc Cocco, S., Leibler, S., and Monasson, R.}
\newblock Neuronal couplings between retinal ganglion cells inferred by
  efficient inverse statistical physics methods.
\newblock {\em Proc. Natl. Acad. Sci. U.S.A. 106\/} (2009), 14058.

\bibitem{Cocco:2011fk}
{\sc Cocco, S., and Monasson, R.}
\newblock Adaptive cluster expansion for inferring boltzmann machines with
  noisy data.
\newblock {\em Phys. Rev. Lett. 106}, 9 (2011), 090601.

\bibitem{Cocco:2012uq}
{\sc Cocco, S., and Monasson, R.}
\newblock Adaptive cluster expansion for the inverse ising problem:
  Convergence, algorithm and tests.
\newblock {\em J. Stat. Phys. 147}, 2 (2012), 252--314.

\bibitem{Cover:1991fk}
{\sc Cover, T., Thomas, J., Wiley, J., et~al.}
\newblock {\em Elements of information theory}, vol.~6.
\newblock Wiley Online Library, 1991.

\bibitem{Dacorogna:2001uf}
{\sc Dacorogna, M., Gen{\c c}lay, R., M{\"u}ller, U., Olsen, R., and Pictet,
  O.}
\newblock {\em An Introduction to High-Frequency Finance}.
\newblock Academic Press, 2001.

\bibitem{Lachapelle:2010kv}
{\sc de~Lachapelle, D., and Challet, D.}
\newblock Turnover, account value and diversification of real traders: evidence
  of collective portfolio optimizing behavior.
\newblock {\em New J. Phys. 12\/} (2010), 075039.

\bibitem{De-Martino:2006os}
{\sc De~Martino, A., and Marsili, M.}
\newblock Statistical mechanics of socio-economic systems with heterogeneous
  agents.
\newblock {\em J. Phys. A-Math. Gen. 39\/} (2006), R465.

\bibitem{Donoho:2006ly}
{\sc Donoho, D.}
\newblock Compressed sensing.
\newblock {\em IEEE T. Inform Theor. 52}, 4 (2006), 1289--1306.

\bibitem{Epps:1979rp}
{\sc Epps, T.}
\newblock Comovements in stock prices in the very short run.
\newblock {\em J. Amer. Stat. Ass. 74\/} (1979), 291--298.

\bibitem{Fama:1970ly}
{\sc Fama, E.}
\newblock Efficient capital markets: A review of theory and empirical work.
\newblock {\em J. Financ. 25}, 2 (1970), 383--417.

\bibitem{Feller:1950zr}
{\sc Feller, W.}
\newblock {\em An introduction to probability theory and its applications}.
\newblock John Wiley \& Sons, 1950.

\bibitem{Gori:2011ly}
{\sc Gori, G., and Trombettoni, A.}
\newblock The inverse ising problem for one-dimensional chains with arbitrary
  finite-range couplings.
\newblock {\em J. Stat. Mech. 2011\/} (2011), P10021.

\bibitem{Hastings:1970vn}
{\sc Hastings, W.}
\newblock Monte carlo sampling methods using markov chains and their
  applications.
\newblock {\em Biometrika 57}, 1 (1970), 97--109.

\bibitem{Hawkes:1971lc}
{\sc Hawkes, A.}
\newblock Point spectra of some mutually exciting point processes.
\newblock {\em J. R. Statist. Soc. B 33\/} (1971), 438--443.

\bibitem{Hawkes:1971nq}
{\sc Hawkes, A.}
\newblock Spectra of some self-exciting and mutually exciting point processes.
\newblock {\em Biometrika 58}, 1 (1971), 83--90.

\bibitem{Hinton:2010vn}
{\sc Hinton, G.}
\newblock A practical guide to training restricted boltzmann machines.
\newblock Tech. rep., Univ. Toronto, 2010.

\bibitem{Huang:1987nx}
{\sc Huang, K.}
\newblock {\em Statistical Mechanics}.
\newblock John Wiley \& Sons, 1987.

\bibitem{Ising:1925kx}
{\sc Ising, E.}
\newblock Beitrag zur theorie des ferromagnetismus.
\newblock {\em Z. Phys. A-Hadron. Nucl. 31}, 1 (1925), 253--258.

\bibitem{Jaeger:1990ys}
{\sc Jaeger, F., Vertigan, D., and Welsh, D.}
\newblock On the computational complexity of the jones and tutte polynomials.
\newblock {\em Math. Proc. Cambridge 108}, 01 (1990), 35--53.

\bibitem{Jerrum:1990zr}
{\sc Jerrum, M., and Sinclair, A.}
\newblock Polynomial-time approximation algorithms for the ising model.
\newblock {\em Lect. Notes. Comput. Sc.\/} (1990), 462--475.

\bibitem{Joulin:2008bv}
{\sc Joulin, A., Lefevre, A., Grunberg, D., and Bouchaud, J.}
\newblock Stock price jumps: news and volume play a minor role.
\newblock {\em Arxiv preprint arxiv:0803.1769\/} (2008).

\bibitem{Kappen:1998dt}
{\sc Kappen, H., and Rodriguez, F.}
\newblock Efficient learning in boltzmann machines using linear response
  theory.
\newblock {\em Neural. Comput. 10\/} (1998), 1137--1156.

\bibitem{Kirman:2010fk}
{\sc Kirman, A.}
\newblock {\em Complex economics: individual and collective rationality}.
\newblock Routledge, 2010.

\bibitem{Krauth:1998fk}
{\sc Krauth, W.}
\newblock Introduction to monte carlo algorithms.
\newblock {\em Lect. Notes. Phys.\/} (1998), 1--35.

\bibitem{Lillo:2008ht}
{\sc Lillo, F., Moro, E., Vaglica, G., and Mantegna, N.}
\newblock Specialization and herding behavior of trading firms in a financial
  market.
\newblock {\em New J. Phys. 10\/} (2008), 043019.

\bibitem{Liu:1989uq}
{\sc Liu, D., and Nocedal, J.}
\newblock On the limited memory bfgs method for large scale optimization.
\newblock {\em Math. Program. 45}, 1 (1989), 503--528.

\bibitem{MacKay:2003vn}
{\sc MacKay, D.}
\newblock {\em Information theory, inference, and learning algorithms}.
\newblock Cambridge University Press, 2003.

\bibitem{Mantegna:1999ay}
{\sc Mantegna, N., and Stanley, E.}
\newblock {\em An Introduction to Econophysics: Correlations and Complexity in
  Finance}.
\newblock Cambridge University Press, 1999.

\bibitem{Marinari:2010vn}
{\sc Marinari, E., and Van~Kerrebroeck, V.}
\newblock Intrinsic limitations of the susceptibility propagation inverse
  inference for the mean field ising spin glass.
\newblock {\em J. Stat. Mech. 2010\/} (2010), P02008.

\bibitem{Mastromatteo:2011ly}
{\sc Mastromatteo, I., and Marsili, M.}
\newblock On the criticality of inferred models.
\newblock {\em J. Stat. Mech. 2011\/} (2011), P10012.

\bibitem{Mezard:2009ko}
{\sc M{\'e}zard, M., and Montanari, A.}
\newblock {\em Information, Physics and Computation}.
\newblock Oxford University Press, 2009.

\bibitem{Mezard:2009ul}
{\sc M{\'e}zard, M., and Mora, T.}
\newblock Constraint satisfaction problems and neural networks: a statistical
  physics perspective.
\newblock {\em J. Physiol. Paris 103\/} (2009), 107--113.

\bibitem{Mezard:1987fk}
{\sc M{\'e}zard, M., Parisi, G., and Virasoro, M.}
\newblock {\em Spin glass theory and beyond}.
\newblock World scientific Singapore, 1987.

\bibitem{Monasson:2010hc}
{\sc Monasson, R.}
\newblock The mean-field ising model.
\newblock {\em Lecture notes, \url{http://www.phys.ens.fr/~monasson/}\/}
  (2010).

\bibitem{Mora:2011kq}
{\sc Mora, T., and Bialek, W.}
\newblock Are biological systems poised at criticality?
\newblock {\em J. Stat. Phys.\/} (2011), 1--35.

\bibitem{Moro:2009wy}
{\sc Moro, E., Vicente, J., Moyano, L., Gerig, A., Farmer, J.~D., Vaglica, G.,
  Lillo, F., and Mantegna, N.}
\newblock Market impact and trading profile of hidden orders in stock markets.
\newblock {\em Phys. Rev. E 80\/} (2009), 066102.

\bibitem{Morters:2008fv}
{\sc M\"orters, P.}
\newblock Large deviation theory and applications.
\newblock {\em Lecture notes, \url{http://people.bath.ac.uk/maspm/}\/} (2008).

\bibitem{Myung:2000jy}
{\sc Myung, I., and Balasubramanian, V.}
\newblock Counting probability distributions: Differential geometry and model
  selection.
\newblock {\em Proc. Natl. Acad. Sci. U.S.A. 97\/} (2000), 11170.

\bibitem{Plefka:1982vn}
{\sc Plefka, T.}
\newblock Convergence condition of the tap equation for the infinite-ranged
  ising spin glass model.
\newblock {\em J. Phys. A-Math. Gen. 15\/} (1982), 1971--1978.

\bibitem{Ravikumar:2010ys}
{\sc Ravikumar, P., Wainwright, M., and Lafferty, J.}
\newblock High-dimensional ising model selection using $\ell$1-regularized
  logistic regression.
\newblock {\em Ann. Stat. 38}, 3 (2010), 1287--1319.

\bibitem{Ricci-Tersenghi:2011uq}
{\sc Ricci-Tersenghi, F.}
\newblock The bethe approximation for solving the inverse ising problem: a
  comparison with other inference methods.
\newblock {\em J. Stat. Mech. 2012}, 08 (2012), P08015.

\bibitem{Rissanen:1984ix}
{\sc Rissanen, J.}
\newblock Universal coding, information, prediction, and estimation.
\newblock {\em IEEE T. Inform Theor. 30}, 4 (1984), 629--636.

\bibitem{Rissanen:1986qz}
{\sc Rissanen, J.}
\newblock Stochastic complexity and modelling.
\newblock {\em Ann. Stat. 14\/} (1986), 1080.

\bibitem{Roudi:2009qm}
{\sc Roudi, Y., Aurell, E., and Hertz, J.}
\newblock Statistical physics of pairwise probability models.
\newblock {\em Front. Comput. Neurosci. 3}, 22 (2009), 1--15.

\bibitem{Roudi:2009li}
{\sc Roudi, Y., Tyrcha, J., and Hertz, J.}
\newblock The ising model for neural data: Model quality and approximate
  methods for extracting functional connectivity.
\newblock {\em Phys. Rev. E 79\/} (2009), 051915.

\bibitem{Rual:2005bh}
{\sc Rual, J., Venkatesan, K., Hao, T., Hirozane-Kishikawa, T., Dricot, A., Li,
  N., Berriz, G., Gibbons, F., Dreze, M., Ayivi-Guedehoussou, N., et~al.}
\newblock Towards a proteome-scale map of the human protein--protein
  interaction network.
\newblock {\em Nature 437}, 7062 (2005), 1173--1178.

\bibitem{Russell:2010ve}
{\sc Russell, S., and Norvig, P.}
\newblock {\em Artificial intelligence: a modern approach}.
\newblock Prentice Hall, 2010.

\bibitem{Schmidt:2010hc}
{\sc Schmidt, M., and Murphy, K.}
\newblock Convex structure learning in log-linear models: Beyond pairwise
  potentials.
\newblock In {\em Proceedings of the International Conference on Artificial
  Intelligence and Statistics (AISTATS)\/} (2010).

\bibitem{Schneidman:2006vg}
{\sc Schneidman, E., Berry~II, M., Segev, R., and Bialek, W.}
\newblock Weak pairwise correlations imply strongly correlated network states
  in a neural population.
\newblock {\em Nature 440\/} (2006), 1007--1012.

\bibitem{Schwarz:1978kx}
{\sc Schwarz, G.}
\newblock Estimating the dimension of a model.
\newblock {\em Ann. Stat. 6}, 2 (1978), 461--464.

\bibitem{Sessak:2009lf}
{\sc Sessak, V., and Monasson, R.}
\newblock Small-correlation expansions for the inverse ising problem.
\newblock {\em J. Phys. A-Math. Theor. 42\/} (2009), 055001.

\bibitem{Shendure:2008qf}
{\sc Shendure, J., and Ji, H.}
\newblock Next-generation dna sequencing.
\newblock {\em Nat. Biotechnol. 26}, 10 (2008), 1135--1145.

\bibitem{Shlens:2006uq}
{\sc Shlens, J., Field, G., Gauthier, J., Grivich, M., Petrusca, D., Sher, A.,
  Litke, A., and Chichilnisky, E.}
\newblock The structure of multi-neuron firing patterns in primate retina.
\newblock {\em J. Neurosci. 26}, 32 (2006), 8254--8266.

\bibitem{Socolich:2005vy}
{\sc Socolich, M., Lockless, S., Russ, W., Lee, H., Gardner, K., and
  Ranganathan, R.}
\newblock Evolutionary information for specifying a protein fold.
\newblock {\em Nature 437\/} (2005), 512--518.

\bibitem{Stephens:2008cg}
{\sc Stephens, G., Mora, T., Tkacik, G., and Bialek, W.}
\newblock Thermodynamics of natural images.
\newblock {\em Arxiv preprint arXiv:0806.2694\/} (2008).

\bibitem{Tanaka:1998tg}
{\sc Tanaka, T.}
\newblock Mean field theory of boltzmann machine learning.
\newblock {\em Phys. Rev. E 58\/} (1998), 2302.

\bibitem{Teller:1953kx}
{\sc Teller, E., Metropolis, N., and Rosenbluth, A.}
\newblock Equation of state calculations by fast computing machines.
\newblock {\em J. Chem. Phys 21}, 13 (1953), 1087--1092.

\bibitem{Thouless:1977kx}
{\sc Thouless, D., Anderson, P., and Palmer, R.}
\newblock Solution of'solvable model of a spin glass'.
\newblock {\em Philos. Mag. 35}, 3 (1977), 593--601.

\bibitem{Tkacik2006:mm}
{\sc Tkacik, G., Schneidman, E., Berry~II, M., and Bialek, W.}
\newblock Ising models for networks of real neurons.
\newblock {\em Arxiv preprint arXiv:q-bio/0611072v1\/} (2006).

\bibitem{Trichet:2010la}
{\sc Trichet, J.}
\newblock Reflections on the nature of monetary policy non-standard measures
  and finance theory.

\bibitem{Tyrcha:2012fk}
{\sc Tyrcha, J., Roudi, Y., Marsili, M., and Hertz, J.}
\newblock Effect of nonstationarity on models inferred from neural data.
\newblock {\em Arxiv preprint arXiv:1203.5673\/} (2012).

\bibitem{Wainwright:2008kx}
{\sc Wainwright, M., and Jordan, M.}
\newblock Graphical models, exponential families, and variational inference.
\newblock {\em Foundations and Trends in Machine Learning 1}, 1-2 (2008),
  1--305.

\bibitem{Wainwright:2007zr}
{\sc Wainwright, M., Ravikumar, P., and Lafferty, J.}
\newblock High-dimensional graphical model selection using $\ell$1-regularized
  logistic regression.
\newblock {\em Adv. Neur. In. 19\/} (2006), 1465--1472.

\bibitem{Weigt:2009on}
{\sc Weigt, M., White, R., Szurmant, H., Hoch, J., and Hwa, T.}
\newblock Identification of direct residue contacts in protein-protein
  interaction by message passing.
\newblock {\em Proc. Natl. Acad. Sci. U.S.A. 106\/} (2009), 67.

\bibitem{Wheeler:2008ve}
{\sc Wheeler, D., Srinivasan, M., Egholm, M., Shen, Y., Chen, L., McGuire, A.,
  He, W., Chen, Y., Makhijani, V., Roth, G., et~al.}
\newblock The complete genome of an individual by massively parallel dna
  sequencing.
\newblock {\em Nature 452}, 7189 (2008), 872--876.

\bibitem{Zamponi:2010fk}
{\sc Zamponi, F.}
\newblock Mean field theory of spin glasses.
\newblock {\em Arxiv preprint arXiv:1008.4844\/} (2010).

\end{thebibliography}

% add the Notation part to the Table of Contents
\cleardoublepage
\ifdefined\phantomsection
  \phantomsection  % makes hyperref recognize this section properly for pdf link
\else
\fi
\addcontentsline{toc}{chapter}{Notation}

% Notation section
\pagestyle{plain}
%\listofboth
\includegraphics[width=\textwidth]{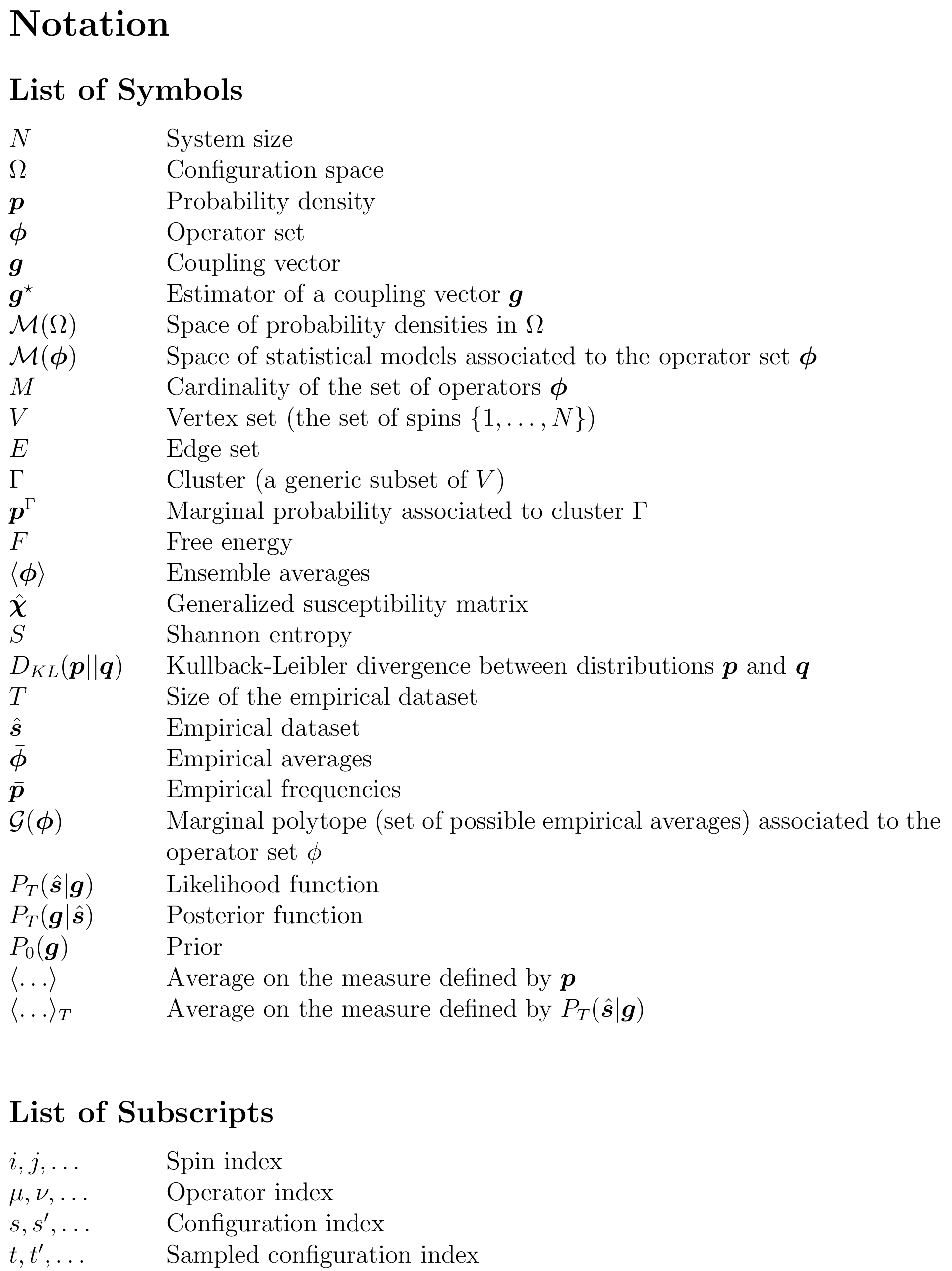}

\end{document}